\pdfoutput=1

\documentclass[12pt,a4paper]{article}

\usepackage{ifthen} 
\newboolean{pdflatex}
\setboolean{pdflatex}{true} 

\newboolean{articletitles}
\setboolean{articletitles}{true} 

\newboolean{uprightparticles}
\setboolean{uprightparticles}{false} 

\newboolean{inbibliography}
\setboolean{inbibliography}{false} 

\usepackage{microtype}
\usepackage{lineno}  
\usepackage{xspace} 
\usepackage{caption} 

\usepackage{graphicx}  
\usepackage{color}
\usepackage{colortbl}
\graphicspath{{./figs/}} 

\usepackage{amsmath} 
\usepackage{amssymb}
\usepackage{amsfonts}
\usepackage{upgreek} 

\usepackage{calc}
\newcommand*\patchAmsMathEnvironmentForLineno[1]{%
\expandafter\let\csname old#1\expandafter\endcsname\csname #1\endcsname
\expandafter\let\csname oldend#1\expandafter\endcsname\csname
end#1\endcsname
 \renewenvironment{#1}%
   {\linenomath\csname old#1\endcsname}%
   {\csname oldend#1\endcsname\endlinenomath}%
}
\newcommand*\patchBothAmsMathEnvironmentsForLineno[1]{%
  \patchAmsMathEnvironmentForLineno{#1}%
  \patchAmsMathEnvironmentForLineno{#1*}%
}
\AtBeginDocument{%
\patchBothAmsMathEnvironmentsForLineno{equation}%
\patchBothAmsMathEnvironmentsForLineno{align}%
\patchBothAmsMathEnvironmentsForLineno{flalign}%
\patchBothAmsMathEnvironmentsForLineno{alignat}%
\patchBothAmsMathEnvironmentsForLineno{gather}%
\patchBothAmsMathEnvironmentsForLineno{multline}%
\patchBothAmsMathEnvironmentsForLineno{eqnarray}%
}

\usepackage{hyperref}    
\usepackage[all]{hypcap} 

\def\lhcb {\mbox{LHCb}\xspace}

\def\MagUp {\mbox{\em Mag\kern -0.05em Up}\xspace}

\ifthenelse{\boolean{uprightparticles}}%
{

 \def\Pmu         {\ensuremath{\upmu}\xspace}

 \def\Ppi         {\ensuremath{\uppi}\xspace}

 \def\Ppsi        {\ensuremath{\uppsi}\xspace}

 \def\PDelta      {\ensuremath{\Delta}\xspace}                 
 \def\PXi      {\ensuremath{\Xi}\xspace}                 
 \def\PLambda      {\ensuremath{\Lambda}\xspace}                 
 \def\PSigma      {\ensuremath{\Sigma}\xspace}                 
 \def\POmega      {\ensuremath{\Omega}\xspace}                 
 \def\PUpsilon      {\ensuremath{\Upsilon}\xspace}                 
 
 \def\PB      {\ensuremath{\mathrm{B}}\xspace}                 
                  
 \def\PD      {\ensuremath{\mathrm{D}}\xspace}

 \def\PJ      {\ensuremath{\mathrm{J}}\xspace}                 
 \def\PK      {\ensuremath{\mathrm{K}}\xspace}

 \def\Pb      {\ensuremath{\mathrm{b}}\xspace}                 
 \def\Pc      {\ensuremath{\mathrm{c}}\xspace}

 \def\Pi      {\ensuremath{\mathrm{i}}\xspace}

 \def\Pp      {\ensuremath{\mathrm{p}}\xspace}

 \def\Ps      {\ensuremath{\mathrm{s}}\xspace}

}
{

 \def\Pmu         {\ensuremath{\mu}\xspace}

 \def\Ppi         {\ensuremath{\pi}\xspace}

 \def\Ppsi        {\ensuremath{\psi}\xspace}                 
                  
 \mathchardef\PDelta="7101
 \mathchardef\PXi="7104
 \mathchardef\PLambda="7103
 \mathchardef\PSigma="7106
 \mathchardef\POmega="710A
 \mathchardef\PUpsilon="7107
                  
 \def\PB      {\ensuremath{B}\xspace}                 
                  
 \def\PD      {\ensuremath{D}\xspace}

 \def\PJ      {\ensuremath{J}\xspace}                 
 \def\PK      {\ensuremath{K}\xspace}

 \def\Pb      {\ensuremath{b}\xspace}                 
 \def\Pc      {\ensuremath{c}\xspace}

 \def\Pi      {\ensuremath{i}\xspace}

 \def\Pp      {\ensuremath{p}\xspace}

 \def\Ps      {\ensuremath{s}\xspace}

}

\makeatletter
\ifcase \@ptsize \relax
  \newcommand{\miniscule}{\@setfontsize\miniscule{4}{5}}
\or
  \newcommand{\miniscule}{\@setfontsize\miniscule{5}{6}}
\or
  \newcommand{\miniscule}{\@setfontsize\miniscule{5}{6}}
\fi
\makeatother

\DeclareRobustCommand{\optbar}[1]{\shortstack{{\miniscule (\rule[.5ex]{1.25em}{.18mm})}
  \\ [-.7ex] $#1$}}

\def\mup        {{\ensuremath{\Pmu^+}}\xspace}
\def\mun        {{\ensuremath{\Pmu^-}}\xspace} 
\def\mumu       {{\ensuremath{\Pmu^+\Pmu^-}}\xspace}

\def\squark    {{\ensuremath{\Ps}}\xspace}

\def\cquark    {{\ensuremath{\Pc}}\xspace}

\def\bquark    {{\ensuremath{\Pb}}\xspace}

\def\pion   {{\ensuremath{\Ppi}}\xspace}

\def\pip    {{\ensuremath{\pion^+}}\xspace}
\def\pim    {{\ensuremath{\pion^-}}\xspace}

\def\kaon    {{\ensuremath{\PK}}\xspace}
  \def\Kbar    {{\kern 0.2em\overline{\kern -0.2em \PK}{}}\xspace}

\def\KorKbar    {\kern 0.18em\optbar{\kern -0.18em K}{}\xspace}

\def\Kp      {{\ensuremath{\kaon^+}}\xspace}
\def\Km      {{\ensuremath{\kaon^-}}\xspace}

\def\Kstarz  {{\ensuremath{\kaon^{*0}}}\xspace}

  \def\Dbar    {{\kern 0.2em\overline{\kern -0.2em \PD}{}}\xspace}
\def\D       {{\ensuremath{\PD}}\xspace}

\def\DorDbar    {\kern 0.18em\optbar{\kern -0.18em D}{}\xspace}

\def\Dsm     {{\ensuremath{\D^-_\squark}}\xspace}

\def\B       {{\ensuremath{\PB}}\xspace}
\def\Bbar    {{\ensuremath{\kern 0.18em\overline{\kern -0.18em \PB}{}}}\xspace}

\def\BorBbar    {\kern 0.18em\optbar{\kern -0.18em B}{}\xspace}

\def\Bd      {{\ensuremath{\B^0}}\xspace}
\def\Bs      {{\ensuremath{\B^0_\squark}}\xspace}

\def\jpsi     {{\ensuremath{{\PJ\mskip -3mu/\mskip -2mu\Ppsi\mskip 2mu}}}\xspace}
\def\psitwos  {{\ensuremath{\Ppsi{(2S)}}}\xspace}

  \def\Y#1S{\ensuremath{\PUpsilon{(#1S)}}\xspace}

\def\proton      {{\ensuremath{\Pp}}\xspace}

\def\Lz          {{\ensuremath{\PLambda}}\xspace}
\def\Lbar        {{\ensuremath{\kern 0.1em\overline{\kern -0.1em\PLambda}}}\xspace}
\def\LorLbar    {\kern 0.18em\optbar{\kern -0.18em \PLambda}{}\xspace}

\def\Lb      {{\ensuremath{\Lz^0_\bquark}}\xspace}

\def\BF         {{\ensuremath{\cal B}}\xspace}

\def\BR         {\BF}
\newcommand{\decay}[2]{\ensuremath{#1\!\to #2}\xspace}         

\def\to                 {\ensuremath{\rightarrow}\xspace}

\def\qsq       {{\ensuremath{q^2}}\xspace}

\def\CP                {{\ensuremath{C\!P}}\xspace}

\def\AT#1     {\ensuremath{A_{\mathrm{T}}^{#1}}\xspace}           

\def\C#1      {\ensuremath{\mathcal{C}_{#1}}\xspace}                       
\def\Cp#1     {\ensuremath{\mathcal{C}_{#1}^{'}}\xspace}                    
\def\Ceff#1   {\ensuremath{\mathcal{C}_{#1}^{\mathrm{(eff)}}}\xspace}        
\def\Cpeff#1  {\ensuremath{\mathcal{C}_{#1}^{'\mathrm{(eff)}}}\xspace}       
\def\Ope#1    {\ensuremath{\mathcal{O}_{#1}}\xspace}                       
\def\Opep#1   {\ensuremath{\mathcal{O}_{#1}^{'}}\xspace}                    

\newcommand{\tev}{\ifthenelse{\boolean{inbibliography}}{\ensuremath{~T\kern -0.05em eV}\xspace}{\ensuremath{\mathrm{\,Te\kern -0.1em V}}}\xspace}
\newcommand{\gev}{\ensuremath{\mathrm{\,Ge\kern -0.1em V}}\xspace}
\newcommand{\mev}{\ensuremath{\mathrm{\,Me\kern -0.1em V}}\xspace}
\newcommand{\kev}{\ensuremath{\mathrm{\,ke\kern -0.1em V}}\xspace}
\newcommand{\ev}{\ensuremath{\mathrm{\,e\kern -0.1em V}}\xspace}
\newcommand{\gevc}{\ensuremath{{\mathrm{\,Ge\kern -0.1em V\!/}c}}\xspace}
\newcommand{\mevc}{\ensuremath{{\mathrm{\,Me\kern -0.1em V\!/}c}}\xspace}
\newcommand{\gevcc}{\ensuremath{{\mathrm{\,Ge\kern -0.1em V\!/}c^2}}\xspace}
\newcommand{\gevgevcccc}{\ensuremath{{\mathrm{\,Ge\kern -0.1em V^2\!/}c^4}}\xspace}
\newcommand{\mevcc}{\ensuremath{{\mathrm{\,Me\kern -0.1em V\!/}c^2}}\xspace}

\def\mum  {\ensuremath{{\,\upmu\rm m}}\xspace}

\def\invfb   {\ensuremath{\mbox{\,fb}^{-1}}\xspace}

\newcommand{\chisq}{\ensuremath{\chi^2}\xspace}

\def\gsim{{~\raise.15em\hbox{$>$}\kern-.85em
          \lower.35em\hbox{$\sim$}~}\xspace}
\def\lsim{{~\raise.15em\hbox{$<$}\kern-.85em
          \lower.35em\hbox{$\sim$}~}\xspace}

\def\sqs   {\ensuremath{\protect\sqrt{s}}\xspace}

\def\ptot       {\mbox{$p$}\xspace}
\def\pt         {\mbox{$p_{\rm T}$}\xspace}

\def\evtgen     {\mbox{\textsc{EvtGen}}\xspace}

\def\geant      {\mbox{\textsc{Geant4}}\xspace}

\def\photos     {\mbox{\textsc{Photos}}\xspace}

\def\pythia     {\mbox{\textsc{Pythia}}\xspace}

\def\tell1  {TELL1\xspace}
\def\ukl1   {UKL1\xspace}

\newcommand{\comment}[1]{}

\def\thetal   {{\ensuremath{\theta_{l}}}\xspace}
\def\thetak   {{\ensuremath{\theta_{K}}}\xspace}

\usepackage{cite} 
\usepackage{mciteplus}

\usepackage{longtable}
\usepackage{multirow}

\usepackage{tikz}
\usetikzlibrary{decorations.pathmorphing}
\usetikzlibrary{decorations.markings}
\usepackage{subfig}

\begin{document}

\renewcommand{\thefootnote}{\fnsymbol{footnote}}
\setcounter{footnote}{1}

\begin{titlepage}
\pagenumbering{roman}

\vspace*{-1.5cm}
\centerline{\large EUROPEAN ORGANIZATION FOR NUCLEAR RESEARCH (CERN)}
\vspace*{1.5cm}
\noindent
\begin{tabular*}{\linewidth}{lc@{\extracolsep{\fill}}r@{\extracolsep{0pt}}}
\ifthenelse{\boolean{pdflatex}}
{\vspace*{-2.7cm}\mbox{\!\!\!\includegraphics[width=.14\textwidth]{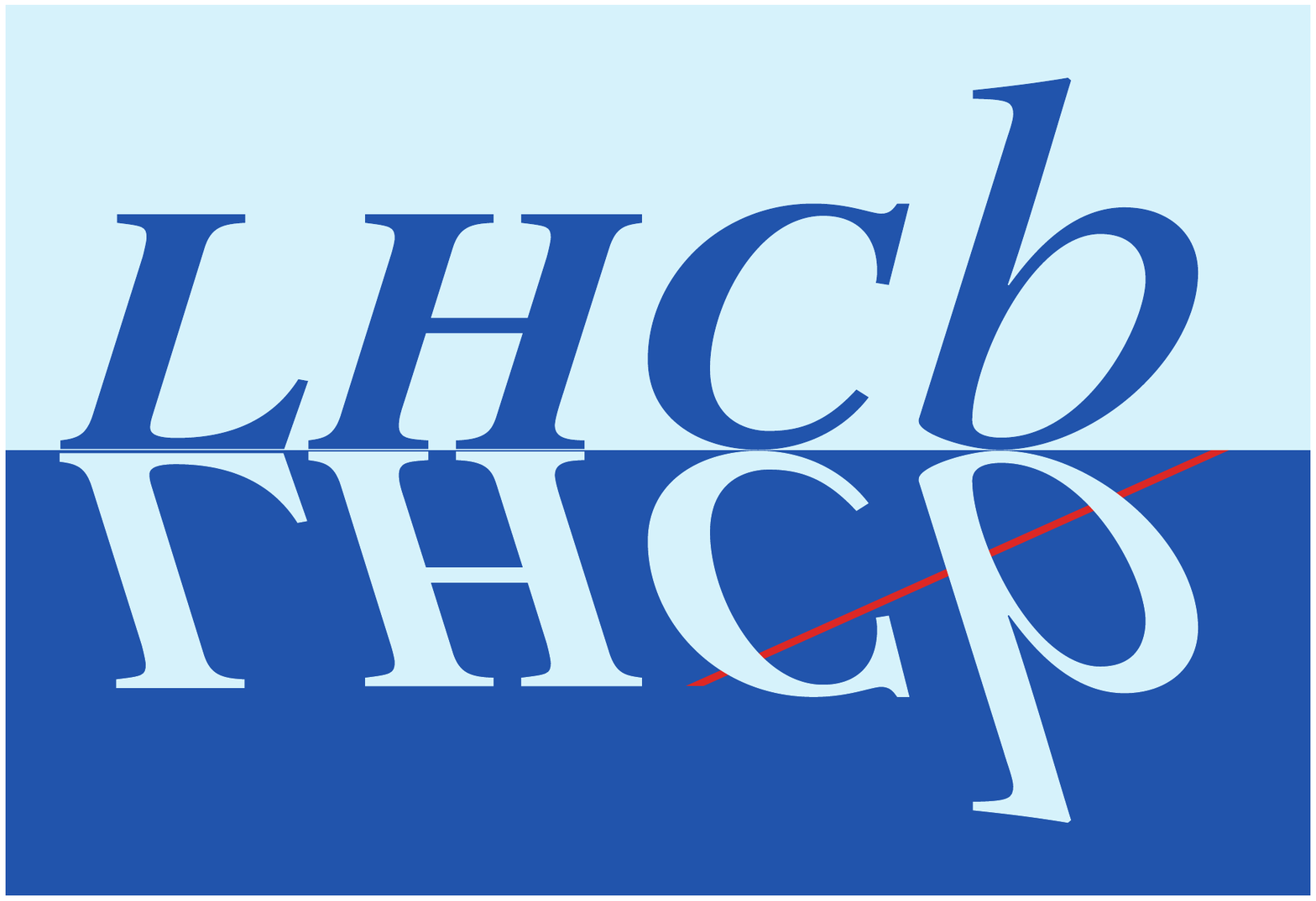}} & &}%
{\vspace*{-1.2cm}\mbox{\!\!\!\includegraphics[width=.12\textwidth]{lhcb-logo.eps}} & &}%
\\
 & & CERN-PH-EP-2015-145 \\  
 & & LHCb-PAPER-2015-023 \\  
 & & 30 September 2015 \\ 
 & & \\

\end{tabular*}

\vspace*{2.5cm}

{\bf\boldmath\huge
\begin{center}
Angular analysis and differential branching fraction of the decay $B^0_s\to\phi\mu^+\mu^-$
\end{center}
}

\vspace*{1.0cm}

\begin{center}
The LHCb collaboration\footnote{Authors are listed at the end of this paper.}
\end{center}

\vspace{\fill}

\begin{abstract}
  \noindent
  An angular analysis and a measurement of the differential branching fraction of the decay $B^0_s\to\phi\mu^+\mu^-$ are presented, 
  using data corresponding to an integrated luminosity of $3.0\invfb$ of $pp$ collisions recorded by the LHCb experiment at $\sqs = 7$ and $8\tev$. 
  Measurements are reported as a function of \qsq, the square of the dimuon invariant mass and results of the angular analysis are found to be consistent with the Standard Model.  
  In the range $1<q^2<6\gevgevcccc$, where precise theoretical calculations are available, the differential branching fraction is found to be more than $3\,\sigma$ below the Standard Model predictions. 
 
\end{abstract}

\vspace*{2.0cm}

\begin{center}
  Published as JHEP 09 (2015) 179
\end{center}

\vspace{\fill}

{\footnotesize 
\centerline{\copyright~CERN on behalf of the \lhcb collaboration, licence \href{http://creativecommons.org/licenses/by/4.0/}{CC-BY-4.0}.}}
\vspace*{2mm}

\end{titlepage}

\newpage
\setcounter{page}{2}
\mbox{~}

\cleardoublepage

\renewcommand{\thefootnote}{\arabic{footnote}}
\setcounter{footnote}{0}

\pagestyle{plain} 
\setcounter{page}{1}
\pagenumbering{arabic}

\section{Introduction}
\label{sec:Introduction}
\enlargethispage{\baselineskip}
The decay $\Bs\to\phi\mup\mun$ 
is mediated by a $b\to s$ flavour changing neutral current (FCNC) transition.
In the Standard Model (SM) it is forbidden at tree-level and proceeds via loop diagrams as shown in Fig.~\ref{fig:feynman}.
In extensions of the SM, new heavy particles can appear in competing diagrams and affect both the branching fraction of the decay
and the angular distributions of the final-state particles. 

This decay channel was first observed and studied by the CDF collaboration~\cite{Aaltonen:2011cn,Aaltonen:2011qs} and subsequently
studied by the LHCb collaboration using data collected during 2011, 
corresponding to an integrated luminosity of $1.0\invfb$~\cite{LHCb-PAPER-2013-017}. 
While the angular distributions were found to be in good agreement with SM expectations, 
the measured branching fraction differs from the recently updated SM prediction by $3.1\,\sigma$~\cite{Altmannshofer:2014rta,Straub:2015ica}. 
A similar trend is also seen for the branching fractions of other $b\to s\mumu$ processes, which tend to be lower than SM predictions~\cite{LHCb-PAPER-2013-019,LHCb-PAPER-2014-006,LHCb-PAPER-2014-030}. 

This paper presents an updated analysis of the decay $\Bs\to\phi(\to\Kp\Km)\mumu$ using data accumulated by LHCb in \proton\proton collisions,
corresponding to an integrated luminosity of $1.0\invfb$ collected during 2011 at $7\tev$ and $2.0\invfb$ collected during 2012 at $8\tev$ centre-of-mass energy. 
The differential branching fraction ${\rm d}{\cal B}(\Bs\to\phi\mumu)/{\rm d}q^2$  
is determined as a function of \qsq, the square of the dimuon invariant mass. 
In addition, a three-dimensional angular analysis in 
$\cos\thetal$, $\cos\thetak$ and $\Phi$ is performed in bins of $q^2$.
Here, the angle $\thetak$ ($\thetal$) denotes the angle of the $\Km$ ($\mun$) with respect to the 
direction of flight of the $\Bs$ meson in the $\Kp\Km$ ($\mup\mun$) centre-of-mass frame,
and $\Phi$ denotes the angle between the $\mup\mun$ and the $\Kp\Km$ decay planes in the $\Bs$ meson 
centre-of-mass frame.
Compared to the previously published fit of the one-dimensional projections of the decay angles~\cite{LHCb-PAPER-2013-017},
the full three-dimensional angular fit gives improved sensitivity and allows access to more angular observables. 

The decay $\decay{\Bs}{\phi\mup\mun}$ is closely related to the decay $\decay{\Bd}{\Kstarz\mup\mun}$, 
which has been studied extensively by LHCb~\cite{LHCb-PAPER-2013-019,LHCb-PAPER-2013-037,LHCb-CONF-2015-002}. 
Although $\Bs$ meson production is suppressed with respect to the $\Bd$ meson by the fragmentation fraction ratio $f_s/f_d\sim 1/4$, 
the narrow $\phi$ resonance allows a clean selection with low background levels. 
Furthermore, the contribution from the S wave, where the $\Kp\Km$ system is in a spin-0 configuration, is expected to be low~\cite{LHCb-PAPER-2012-040}. 
Since the $\Kp\Km\mup\mun$ final state is not flavour-specific,
the angular observables accessible in the decay $\decay{\Bs}{\phi\mup\mun}$ are the \CP averages $F_{\rm L}$, $S_{3,4,7}$ and the \CP asymmetries $A_{5,6,8,9}$~\cite{Bobeth:2008ij}. 
The flavour-averaged differential decay rate,
as a function of the decay angles 
in bins of $q^2$, is given by
\begin{align}
  \frac{1}{{\rm d}\Gamma/{\rm d}q^2}
  \frac{{\rm d}^3\Gamma}{{\rm dcos}\thetal\,{\rm dcos}\thetak\,{\rm d}\Phi}
&= \frac{9}{32\pi}
  \bigl[ \tfrac{3}{4} (1-{F_{\rm L}})\sin^2\thetak\nonumber+ {F_{\rm L}}\cos^2\thetak\nonumber\\
    &+ \tfrac{1}{4}(1-{F_{\rm L}})\sin^2\thetak\cos 2\thetal - {F_{\rm L}} \cos^2\thetak\cos 2\thetal\nonumber\\
    &+ {S_3}\sin^2\thetak \sin^2\thetal \cos 2\Phi + {S_4} \sin 2\thetak \sin 2\thetal \cos\Phi\nonumber\\
    &+ {A_5}\sin 2\thetak \sin \thetal \cos \Phi+ A_{6} \sin^2\thetak \cos\thetal\nonumber\\
    &+ {S_7} \sin 2\thetak \sin\thetal \sin\Phi + {A_8} \sin 2\thetak \sin 2\thetal \sin\Phi\nonumber\\ 
    &+ {A_9}\sin^2\thetak \sin^2\thetal \sin 2\Phi 
    \bigr].\label{eq:untagged}
\end{align}
The {\it T}-odd \CP asymmetries $A_8$ and $A_9$ are predicted to be close to zero in the SM 
and are of particular interest, as they can be large in the presence of contributions beyond the SM~\cite{Bobeth:2008ij}. 

\begin{figure}
\tikzset{
  photon/.style={decorate, decoration={snake,amplitude=.5mm,segment length=1.5mm,post length=0mm}},
  wboson/.style={dashed},
  fermion/.style={solid, decoration={markings,mark=at position 0.5 with {\arrow{>}}},postaction={decorate}},
  antifermion/.style={solid, decoration={markings,mark=at position 0.5 with {\arrow{<}}},postaction={decorate}},
  gluon/.style={decorate,decoration={coil,amplitude=1mm,segment length=1.25mm}}
}
\begin{center}
\begin{tikzpicture}
\coordinate [label=left:$s$] (sin) at (-2.5,0);
\coordinate [label=left:$\bar{b}$] (bin) at (-2.5,-1);
\coordinate [label=right:$s$] (sout) at (2.5,0);
\coordinate [label=right:$\bar{s}$] (bout) at (2.5,-1);
\coordinate [label=left:$\Bs$] (bs) at (-2.75,-0.5);
\coordinate [label=right:$\phi$] (phi) at (2.75,-0.5);
\coordinate (vtb) at (-1,-1);
\coordinate (vts) at (1,-1);
\coordinate (gammafrom) at (0.7107, -1.7107);
\coordinate (gammato) at (1.5,-2.5);
\coordinate [label=right:$\mu^-$] (muminusout) at (2.5,-2.0);
\coordinate [label=right:$\mu^+$] (muplusout) at (2.5,-3.0);
\draw [fermion] (sin) -- (sout);
\draw [antifermion] (bin) -- (vtb);
\draw [antifermion] (vts) -- (bout);
\draw [antifermion] (vtb) -- (vts);
\node (top) at (0,-0.7) {${\bar{t}}$};
\node (w) at (0,-1.7) {$W$};
\draw [wboson] (-1.0,-1.0) arc (180:360:1.0); 
\draw [photon] (gammafrom) -- (gammato); 
\draw [antifermion] (gammato) -- (muplusout);
\draw [fermion] (gammato) -- (muminusout);
\node (photonz) at (0.65,-2.3) {${Z^0,\gamma}$};
\end{tikzpicture}
\hspace{1cm}
\begin{tikzpicture}
\coordinate [label=left:$s$] (sin) at (-2.5,0);
\coordinate [label=left:$\bar{b}$] (bin) at (-2.5,-1);
\coordinate [label=right:$s$] (sout) at (2.5,0);
\coordinate [label=right:$\bar{s}$] (bout) at (2.5,-1);
\coordinate [label=left:$\Bs$] (bs) at (-2.75,-0.5);
\coordinate [label=right:$\phi$] (phi) at (2.75,-0.5);
\coordinate (vtb) at (-0.5,-1);
\coordinate (vts) at (0.5,-1);
\coordinate (gammafrom) at (0.7107, -1.7107);
\coordinate (gammato) at (1.5,-2.5);
\coordinate [label=right:$\mu^-$] (muminusout) at (2.5,-2.0);
\coordinate [label=right:$\mu^+$] (muplusout) at (2.5,-3.0);
\coordinate (muplusin) at (1.5,-3.0);
\coordinate (muminusin) at (1.5,-2.0);
\draw [fermion] (sin) -- (sout);
\draw [antifermion] (bin) -- (vtb);
\draw [antifermion] (vts) -- (bout);
\draw [antifermion] (vtb) -- (vts);
\node (top) at (0,-0.7) {${\bar{t}}$};
\draw [antifermion] (muplusin) -- (muplusout);
\draw [fermion] (muminusin) -- (muminusout);
\draw [fermion] (muplusin) -- (muminusin);
\draw [wboson] (vtb) -- (muplusin);
\draw [wboson] (vts) -- (muminusin);
\node at (1.8,-2.5) {$\nu_\mu$};
\node at (1.5,-1.5) {$W$};
\node at (0.0,-2.0) {$W$};
\end{tikzpicture}
\end{center}
\caption{
Examples of $b\to s$ loop diagrams contributing to the decay $\Bs\to\phi\mumu$ in the SM. 
}
\label{fig:feynman}
\end{figure}

\section{Detector and simulation}
\label{sec:detector}
The \lhcb detector~\cite{Alves:2008zz,LHCb-DP-2014-002} is a single-arm forward
spectrometer covering the \mbox{pseudorapidity} range $2<\eta <5$,
designed for the study of particles containing \bquark or \cquark
quarks. The detector includes a high-precision tracking system
consisting of a silicon-strip vertex detector surrounding the $pp$
interaction region, a large-area silicon-strip detector located
upstream of a dipole magnet with a bending power of about
$4{\rm\,Tm}$, and three stations of silicon-strip detectors and straw
drift tubes placed downstream of the magnet.
The tracking system provides a measurement of momentum, \ptot, of charged particles with
a relative uncertainty that varies from 0.5\% at low momentum to 1.0\% at 200\gevc.
The minimum distance of a track to a primary vertex, the impact parameter (IP), is measured with a resolution of $(15+29/\pt)\mum$,
where \pt is the component of the momentum transverse to the beam, in\,\gevc.
Different types of charged hadrons are distinguished using information
from two ring-imaging Cherenkov detectors. 
Photons, electrons and hadrons are identified by a calorimeter system consisting of
scintillating-pad and preshower detectors, an electromagnetic
calorimeter and a hadronic calorimeter. Muons are identified by a
system composed of alternating layers of iron and multiwire
proportional chambers.
The online event selection is performed by a trigger~\cite{LHCb-DP-2012-004}, 
which consists of a hardware stage, based on information from the calorimeter and muon
systems, followed by a software stage, which applies a full event
reconstruction.

Simulated signal samples are used to determine the effect of the detector geometry,
trigger, reconstruction and selection on the signal efficiency. 
In addition, simulated background samples are used to determine the pollution from specific background processes. 
In the simulation, $pp$ collisions are generated using
\pythia~\cite{Sjostrand:2006za,Sjostrand:2007gs} 
 with a specific \lhcb
configuration~\cite{LHCb-PROC-2010-056}.  Decays of hadronic particles
are described by \evtgen~\cite{Lange:2001uf}, in which final-state
radiation is generated using \photos~\cite{Golonka:2005pn}. The
interaction of the generated particles with the detector, and its response,
are implemented using the \geant
toolkit~\cite{Allison:2006ve, Agostinelli:2002hh} as described in
Ref.~\cite{LHCb-PROC-2011-006}. 
Data-driven corrections are applied to the simulated samples to account 
for imperfect modelling of particle identification performance, 
the $\Bs$ meson transverse momentum spectrum and $\Bs$ vertexing quality, as well as track multiplicity.

\section{Selection of signal candidates}
\label{sec:selection}
The $\Bs\to\phi\mup\mun$ signal candidates are required to satisfy the hardware trigger requirement,
which selects muons with 
$\pt>1.48\gevc$ 
in the $7\tev$ data and $\pt>1.76\gevc$ in the $8\tev$ data. 
In the subsequent software trigger, at least
one of the final-state particles is required to have both
$\pt>0.8\gevc$ and IP larger than $100\mum$ with respect to all
of the primary $pp$ interaction vertices~(PVs) in the
event. The tracks of two or more of the final-state
particles are also required to form a vertex that is significantly
displaced from any PV.

Signal candidates are accepted if their reconstructed invariant mass is in the range $5267<m(\Kp\Km\mup\mun)<5800\mevcc$ and the invariant mass of the $\Kp\Km$ system is within $12\mevcc$ of the known $\phi$ mass~\cite{PDG2014}. The mass resolutions are $19\mevcc$ for the invariant $\Kp\Km\mup\mun$ mass and $4\mevcc$ for the $\Kp\Km$ invariant mass. 
The final-state particles are required to have significant $\chi^2_{\rm IP}$ 
with respect to any PV in the event, 
where $\chi^2_{\rm IP}$ denotes the change in the $\chi^2$ of the PV when reconstructed with or without the considered track.
The four final-state tracks are then fitted to a common vertex which is required to be of good quality 
and significantly displaced from any PV in the event. 
The signal candidate is required to have small $\chi^2_{\rm IP}$ 
with respect to a PV in the event. 
Furthermore, the angle $\theta_{\rm DIRA}$ between the reconstructed $\Bs$ momentum and the vector connecting the PV with the decay vertex is required to be small.

To further reduce the combinatorial background, a boosted decision tree (BDT)~\cite{Breiman} using the AdaBoost algorithm~\cite{AdaBoost} is employed. 
The BDT is trained with a sample of $\Bs\to\jpsi(\to\mup\mun)\phi(\to\Kp\Km)$ decays as a signal proxy and events from the upper mass sideband,
$5567<m(\Kp\Km\mup\mun)<5800\mevcc$, as a proxy for the background. 
The discriminating variables of the BDT are 
the $\chi^2_{\rm IP}$ of the $\Bs$ signal candidate and all final-state tracks, 
the $\Bs$ transverse momentum, 
the \chisq of the vertex fit ($\chi^2_{\rm Vtx}$), 
the flight distance significance 
of the signal candidate, 
and particle identification information for the final-state particles. 
The BDT selection has an efficiency of $96\%$ for signal events with a background rejection of $95\%$. 
The total efficiency for signal events, including detector geometry, trigger and reconstruction effects is $1.1\%$.

\begin{figure}
  \centering
  \includegraphics[width=12cm]{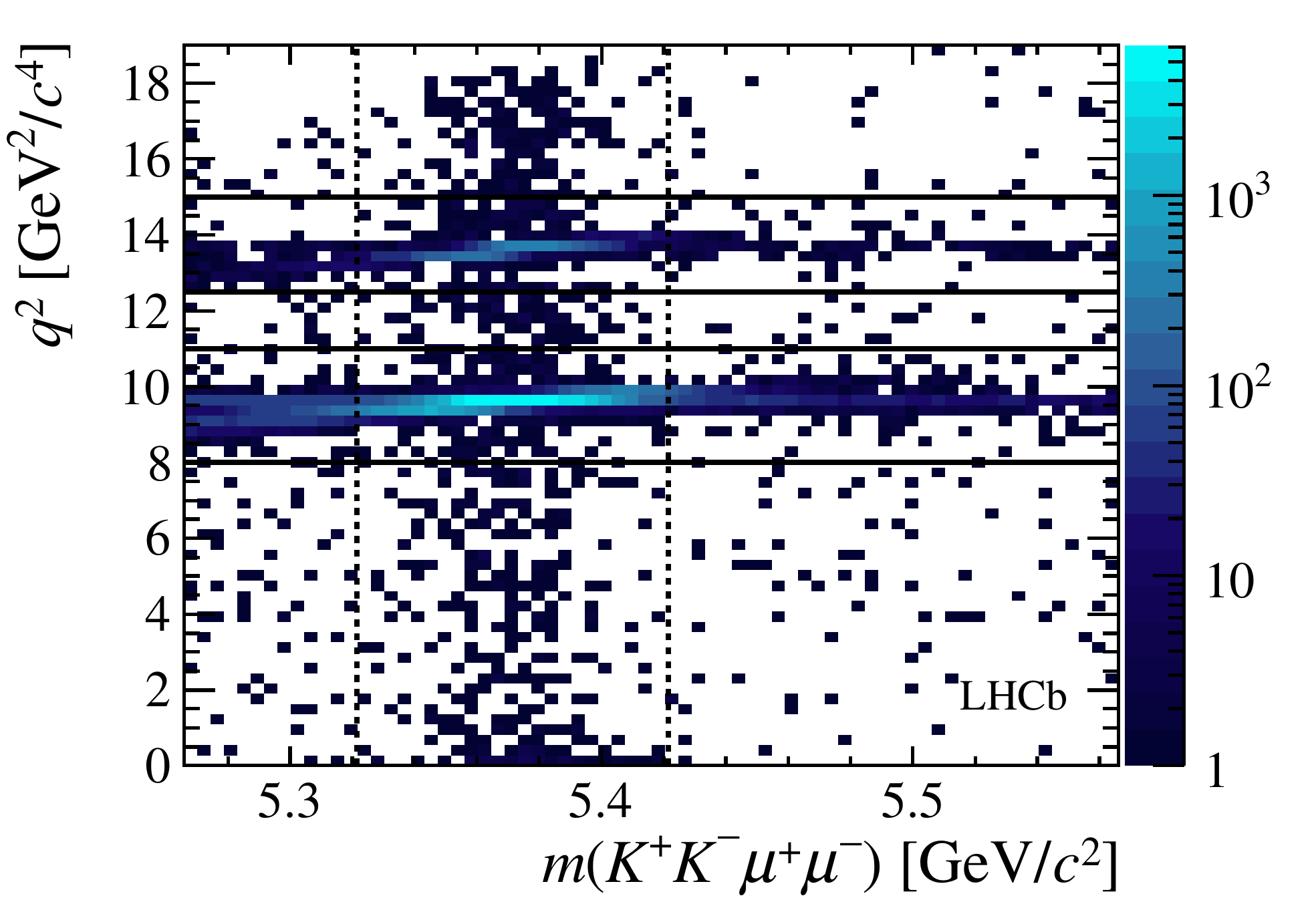}
  \caption{Two-dimensional distribution of \qsq versus the invariant mass of the $\Kp\Km\mumu$ system.
    The signal decay $\decay{\Bs}{\phi\mumu}$ is clearly visible within the $\pm50\mevcc$ interval around the $\Bs$ mass, indicated by the dashed vertical lines. 
    The horizontal lines denote the charmonium regions, where the tree-level decays $\decay{\Bs}{\jpsi\phi}$ and $\decay{\Bs}{\psitwos\phi}$ dominate.\label{fig:mbvsq2}}
\end{figure}

The reconstructed $\Bs$ mass versus $q^2$ for signal candidates after the full selection is given in Fig.~\ref{fig:mbvsq2}. 
The signal decay $\Bs\to\phi\mup\mun$ is clearly visible as a vertical band. 
In the $q^2$ region around the $\jpsi$ and $\psitwos$ masses, the tree-level charmonium decays $\Bs\to\jpsi\phi$ and $\Bs\to\psitwos\phi$ dominate.
The decay $\Bs\to\jpsi\phi$ is used as a control mode throughout the analysis. 

\subsection{Backgrounds}
The decays $\Bs\to\jpsi\phi$ and $\Bs\to\psitwos\phi$, primarily originating from $b\to c\bar{c}s$ tree-level processes, are vetoed by rejecting
candidates in the $q^2$ regions $8.0<q^2<11.0\gevgevcccc$ and $12.5<q^2<15.0\gevgevcccc$.
The $\Bs\to\jpsi\phi$ decay can also constitute a peaking background if one of the final-state muons is misidentified as a kaon and vice-versa. 
This background is vetoed by rejecting candidates for which the invariant mass of the $K^\pm\mu^\mp$ system, 
with the kaon reconstructed under the muon mass hypothesis, is within $45\mevcc$ of the known $\jpsi$ meson mass~\cite{PDG2014},
unless the final-state particles fulfil stringent particle identification requirements. 
After the veto is applied, this background contribution is found to be negligible. 

The rare baryonic decay $\decay{\Lb}{\Lz(1520)(\to p\Km)\mup\mun}$ can mimic the signal decay if the proton in the final state is misidentified as a kaon.
This potential background is vetoed by rejecting events with invariant mass close to the known $\Lb$ baryon mass~\cite{PDG2014} where one kaon has the proton mass
hypothesis assigned,
unless the kaon passes stringent particle identification requirements. 
Assuming a $\qsq$ dependence following Ref.~\cite{Mott:2011cx} and using $\BR(\decay{\Lb}{\Lz\mup\mun}) = (0.96\pm 0.29)\times 10^{-6}$ \cite{LHCb-PAPER-2013-025} as
an estimate for the unknown $\decay{\Lb}{\Lz(1520)\mup\mun}$ branching fraction, a yield of $2.0\pm 0.8$ $\decay{\Lb}{\Lz(1520)\mup\mun}$ background events is
expected in the signal region, within $50\mevcc$ of the known $\Bs$ mass~\cite{PDG2014}, 
after the veto. 
The rare decay $\decay{\Bd}{\Kstarz\mup\mun}$ can be a peaking background if the pion in the final state is reconstructed as a kaon.
After suppressing this background using particle identification information,  
a yield of $1.7\pm 0.4$ events is expected in the signal region.
The background pollution from $\decay{\Lb}{\Lz(1520)\mup\mun}$ and $\decay{\Bd}{\Kstarz\mup\mun}$ decays is neglected in the fit and treated as a systematic uncertainty. 
Backgrounds from semileptonic $b\to c(\to s\mun\bar{\nu}_\mu)\mup\nu_\mu$ cascade decays and fully hadronic decays such as $\Bs\to\Dsm(\to \Kp\Km\pim)\pip$,
where hadrons are misidentified as muons, are estimated to be small and can therefore be neglected.

\section{Differential branching fraction}
\label{sec:branchingfraction}
\begin{figure}
  \centering
  \includegraphics[width=0.49\textwidth]{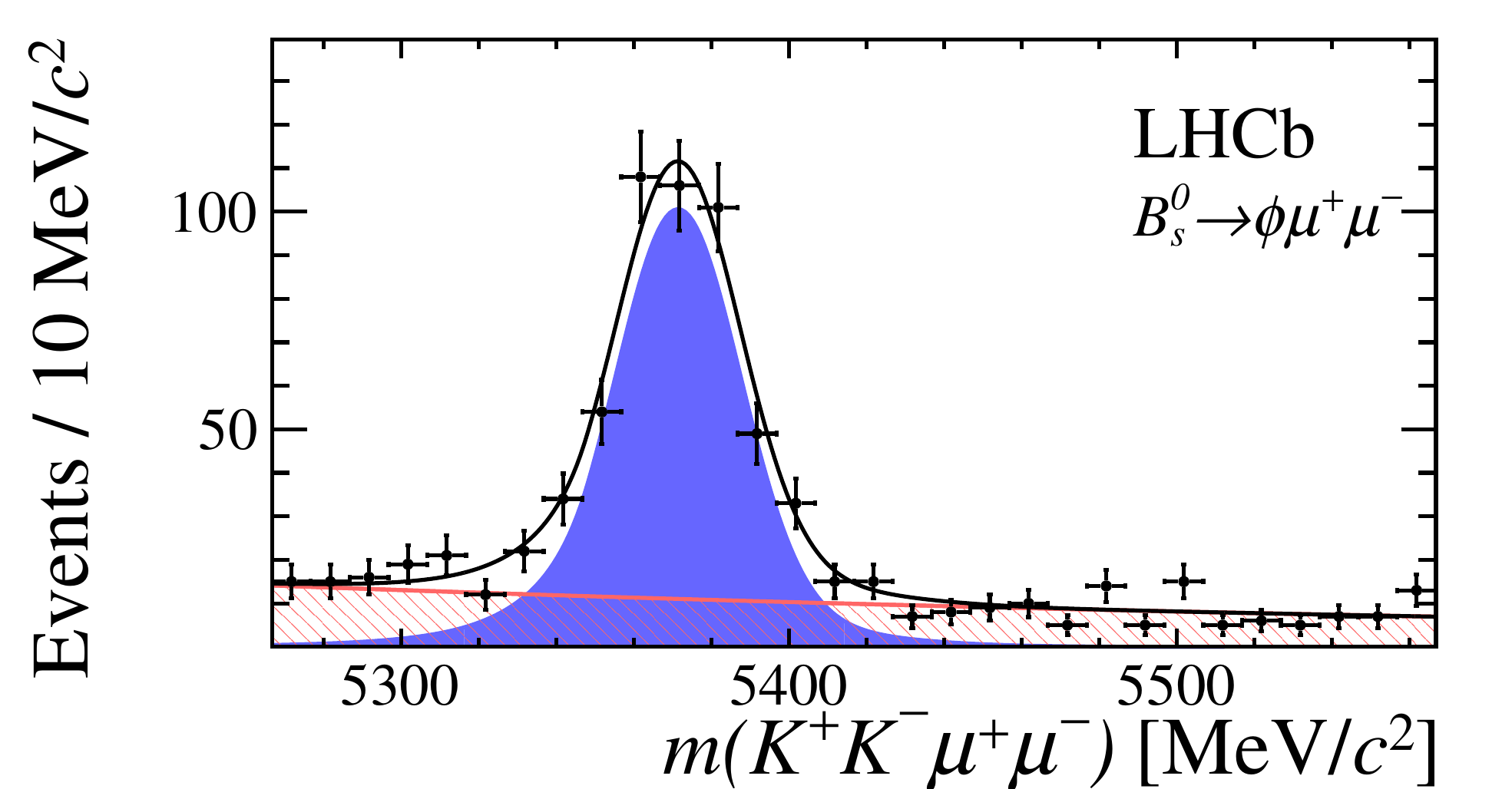}
  \hfill\includegraphics[width=0.49\textwidth]{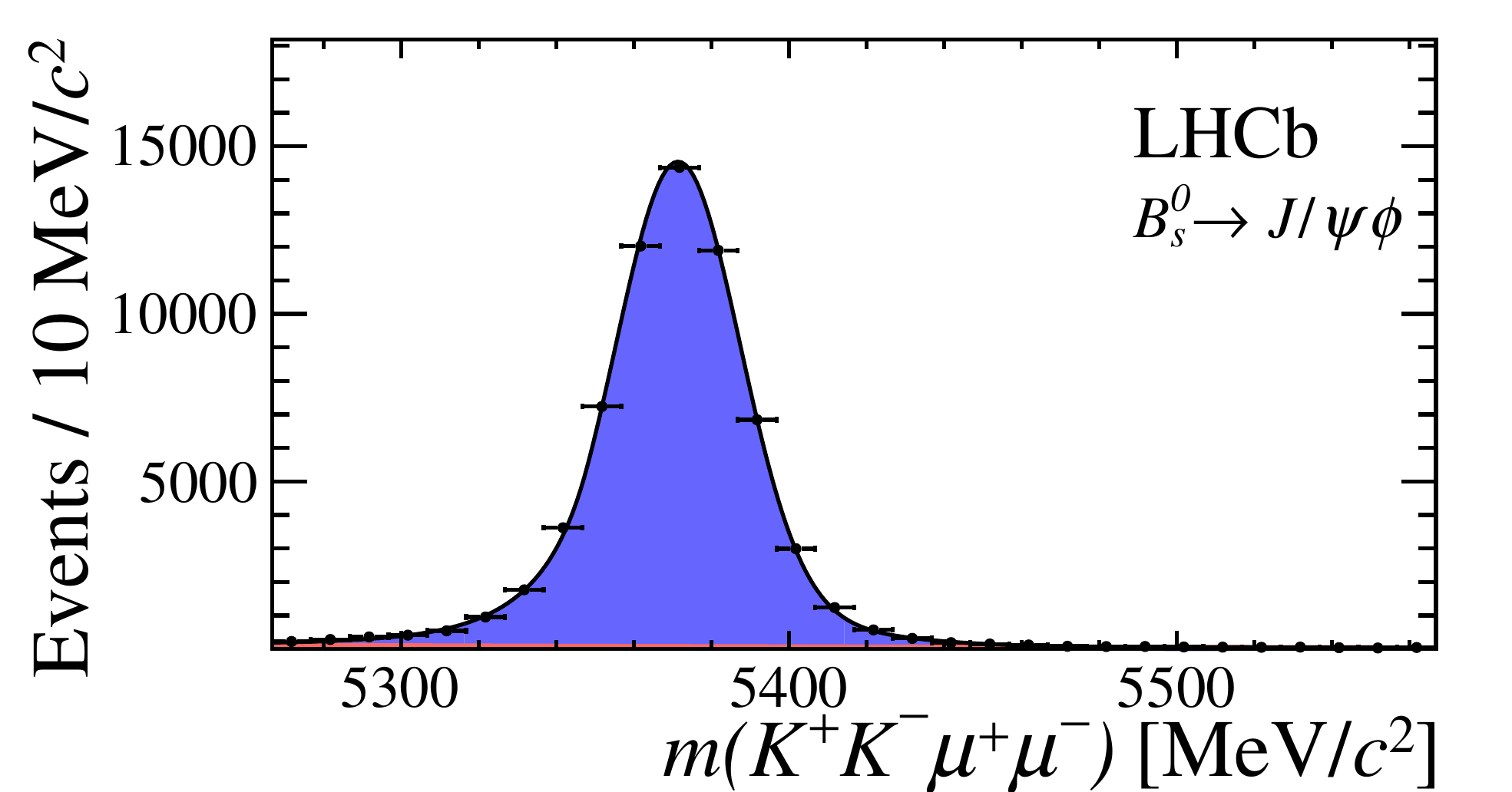}
  \caption{
    Invariant mass distribution for (left) $\decay{\Bs}{\phi\mumu}$ signal decays,
    integrated over the $q^2$ bins used,
    and for (right) the control mode $\decay{\Bs}{\jpsi\phi}$ in the $\Kp\Km\mumu$ final state.
    The signal component is given by the solid blue area, the background component by the shaded red area.\label{fig:massmodel}}
\end{figure}

Figure~\ref{fig:massmodel} shows the $\Kp\Km\mup\mun$ invariant mass distribution 
for the $\decay{\Bs}{\phi\mumu}$ signal decay integrated over $q^2$,
as well as for the control mode $\decay{\Bs}{\jpsi\phi}$. 
To determine the $\decay{\Bs}{\phi\mumu}$ signal yields in bins of $q^2$, 
extended maximum likelihood fits are performed. 
The combinatorial background is described by an exponential function,
whilst the signal component is modelled with the sum of two Gaussian functions with a common mean and a radiative power-law tail toward smaller invariant mass values.
The parameters describing the signal mass shape are determined from a fit to the $\decay{\Bs}{\jpsi\phi}$ control mode.
The \qsq dependence of the signal mass resolution is accounted for by using scale factors, which are determined from simulation. 
The $\Kp\Km\mumu$ invariant mass distributions for the signal decay in bins of $q^2$ are given in Appendix~\ref{app:mass};
the yields and corresponding uncertainties 
are listed in Table~\ref{tab:diffbr}.  
Integrating over the $q^2$ bins, the signal yield is found to be $432\pm 24$. 
A fit to the control mode $\decay{\Bs}{\jpsi\phi}$, which is used for normalisation, gives $N_{\jpsi\phi}=62\,033\pm 260$ decays. 

The differential branching fraction for a given $q^2$ bin 
$[q^2_{\rm min},q^2_{\rm max}]$ is calculated according to 
\begin{align}
  \frac{{\rm d}{\cal B}(\decay{\Bs}{\phi\mumu})}{{\rm d}q^2} &=
  \frac{1}{q^2_{\rm max}-q^2_{\rm min}}\cdot\frac{N_{\phi\mu\mu}}{N_{\jpsi\phi}}\cdot\frac{\epsilon_{\jpsi\phi}}{\epsilon_{\phi\mu\mu}}\cdot
       {\cal B}(\decay{\Bs}{\jpsi\phi}) {\cal B}(\decay{\jpsi}{\mup\mun}),~\label{eq:diffbr}
\end{align}
where $N_{\rm \phi\mu\mu}$ and $N_{\jpsi\phi}$ denote the yield of the signal and normalisation mode,
and $\epsilon_{\phi\mu\mu}$ and $\epsilon_{\jpsi\phi}$ their respective efficiencies.
The branching fractions are given by ${\cal B}(\decay{\jpsi}{\mup\mun})=(5.961\pm0.033)\times 10^{-2}$~\cite{PDG2014} and ${\cal B}(\decay{\Bs}{\jpsi\phi})=(10.76\pm 0.81)\times 10^{-4}$.
For the branching fraction of the normalisation channel $\decay{\Bs}{\jpsi\phi}$ the LHCb measurement~\cite{LHCb-PAPER-2012-040} is recalculated using an updated
measurement of $f_s/f_d=0.259\pm0.015$~\cite{fsfd}. 
A weighted average is calculated by combining this updated measurement with the measurements by Belle~\cite{Thorne:2013llu} and CDF~\cite{Abe:1996kc}. 
The resulting relative and absolute differential branching fractions are given in Table~\ref{tab:diffbr}. 

\begin{figure}
  \centering
  \includegraphics[width=12cm]{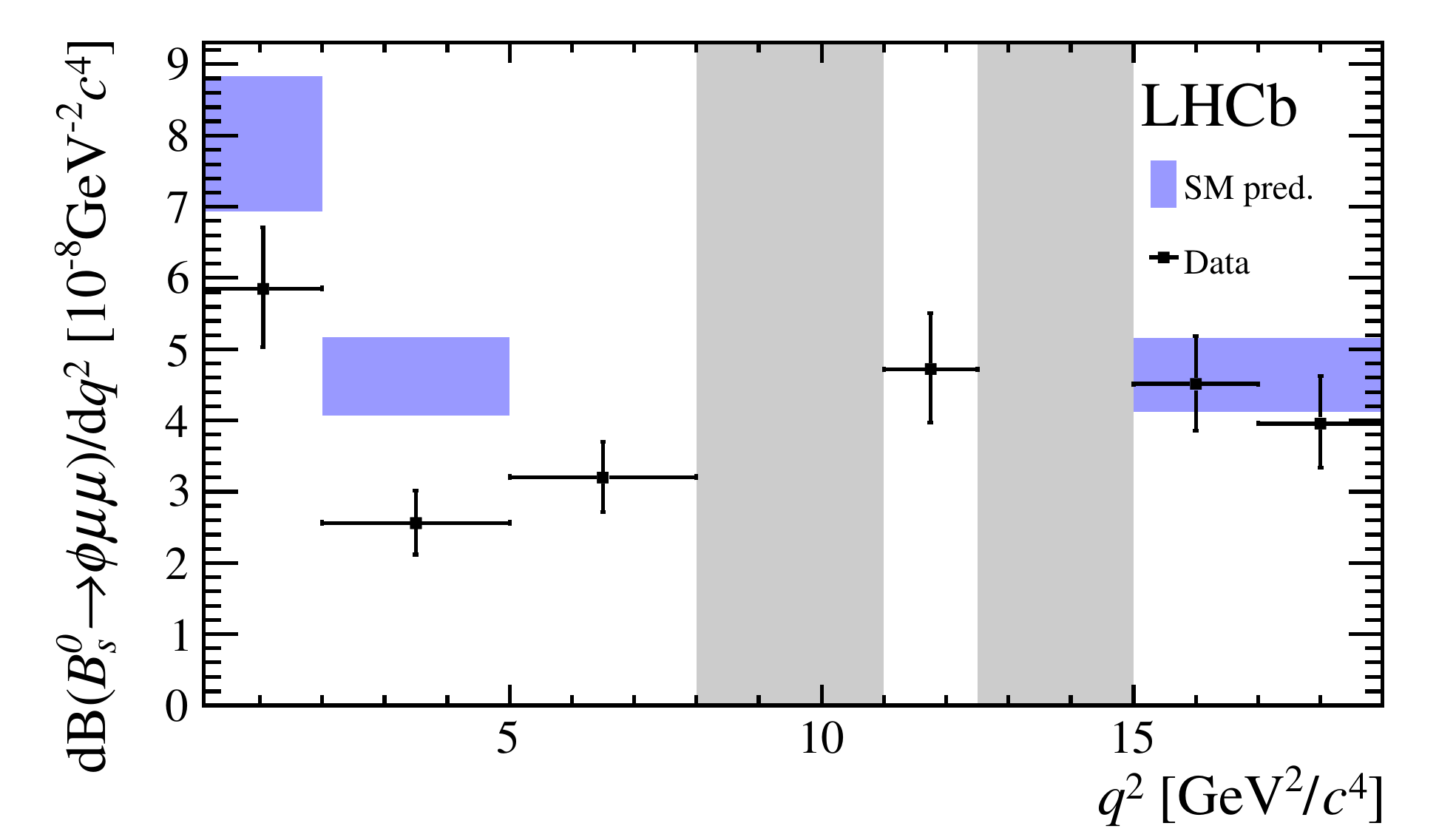}\\
  \caption{
    Differential branching fraction of the decay $\decay{\Bs}{\phi\mumu}$, 
    overlaid with SM predictions~\cite{Altmannshofer:2014rta,Straub:2015ica} indicated by blue shaded boxes. The vetoes excluding the charmonium resonances are indicated by
grey areas.\label{fig:diffbr}}
\end{figure}

The differential branching fraction is also shown in Fig.~\ref{fig:diffbr},
overlaid with SM predictions from Refs.~\cite{Altmannshofer:2014rta,Straub:2015ica}.
In the \qsq region $1.0<q^2<6.0\gevgevcccc$ the measured differential branching fraction lies $3.3\,\sigma$ below the SM expectation of $(4.81\pm  0.56)\times 10^{-8}\gev^{-2}c^{4}$~\cite{Altmannshofer:2014rta,Altmannshofer:2015sma}. 
For the SM predictions, the form factors are determined in a combined fit to the results of
light-cone sum rule calculations at low $q^2$~\cite{Straub:2015ica} and lattice QCD calculations at high $q^2$~\cite{Horgan:2013hoa,Horgan:2015vla}. 
Standard Model predictions for the branching fraction at high $q^2$ that exclusively use the results from lattice calculations~\cite{Horgan:2013pva} are found to be larger than the results from the combined fit. 
Owing to their proximity to the charmonium resonances, no predictions are available corresponding to the $q^2$ bins $5.0 < q^2 < 8.0 \gevgevcccc$ and $11.0 < q^2 < 12.5\gevgevcccc$. 

The total branching fraction of the signal decay is given by the integral over the six $q^2$ bins. 
To account for the fraction of signal events 
in the vetoed \qsq\ regions, 
a correction factor $f_{\rm veto}=1.520\pm 0.003\pm 0.043$ is applied, 
which is determined using the calculation in Ref.~\cite{Ali:2002jg} with updated form factors from Ref.~\cite{PhysRevD.71.014029}.
The first given uncertainty is statistical, the second is systematic.

The resulting relative and total branching fractions are
\begin{align*}
  \frac{{\cal B}(\decay{\Bs}{\phi\mup\mun})}{{\cal B}(\decay{\Bs}{\jpsi\phi})}&= (7.41^{+0.42}_{-0.40}\pm 0.20\pm 0.21)\times 10^{-4},\\
{\cal B}(\decay{\Bs}{\phi\mup\mun}) &= (7.97^{+0.45}_{-0.43} \pm 0.22 \pm 0.23 \pm 0.60)\times 10^{-7},
\end{align*}
where the uncertainties are (from left to right) statistical, systematic, and from the extrapolation to the full $q^2$ region. For the total branching fraction, a further
uncertainty originates from the uncertainty on the branching fraction of the normalisation mode.

\bgroup 
\def\arraystretch{1.75}
\begin{table}
\small
  \newlength{\maxwidth}
  \setlength{\maxwidth}{\widthof{19.0}}
  \caption{
    The signal yields for $\decay{\Bs}{\phi\mumu}$ decays, as well as the differential branching fraction relative to the normalisation mode
    and the absolute differential branching fraction, in bins of $q^2$. The given uncertainties are (from left to right) statistical, systematic, and the uncertainty on the branching fraction of the normalisation mode.\label{tab:diffbr}}
  \centering
\begin{tabular}{rrcc}\hline
  $q^2$ bin $[\gevgevcccc]$ & \multicolumn{1}{c}{$N_{\phi\mu\mu}$} & $\frac{{\rm d}{\cal B}(\Bs\to\phi\mu\mu)}{{\cal B}(\Bs\to\jpsi\phi){\rm d}\qsq}$ $[10^{-5}\gev^{-2}c^{4}]$ &
    $\frac{{\rm d}{\cal B}(\Bs\to\phi\mup\mun)}{{\rm d}q^2}$ $[10^{-8}\gev^{-2}c^{4}]$\\ \hline
  $0.1<q^2<\makebox[\maxwidth][r]{2.0}$   & $85^{+11}_{-10}$     & $5.44^{+0.68}_{-0.64}\pm 0.13$ & $5.85 ^{+0.73}_{-0.69}\pm 0.14\pm 0.44$\\
  $2.0<q^2<\makebox[\maxwidth][r]{5.0}$   & $60^{+10}_{-9\hphantom{0}}$       & $2.38^{+0.39}_{-0.37}\pm 0.06$ & $2.56 ^{+0.42}_{-0.39}\pm 0.06\pm 0.19$\\
  $5.0<q^2<\makebox[\maxwidth][r]{8.0}$   & $83^{+12}_{-11}$     & $2.98^{+0.41}_{-0.39}\pm 0.07$ & $3.21 ^{+0.44}_{-0.42}\pm 0.08\pm 0.24$\\
  $11.0<q^2<\makebox[\maxwidth][r]{12.5}$ & $70^{+10}_{-10}$      & $4.37^{+0.64}_{-0.61}\pm 0.14$ & $4.71 ^{+0.69}_{-0.65}\pm 0.15\pm 0.36$\\
  $15.0<q^2<\makebox[\maxwidth][r]{17.0}$ & $83^{+10}_{-10}$      & $4.20^{+0.53}_{-0.50}\pm 0.11$ & $4.52 ^{+0.57}_{-0.54}\pm 0.12\pm 0.34$\\
  $17.0<q^2<\makebox[\maxwidth][r]{19.0}$ & $54^{+8\hphantom{0}}_{-7\hphantom{0}}$       & $3.68^{+0.53}_{-0.50}\pm 0.13$ & $3.96 ^{+0.57}_{-0.54}\pm 0.14\pm 0.30$\\\hline
  $1.0<q^2<\makebox[\maxwidth][r]{6.0}$   & $101^{+13}_{-12}$    & $2.40^{+0.30}_{-0.29}\pm 0.07$ & $2.58 ^{+0.33}_{-0.31}\pm 0.08\pm 0.19$\\
  $15.0<q^2<\makebox[\maxwidth][r]{19.0}$ & $136^{+13}_{-13}$    & $3.75^{+0.37}_{-0.35}\pm 0.12$ & $4.04 ^{+0.39}_{-0.38}\pm 0.13\pm 0.30$\\
\hline\end{tabular}
\end{table}
\egroup

\subsection{Systematic uncertainties}
For the branching fraction ratio ${\cal B}(\decay{\Bs}{\phi\mumu})/{\cal B}(\decay{\Bs}{\jpsi\phi})$, systematic uncertainties are mostly
due to uncertainties on 
the efficiency ratio $\epsilon_{\jpsi\phi}/\epsilon_{\phi\mu\mu}$, 
which is taken from simulation. 
To evaluate the size of uncertainties affecting the efficiency ratio, 
it is recalculated after applying the corresponding systematic variation to the simulated samples. 
The observed deviation is taken as systematic uncertainty. 
The procedure to correct the tracking efficiency in simulation introduces a systematic uncertainty on the efficiency ratio of less than $0.6\%$. 
The correction to particle identification performance in simulation has a systematic uncertainty of $0.5\%$. 
The relative efficiency is further affected by the data-driven corrections 
to the simulation in the distribution of the variables 
$\pt(\Bs)$ and $\chi^2_{\rm Vtx}(\Bs)$, as well as the track multiplicity,
which have a combined systematic effect of~$1.0\%$. 
The non-uniform angular acceptance detailed in Sec.~\ref{sec:angularanalysis} introduces a dependence of the signal efficiency on the underlying physics model. Its
effect on the branching fraction measurement is evaluated by varying the Wilson coefficient \C9 used in the generation of simulated signal events. By allowing a New Physics contribution of $-1.5$, which is motivated by the global fit results in Ref.~\cite{PhysRevD.88.074002}, the resulting systematic uncertainty is found to be less than $1.6\%$. 
The selection requirements introduce a decay-time dependence of the efficiencies which can, 
due to the sizeable lifetime difference in the \Bs\ system \cite{LHCb-PAPER-2014-059}, affect the measured branching fraction \cite{DeBruyn:2012wj}. 
The systematic uncertainty is determined with simulated $\decay{\Bs}{\phi\mumu}$ signal events,
generated using time-dependent decay amplitudes as described in Ref.~\cite{Bobeth:2008ij}.
When varying the Wilson coefficients, the size of the effect is found to be at most $1.6\%$, which is taken as the systematic uncertainty. 
The statistical uncertainty due to the limited size of the simulated signal samples leads to a systematic uncertainty of $1.9\%$. 

The systematic uncertainties due to the parametrisation of the mass shapes are evaluated using pseudoexperiments. 
For the signal mass model, events are generated using a double Gaussian mass shape, 
and then fitted using both the double Gaussian as well as the nominal signal mass shape, taking the observed deviation as the systematic uncertainty. 
For the parametrisation of the combinatorial background, the nominal exponential function is compared with a linear mass model. 
The systematic uncertainties due to the modelling of the signal and background mass shape are $2.1\%$ and $1.6\%$, respectively. 
Peaking backgrounds are neglected in the fit for determination of the signal yields.
The main sources of systematic uncertainty are caused by contributions from the decays $\Lb\to p\Km\mumu$ and $\Bd\to\Kstarz\mumu$,
resulting in systematic uncertainties of $0.2-2.2\%$, depending on the $q^2$ bin.
Finally, the uncertainty on the branching fraction of the decay $\jpsi\to\mumu$ amounts to a systematic uncertainty of $0.6\%$. 
The complete list of systematic uncertainties is given in Table~\ref{tab:syst_BRratio}.

For the total branching fraction of the signal decay, 
the uncertainty on the branching fraction of the normalisation channel 
is the dominant systematic uncertainty, at the level of $7.5\%$. 
The uncertainty on the correction factor $f_{\rm veto}$ to account for signal events that are rejected by the charmonium vetoes is 
estimated by varying the Wilson coefficients and form-factor parameters, leading to a systematic uncertainty of $2.9\%$. 

\bgroup 
\def\arraystretch{1.25}
\begin{table}
\small
\caption{\label{tab:syst_BRratio} Systematic uncertainties $[10^{-5}\gev^{-2}c^{4}]$ on the branching fraction ratio ${\rm d}\BR(\Bs\to\phi\mumu)/\BR(\Bs\to\jpsi\phi){\rm
d}\qsq$ per bin of \qsq $[\gevgevcccc]$.}
\begin{center}
\begin{tabular}{lcccccccc} \hline
Source & $[0.1,2]$      &$[2,5]$        &$[5,8]$        &$[11,12.5]$    &$[15,17]$      &$[17,19]$      &$[1,6]$
    &$[15,19]$   \\ \hline 
Simulation corr.        &$0.01$   &$0.01$   &$0.01$   &$0.01$   &$0.05$   &$0.04$   &$0.00$   &$0.04$      \\
Angular model           &$0.04$   &$0.00$   &$0.01$   &$0.00$   &$0.01$   &$0.06$   &$0.00$   &$0.01$      \\
Efficiency ratio        &$0.06$   &$0.03$   &$0.03$   &$0.06$   &$0.06$   &$0.07$   &$0.02$   &$0.04$      \\
Signal mass model       &$0.02$   &$0.01$   &$0.03$   &$0.03$   &$0.03$   &$0.00$   &$0.05$   &$0.05$      \\
Bkg. mass model         &$0.02$   &$0.02$   &$0.02$   &$0.02$   &$0.03$   &$0.05$   &$0.01$   &$0.06$      \\ 
Time acceptance         &$0.09$   &$0.04$   &$0.05$   &$0.07$   &$0.07$   &$0.06$   &$0.04$   &$0.06$      \\ 
$\BR(\jpsi \to \mup \mun)$ &$0.03$   &$0.01$   &$0.02$   &$0.02$   &$0.02$   &$0.02$   &$0.01$   &$0.02$      \\
Peaking bkg.            &$0.03$   &$0.02$   &$0.02$   &$0.10$   &$0.02$   &$0.01$   &$0.02$   &$0.01$      \\ \hline
Quadratic sum           &$0.13$   &$0.06$   &$0.07$   &$0.14$   &$0.11$   &$0.13$   &$0.07$   &$0.12$      \\
\hline\end{tabular}
\end{center}
\end{table}
\egroup

\section{Angular analysis}
\label{sec:angularanalysis}
For the determination of the four \CP averages $F_{\rm L}$, $S_{3,4,7}$ and the four \CP asymmetries $A_{5,6,8,9}$ an unbinned maximum likelihood fit
to the three-dimensional angular distribution and the $\Kp\Km\mup\mun$ invariant mass distribution is performed in each $q^2$ bin.
The models described in Sec.~\ref{sec:branchingfraction} are used to parametrise
the mass line shapes for signal and background. 
The angular distribution of the signal component is given by Eq.~\ref{eq:untagged}. 
The angular background distribution is described by the product of second-order Chebyshev polynomials in the three decay angles. 

The non-uniform efficiency due to the reconstruction, triggering and selection of signal candidates distorts the angular distributions of the final-state particles, as well as the $q^2$ distribution. 
This acceptance effect is parametrised using Legendre polynomials, according to
\begin{align}
  \epsilon(\cos\thetal,\cos\thetak,\Phi,q^2)&=\sum_{klmn}c_{klmn}P_k(\cos\thetal)P_l(\cos\thetak)P_m(\Phi)P_n(q^2),\label{eq:acceptance}
\end{align}
where $P_i(x)$ denote Legendre polynomials of order $i$ and $c_{klmn}$ the coefficients that are determined by performing a moments analysis
using a large sample of simulated $\decay{\Bs}{\phi\mumu}$ signal events generated according to a phase-space model.  
The maximum order of the polynomials that is included is four for $\cos\thetal$, two for $\cos\thetak$, six for the angle $\Phi$ and five for $q^2$.
In addition, the acceptance is assumed to be symmetric in the decay angles.
This choice corresponds to the lowest orders of polynomials that describe the acceptance effect. 
The acceptance description is cross-checked using the control mode $\decay{\Bs}{\jpsi\phi}$.
An angular analysis of the control mode is performed and the angular observables
are found to be in good agreement with the previous measurement~\cite{LHCb-PAPER-2014-059}. 

\begin{figure}[htbp!]
    \captionsetup[subfigure]{labelformat=empty}
  \centering
\includegraphics[width=0.49\textwidth]{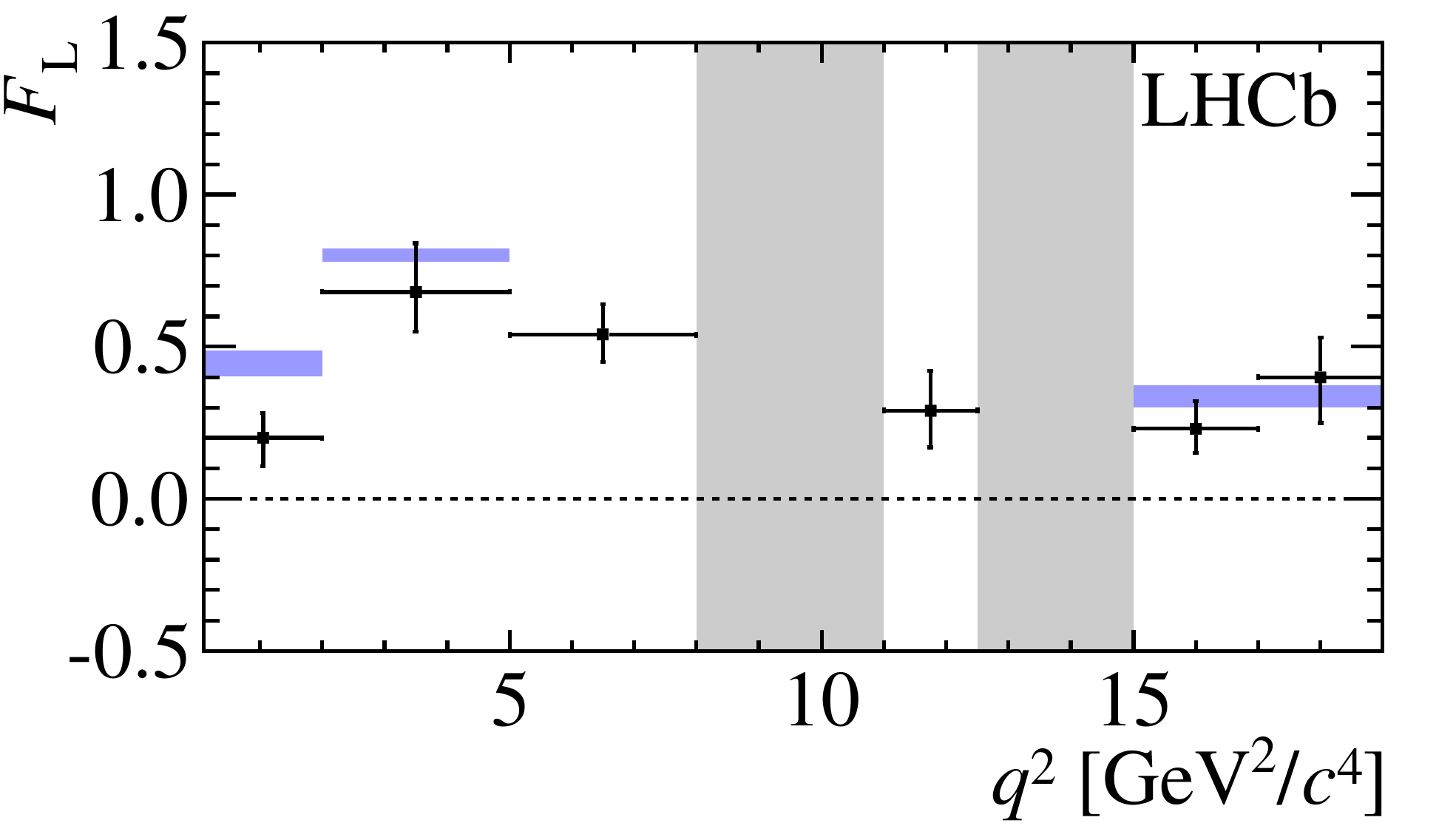}
\hfill
\includegraphics[width=0.49\textwidth]{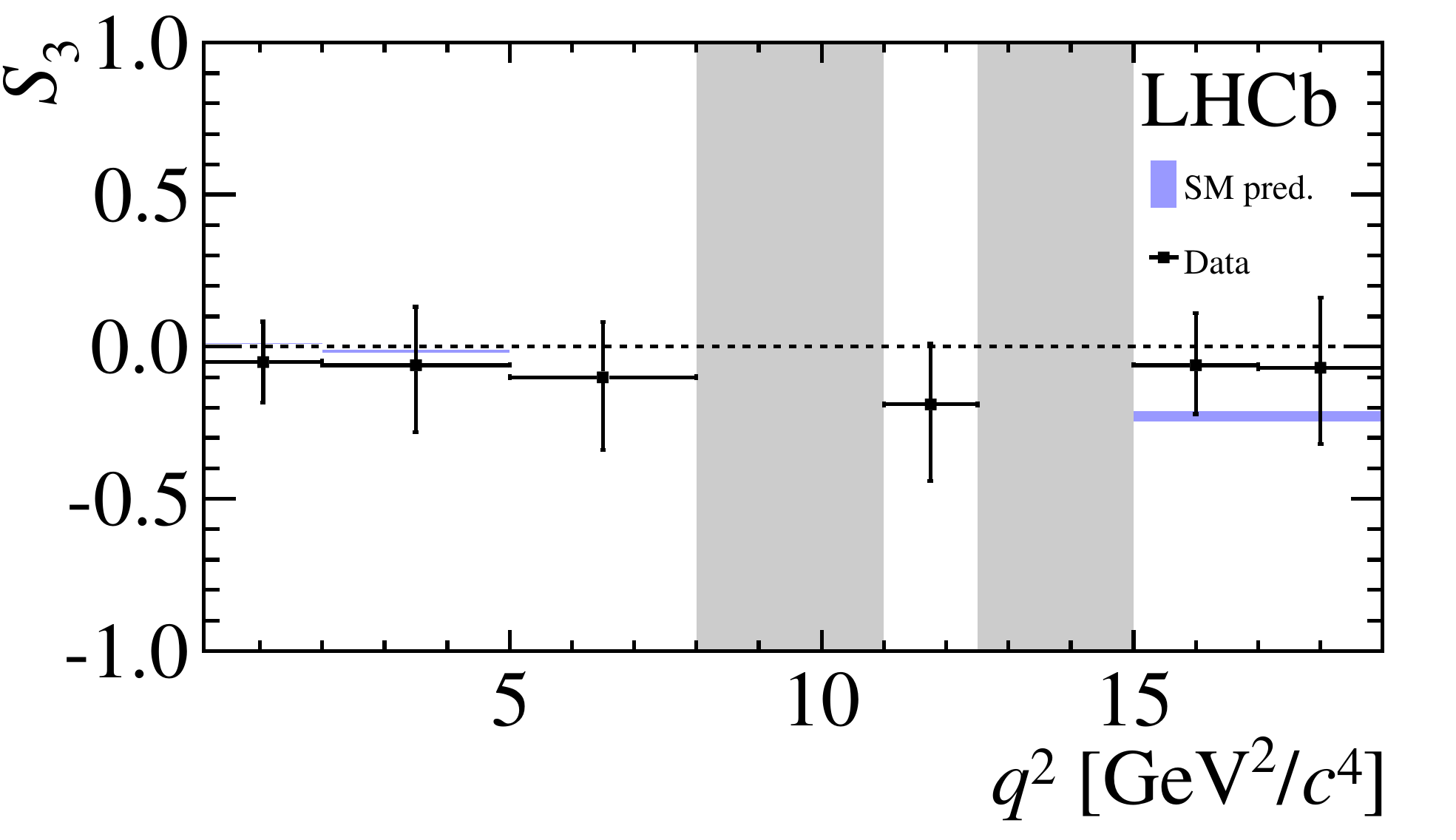}\\[0.2cm]
\includegraphics[width=0.49\textwidth]{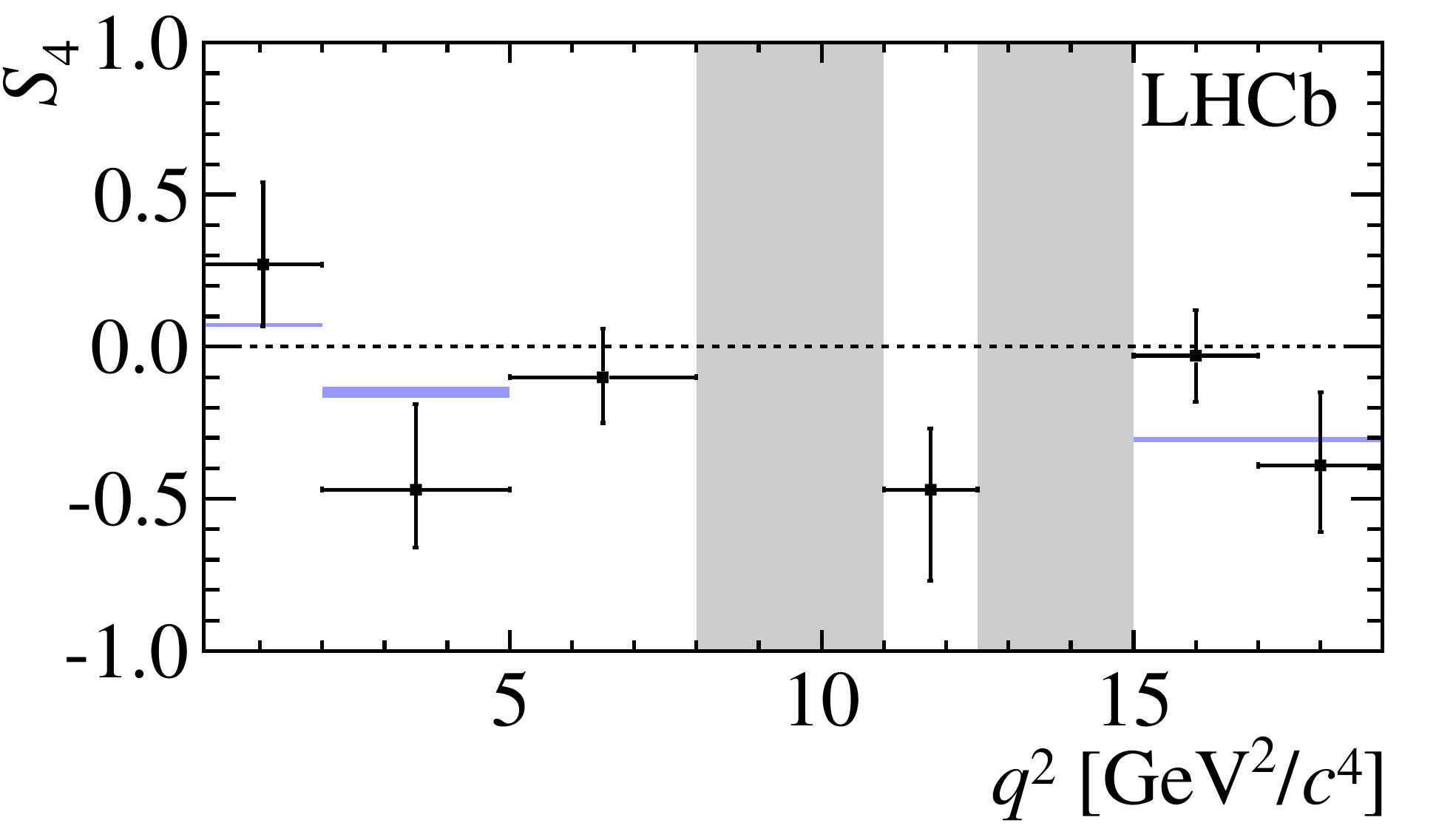}
\hfill
\includegraphics[width=0.49\textwidth]{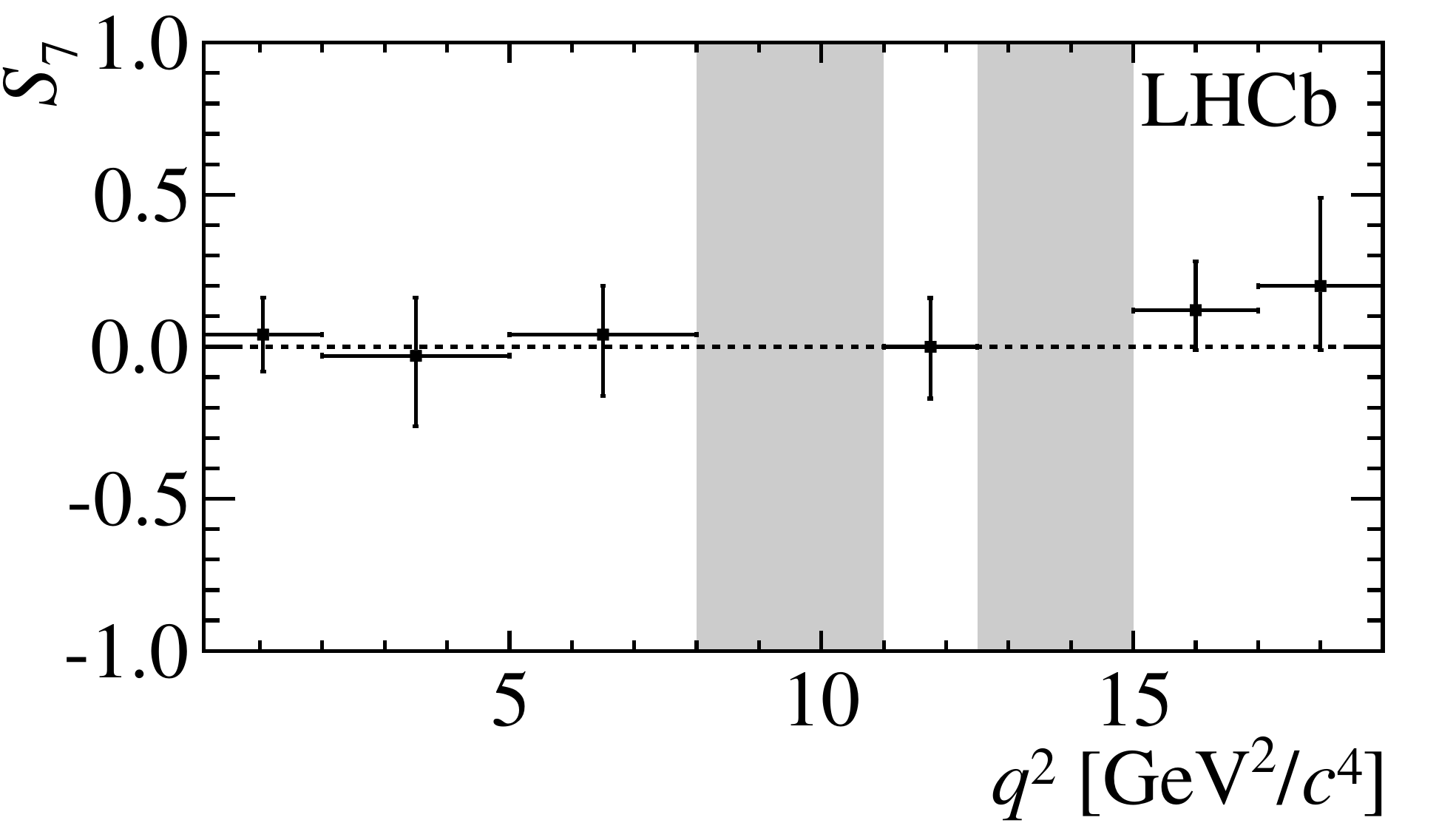}\\[0.2cm]
\includegraphics[width=0.49\textwidth]{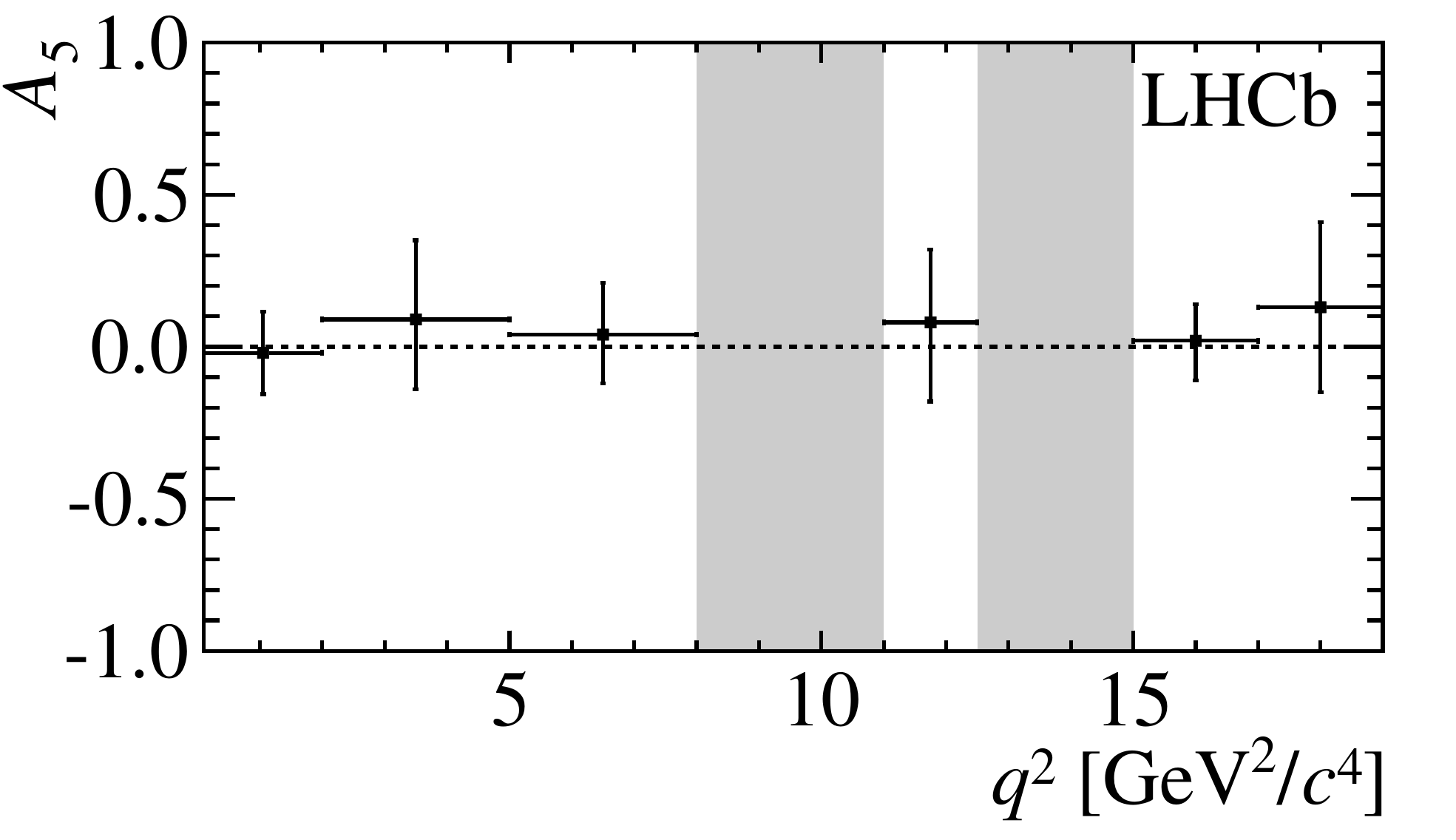}
\hfill
\includegraphics[width=0.49\textwidth]{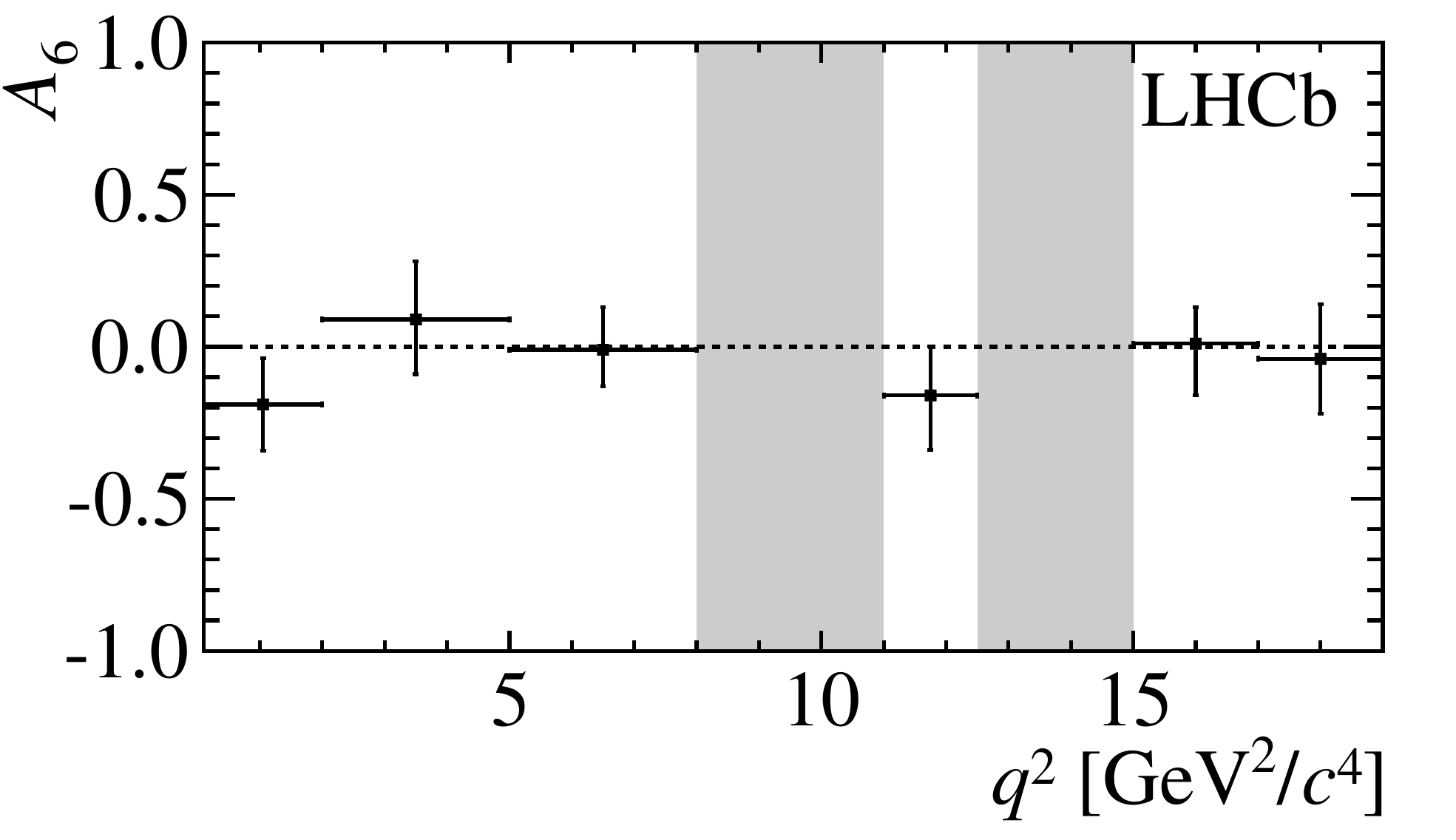}\\[0.2cm]
\includegraphics[width=0.49\textwidth]{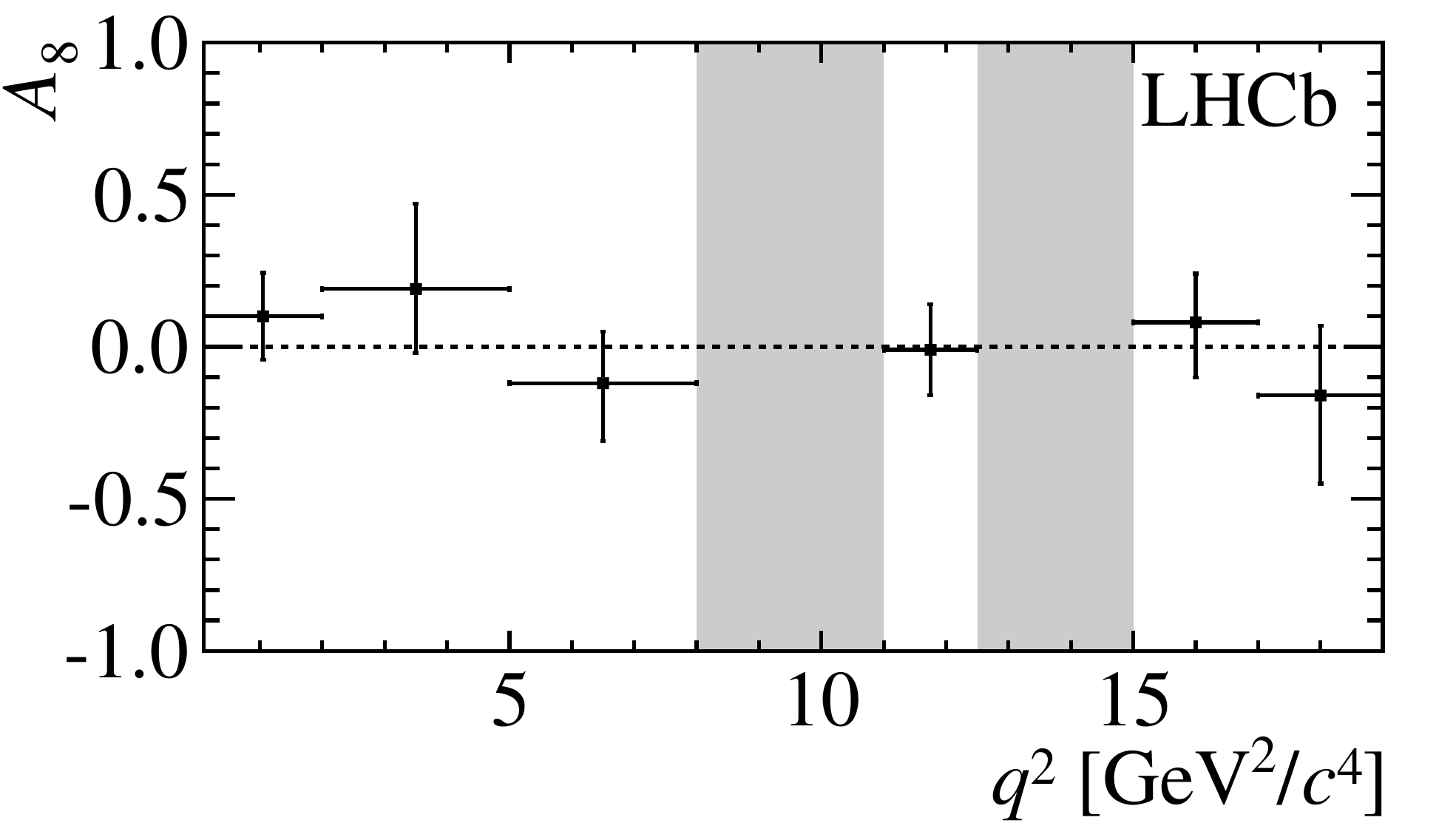}
\hfill
\includegraphics[width=0.49\textwidth]{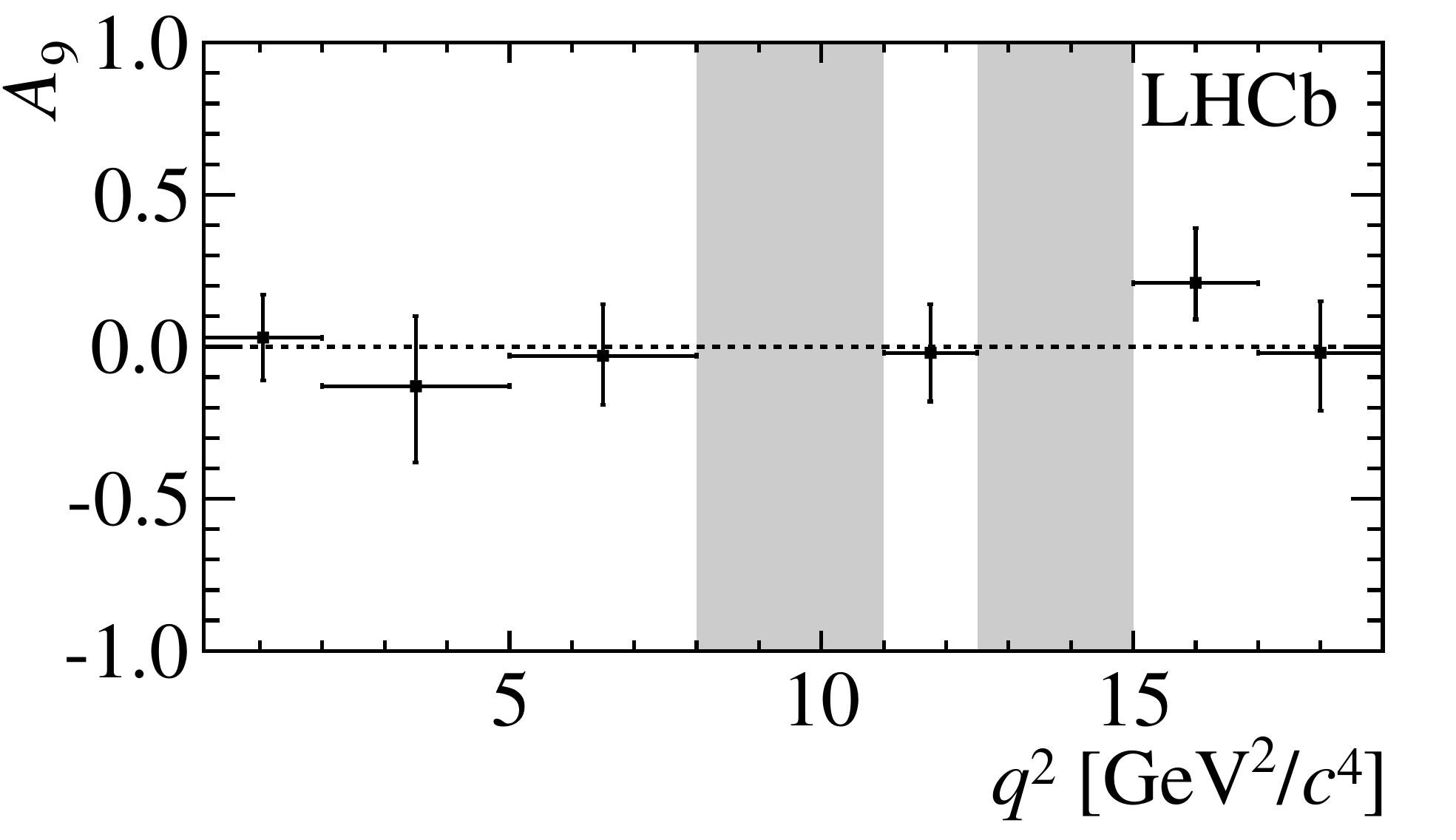}
\caption{
    \CP-averaged angular observables $F_{\rm L}$ and $S_{3,4,7}$ and \CP asymmetries $A_{5,6,8,9}$ shown by black dots,
    overlaid with SM predictions~\cite{Altmannshofer:2014rta,Straub:2015ica}, where available, indicated as blue shaded boxes. The vetoes excluding the charmonium resonances are indicated by
grey areas.\label{fig:angularresults}}
\end{figure}

Appendix~\ref{app:ang-project} gives the one-dimensional angular distributions of the signal decay in each \qsq\ bin, overlaid with the projections of the likelihood fit.
For the \qsq\ bins with the lowest number of signal candidates, pseudoexperiments show the likelihood estimator 
to be biased for certain observables due to boundary effects which arise from the requirement of Eq.~\ref{eq:untagged} being positive for all values of the decay angles. 
Therefore, the Feldman-Cousins method~\cite{1998PhRvD..57.3873F} is used to determine confidence regions for each observable,
which guarantees correct coverage for low signal yields. 
The remaining signal observables are treated as nuisance parameters following the plugin method~\cite{pluginmethod}. 
The Feldman-Cousins scans for the angular observables in bins of $q^2$ are given in Appendix~\ref{app:fc}. 
In some $\qsq$ bins, the shape of the obtained confidence level is dominated by effects arising from the boundary condition. 
Table~\ref{tab:angularresults} gives the minima of the Feldman-Cousins scans and the $68\%$ confidence intervals. 
The linear correlations between the angular observables in the different $q^2$ bins are given in Appendix~\ref{app:correlations}.
The angular observables are shown in Fig.~\ref{fig:angularresults}, overlaid with SM predictions from Refs.~\cite{Altmannshofer:2014rta,Straub:2015ica}. No
predictions are given for $S_7$ and $A_{5,6,8,9}$; they are expected to be close to zero in the SM.

\bgroup 
\def\arraystretch{1.5}
\begin{table}
  \small
\caption{
  (Top) \CP-averaged angular observables $F_{\rm L}$ and $S_{3,4,7}$ and (bottom) \CP asymmetries $A_{5,6,8,9}$ obtained from the unbinned maximum likelihood fit, where
  the first uncertainty is statistical and the second is systematic.\label{tab:angularresults}}
  \centering
\scalebox{0.9}{
  \newlength{\maxscaled}
  \setlength{\maxscaled}{\widthof{19.0}}
  \begin{tabular}{rrrrr}\hline
  $q^2$ bin $[\gevgevcccc]$ & \multicolumn{1}{c}{$F_{\rm L}$} & \multicolumn{1}{c}{$S_3$} & \multicolumn{1}{c}{$S_4$} & \multicolumn{1}{c}{$S_7$}\\ \hline
$0.1 < \qsq < \makebox[\maxscaled][r]{2.0}$  & $\hphantom{-}0.20 ^{+ 0.08 }_{ -0.09 } \pm 0.02 $  & $ -0.05 ^{+ 0.13 }_{ -0.13 } \pm 0.01 $  & $ 0.27 ^{+ 0.28 }_{ -0.18 } \pm 0.01 $  & $ 0.04 ^{+ 0.12 }_{ -0.12 } \pm 0.00 $ \\ 
$2.0 < \qsq < \makebox[\maxscaled][r]{5.0}$ & $ 0.68 ^{+ 0.16 }_{ -0.13 } \pm 0.03 $  & $ -0.06 ^{+ 0.19 }_{ -0.23 } \pm 0.01 $  & $ -0.47 ^{+ 0.30 }_{ -0.44 } \pm 0.01 $  & $ -0.03 ^{+ 0.18 }_{ -0.23 } \pm 0.01 $ \\
$5.0 < \qsq < \makebox[\maxscaled][r]{8.0}$  & $ 0.54 ^{+ 0.10 }_{ -0.09 } \pm 0.02 $  & $ -0.10 ^{+ 0.20 }_{ -0.29 } \pm 0.01 $  & $ -0.10 ^{+ 0.15 }_{ -0.18 } \pm 0.01 $  & $ 0.04 ^{+ 0.16 }_{ -0.20 } \pm 0.01 $ \\
$11.0 < \qsq < \makebox[\maxscaled][r]{12.5}$  & $ 0.29 ^{+ 0.11 }_{ -0.11 } \pm 0.04 $  & $ -0.19 ^{+ 0.20 }_{ -0.23 } \pm 0.01 $  & $ -0.47 ^{+ 0.21 }_{ -0.29 } \pm 0.01 $  & $ 0.00 ^{+ 0.15 }_{ -0.17 } \pm 0.01 $ \\	
$15.0 < \qsq < \makebox[\maxscaled][r]{17.0}$  & $ 0.23 ^{+ 0.09 }_{ -0.08 } \pm 0.02 $  & $ -0.06 ^{+ 0.16 }_{ -0.19 } \pm 0.01 $  & $ -0.03 ^{+ 0.15 }_{ -0.15 } \pm 0.01 $  & $ 0.12 ^{+ 0.16 }_{ -0.13 } \pm 0.01 $ \\
$17.0 < \qsq < \makebox[\maxscaled][r]{19.0}$ & $ 0.40 ^{+ 0.13 }_{ -0.15 } \pm 0.02 $  & $ -0.07 ^{+ 0.23 }_{ -0.27 } \pm 0.02 $  & $ -0.39 ^{+ 0.25 }_{ -0.34 } \pm 0.02 $  & $ 0.20 ^{+ 0.29 }_{ -0.22 } \pm 0.01 $ \\ \hline
$1.0 < \qsq < \makebox[\maxscaled][r]{6.0}$  & $ 0.63 ^{+ 0.09 }_{ -0.09 } \pm 0.03 $  & $ -0.02 ^{+ 0.12 }_{ -0.13 } \pm 0.01 $  & $ -0.19 ^{+ 0.14 }_{ -0.13 } \pm 0.01 $  & $ -0.03 ^{+ 0.14 }_{ -0.14 } \pm 0.00 $ \\
$15.0 < \qsq < \makebox[\maxscaled][r]{19.0}$  & $ 0.29 ^{+ 0.07 }_{ -0.06 } \pm 0.02 $  & $ -0.09 ^{+ 0.11 }_{ -0.12 } \pm 0.01 $  & $ -0.14 ^{+ 0.11 }_{ -0.11 } \pm 0.01 $  & $ 0.13 ^{+ 0.11 }_{ -0.11 } \pm 0.01 $ \\
\hline\end{tabular}}\\[1em]
\scalebox{0.9}{
  \setlength{\maxscaled}{\widthof{19.0}}
\begin{tabular}{rrrrr}\hline
  $q^2$ bin $[\gevgevcccc]$ & \multicolumn{1}{c}{$A_5$} & \multicolumn{1}{c}{$A_6$} & \multicolumn{1}{c}{$A_8$} & \multicolumn{1}{c}{$A_9$}\\ \hline
$0.1 < \qsq < \makebox[\maxscaled][r]{2.0}$ & $ -0.02 ^{+ 0.13 }_{ -0.13 } \pm 0.00 $  & $ -0.19 ^{+ 0.15 }_{ -0.15 } \pm 0.01 $  & $ 0.10 ^{+ 0.14 }_{ -0.14 } \pm 0.00 $  & $ 0.03 ^{+ 0.14 }_{ -0.14 } \pm 0.01 $ \\
$2.0 < \qsq < \makebox[\maxscaled][r]{5.0}$ & $ 0.09 ^{+ 0.28 }_{ -0.22 } \pm 0.01 $  & $ 0.09 ^{+ 0.20 }_{ -0.19 } \pm 0.02 $  & $ 0.19 ^{+ 0.26 }_{ -0.21 } \pm 0.01 $  & $ -0.13 ^{+ 0.24 }_{ -0.30 } \pm 0.01 $ \\
$5.0 < \qsq < \makebox[\maxscaled][r]{8.0}$  & $ 0.04 ^{+ 0.17 }_{ -0.17 } \pm 0.01 $  & $ -0.01 ^{+ 0.14 }_{ -0.12 } \pm 0.01 $  & $ -0.12 ^{+ 0.17 }_{ -0.19 } \pm 0.01 $  & $ -0.03 ^{+ 0.17 }_{ -0.16 } \pm 0.01 $ \\   
$11.0 < \qsq < \makebox[\maxscaled][r]{12.5}$ & $ 0.08 ^{+ 0.21 }_{ -0.21 } \pm 0.01 $  & $ -0.16 ^{+ 0.16 }_{ -0.18 } \pm 0.01 $  & $ -0.01 ^{+ 0.15 }_{ -0.15 } \pm 0.01 $  & $ -0.02 ^{+ 0.16 }_{ -0.15 } \pm 0.01 $ \\
$15.0 < \qsq < \makebox[\maxscaled][r]{17.0}$ & $ 0.02 ^{+ 0.13 }_{ -0.14 } \pm 0.01 $  & $ 0.01 ^{+ 0.12 }_{ -0.17 } \pm 0.01 $  & $ 0.08 ^{+ 0.16 }_{ -0.18 } \pm 0.01 $  & $ 0.21 ^{+ 0.18 }_{ -0.12 } \pm 0.01 $ \\
$17.0 < \qsq < \makebox[\maxscaled][r]{19.0}$ & $ 0.13 ^{+ 0.29 }_{ -0.27 } \pm 0.01 $  & $ -0.04 ^{+ 0.18 }_{ -0.19 } \pm 0.01 $  & $ -0.16 ^{+ 0.24 }_{ -0.29 } \pm 0.01 $  & $ -0.02 ^{+ 0.19 }_{ -0.19 } \pm 0.01 $ \\ \hline
$1.0 < \qsq < \makebox[\maxscaled][r]{6.0}$ & $ 0.20 ^{+ 0.13 }_{ -0.13 } \pm 0.00 $  & $ 0.08 ^{+ 0.12 }_{ -0.11 } \pm 0.01 $  & $ -0.00 ^{+ 0.15 }_{ -0.17 } \pm 0.00 $  & $ -0.01 ^{+ 0.13 }_{ -0.13 } \pm 0.01 $ \\
$15.0 < \qsq < \makebox[\maxscaled][r]{19.0}$  & $ 0.11 ^{+ 0.10 }_{ -0.10 } \pm 0.00 $  & $ 0.00 ^{+ 0.10 }_{ -0.11 } \pm 0.01 $  & $ 0.03 ^{+ 0.12 }_{ -0.12 } \pm 0.00 $  & $ 0.12 ^{+ 0.11 }_{ -0.09 } \pm 0.00 $ \\
  \hline\end{tabular}}
\end{table}
\egroup

\subsection{Systematic uncertainties}
The systematic uncertainties on the angular observables are evaluated using large numbers of pseudoexperiments where simulated events are generated to reflect the measured angular distributions and event yield.
To reduce statistical effects, each sample is fitted twice, once with and once without systematic variations.  

The angular acceptance correction, which is determined from simulation, is a significant source of systematic uncertainties for the angular observables.  
The data-driven corrections of the distributions of $\pt(\Bs)$, $\chi^2_{\rm Vtx}(\Bs)$ and track multiplicity,
as well as particle identification performance and tracking efficiency, amount to a systematic uncertainty of less than $0.01$ in total. 
Furthermore, the kinematic distributions of the final-state particles are cross-checked using the control mode $\decay{\Bs}{\jpsi\phi}$.
Correcting for kinematic differences between data and simulation amounts to a systematic deviation of less than $0.01$. 
The effect of the limited size of the simulated signal samples on the acceptance description is evaluated by varying the Legendre coefficients in Eq.~\ref{eq:acceptance} according to their corresponding covariance matrix.
The resulting systematic uncertainty is smaller than $0.02$ for all observables and $q^2$ bins. 
The four-dimensional acceptance correction is evaluated at the centre of each $q^2$ bin.
To estimate the systematic effect due to this, an alternative acceptance description is used, 
where a separate three-dimensional acceptance is used for each $q^2$ bin. 
The resulting systematic deviation is negligible. 

The systematic effect of neglecting peaking backgrounds is evaluated by performing toy studies, 
where simulated $\decay{\Lb}{p\Km\mumu}$ and $\decay{\Bd}{\Kstarz\mumu}$ background events are added,
according to their expected yields for the specific $q^2$ bin. 
The resulting systematic deviations of the angular observables are smaller than $0.01$ for all observables and $q^2$ bins.
The S-wave pollution for the decay $\decay{\Bs}{\phi\mumu}$ is expected to be similar to that of the $\decay{\Bs}{\jpsi\phi}$ decay, 
at the level of $1.1\%$~\cite{LHCb-PAPER-2012-040}. The effect of the S wave in the \Kp\Km system on the angular observables is determined to be smaller than $0.01$ using toy studies. 
The combinatorial background is described using second-order Chebyshev polynomials determined from the upper mass sideband.
The systematic uncertainty associated with this model choice is estimated by using first-order polynomials as an alternative.
With a systematic effect of up to $0.04$ on the angular observables, depending on $q^2$ bin,
this constitutes the dominant systematic uncertainty for the angular analysis. 
In addition, the effect of fixing the angular background parameters in the nominal fit is evaluated using toy studies.
The systematic deviation is found to be smaller than $0.02$ for all observables and $q^2$ bins. 

The total systematic uncertainty, given by the quadratic sum over all systematic effects,
is found to be small compared to the statistical uncertainties for all angular observables in all \qsq bins.

\section{Conclusions}
\label{sec:conclusions}
Measurements of the differential branching fraction and the first full three-dimensional angular analysis of the decay $\decay{\Bs}{\phi\mumu}$ are presented,
using data collected by the LHCb experiment in $pp$ collisions, corresponding to an integrated luminosity of $3.0\invfb$. 
The results are given in Tables~\ref{tab:diffbr} and~\ref{tab:angularresults} and are the most precise measurements of these quantities to date.
The \CP-averaged angular observables $S_4$ and $S_7$ are determined for the first time for this decay. 
The determination of the \CP asymmetries $A_5$ and $A_8$ constitutes the first measurement of these quantities for any rare $b\to s\ell\ell$ decay, 
providing additional constraints in global fits. 
All angular observables are found to be compatible with SM predictions. 

The $\decay{\Bs}{\phi\mumu}$ branching fraction relative to the normalisation mode $\decay{\Bs}{\jpsi\phi}$ is measured to be 
\begin{align*}
  \frac{{\cal B}(\decay{\Bs}{\phi\mup\mun})}{{\cal B}(\decay{\Bs}{\jpsi\phi})}&= (7.41^{+0.42}_{-0.40}\pm 0.20\pm 0.21)\times 10^{-4},\\
\end{align*}
and the resulting total absolute branching fraction is measured to be 
\begin{align*}
{\cal B}(\decay{\Bs}{\phi\mup\mun}) &= (7.97^{+0.45}_{-0.43} \pm 0.22 \pm 0.23 \pm 0.60)\times 10^{-7}, 
\end{align*}
where the uncertainties are (from left to right) statistical, systematic, and from the extrapolation to the full $q^2$ region. For the total branching fraction, a further
uncertainty originates from the uncertainty on the branching fraction of the normalisation mode.
The measured branching fraction is compatible with the previous measurement~\cite{LHCb-PAPER-2013-017} and lies below SM expectations. 
For the $q^2$ region $1.0<q^2<6.0\gevgevcccc$ the differential branching fraction of $(2.58 ^{+0.33}_{-0.31}\pm 0.08\pm 0.19)\times 10^{-8}\gev^{-2}c^{4}$ is more than $3\,\sigma$ below the SM prediction of $(4.81\pm  0.56)\times 10^{-8}\gev^{-2}c^{4}$~\cite{Altmannshofer:2014rta,Straub:2015ica,Altmannshofer:2015sma}.

\section*{Acknowledgements}

\noindent We express our gratitude to our colleagues in the CERN
accelerator departments for the excellent performance of the LHC. We
thank the technical and administrative staff at the LHCb
institutes. We acknowledge support from CERN and from the national
agencies: CAPES, CNPq, FAPERJ and FINEP (Brazil); NSFC (China);
CNRS/IN2P3 (France); BMBF, DFG, HGF and MPG (Germany); INFN (Italy); 
FOM and NWO (The Netherlands); MNiSW and NCN (Poland); MEN/IFA (Romania); 
MinES and FANO (Russia); MinECo (Spain); SNSF and SER (Switzerland); 
NASU (Ukraine); STFC (United Kingdom); NSF (USA).
The Tier1 computing centres are supported by IN2P3 (France), KIT and BMBF 
(Germany), INFN (Italy), NWO and SURF (The Netherlands), PIC (Spain), GridPP 
(United Kingdom).
We are indebted to the communities behind the multiple open 
source software packages on which we depend. We are also thankful for the 
computing resources and the access to software R\&D tools provided by Yandex LLC (Russia).
Individual groups or members have received support from 
EPLANET, Marie Sk\l{}odowska-Curie Actions and ERC (European Union), 
Conseil g\'{e}n\'{e}ral de Haute-Savoie, Labex ENIGMASS and OCEVU, 
R\'{e}gion Auvergne (France), RFBR (Russia), XuntaGal and GENCAT (Spain), Royal Society and Royal
Commission for the Exhibition of 1851 (United Kingdom).

\clearpage

{\noindent\bf\Large Appendices}

\appendix

\section{Invariant mass distributions}
\label{app:mass}

\begin{figure}[h]
    \captionsetup[subfigure]{labelformat=empty}
  \centering
\includegraphics[width=0.48\textwidth]{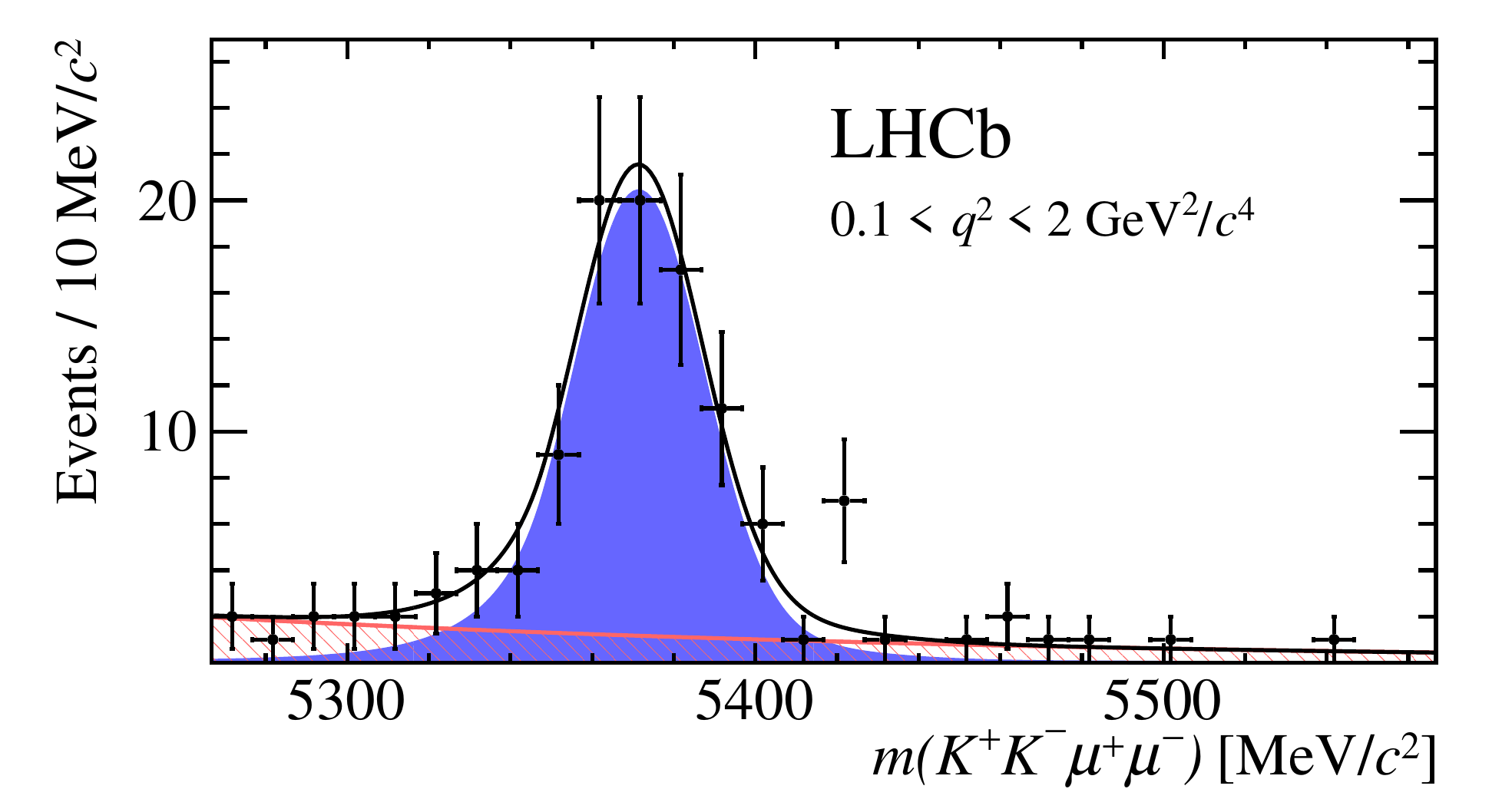}
\hfill\includegraphics[width=0.48\textwidth]{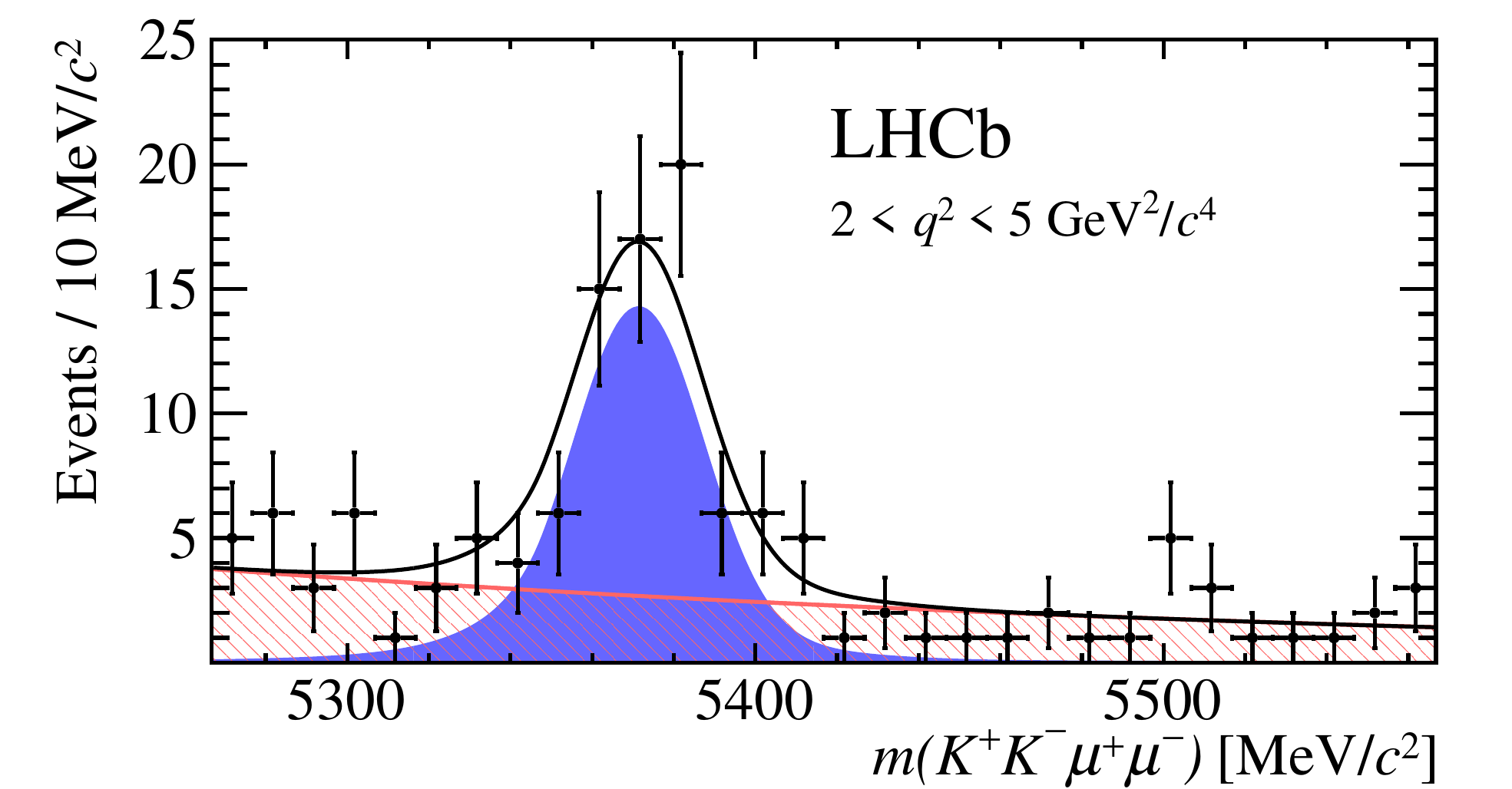}\\[0.2cm]
\includegraphics[width=0.48\textwidth]{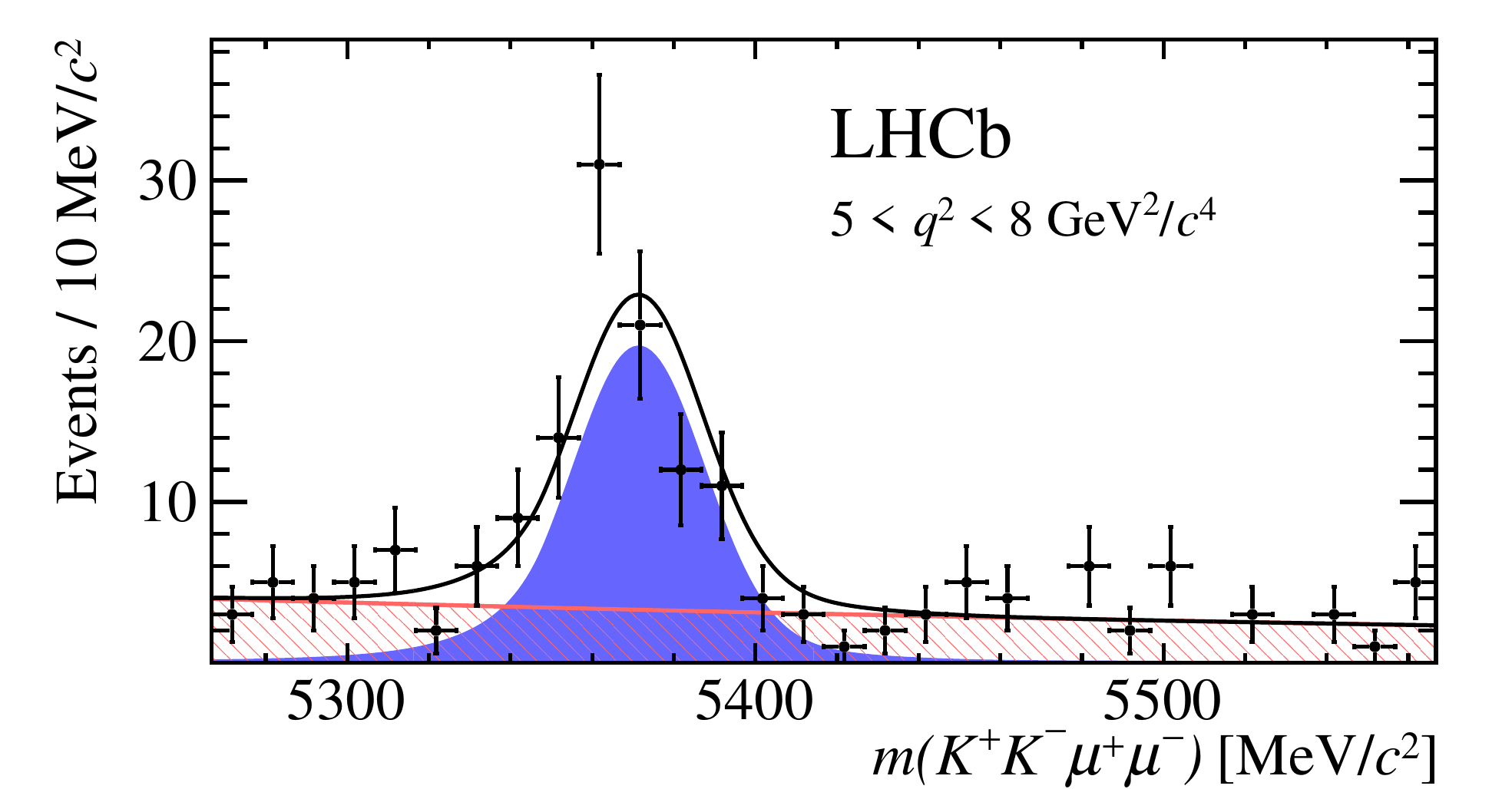}
\hfill\includegraphics[width=0.48\textwidth]{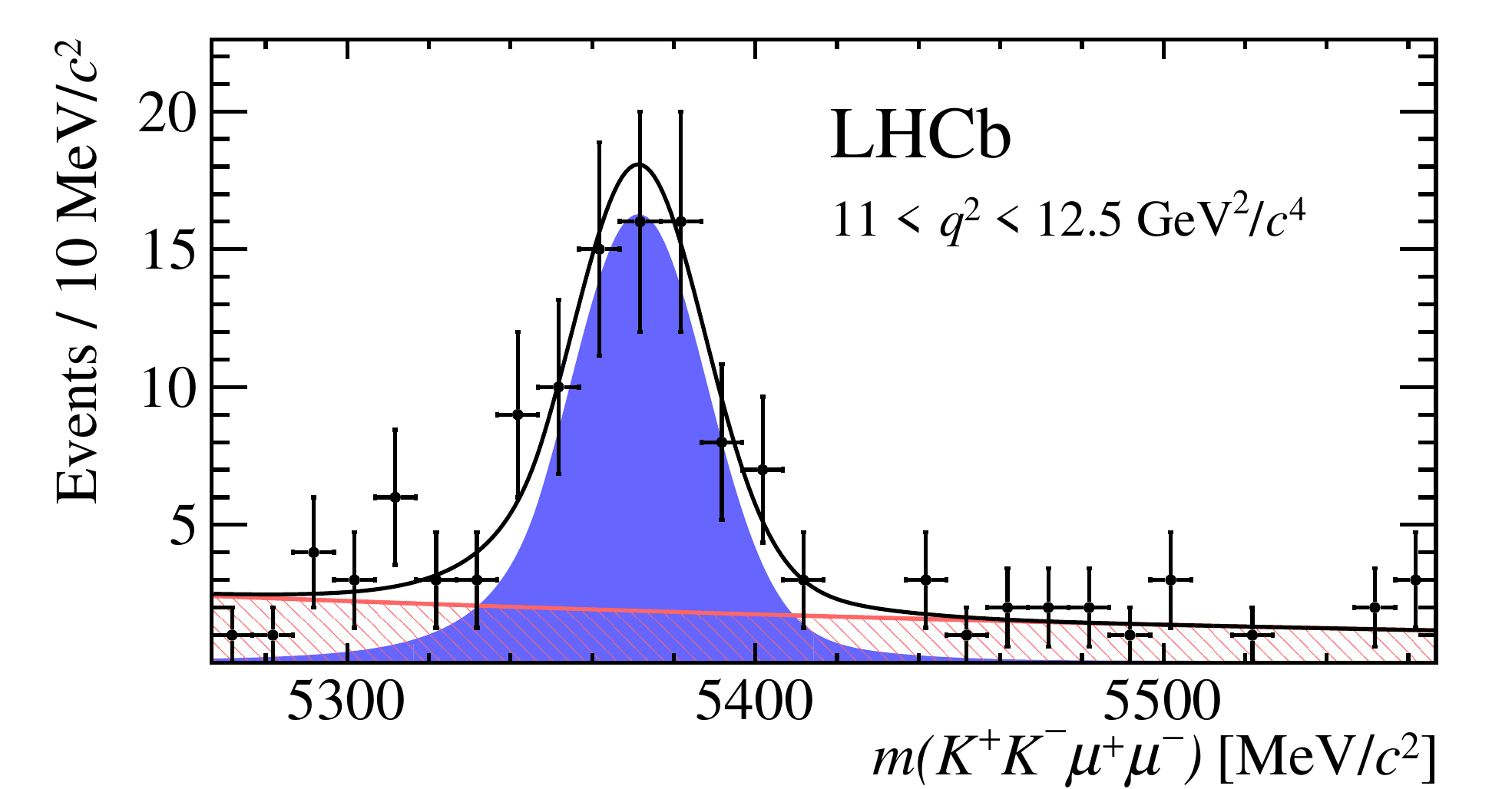}\\[0.2cm]
\includegraphics[width=0.48\textwidth]{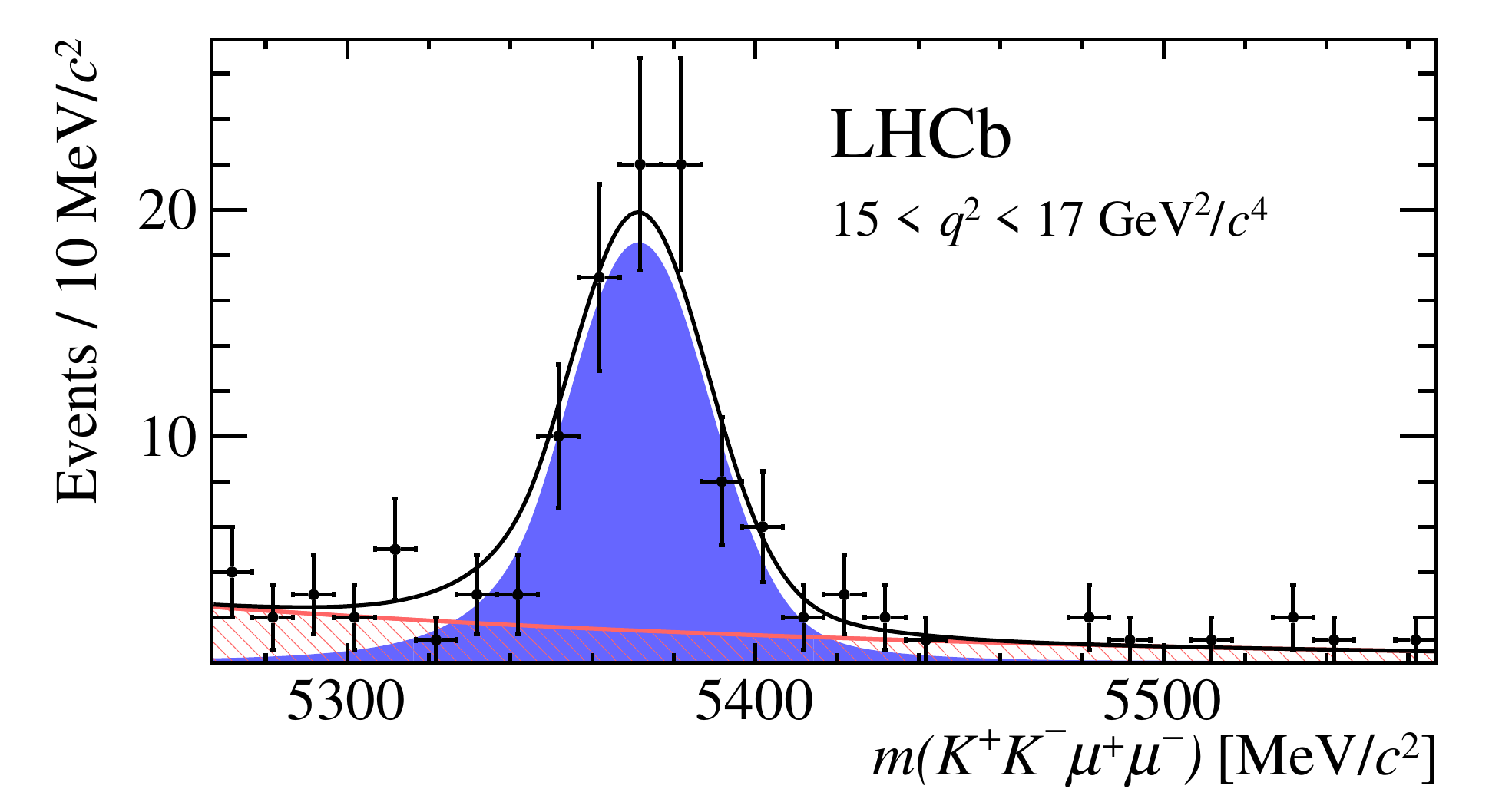}
\hfill\includegraphics[width=0.48\textwidth]{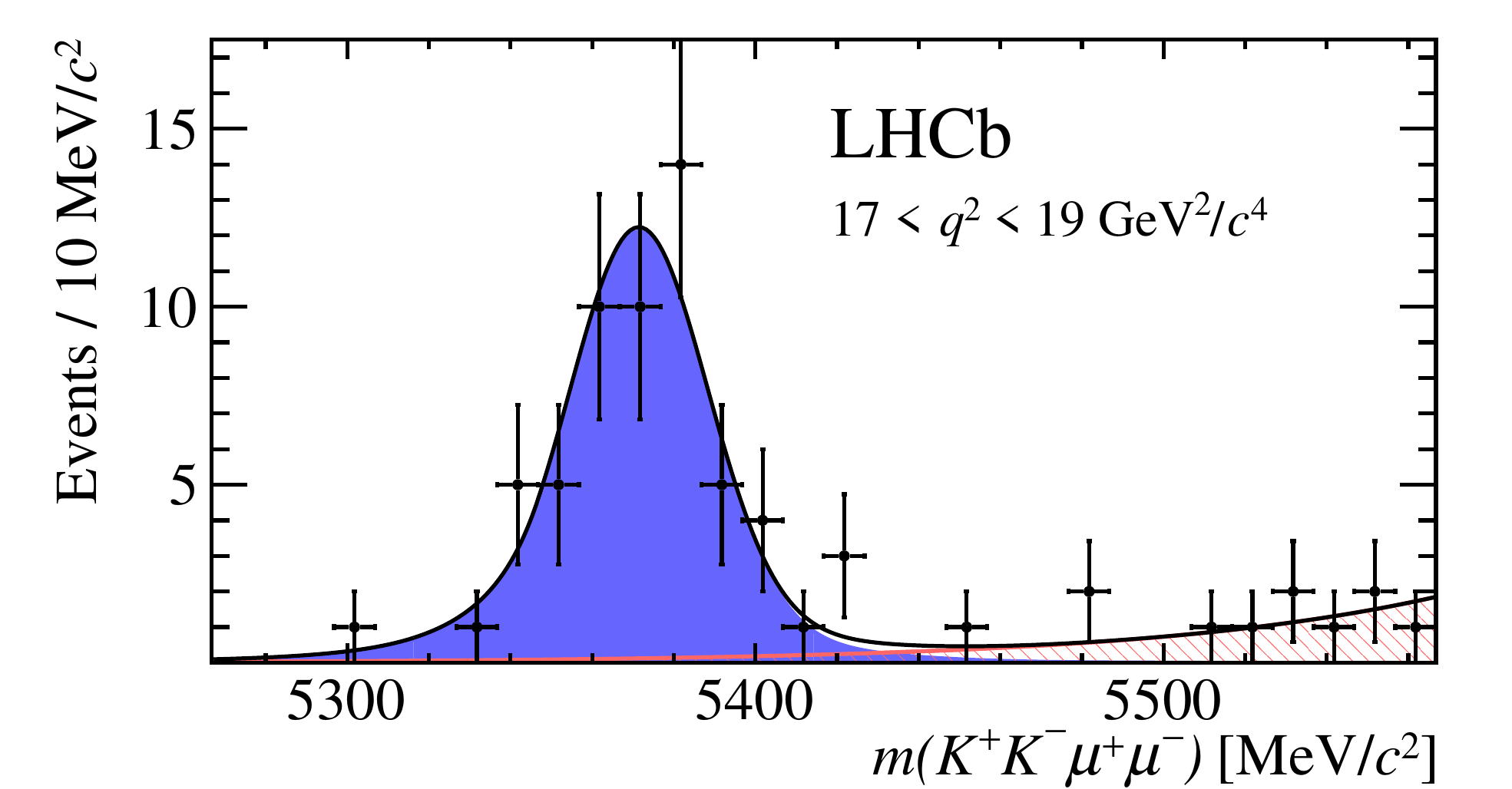}\\[0.2cm]
\includegraphics[width=0.48\textwidth]{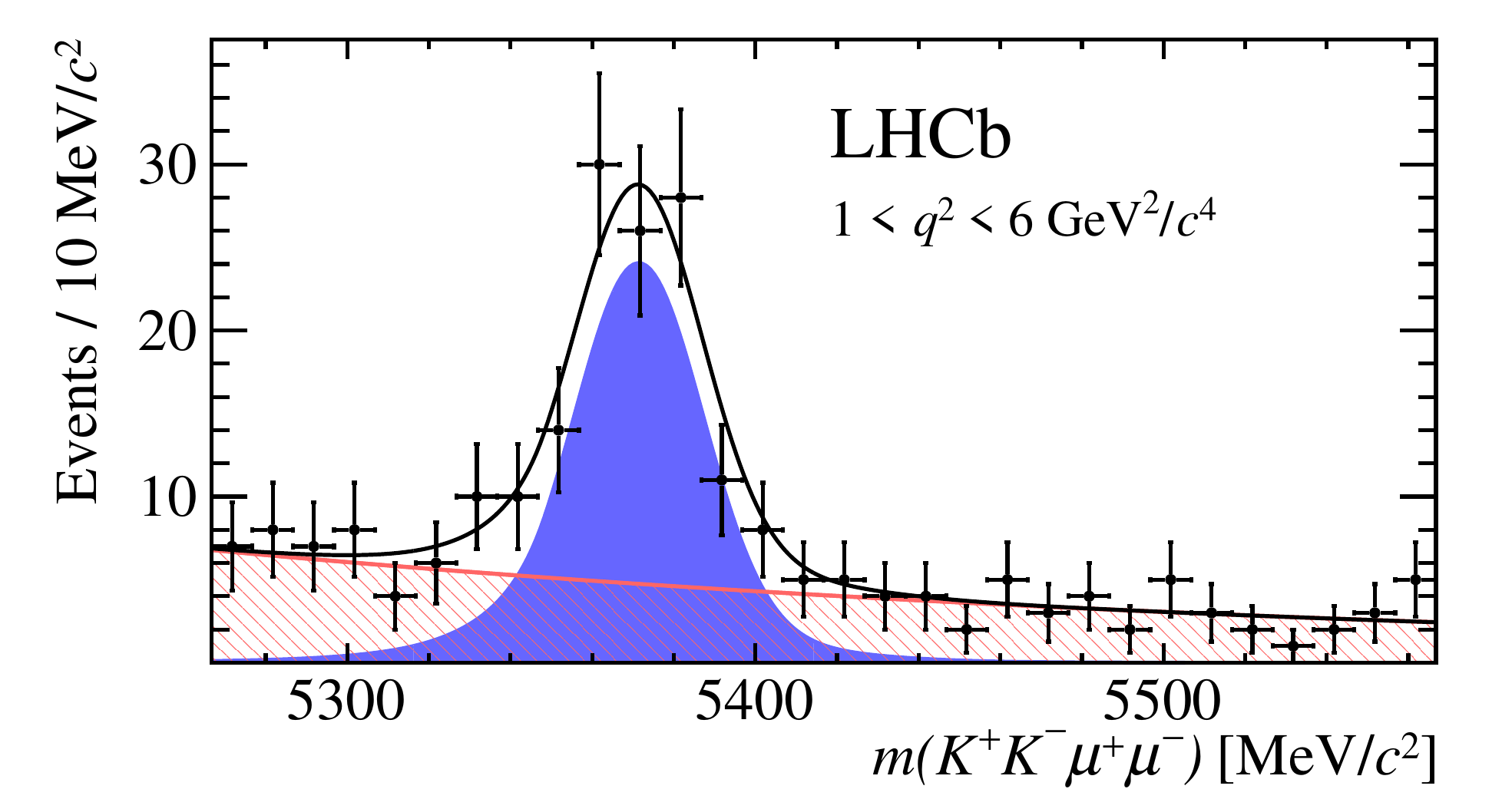}
\hfill\includegraphics[width=0.48\textwidth]{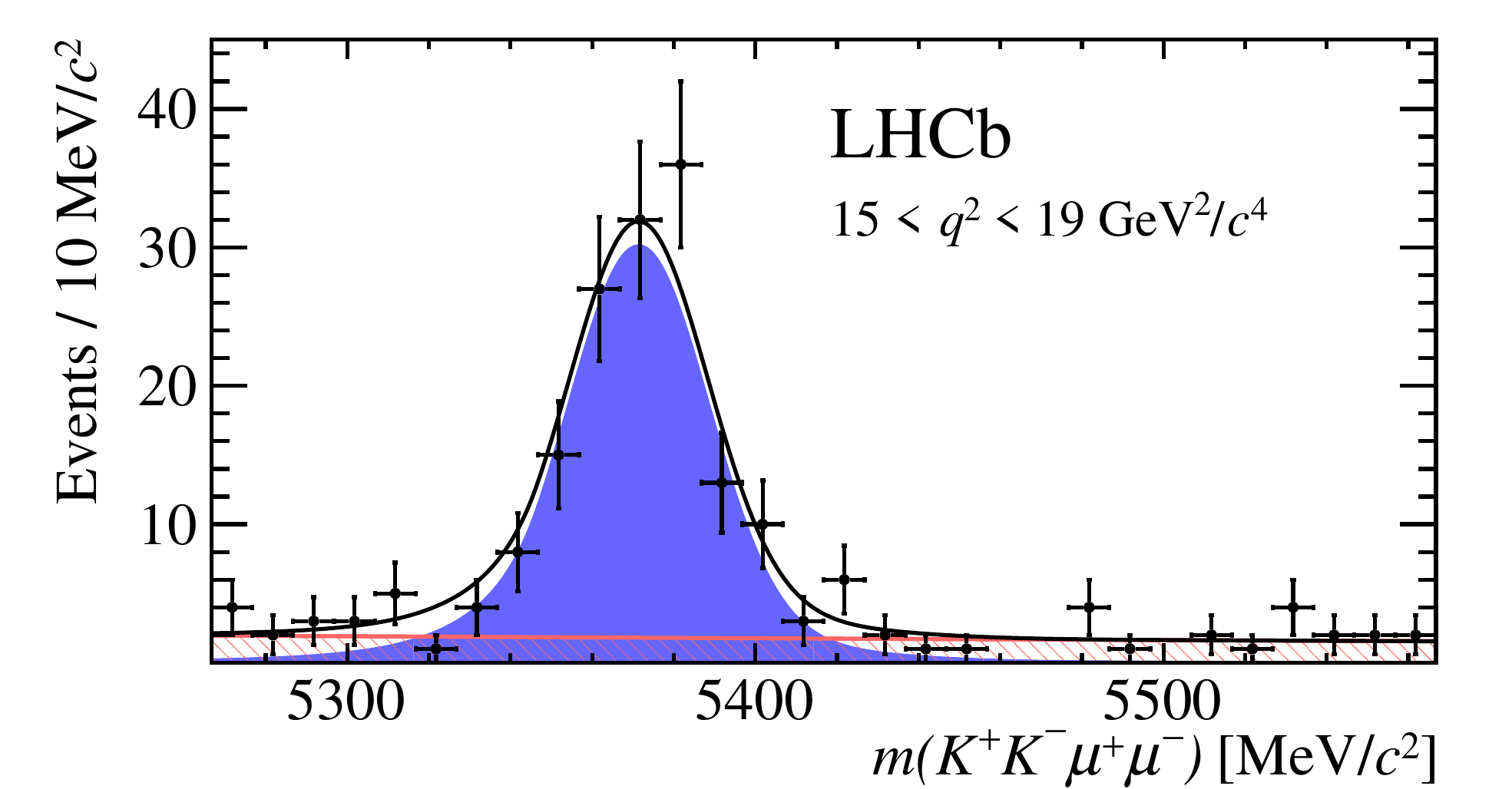}
\caption{
    Invariant mass distributions for $\decay{\Bs}{\phi\mumu}$ signal decays in bins of $q^2$. The signal component is shown by the solid blue area, the background component by the shaded red area.\label{fig:signalyields}}
\end{figure}

\clearpage

\section{Angular fit projections}
\label{app:ang-project}
\begin{figure}[h]
\begin{center}
\includegraphics[width=0.32\textwidth]{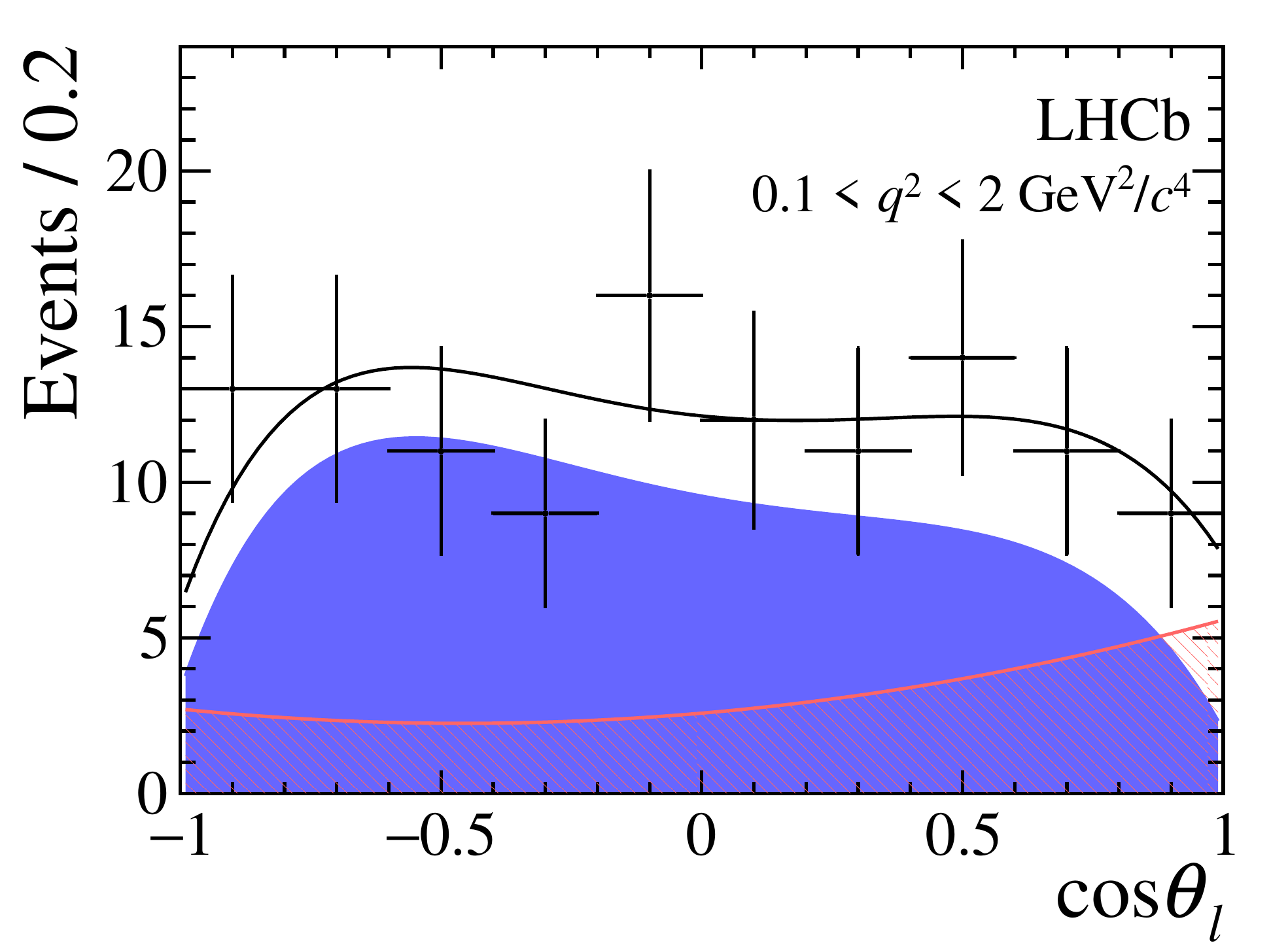}
\includegraphics[width=0.32\textwidth]{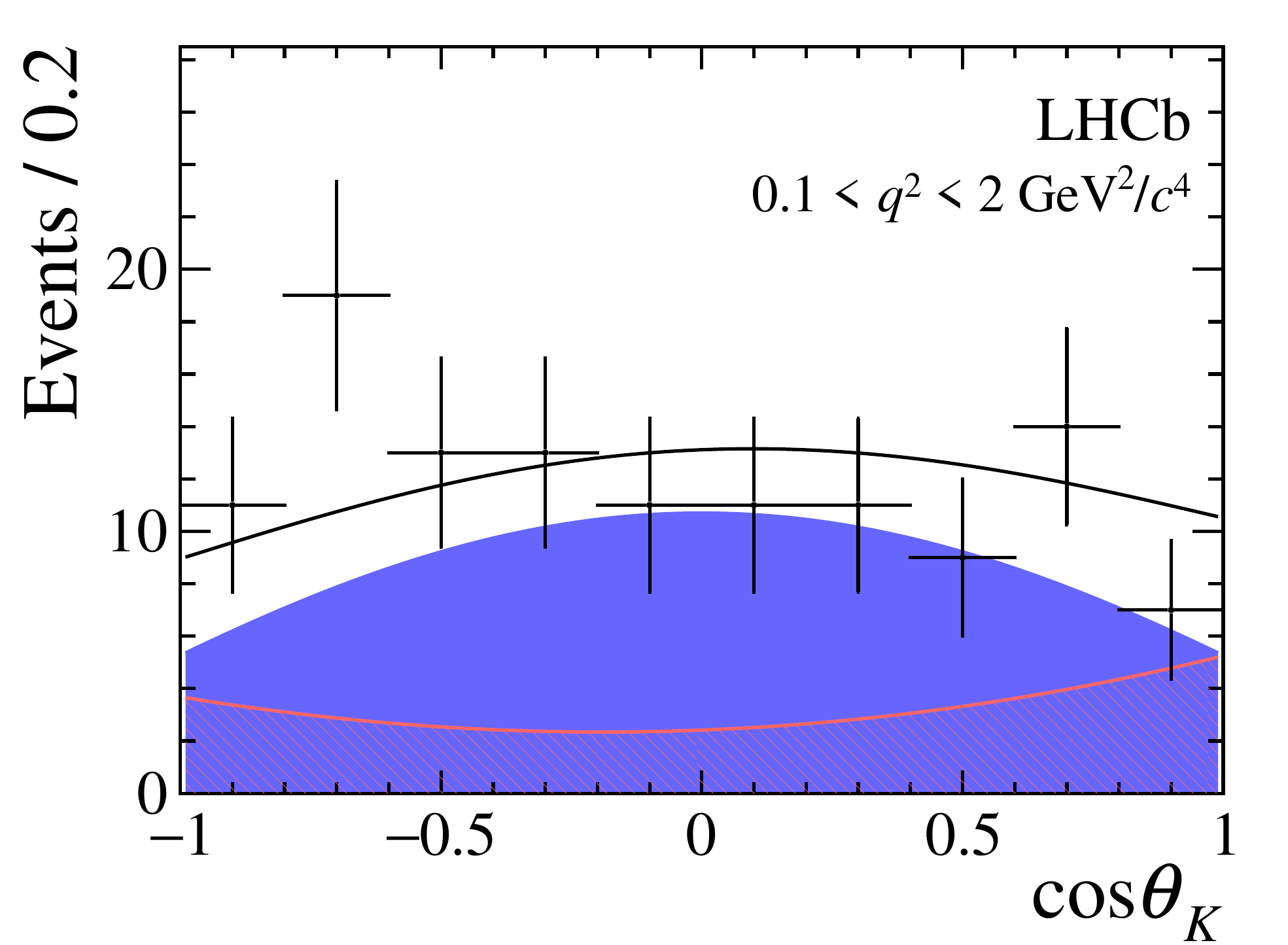}
\includegraphics[width=0.32\textwidth]{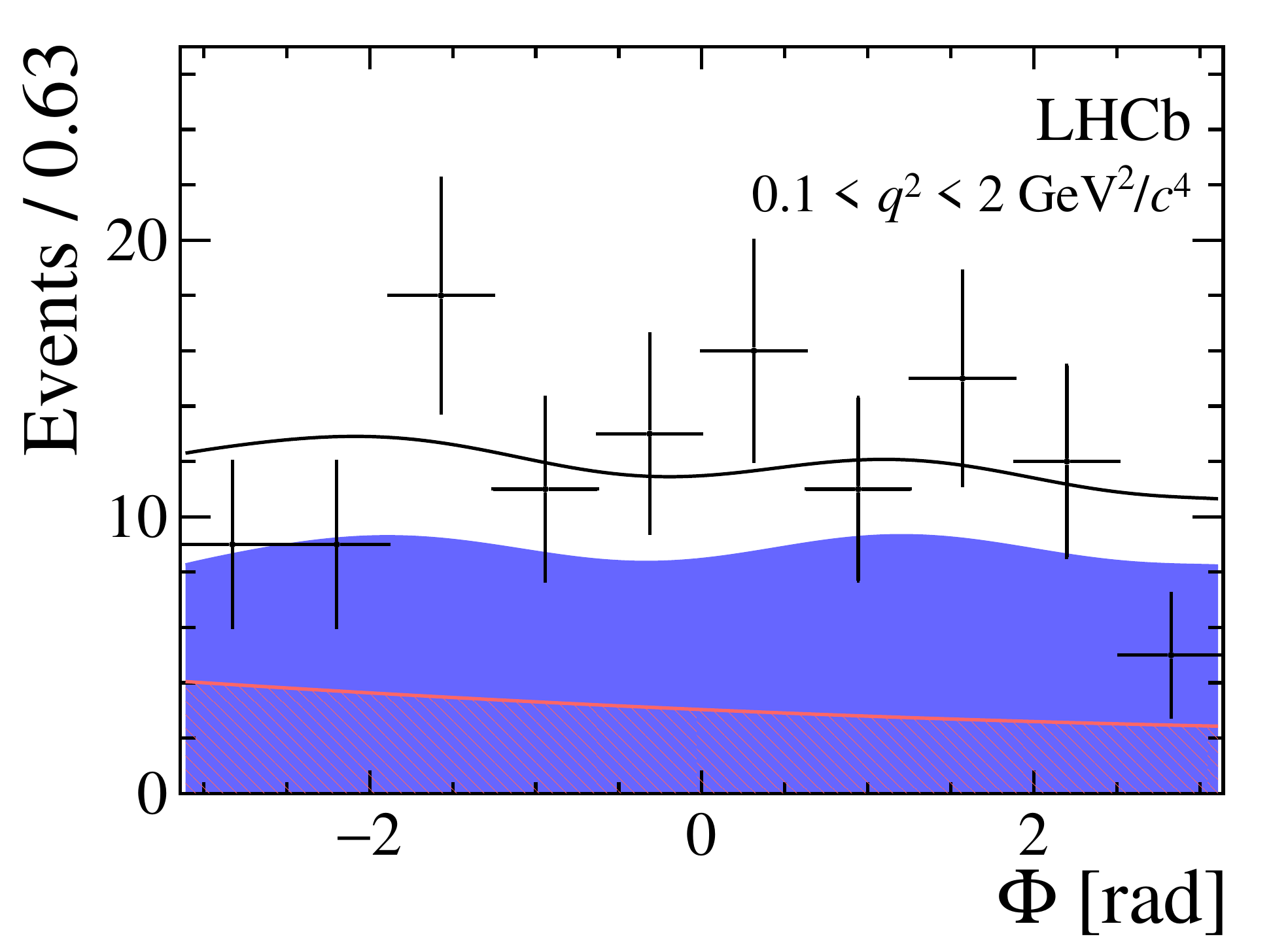}\\
\includegraphics[width=0.32\textwidth]{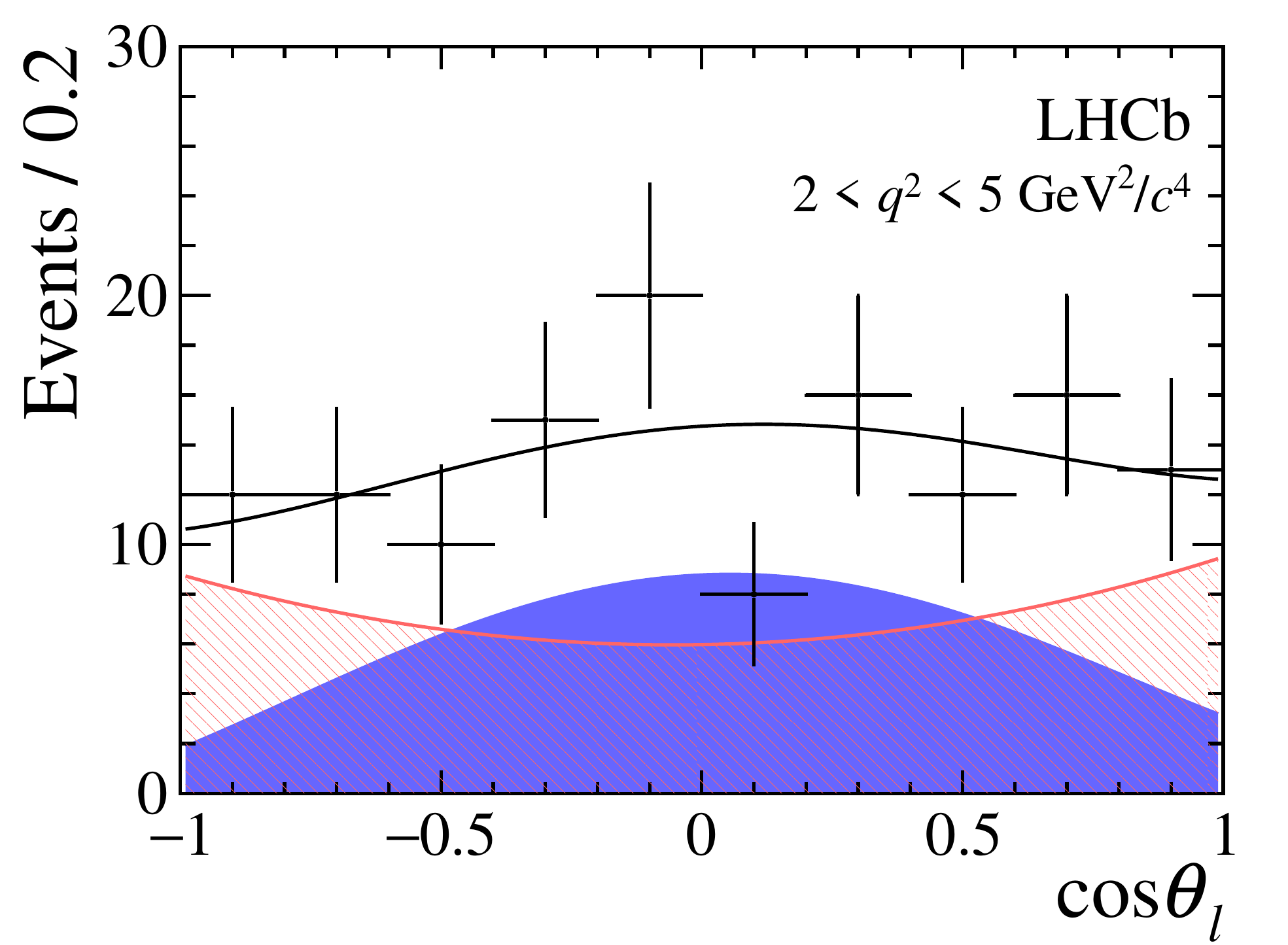}
\includegraphics[width=0.32\textwidth]{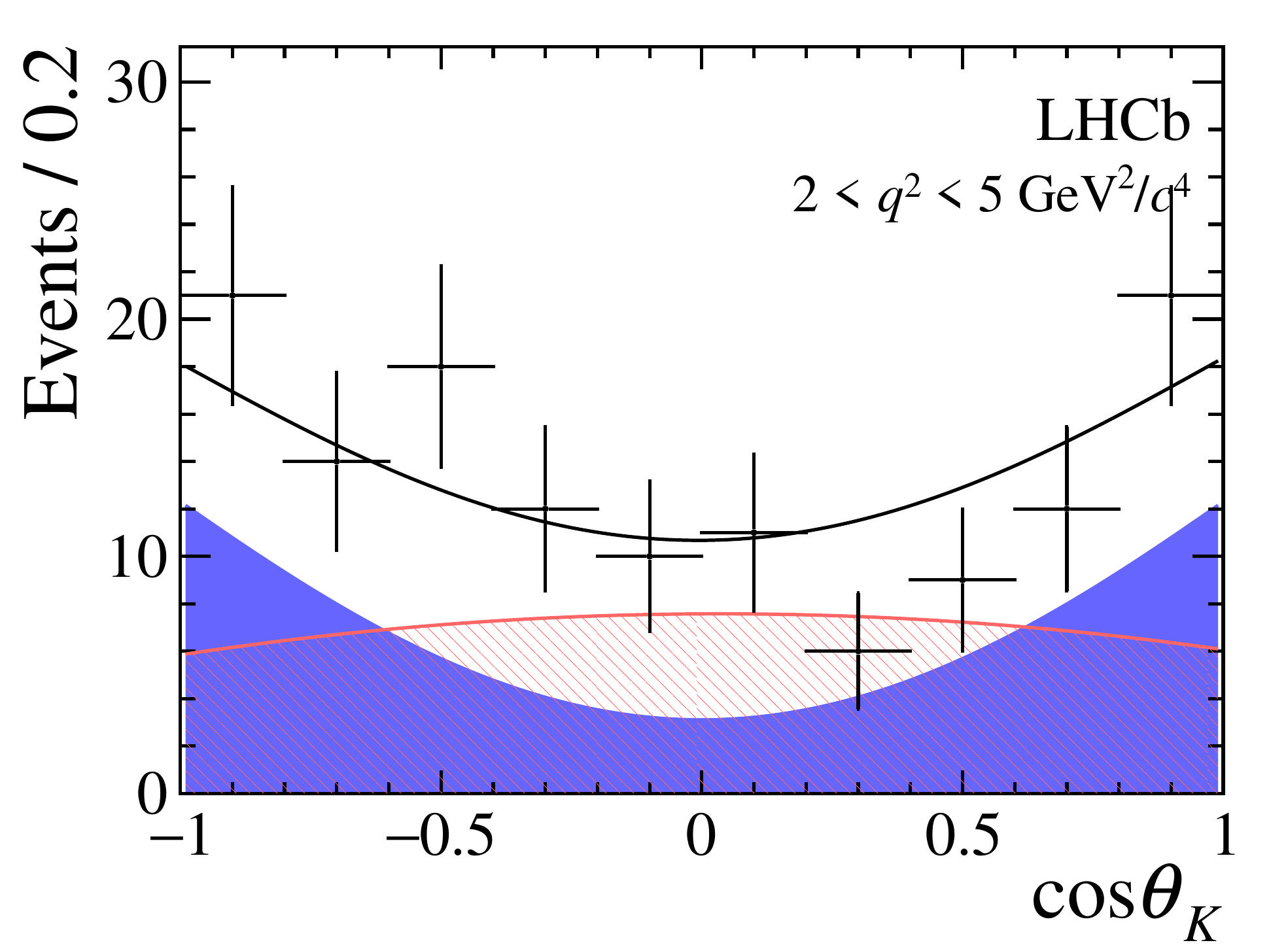}
\includegraphics[width=0.32\textwidth]{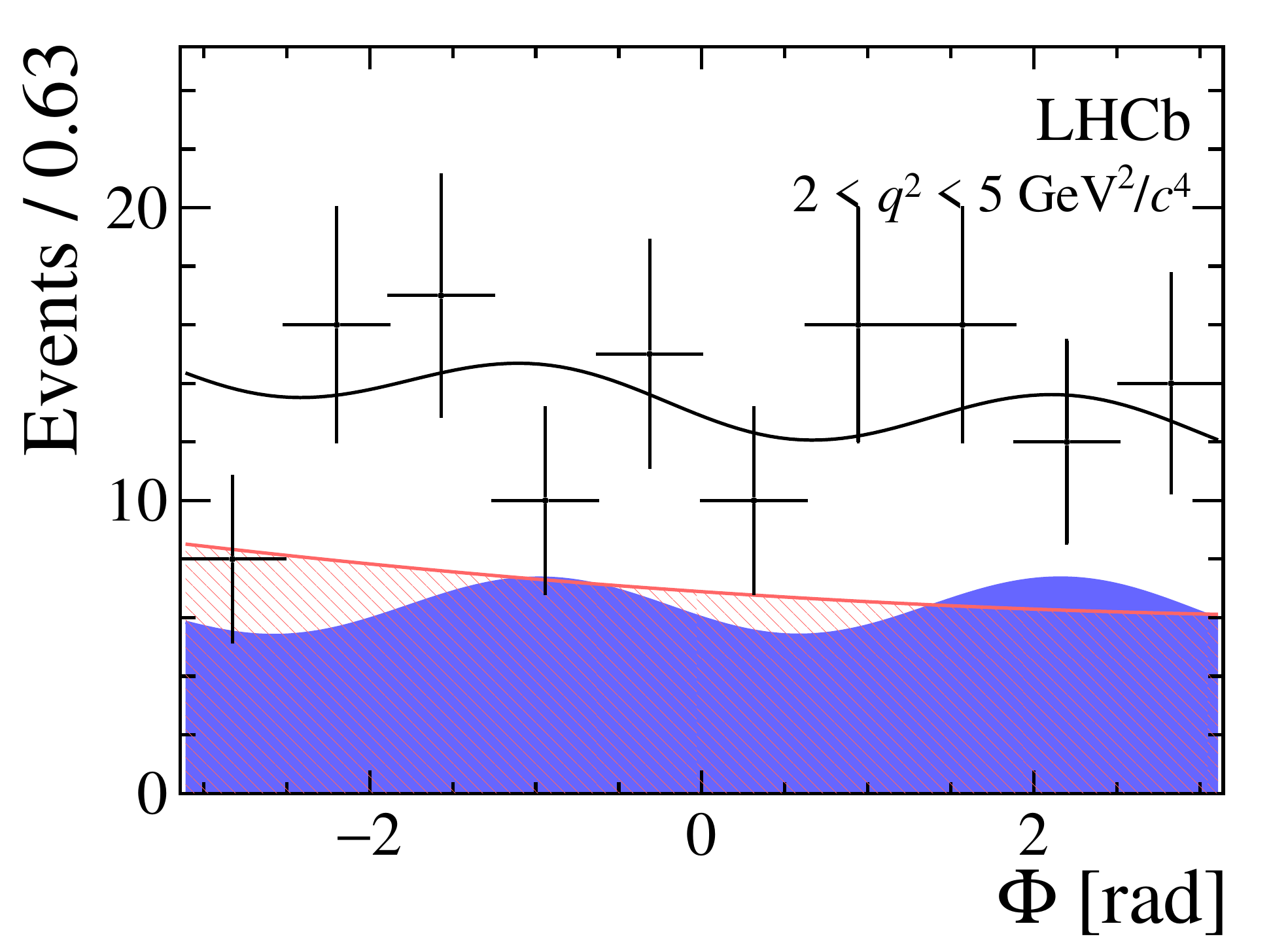}\\
\includegraphics[width=0.32\textwidth]{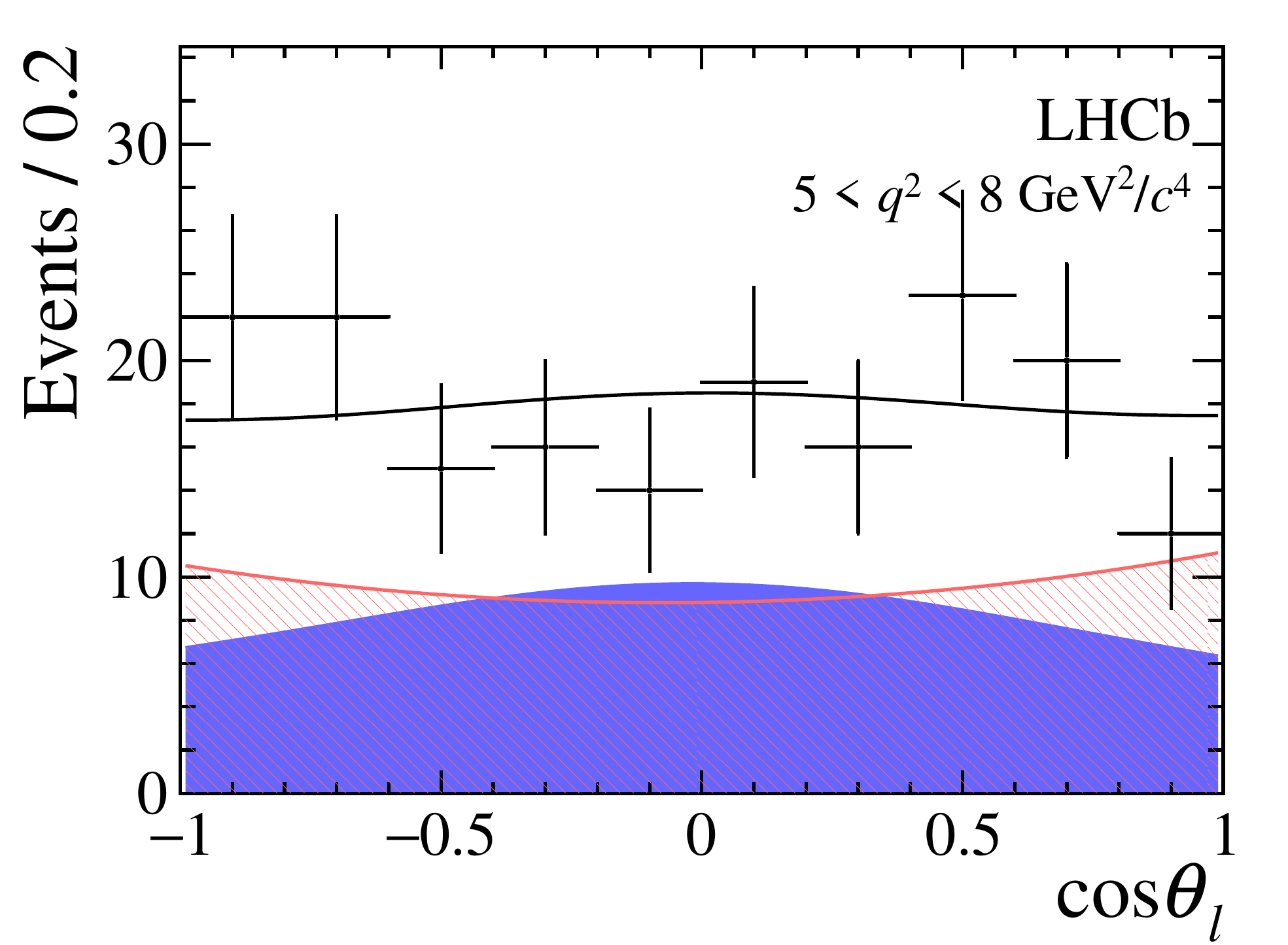}
\includegraphics[width=0.32\textwidth]{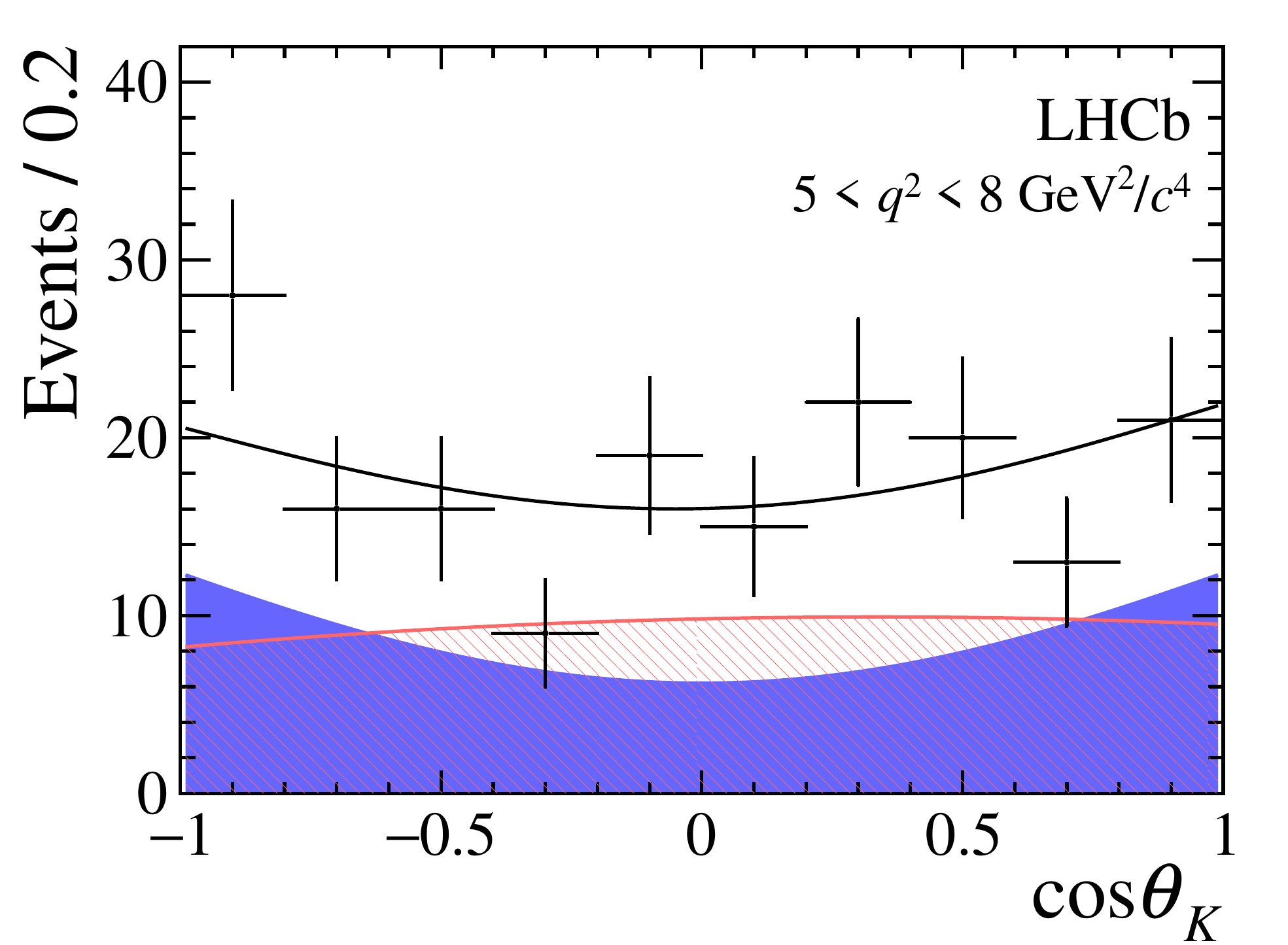}
\includegraphics[width=0.32\textwidth]{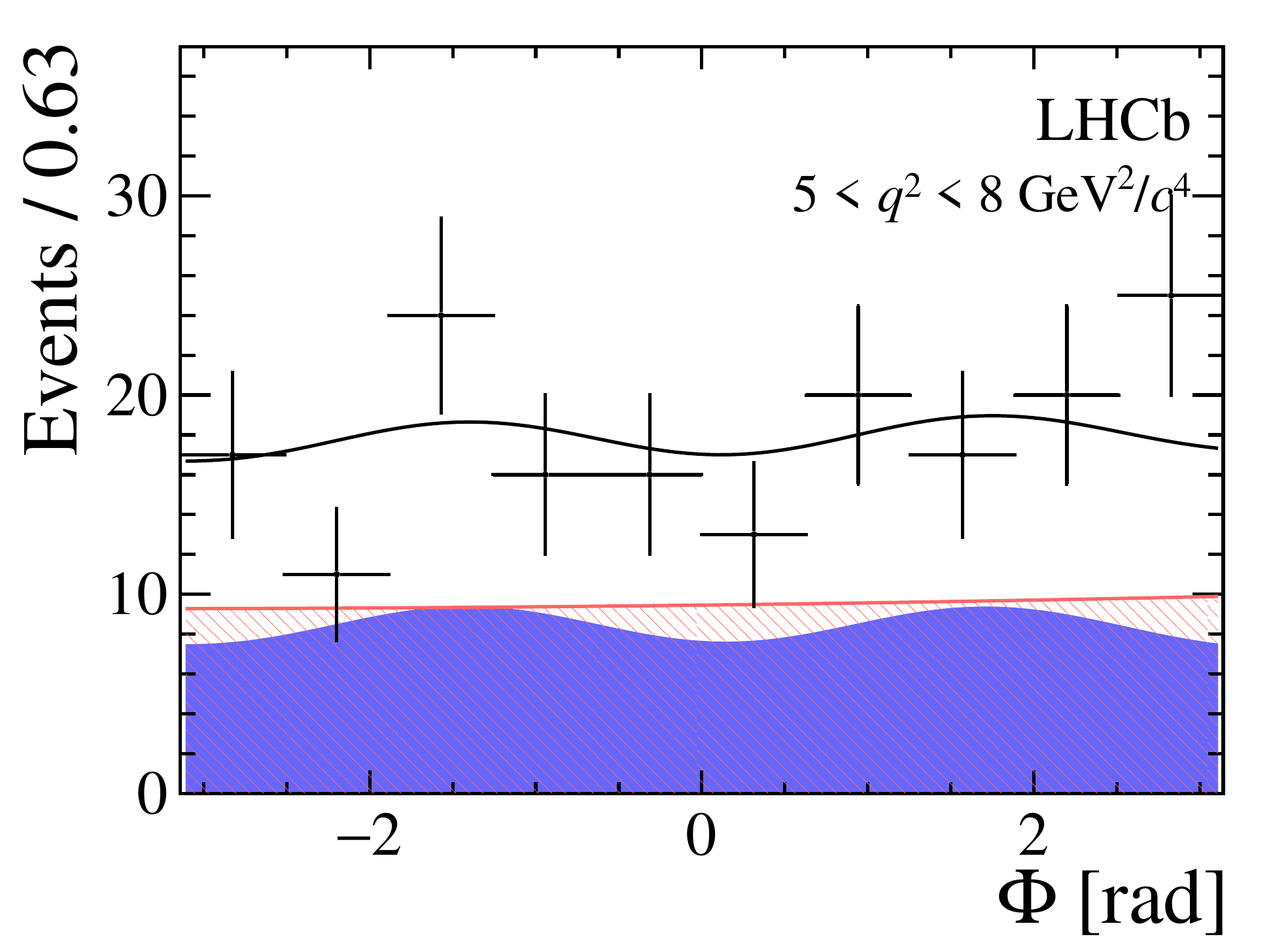}\\
\includegraphics[width=0.32\textwidth]{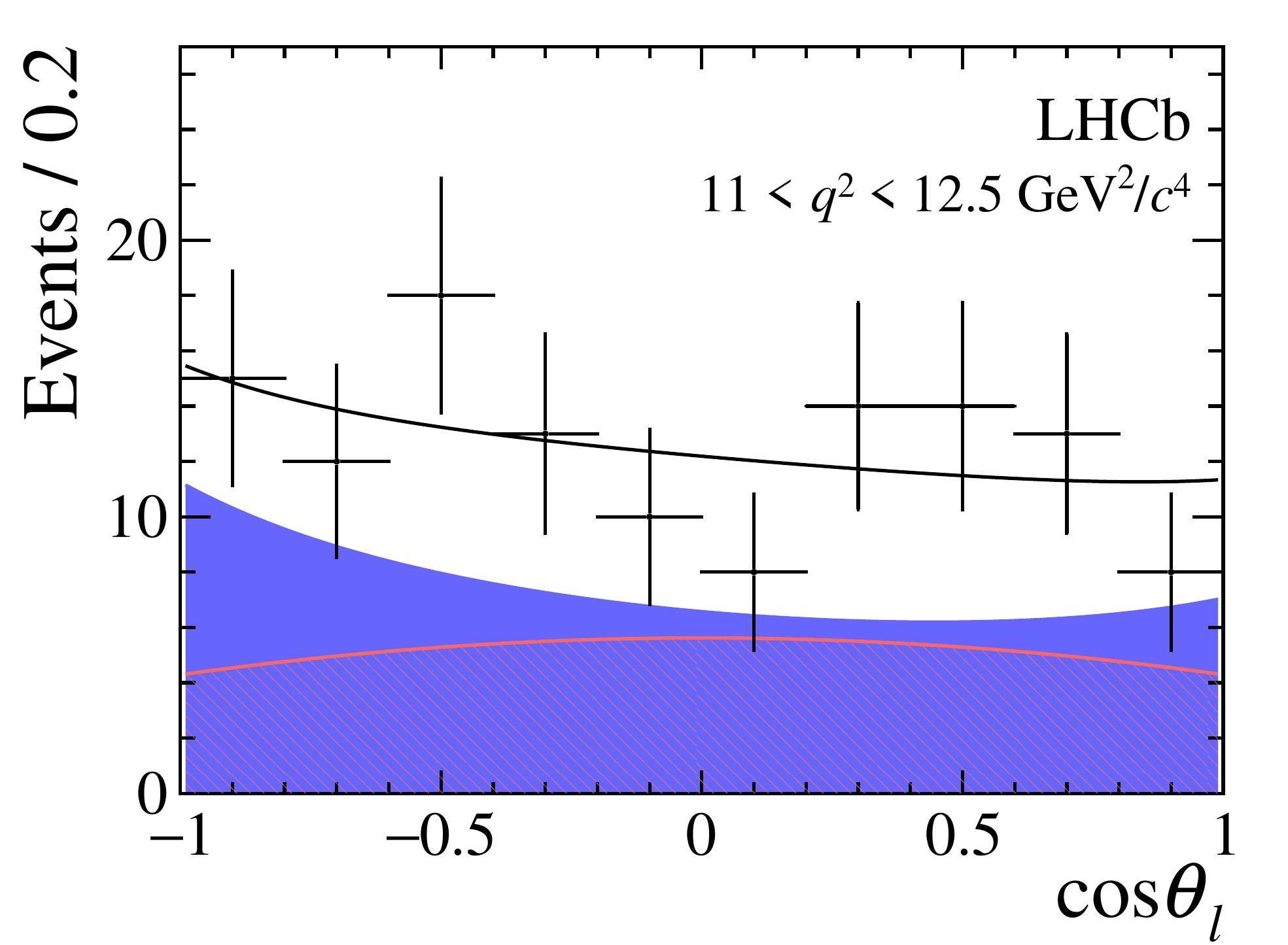}
\includegraphics[width=0.32\textwidth]{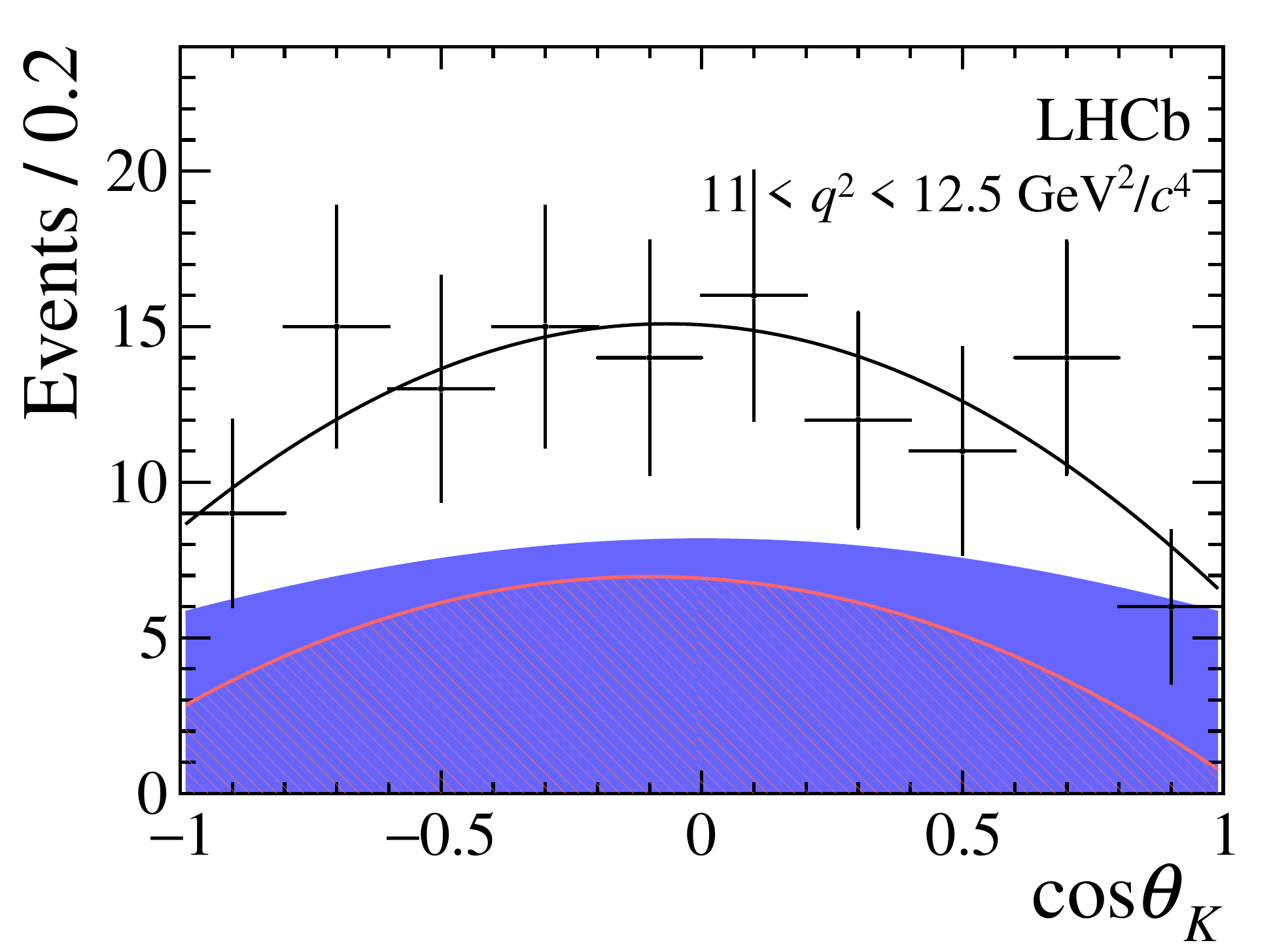}
\includegraphics[width=0.32\textwidth]{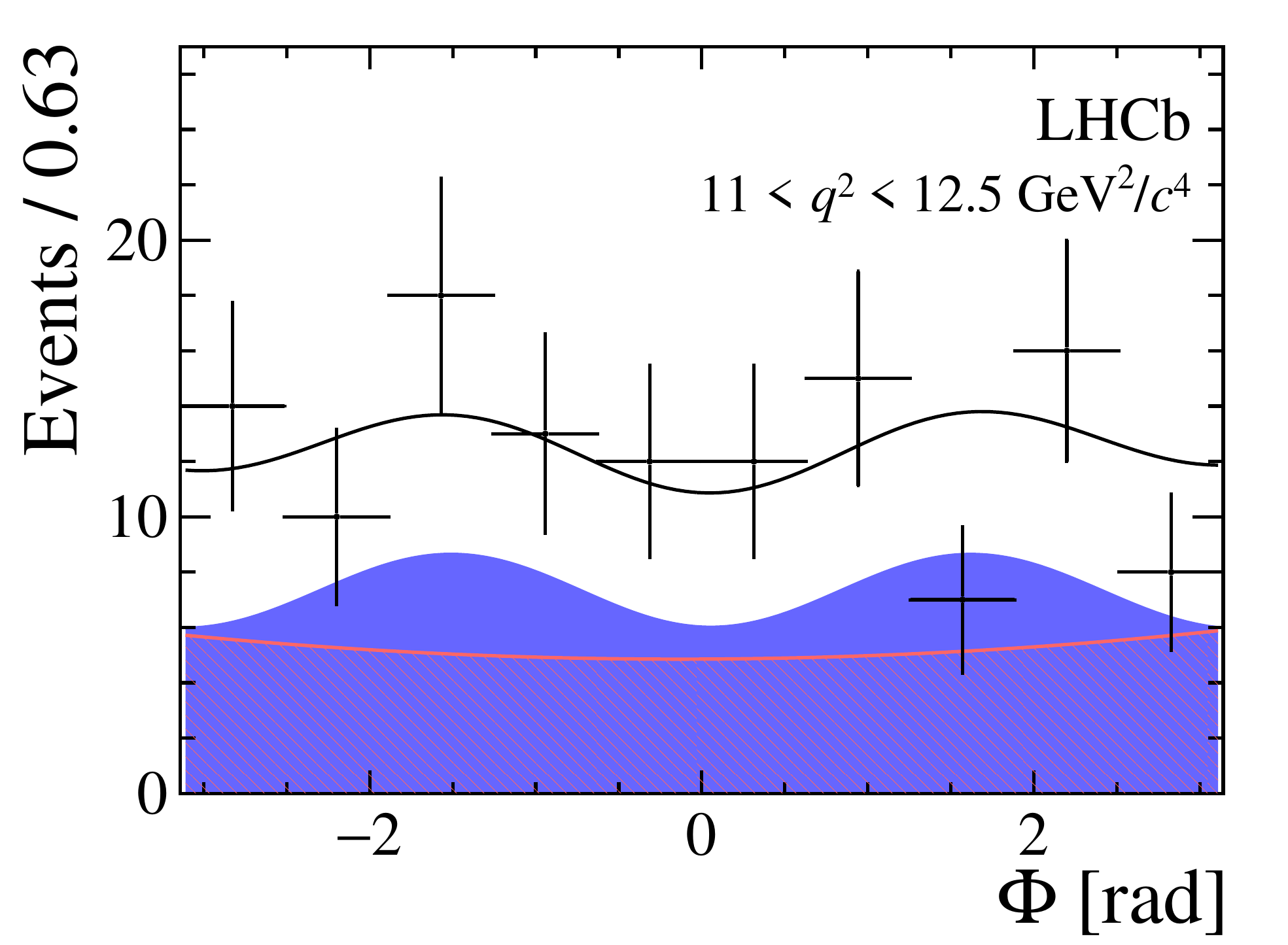}\\
\caption{\label{fig:ang-proj1} One-dimensional projections of the fit to the angles $\cos\theta_l$, $\cos\theta_K$, $\Phi$ in bins of $q^2$. The signal component is shown
by the solid blue area, the background component by the shaded red area.}
\end{center}
\end{figure}

\label{app:ang-project}
\begin{figure}[h]
\begin{center}
\includegraphics[width=0.32\textwidth]{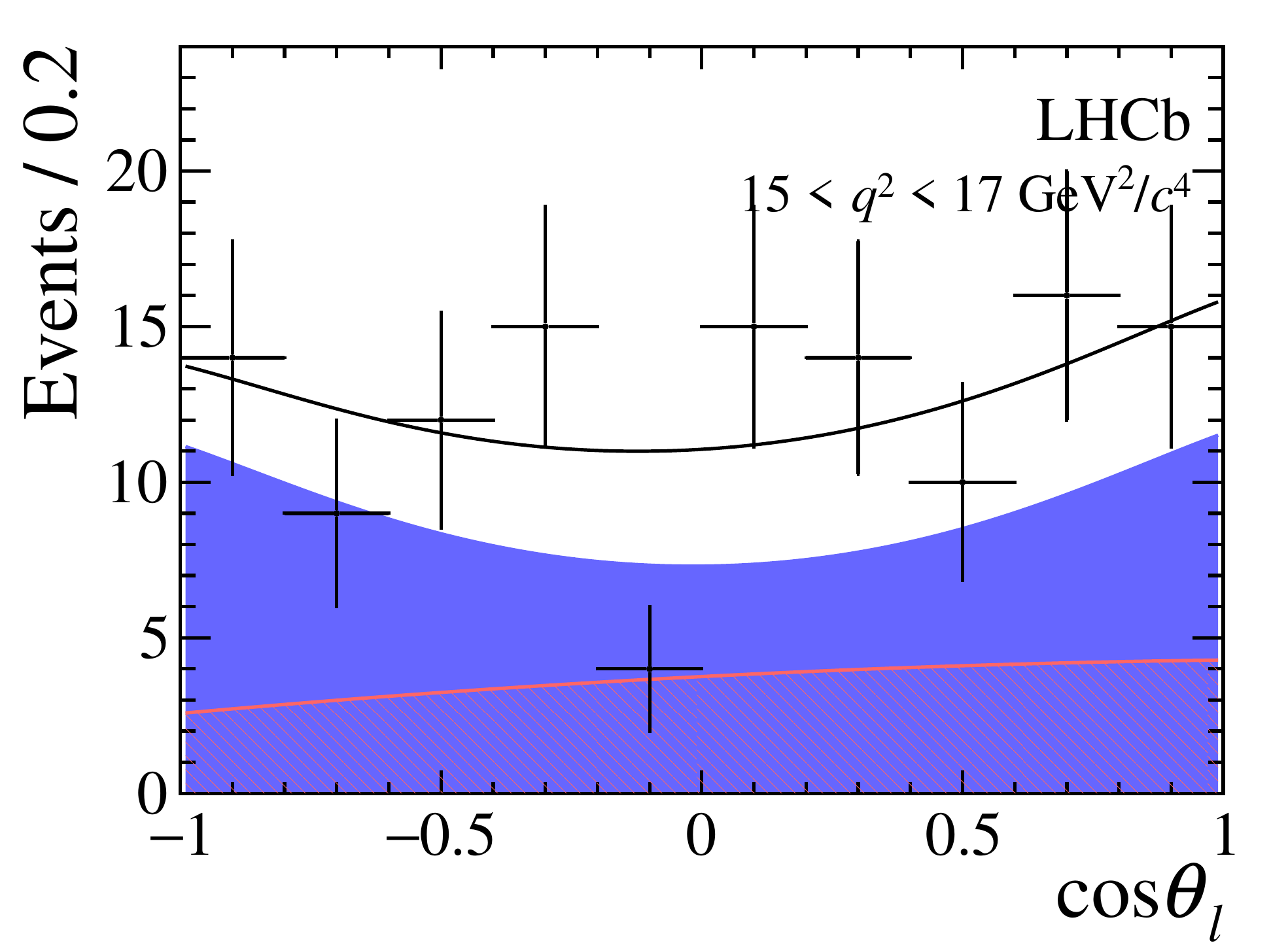}
\includegraphics[width=0.32\textwidth]{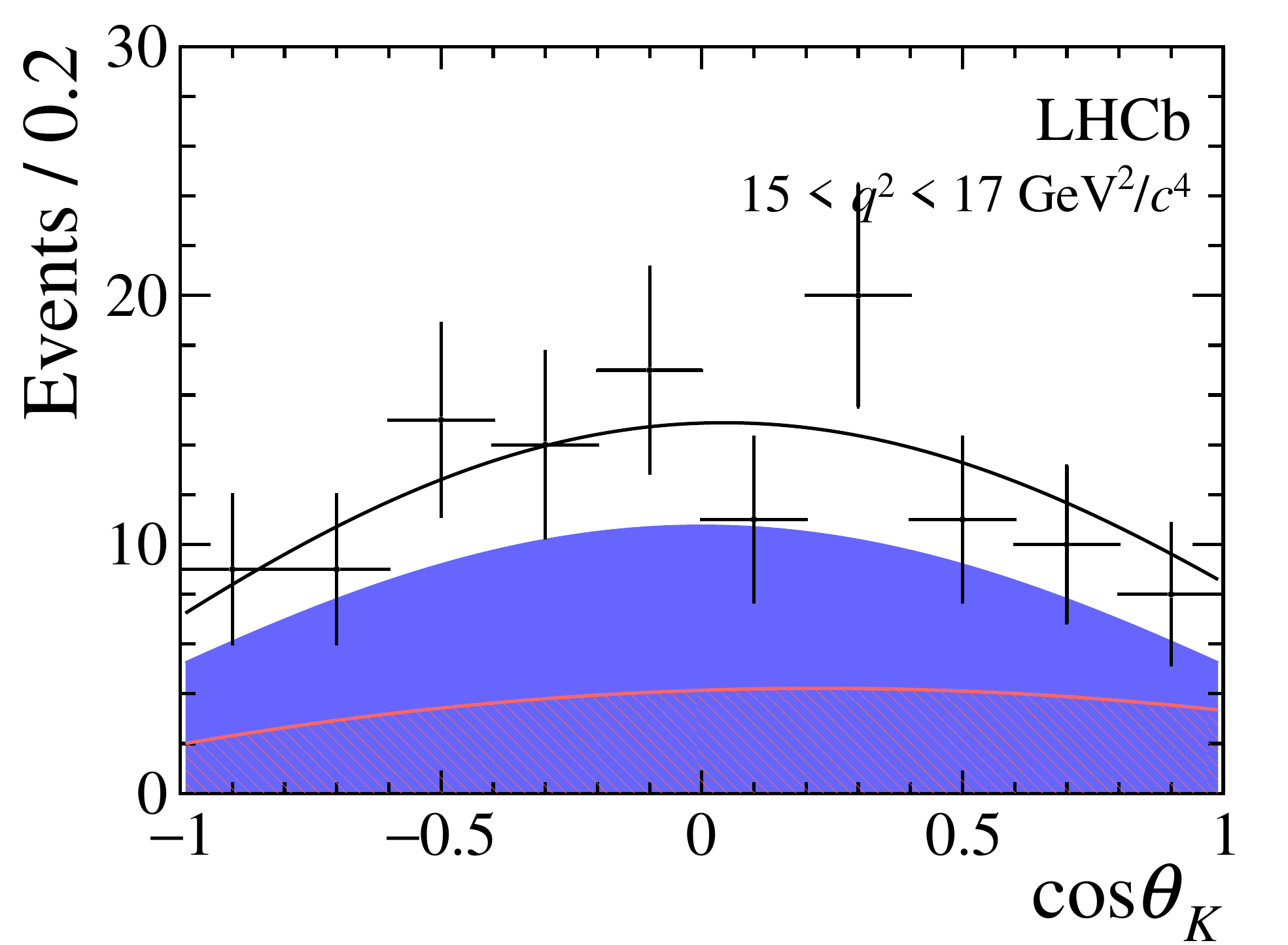}
\includegraphics[width=0.32\textwidth]{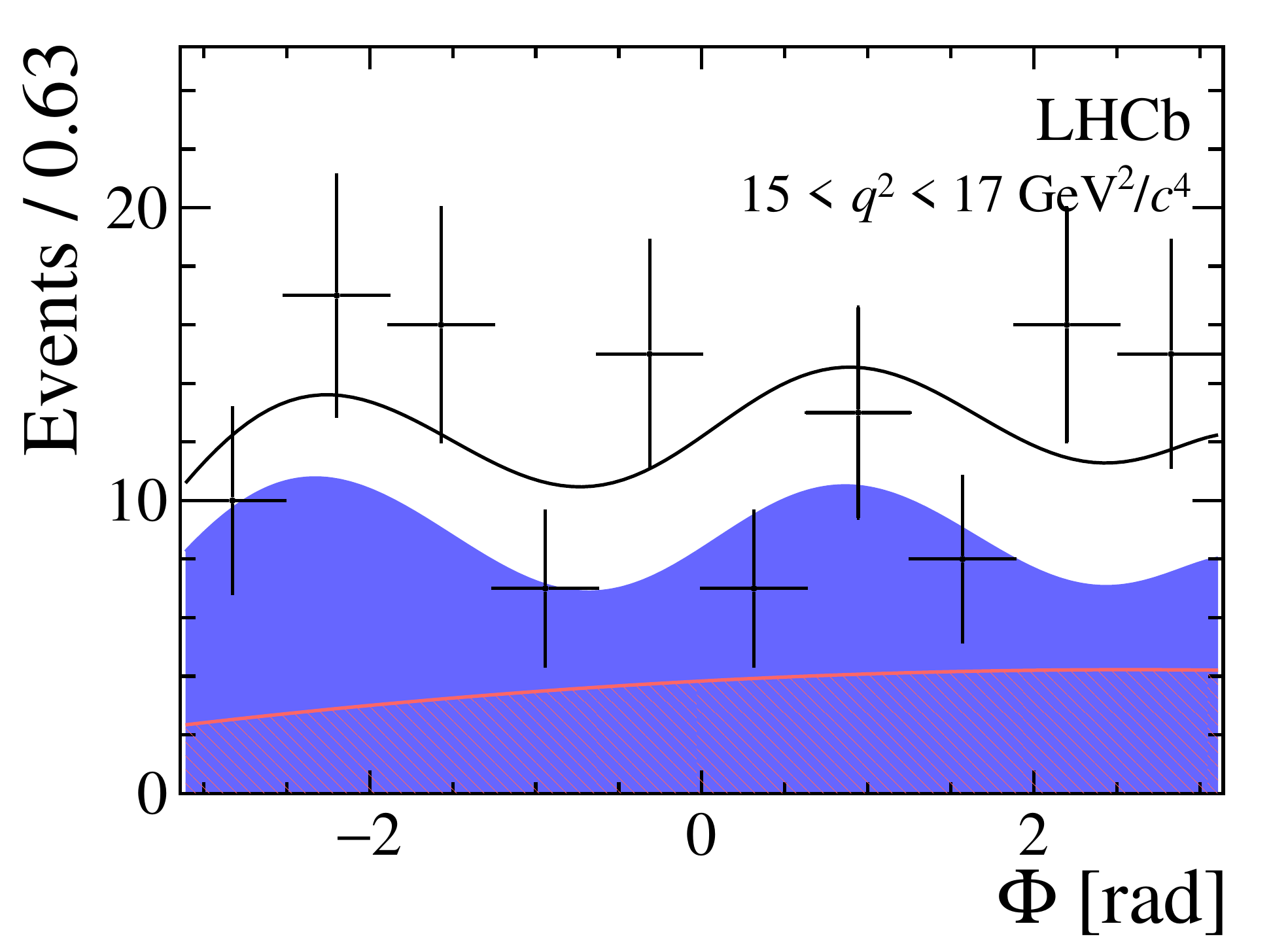}\\
\includegraphics[width=0.32\textwidth]{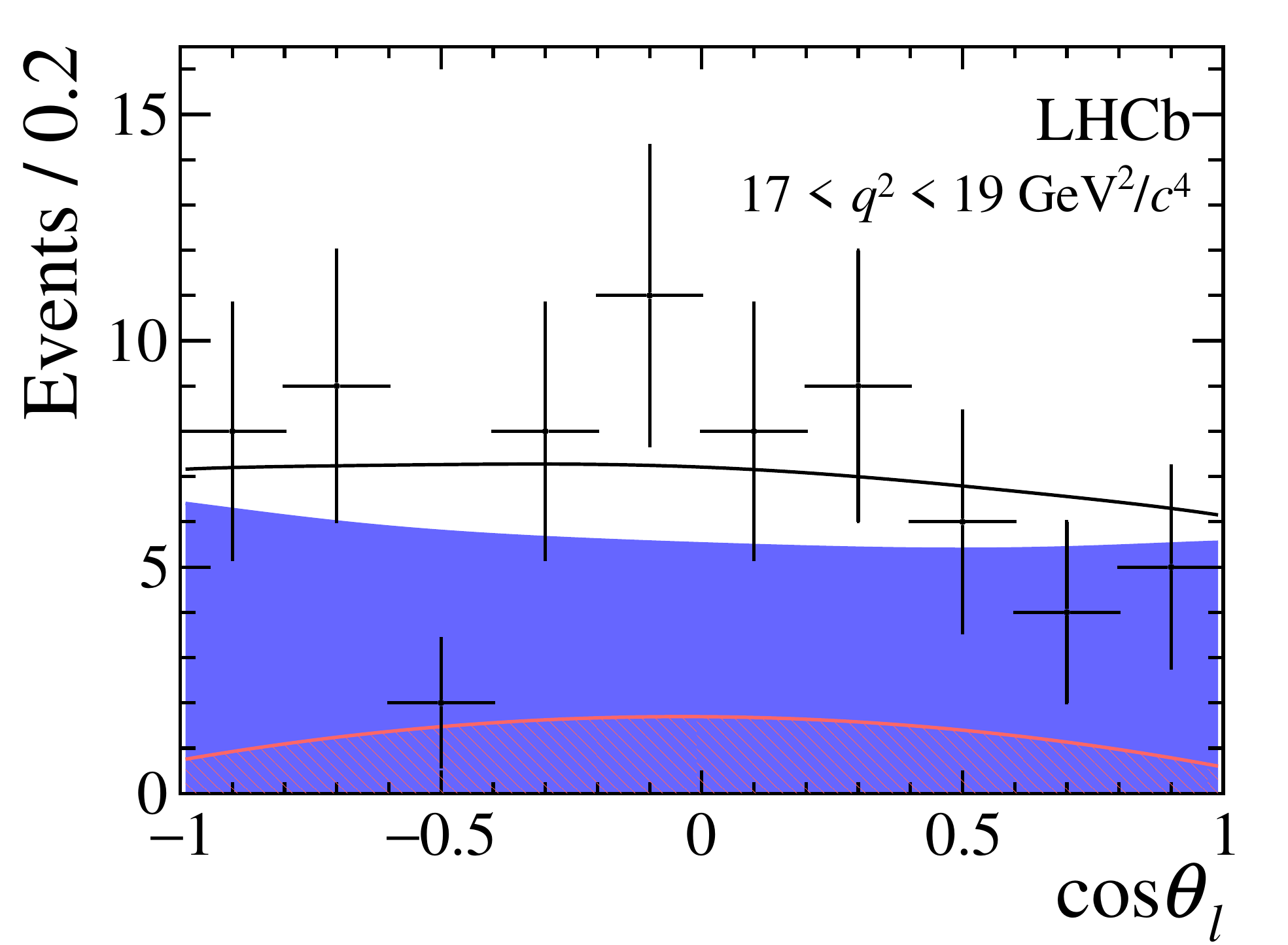}
\includegraphics[width=0.32\textwidth]{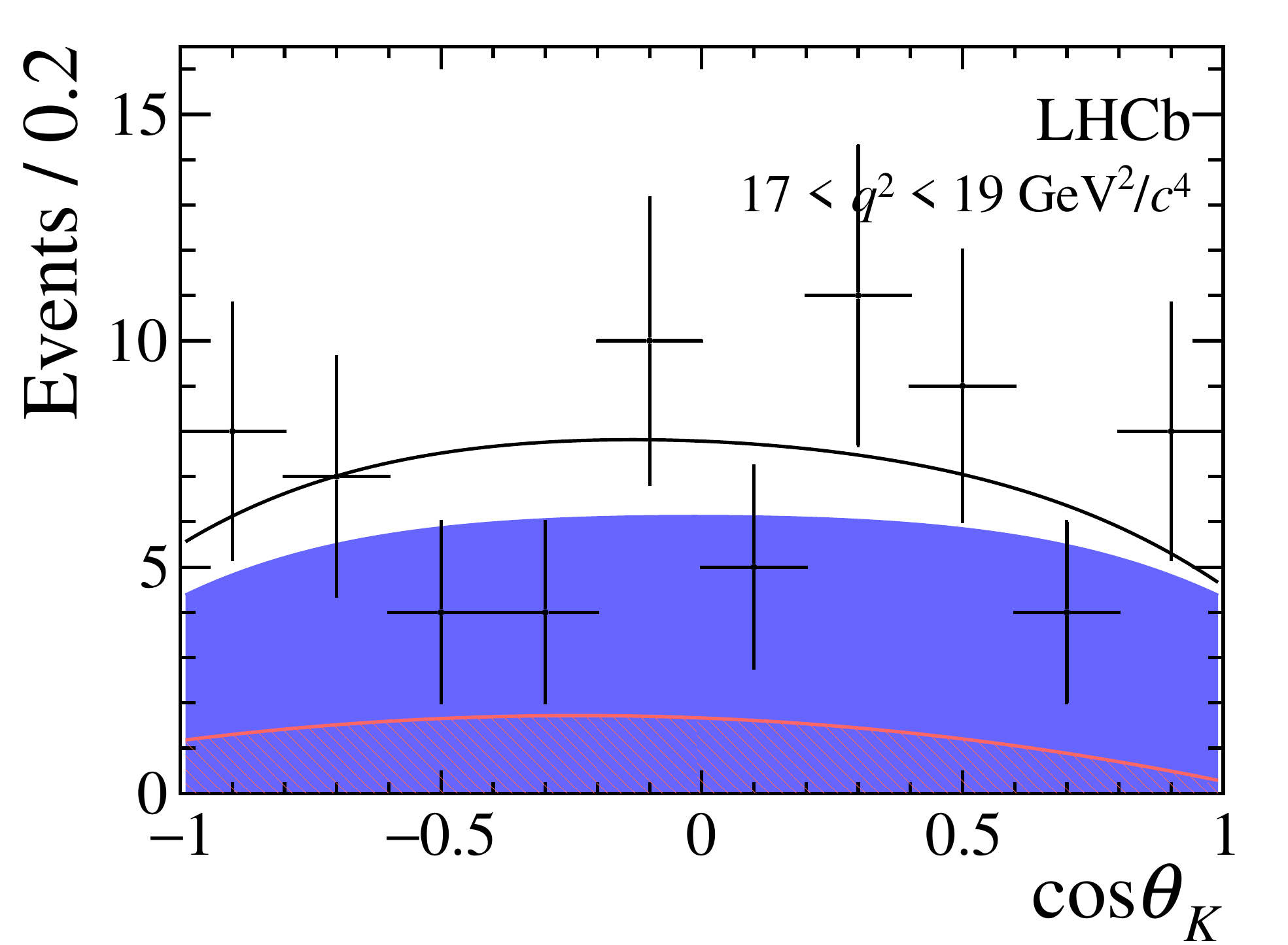}
\includegraphics[width=0.32\textwidth]{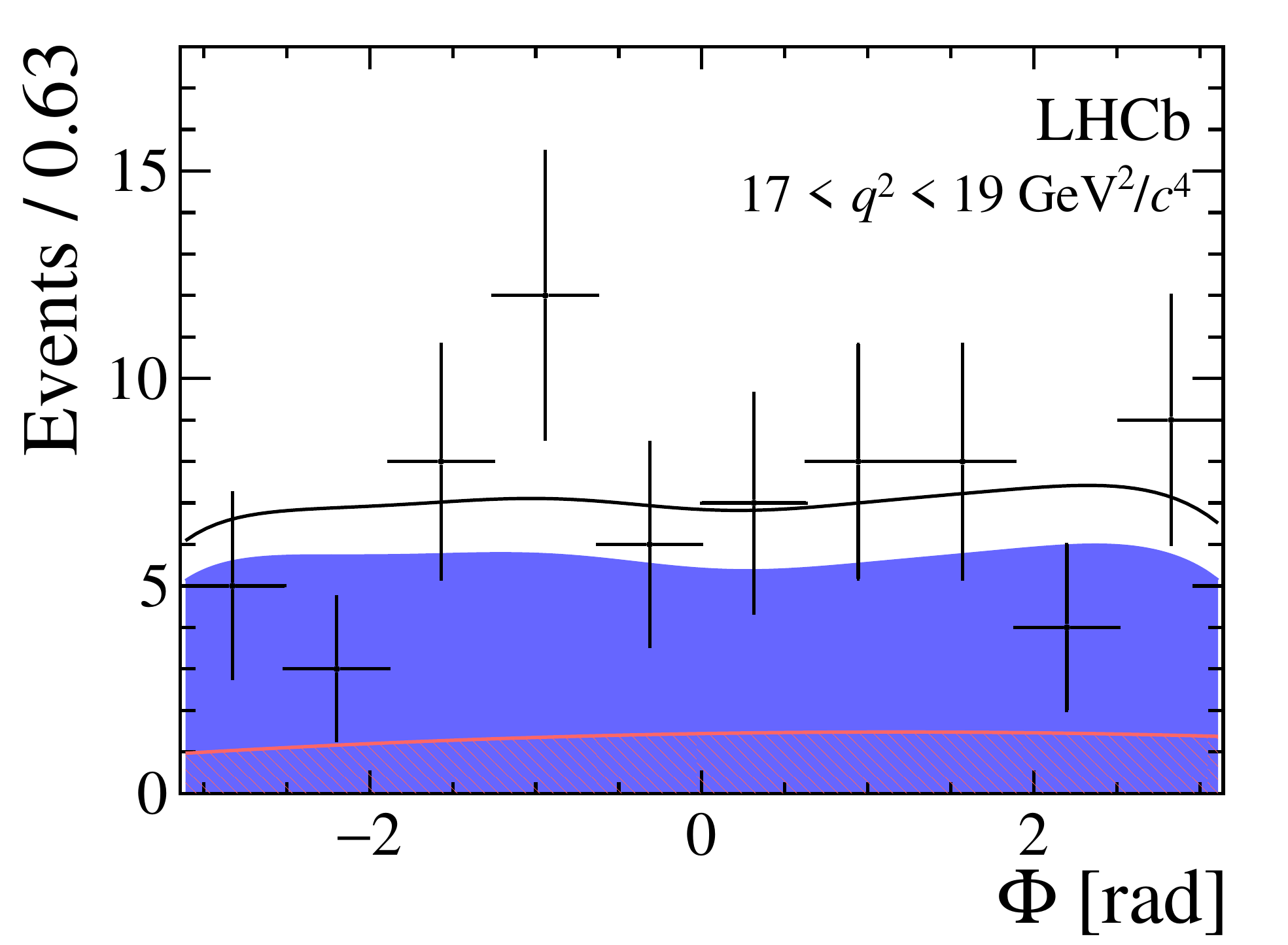}\\
\includegraphics[width=0.32\textwidth]{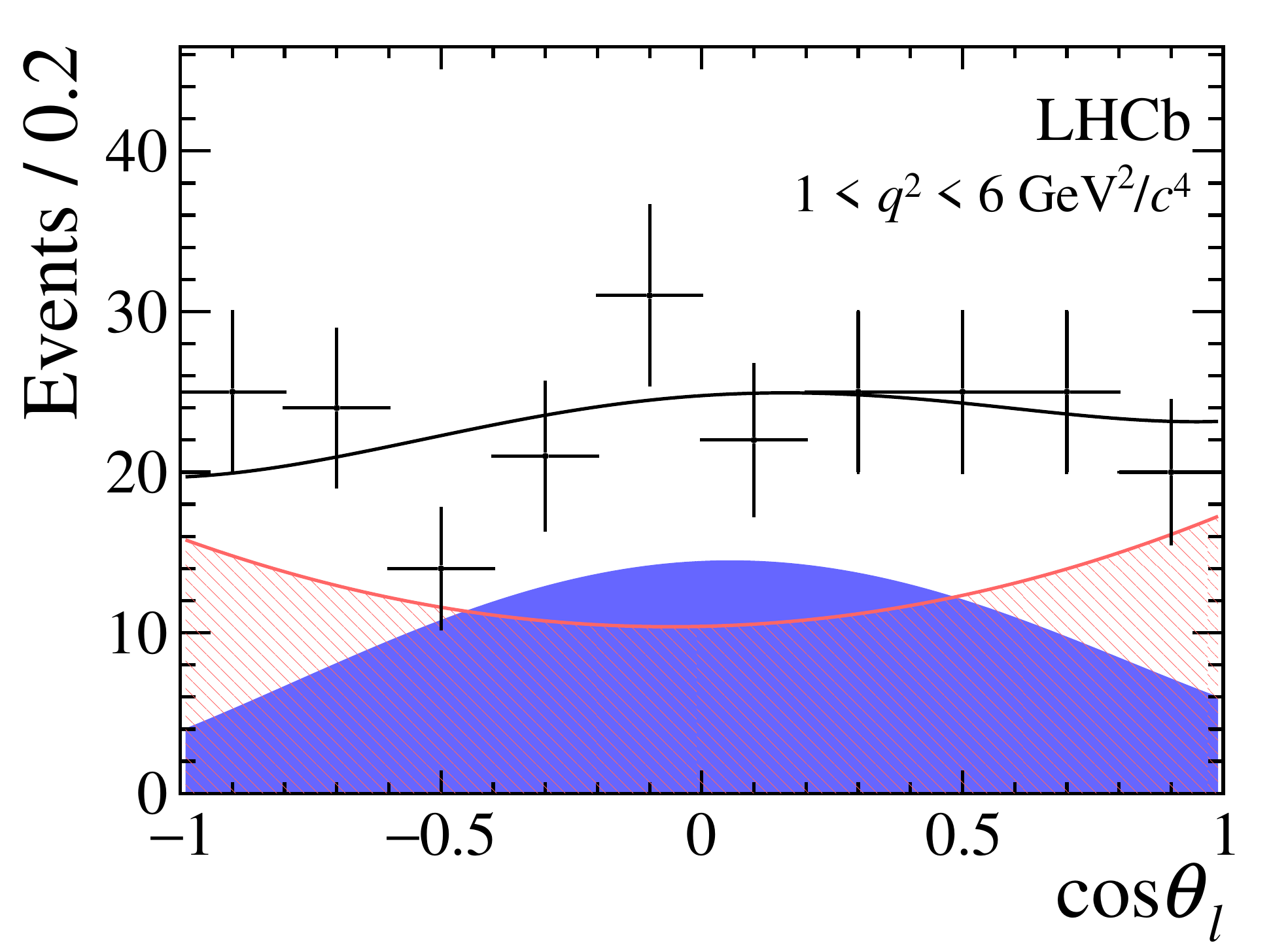}
\includegraphics[width=0.32\textwidth]{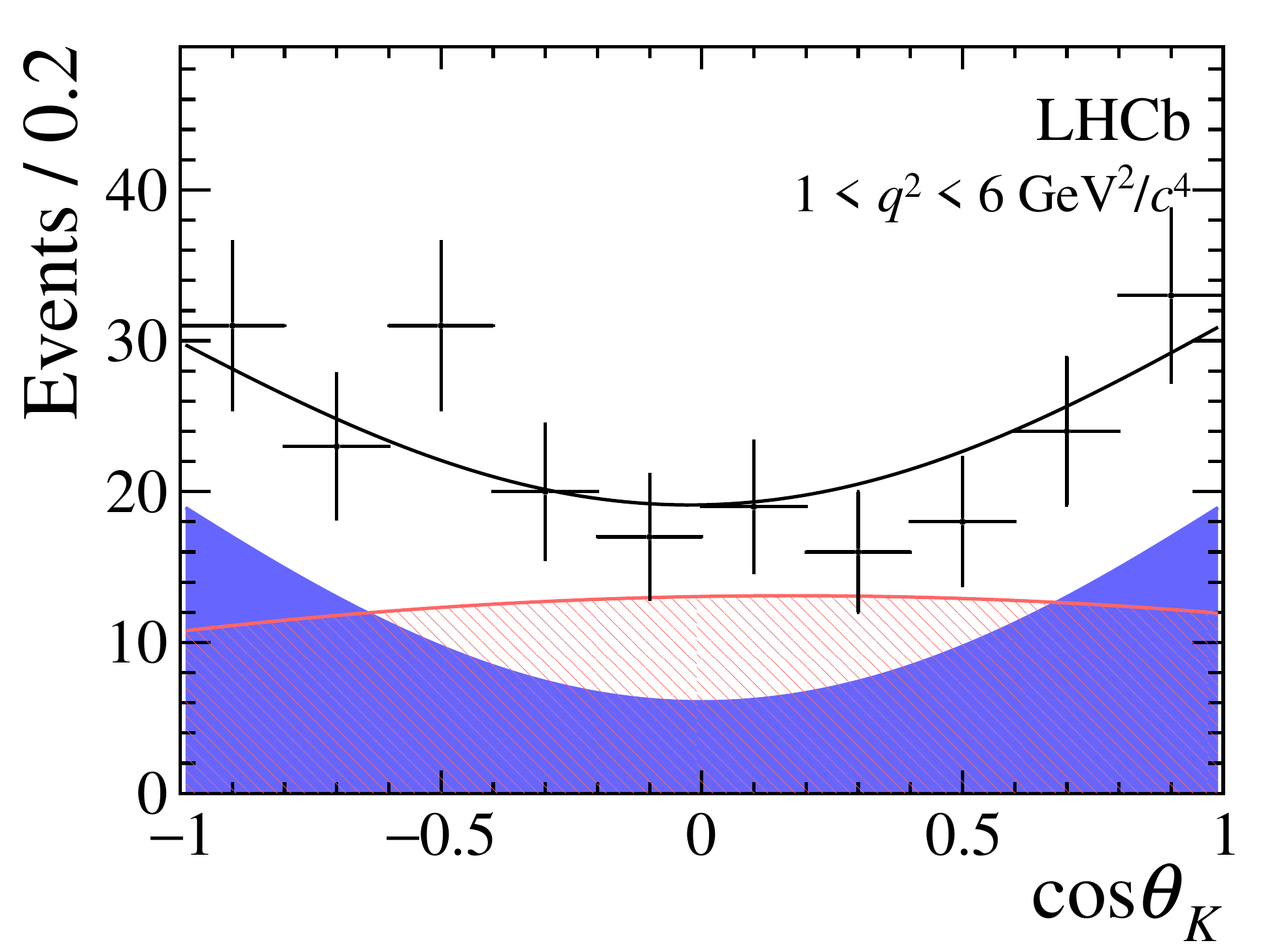}
\includegraphics[width=0.32\textwidth]{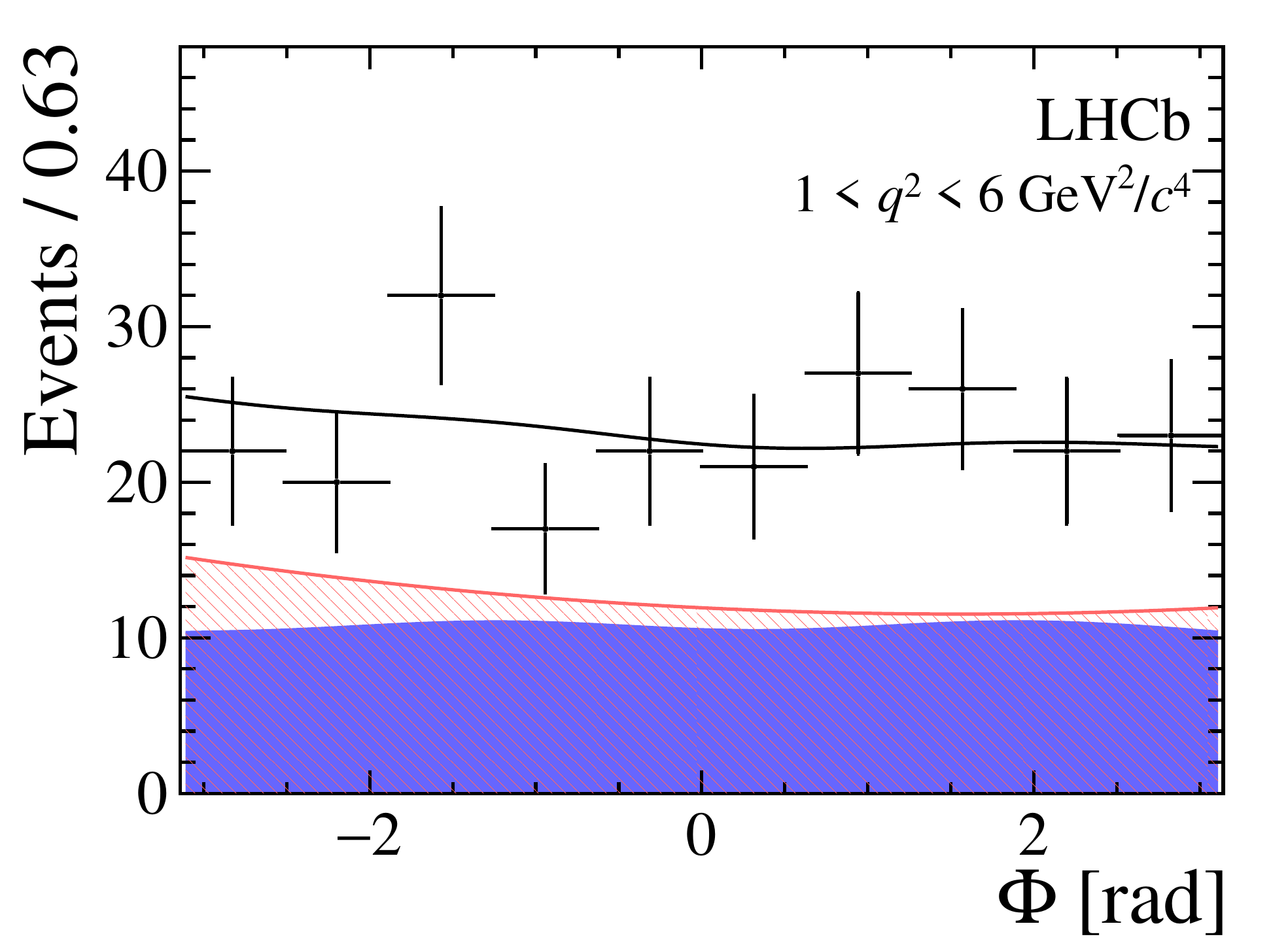}\\
\includegraphics[width=0.32\textwidth]{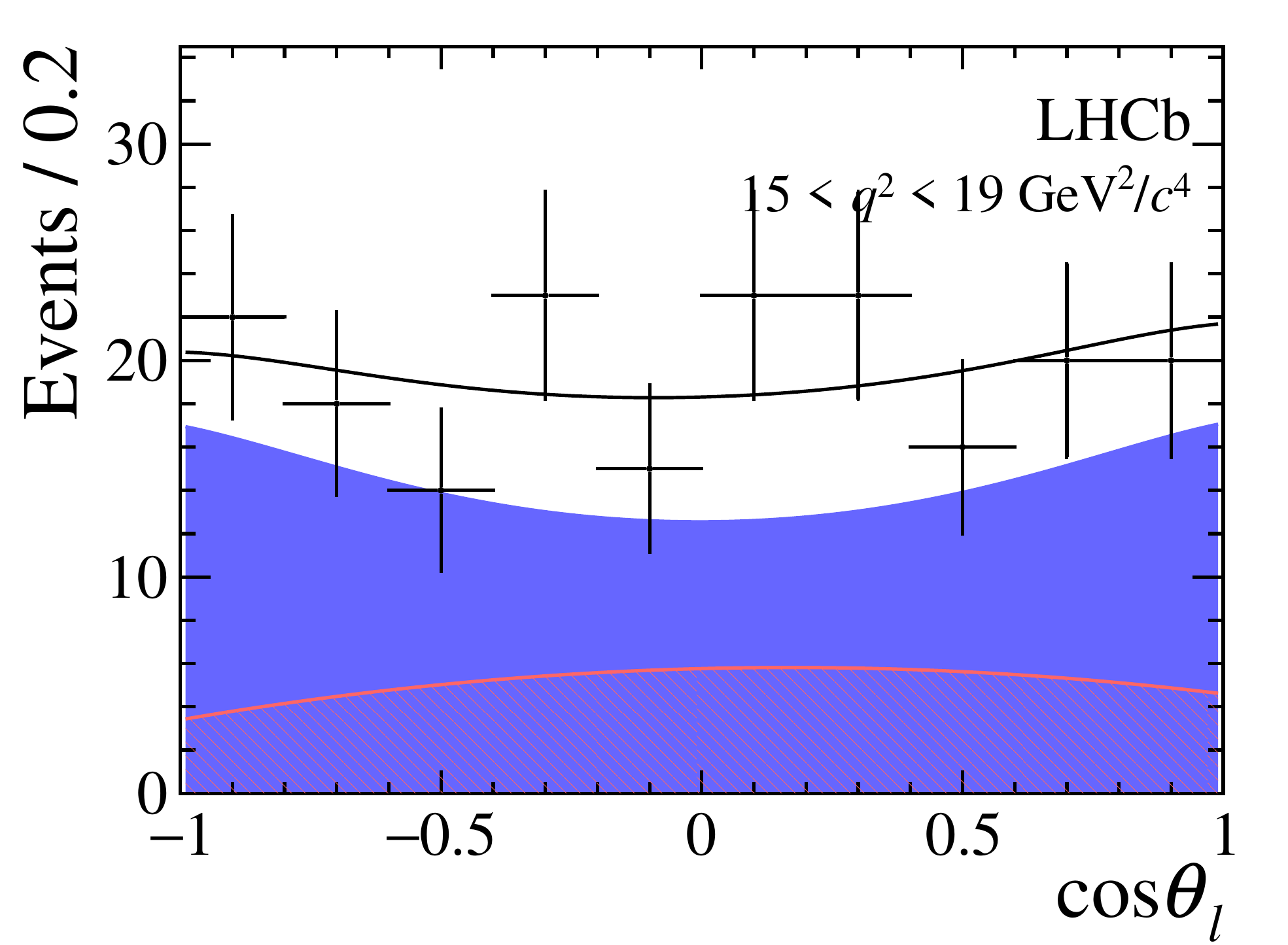}
\includegraphics[width=0.32\textwidth]{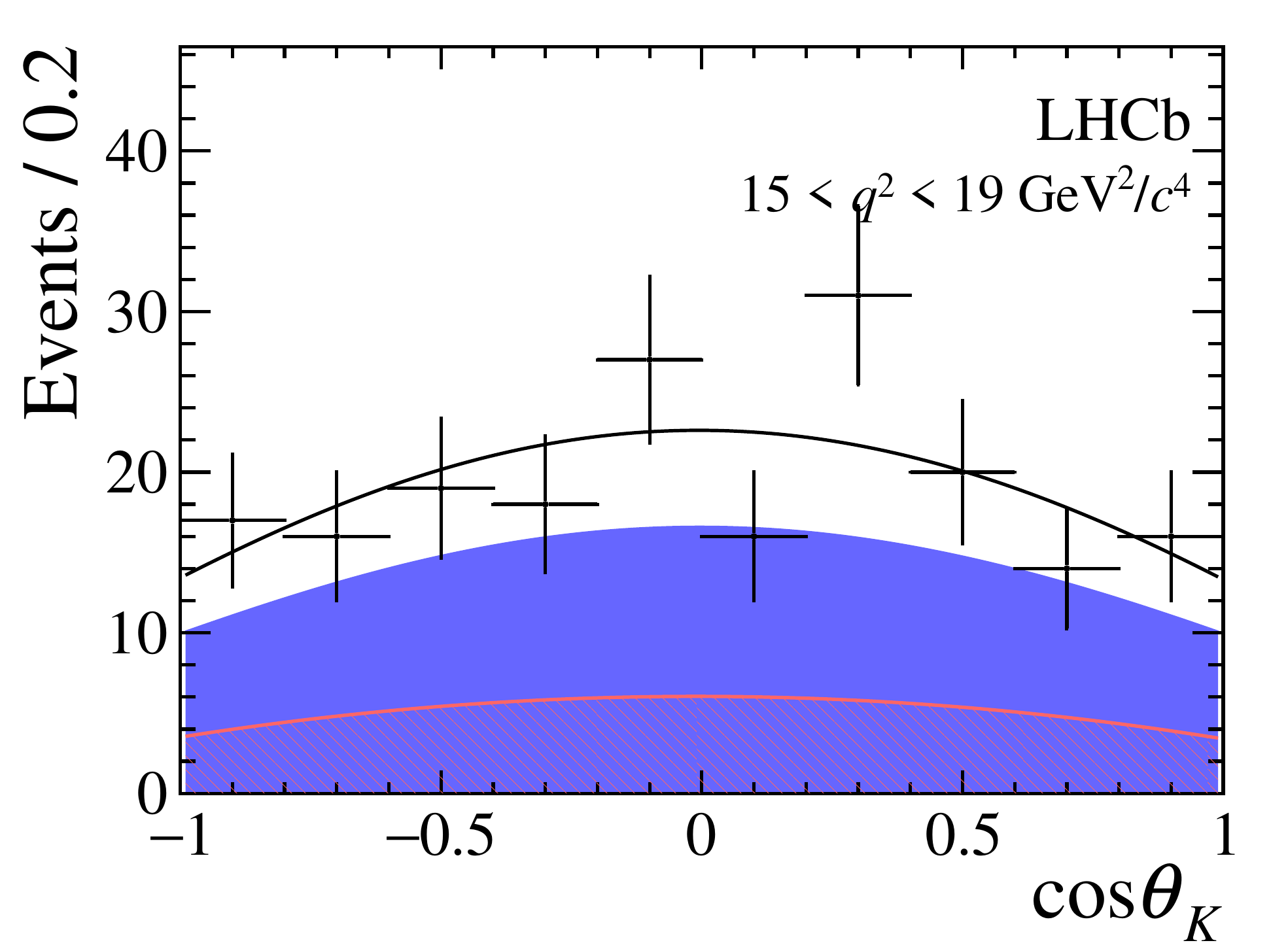}
\includegraphics[width=0.32\textwidth]{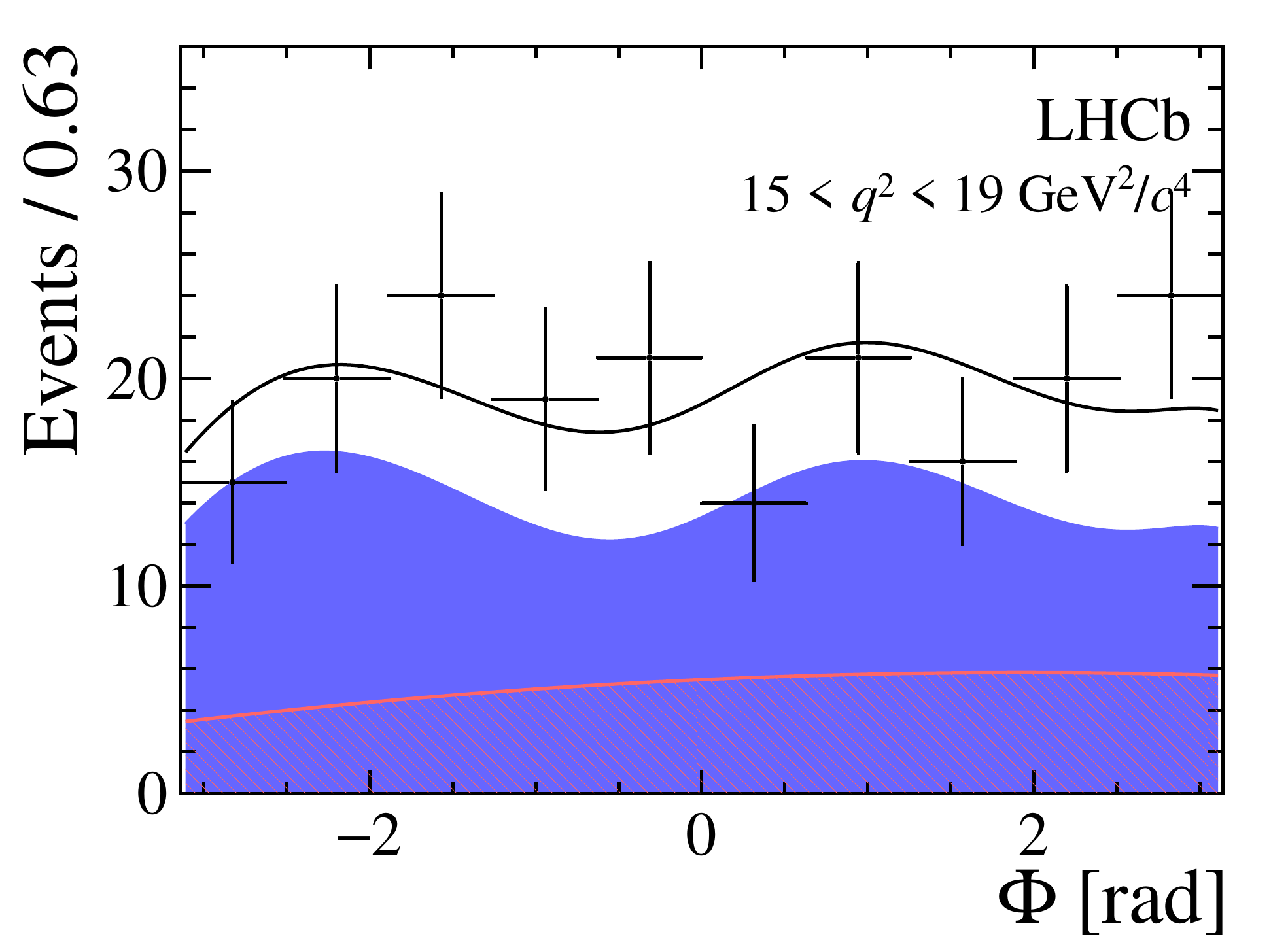}\\
\caption{\label{fig:ang-proj2} One-dimensional projections of the fit to the angles $\cos\theta_l$, $\cos\theta_K$, $\Phi$ in bins of $q^2$. The signal component is shown
by the solid blue area, the background component by the shaded red area.}
\end{center}
\end{figure}

\clearpage
\section{Confidence intervals}
\label{app:fc}
\vspace*{-0.3cm}
\begin{figure}[h]
\begin{center}
\includegraphics[width=0.37\textwidth]{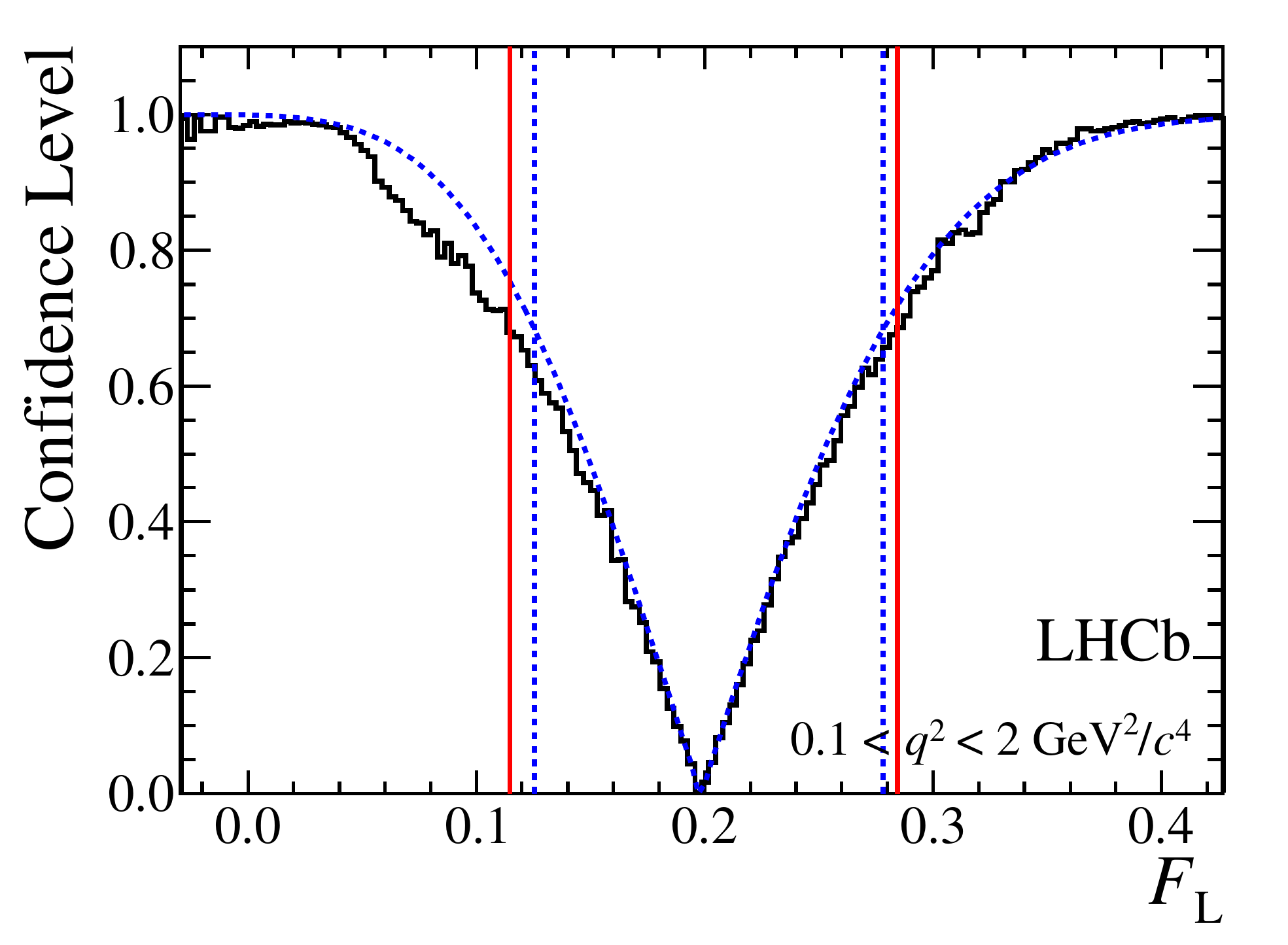}
\includegraphics[width=0.37\textwidth]{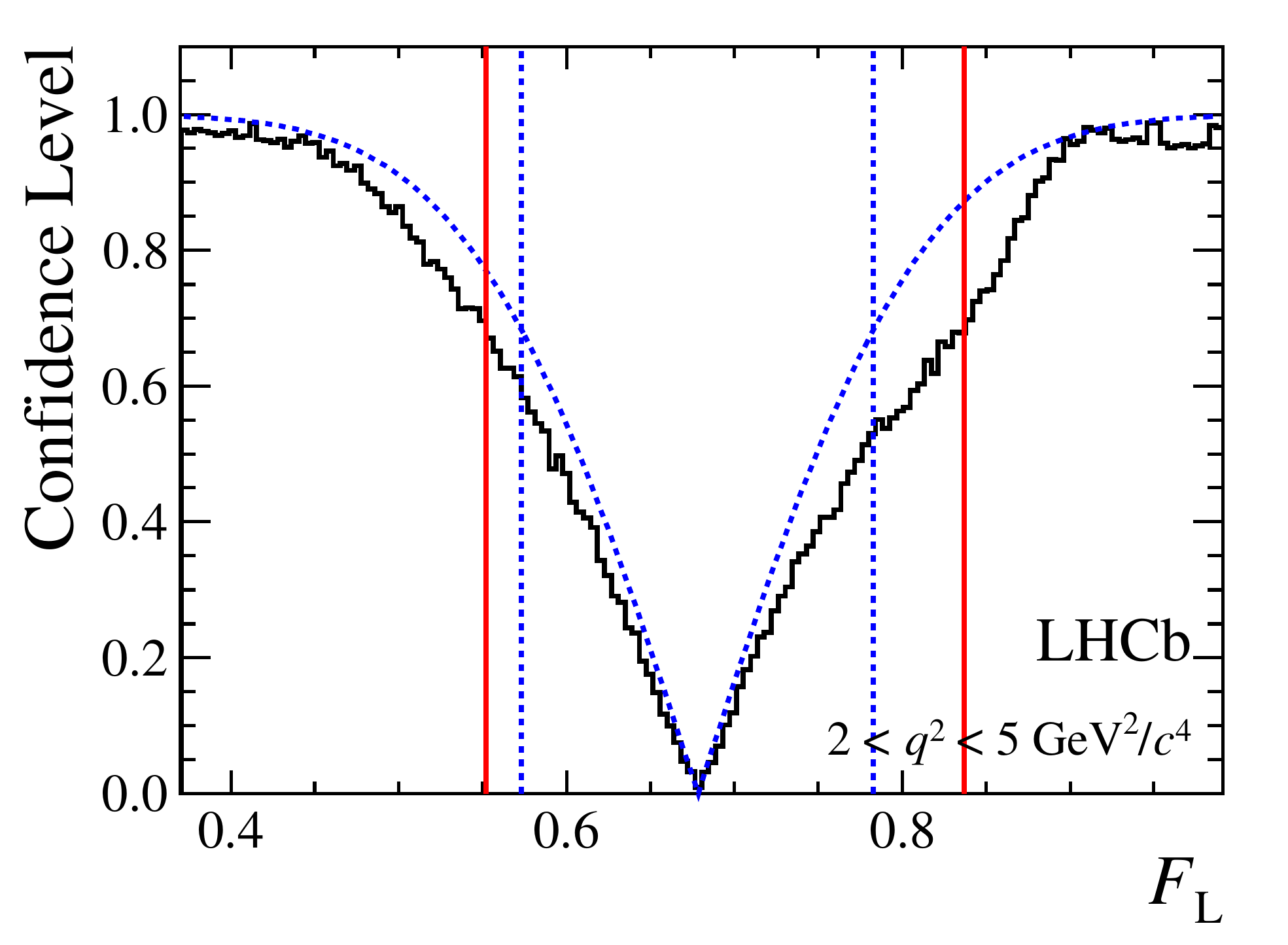}\\
\includegraphics[width=0.37\textwidth]{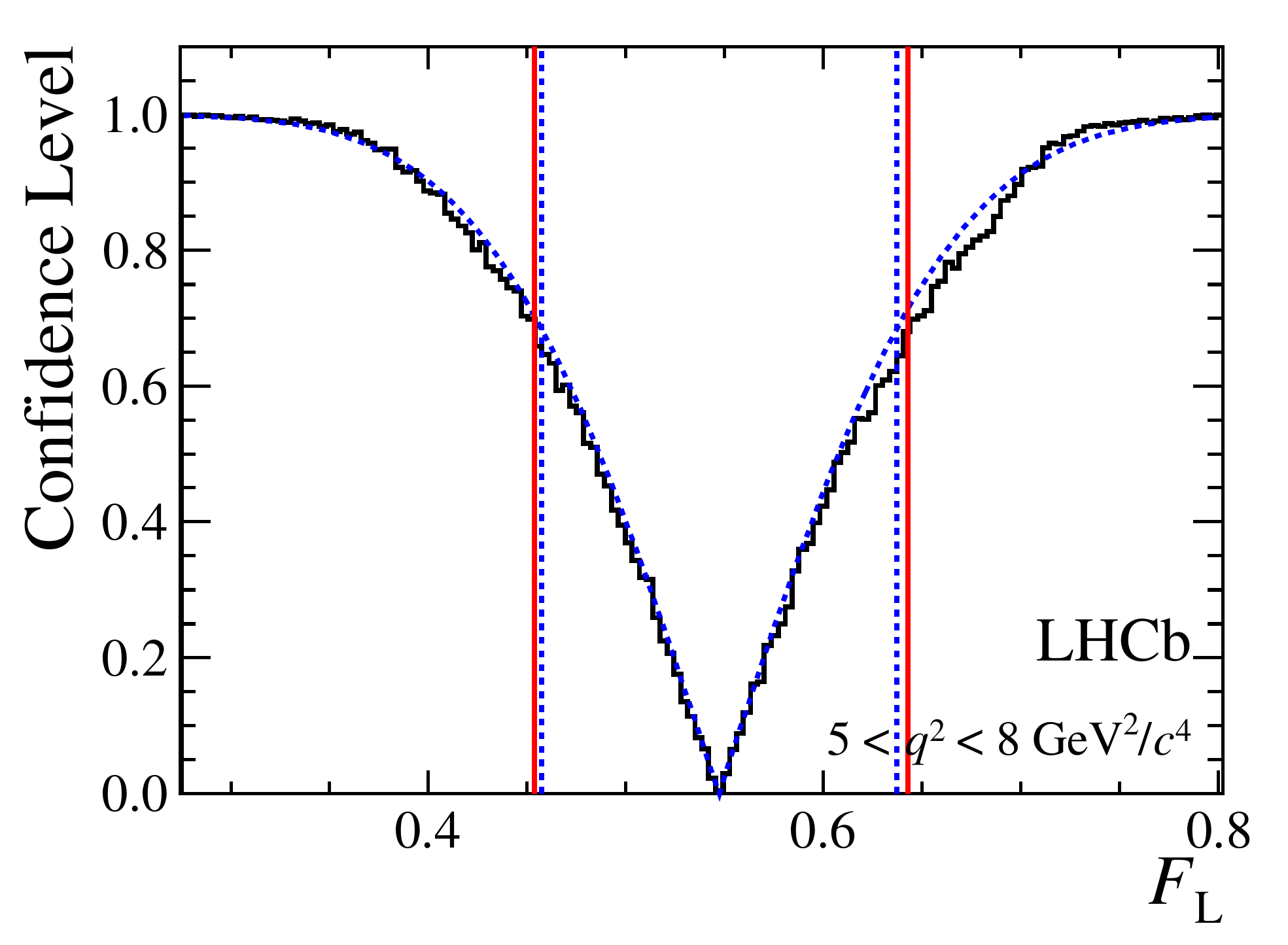}
\includegraphics[width=0.37\textwidth]{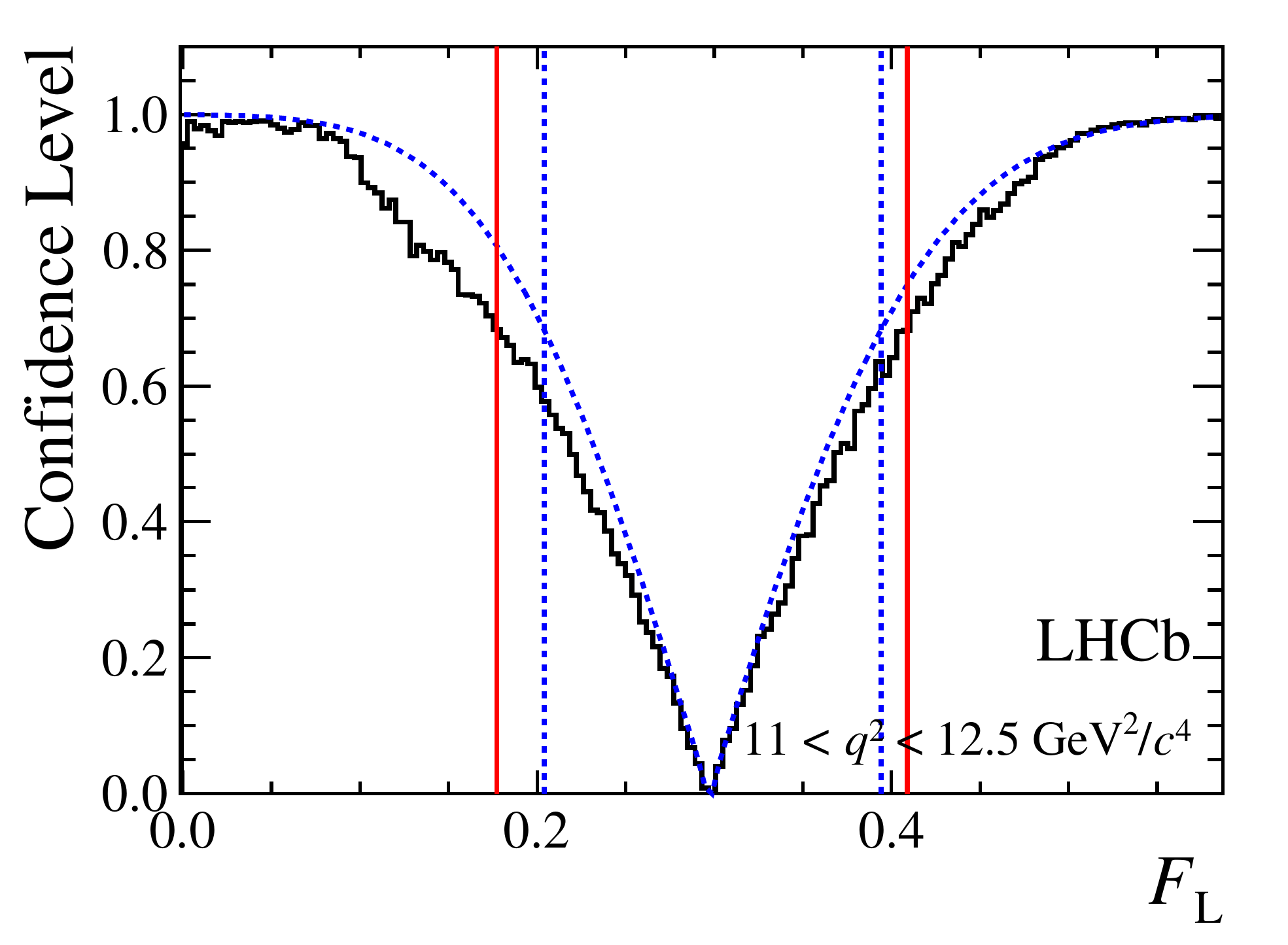}\\
\includegraphics[width=0.37\textwidth]{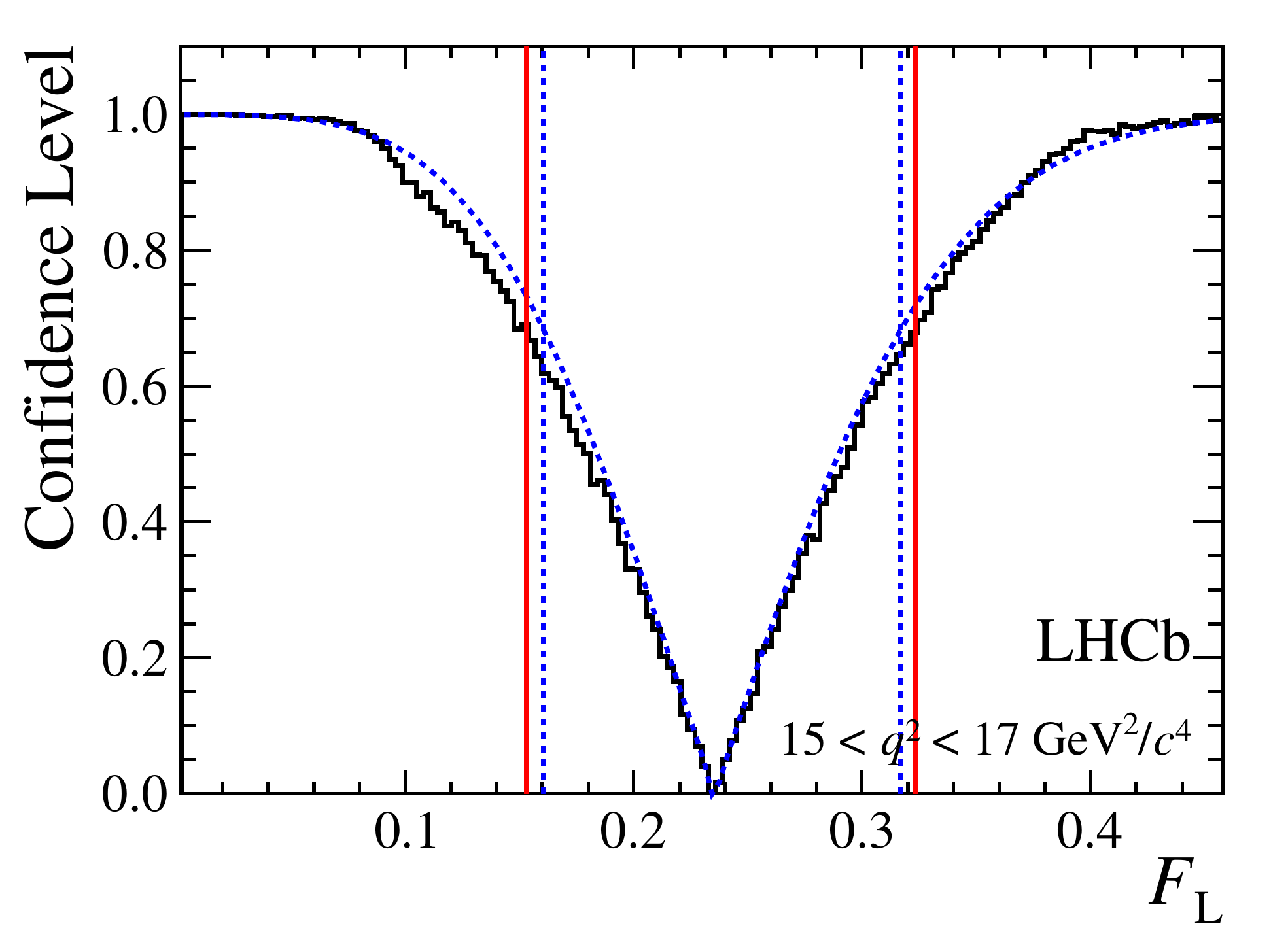}
\includegraphics[width=0.37\textwidth]{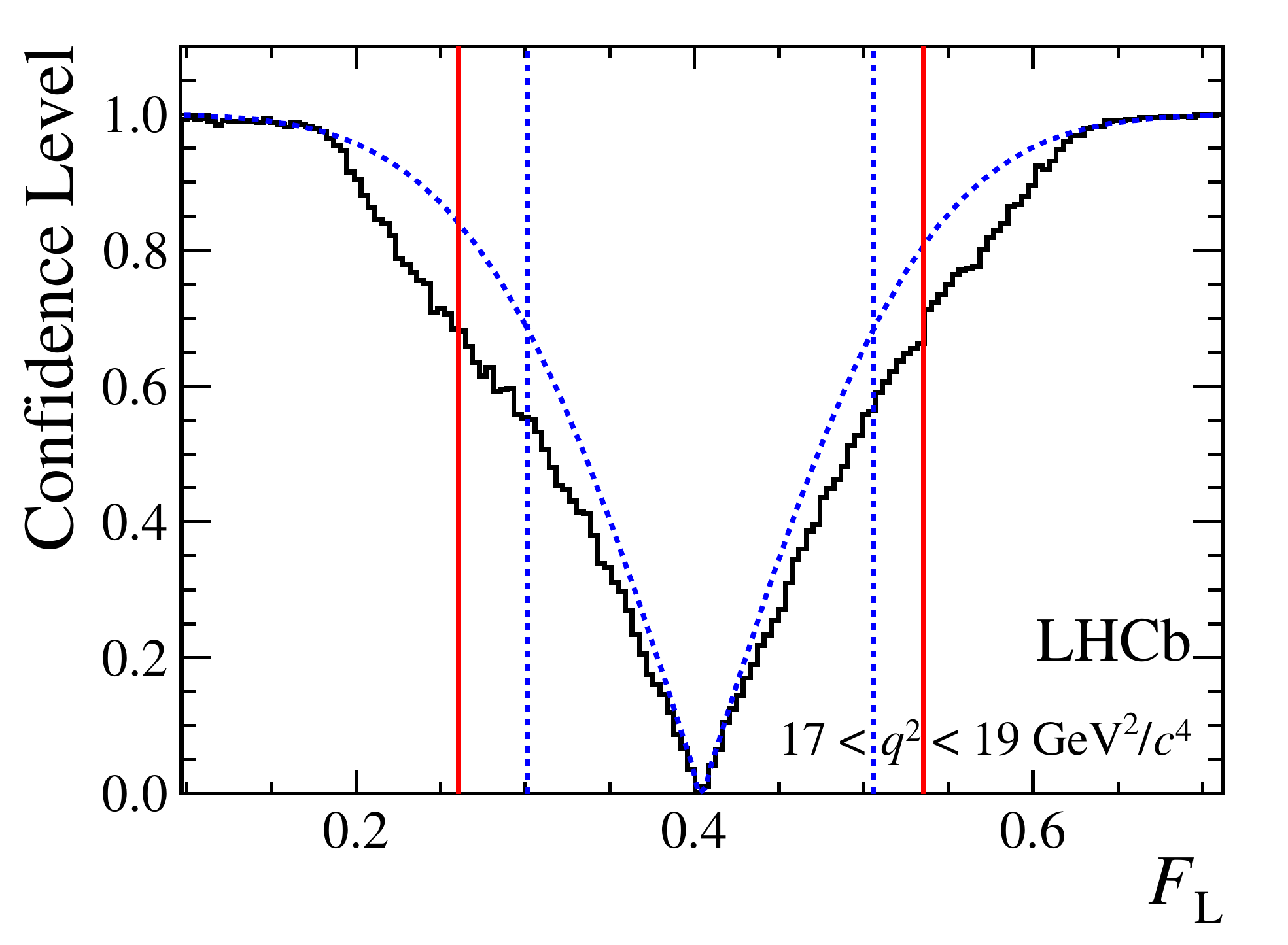}\\
\includegraphics[width=0.37\textwidth]{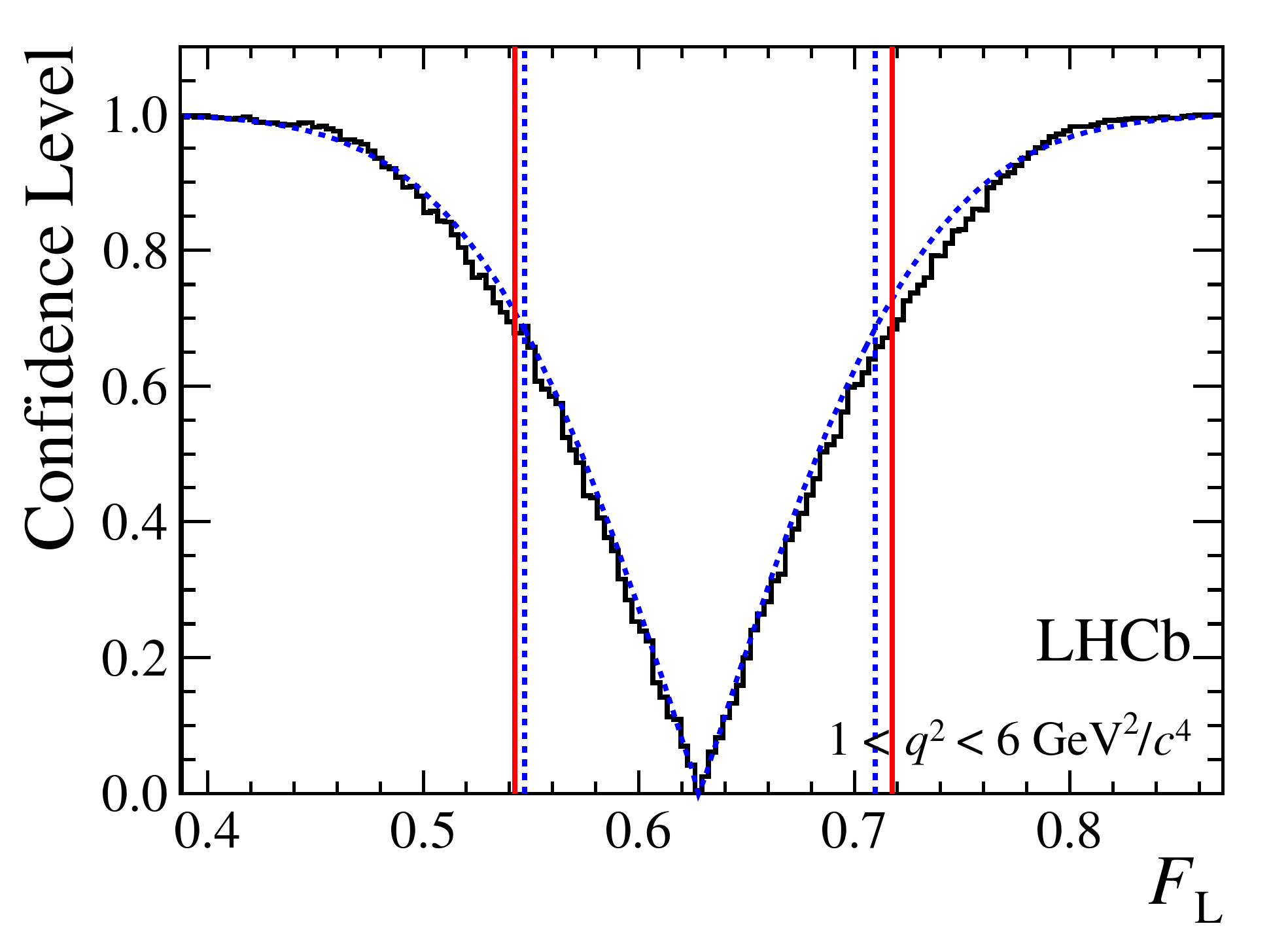}
\includegraphics[width=0.37\textwidth]{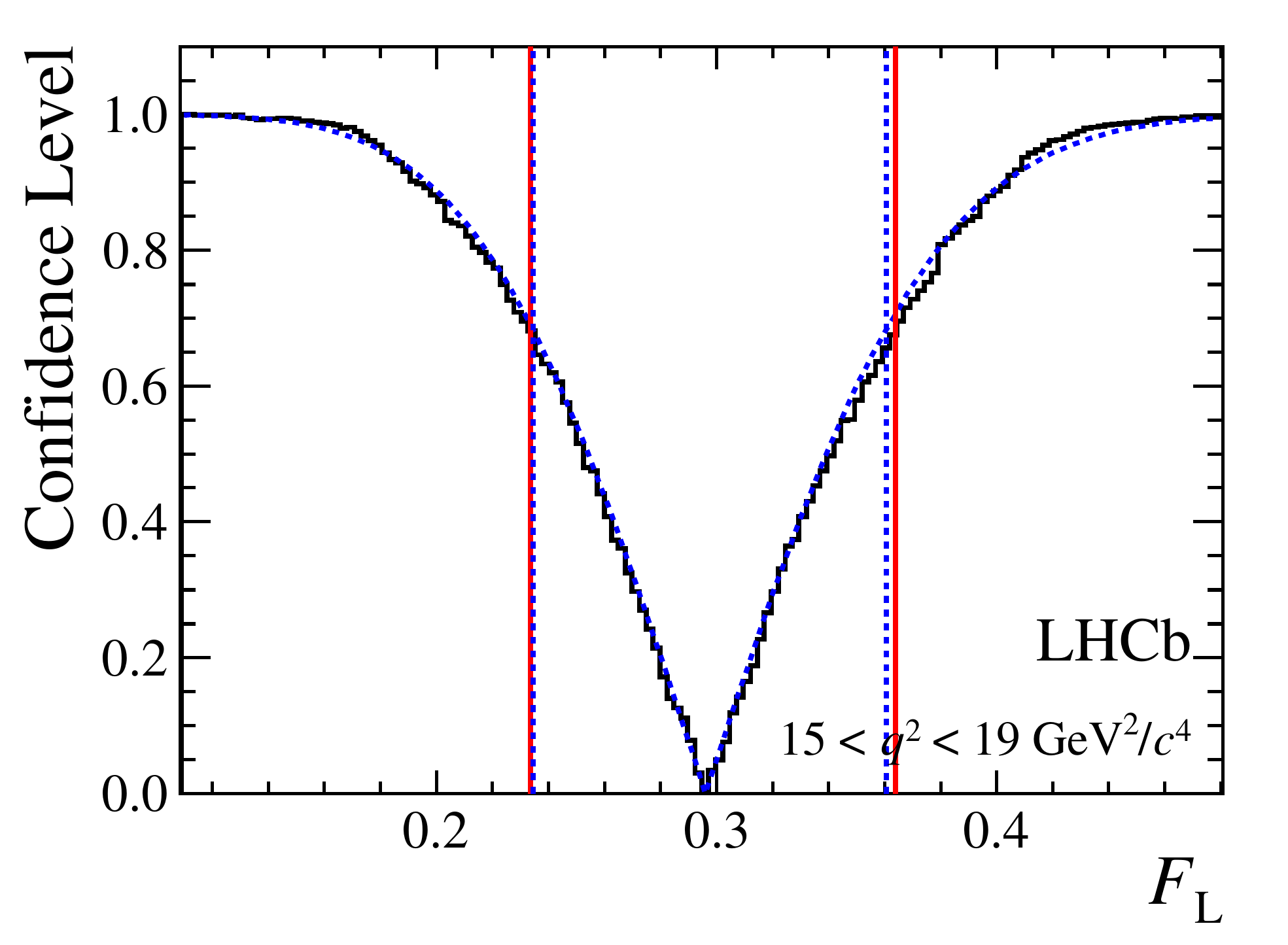}\\
\caption{\label{fig:ang-fc1} Confidence level obtained from a likelihood scan (shaded blue) and from the Feldman-Cousins method (solid black).
The shaded blue and solid red vertical lines indicate the corresponding $68\%$ CL intervals obtained from the likelihood scan and the Feldman-Cousins method, respectively.}
\end{center}
\end{figure}

\begin{figure}[h]
\begin{center}
\includegraphics[width=0.37\textwidth]{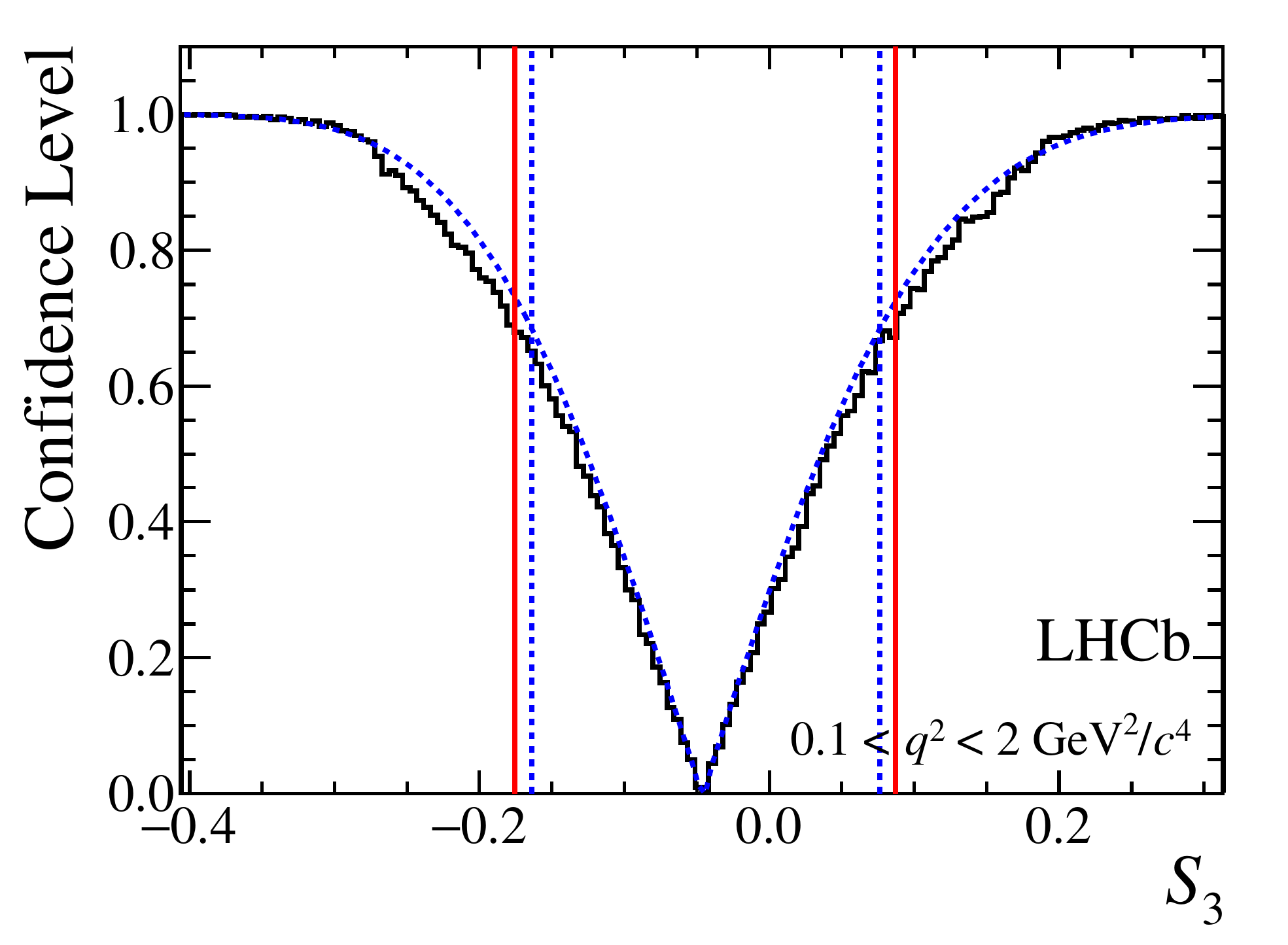}
\includegraphics[width=0.37\textwidth]{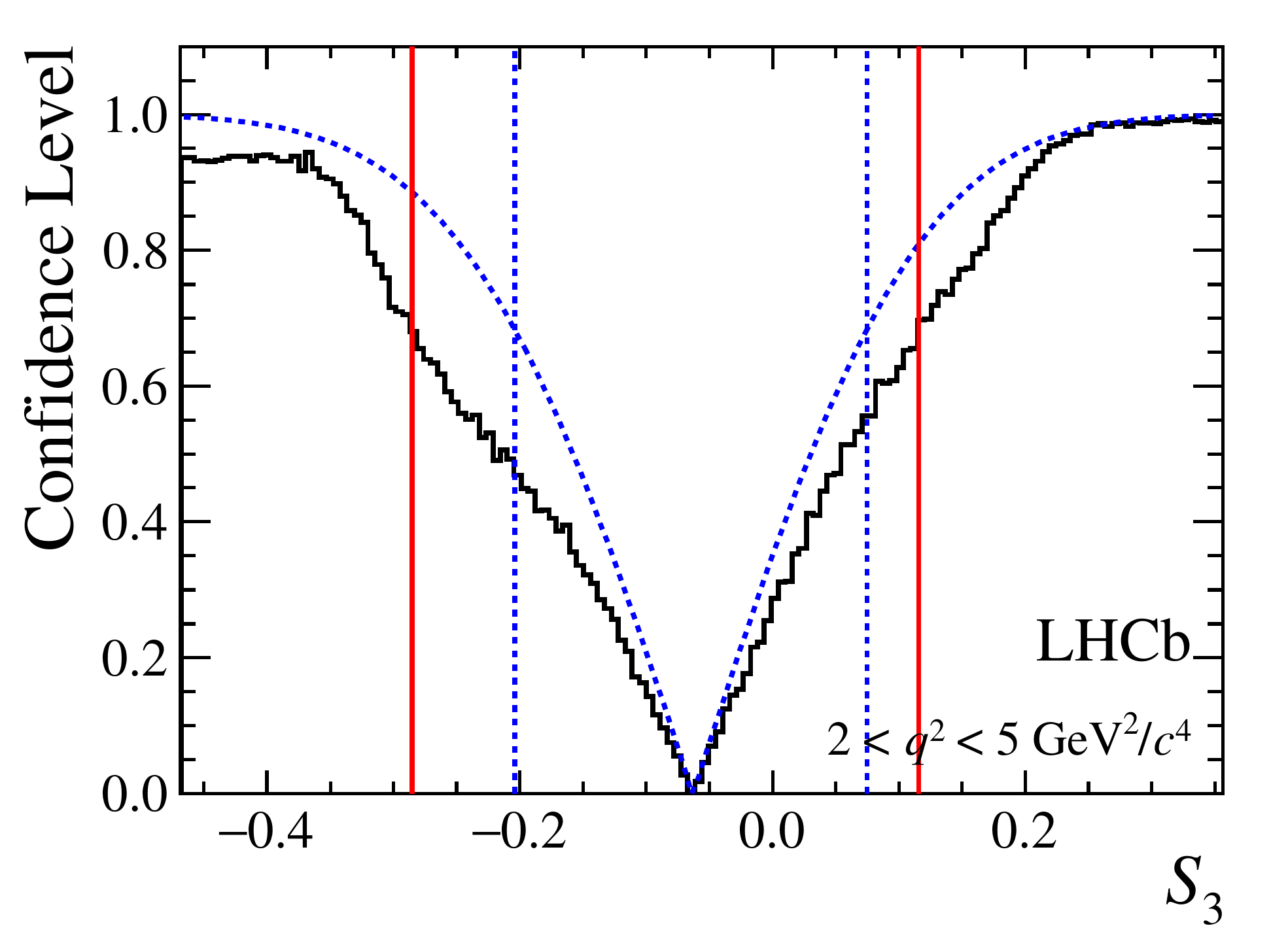}\\
\includegraphics[width=0.37\textwidth]{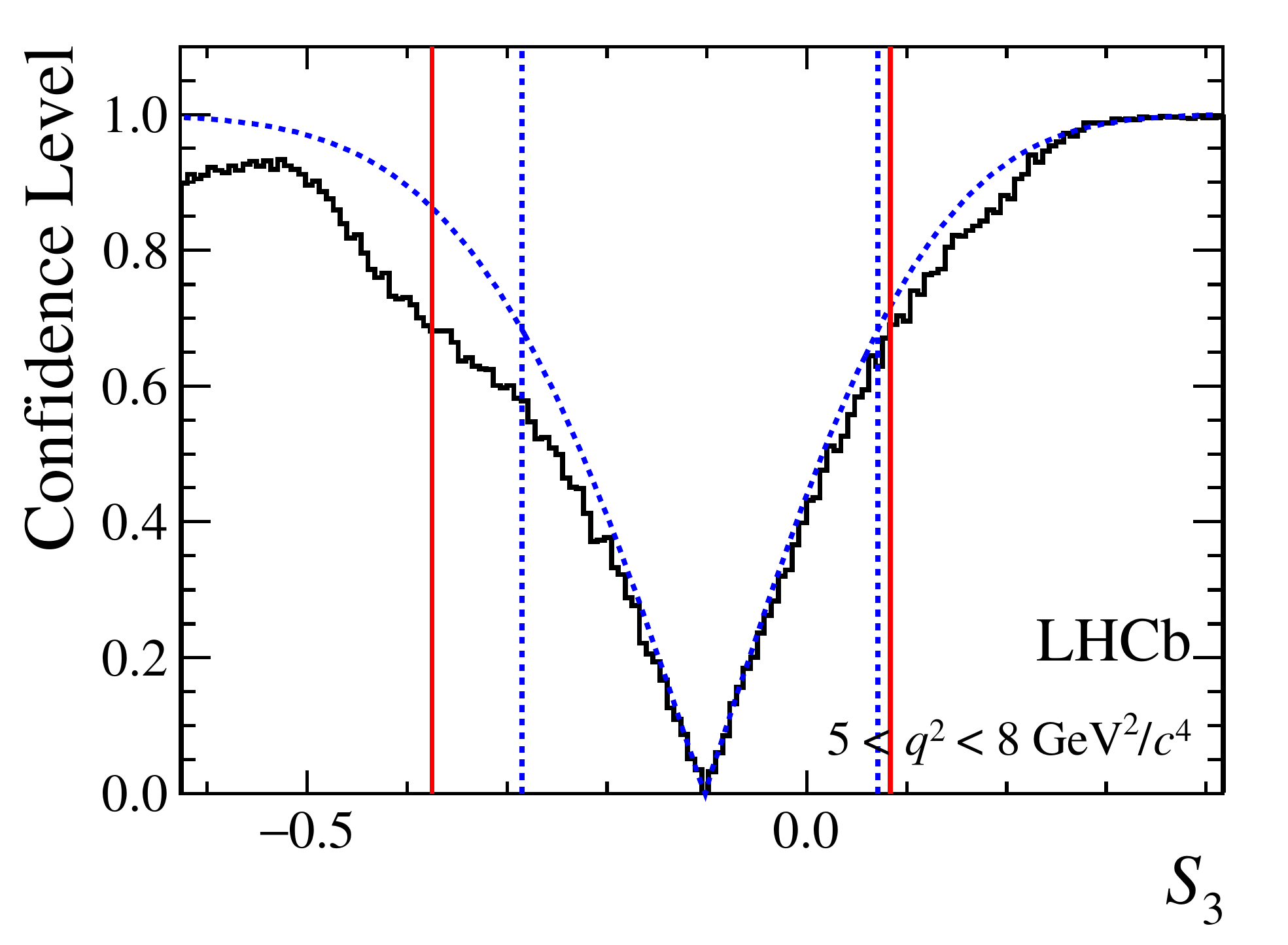}
\includegraphics[width=0.37\textwidth]{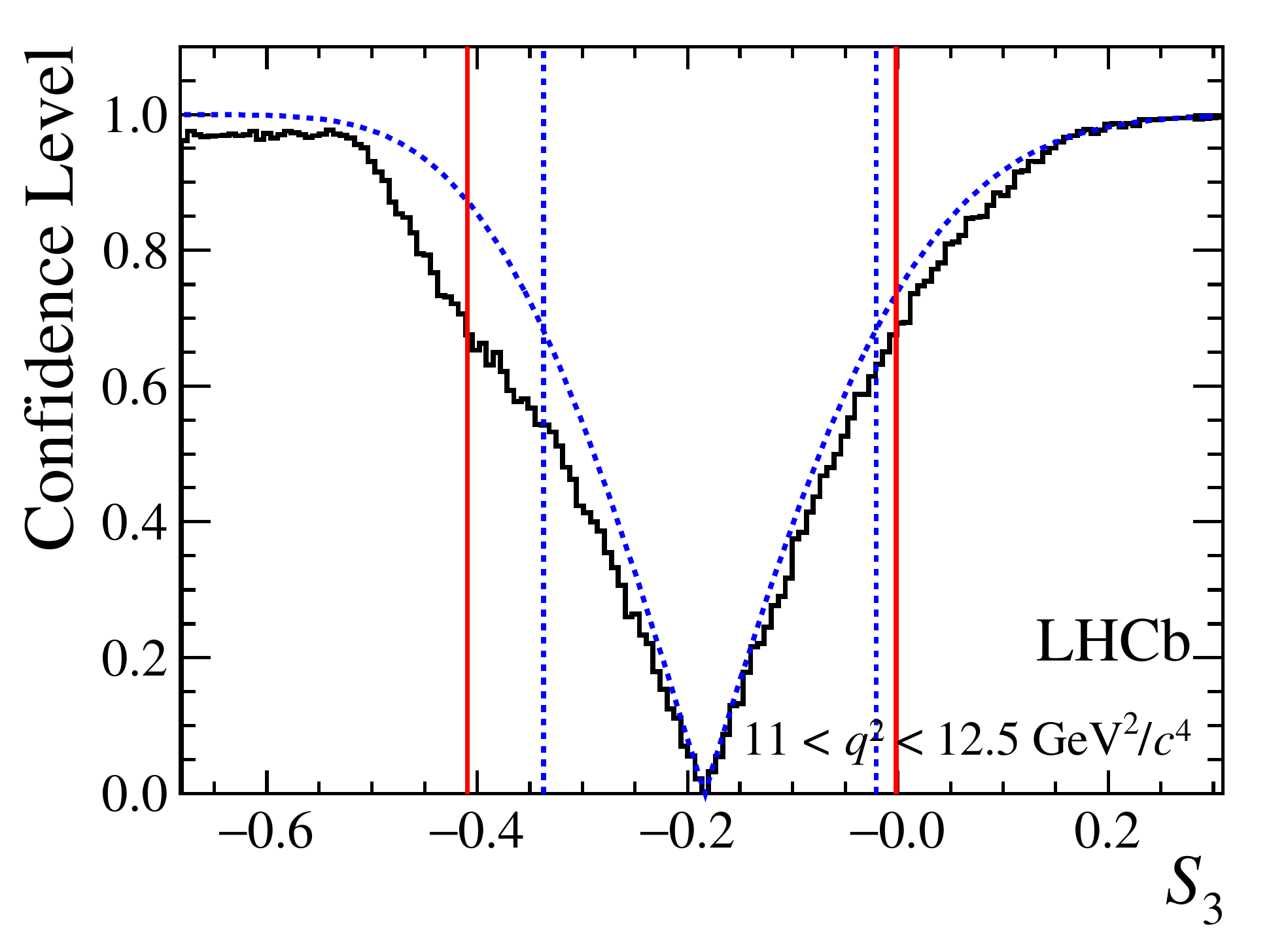}\\
\includegraphics[width=0.37\textwidth]{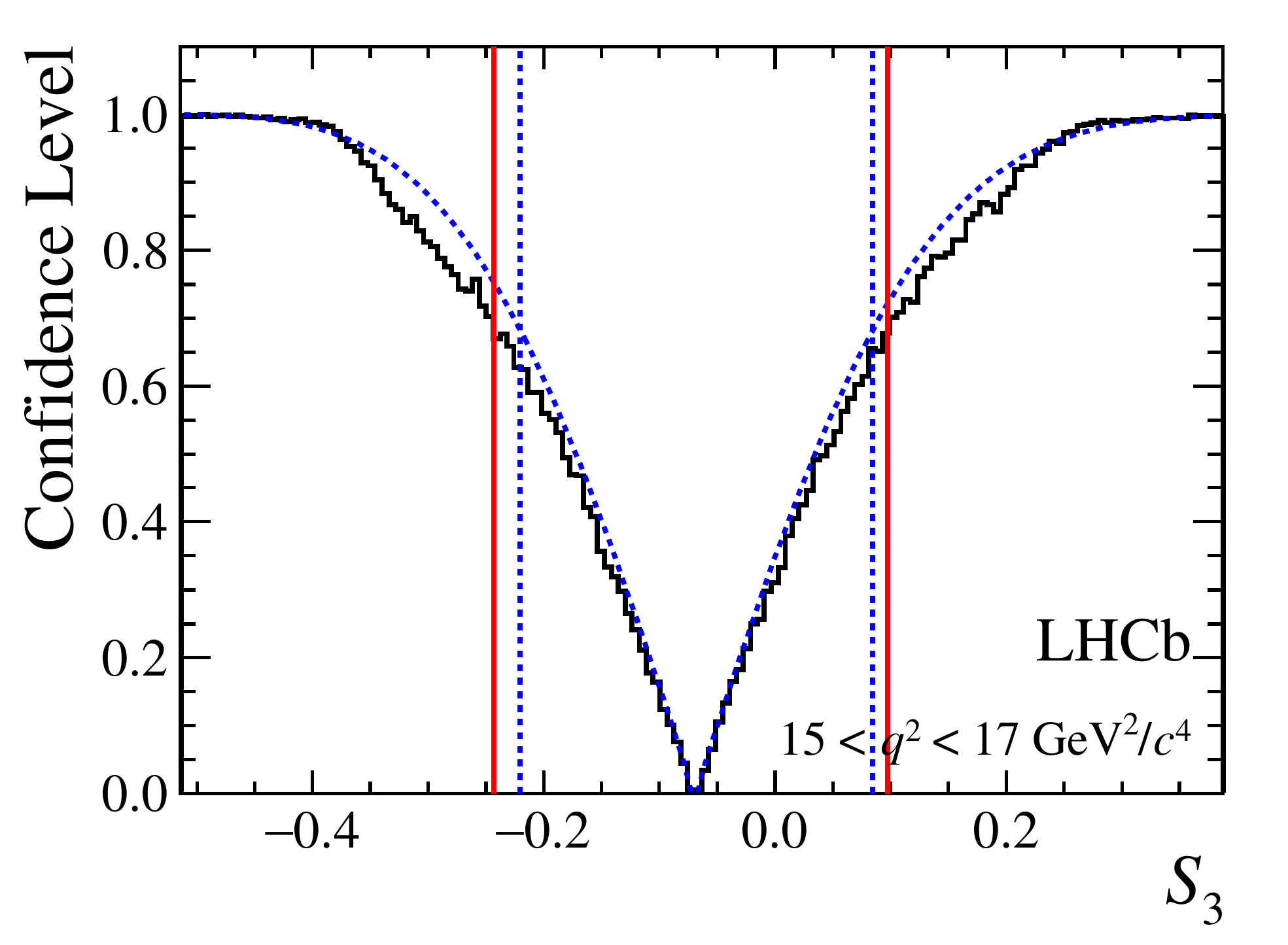}
\includegraphics[width=0.37\textwidth]{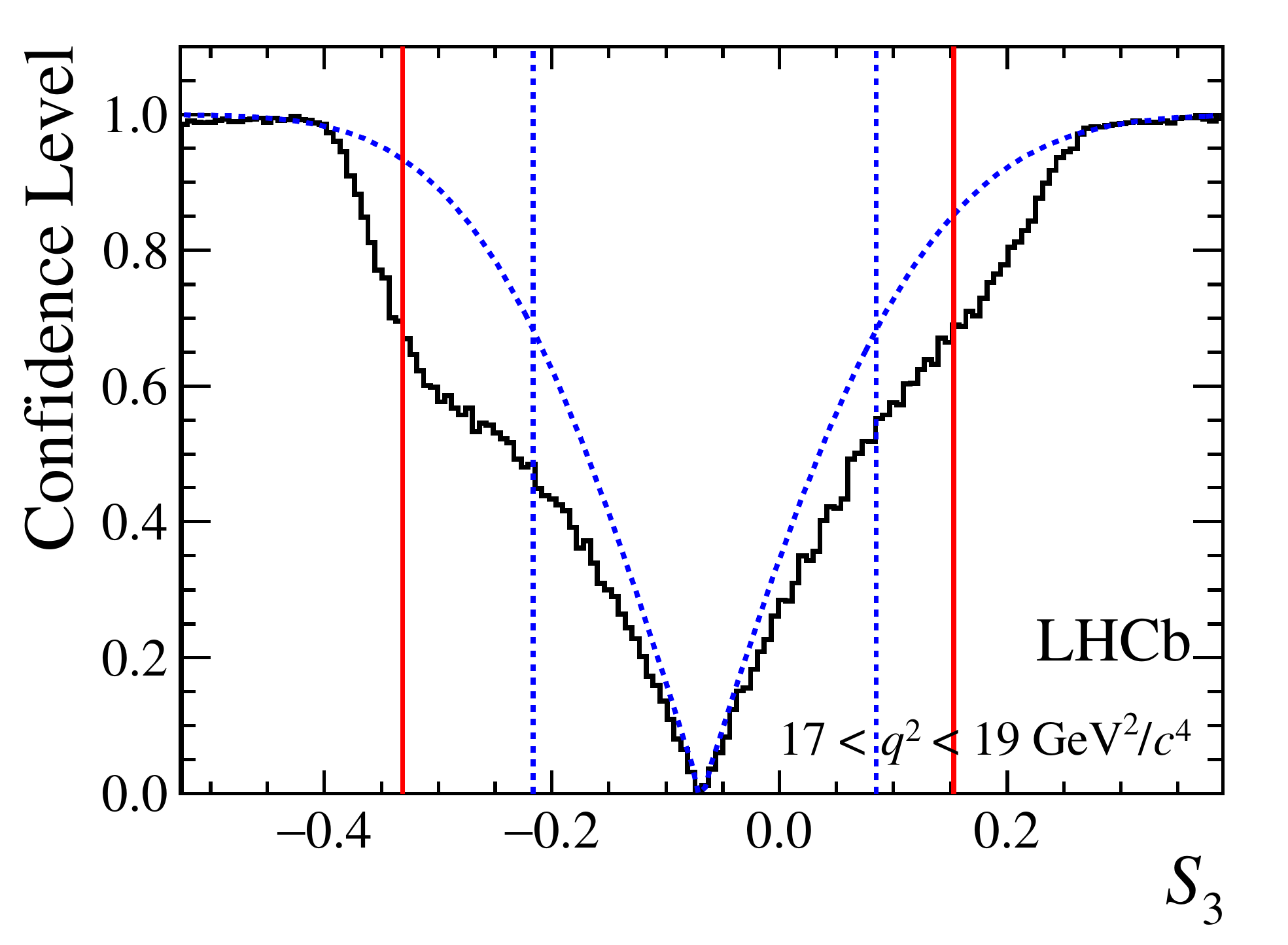}\\
\includegraphics[width=0.37\textwidth]{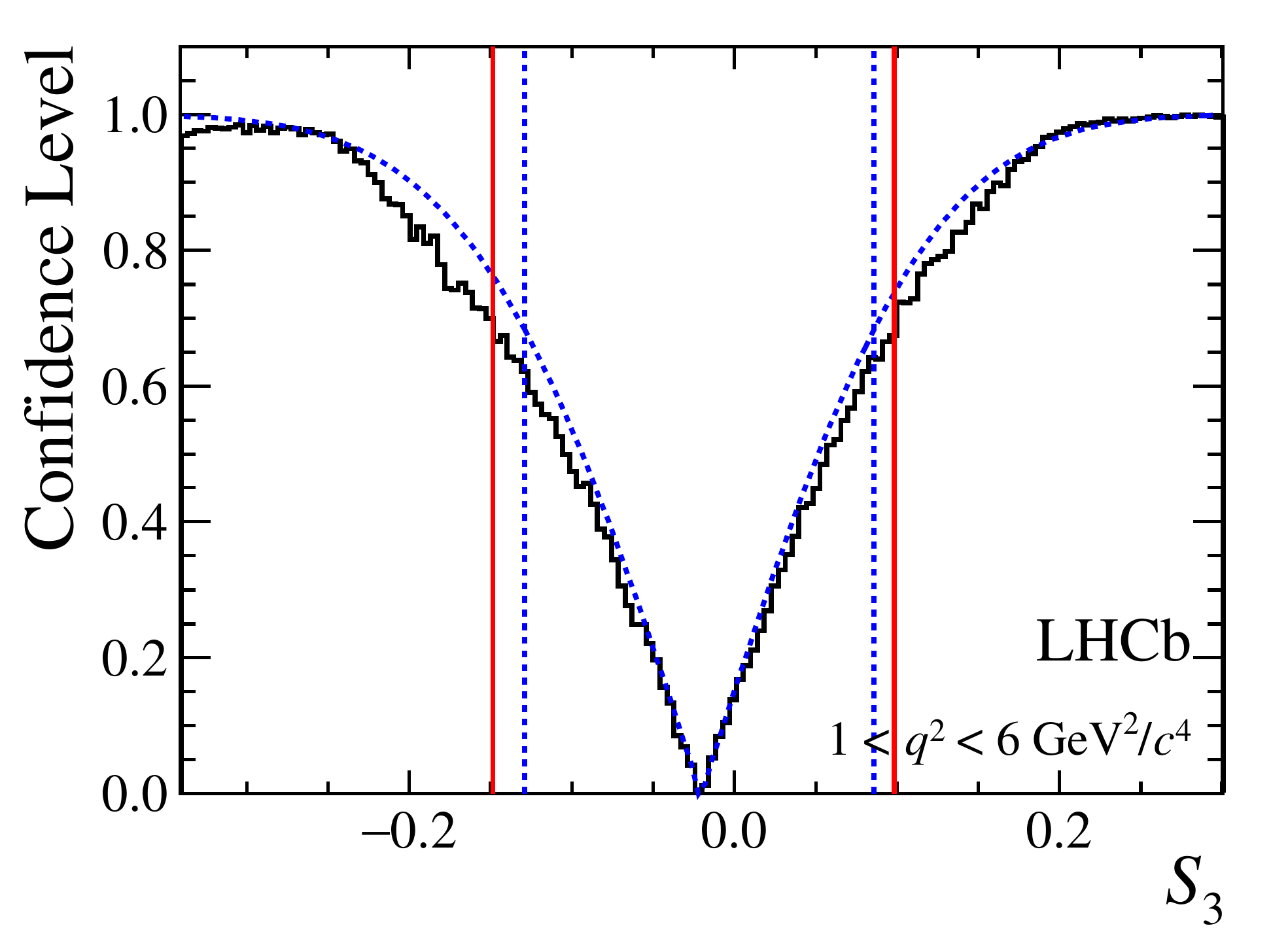}
\includegraphics[width=0.37\textwidth]{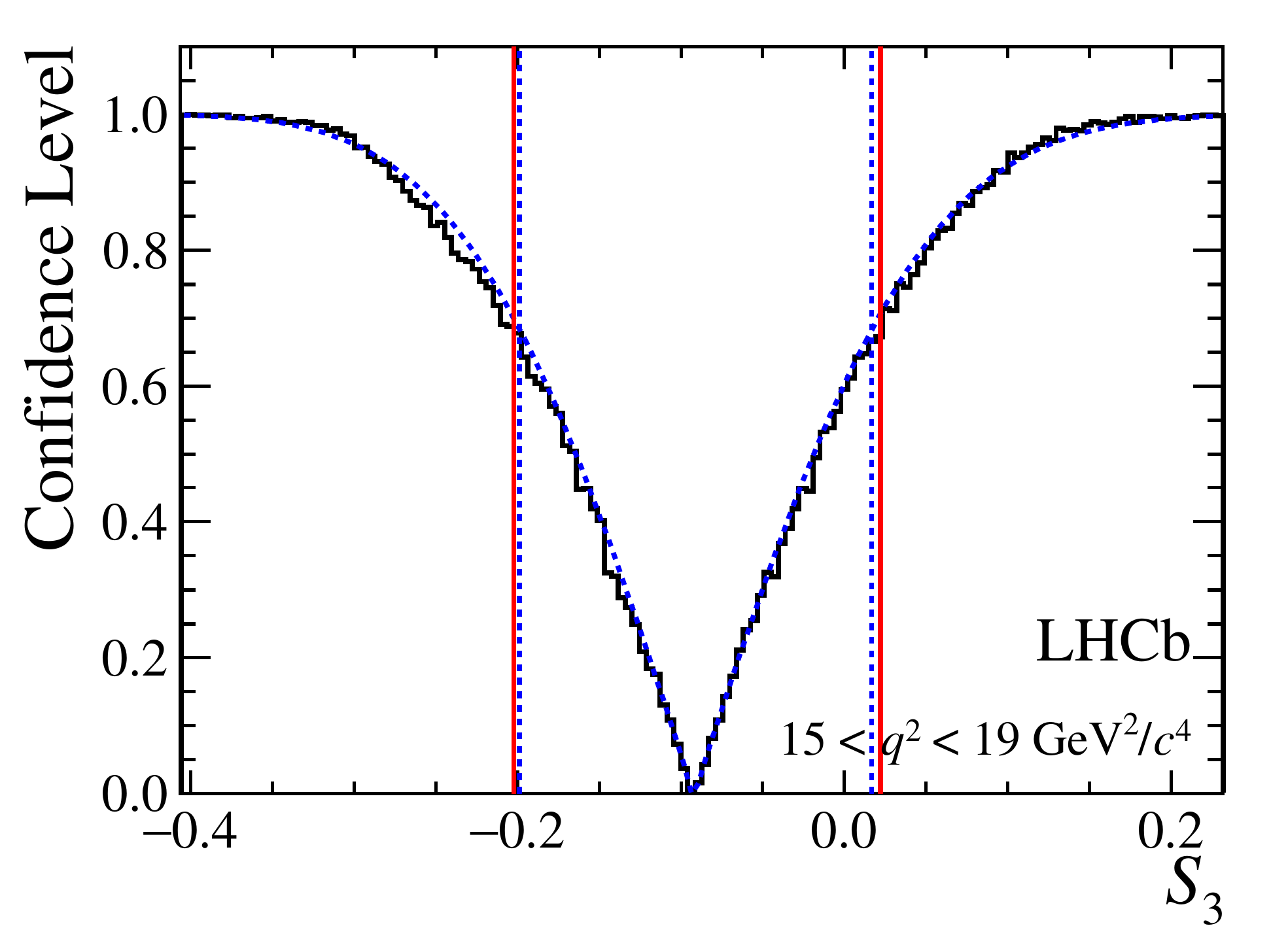}\\
\caption{\label{fig:ang-fc3} Confidence level obtained from a likelihood scan (shaded blue) and from a Feldman-Cousins method (solid black).
The shaded blue and solid red vertical lines indicate the corresponding $68\%$ CL intervals obtained from the likelihood scan and the Feldman-Cousins method, respectively.}
\end{center}
\end{figure}

\begin{figure}[h]
\begin{center}
\includegraphics[width=0.37\textwidth]{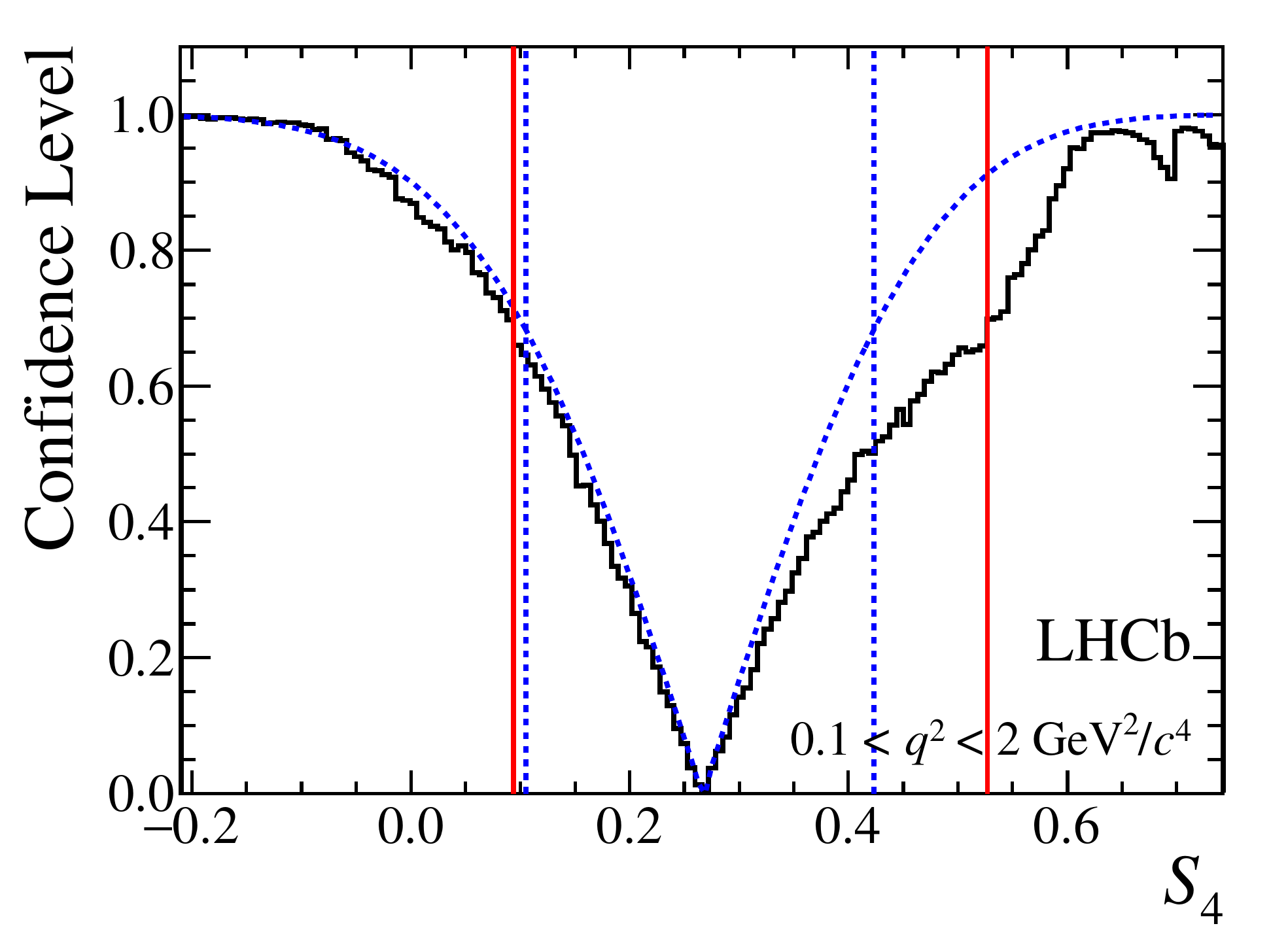}
\includegraphics[width=0.37\textwidth]{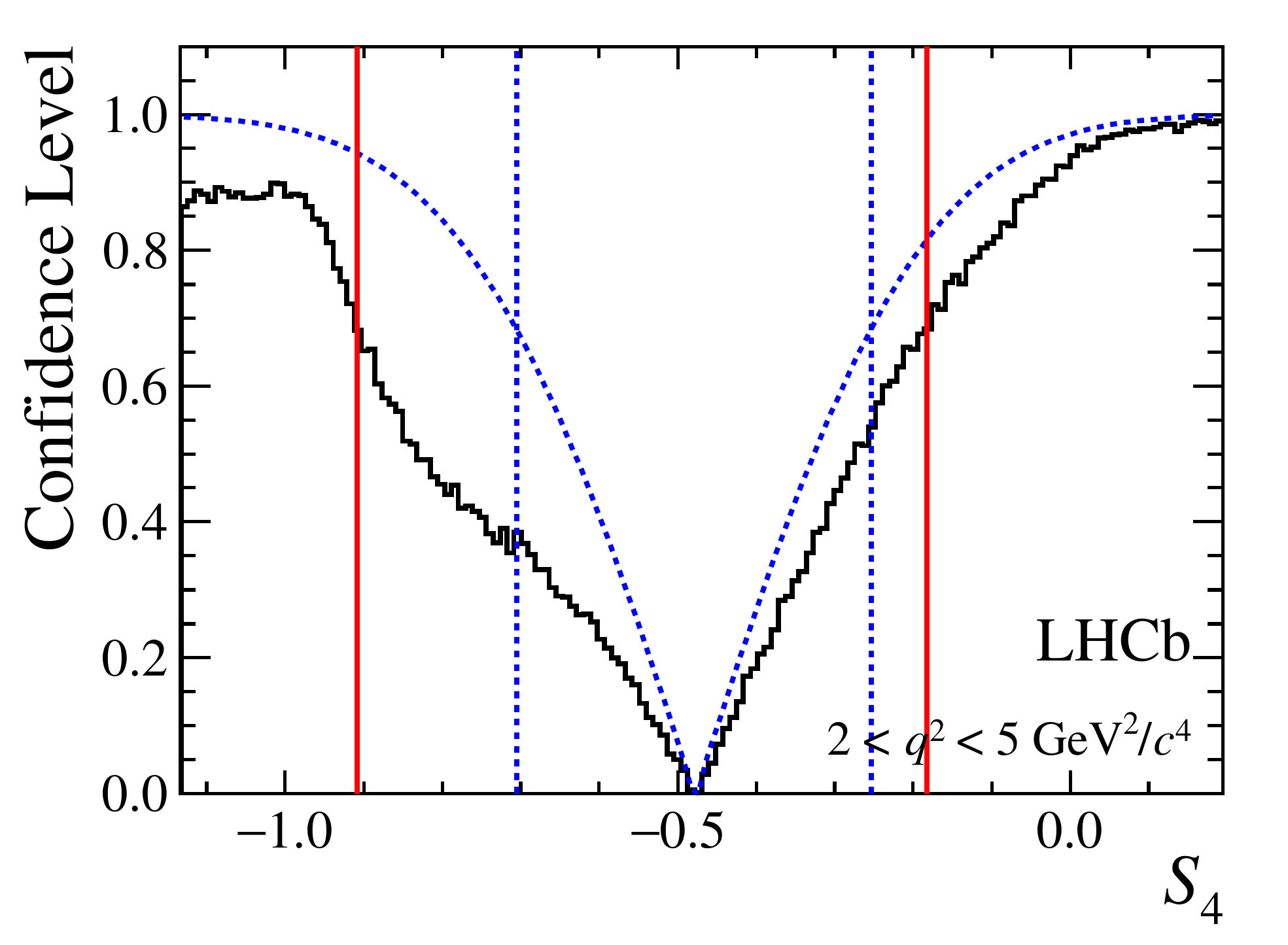}\\
\includegraphics[width=0.37\textwidth]{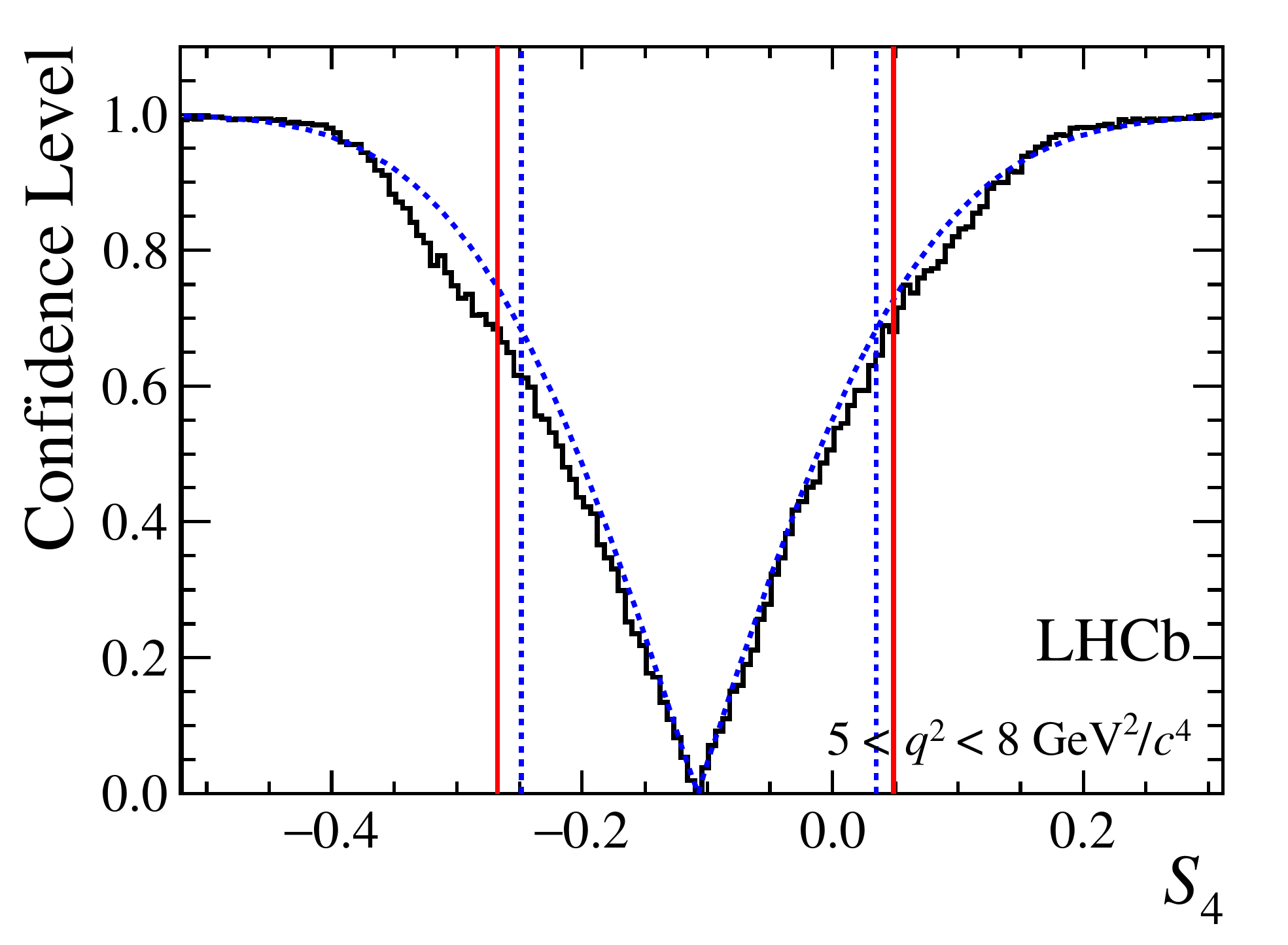}
\includegraphics[width=0.37\textwidth]{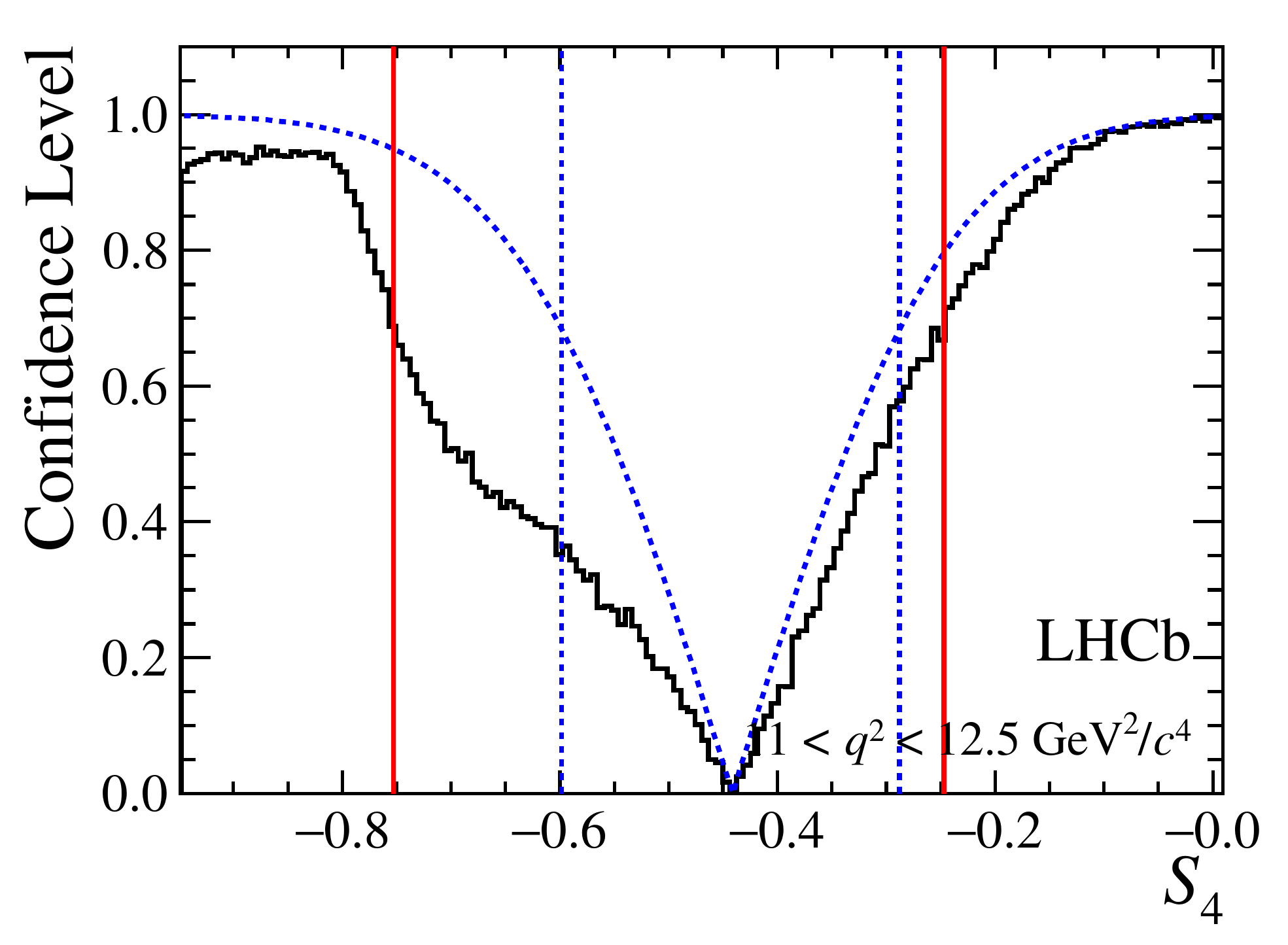}\\
\includegraphics[width=0.37\textwidth]{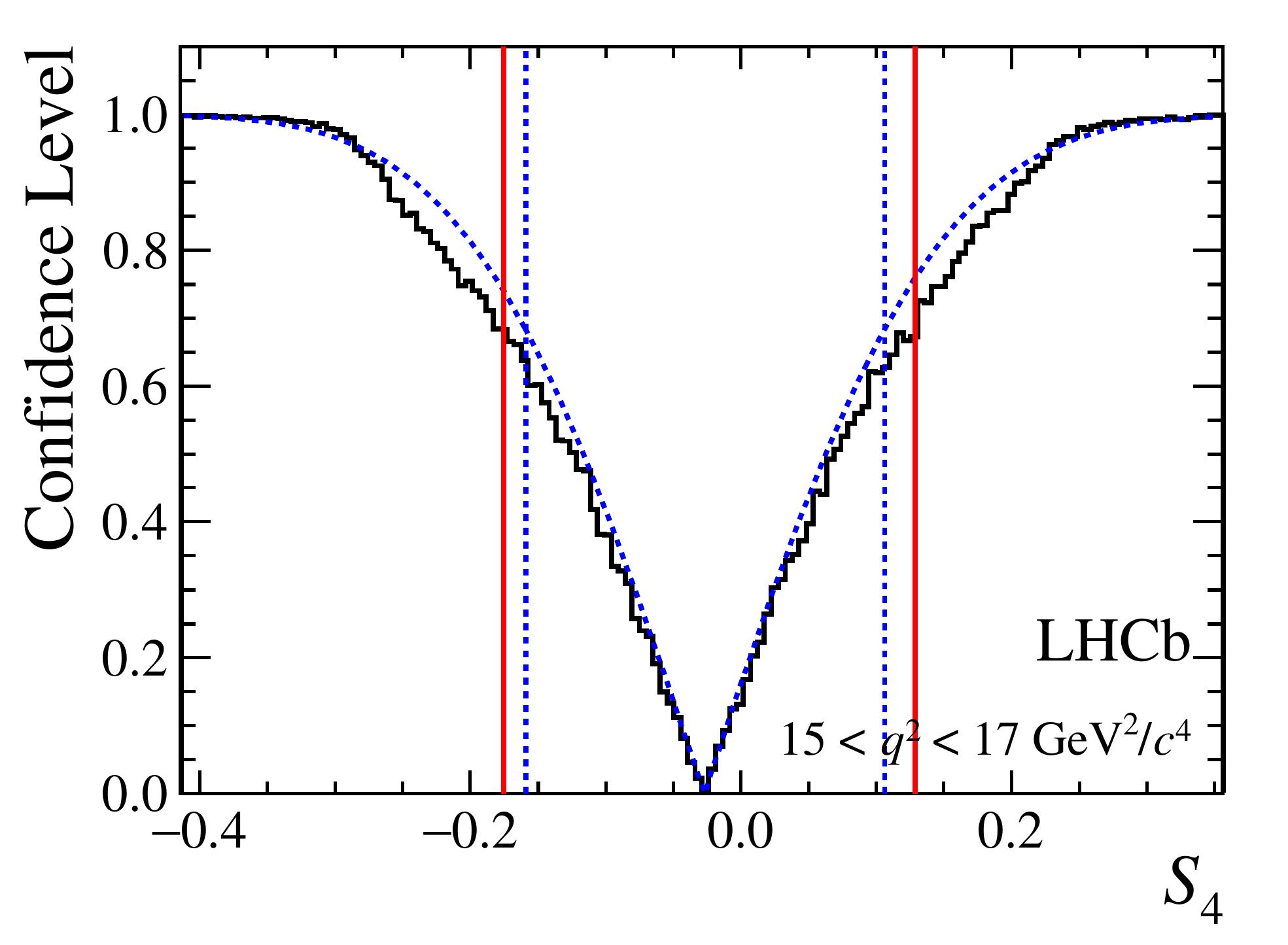}
\includegraphics[width=0.37\textwidth]{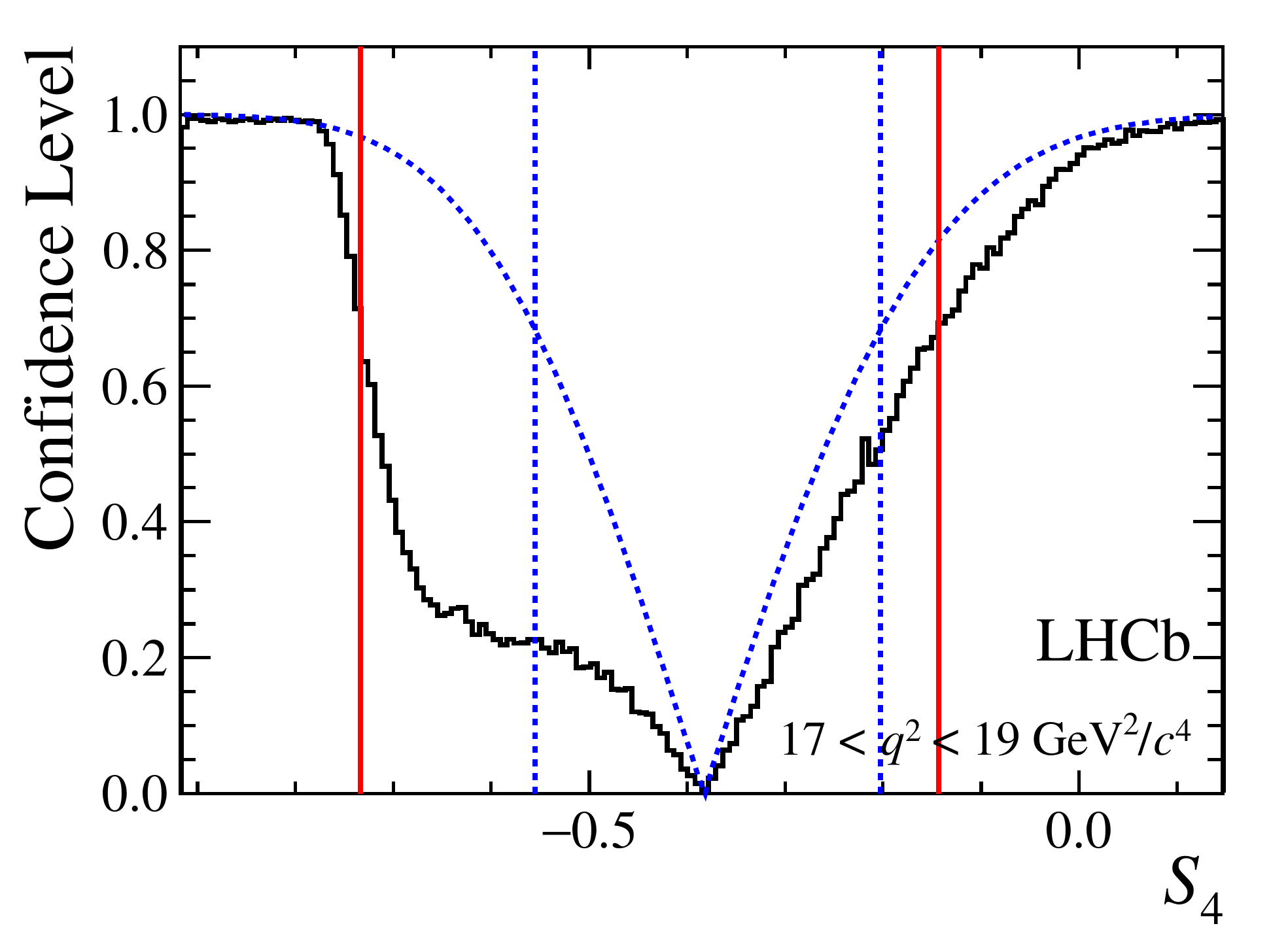}\\
\includegraphics[width=0.37\textwidth]{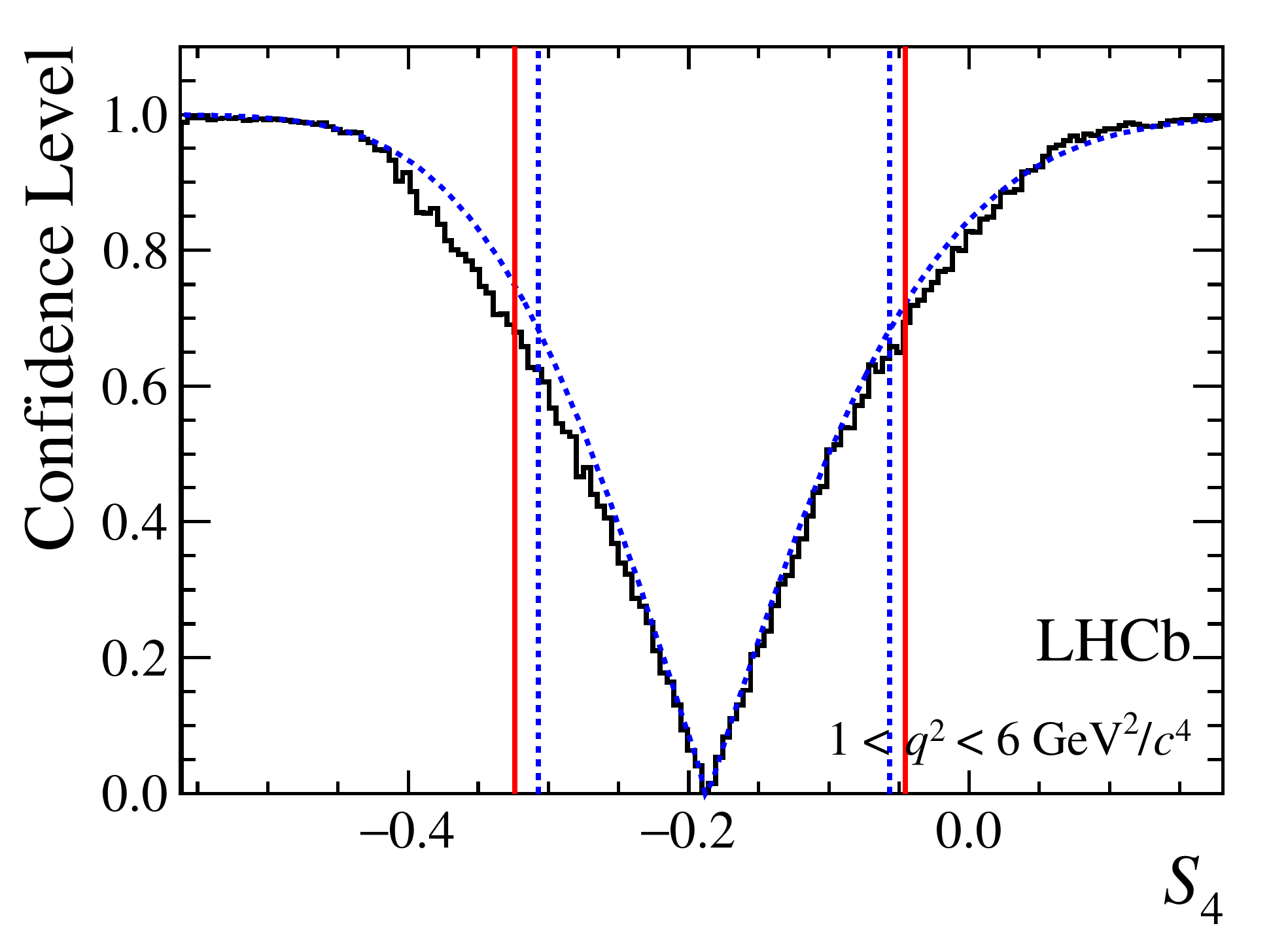}
\includegraphics[width=0.37\textwidth]{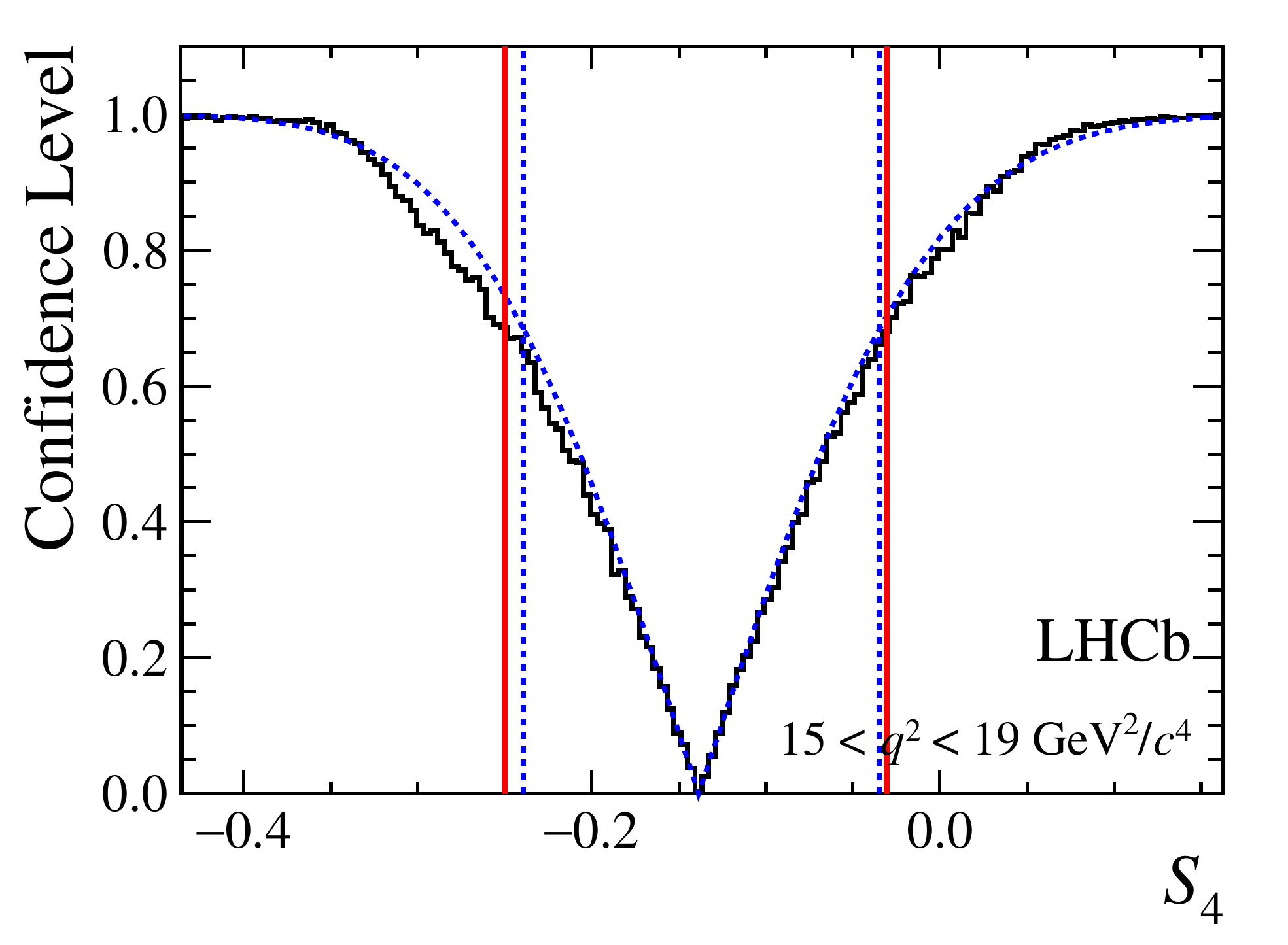}\\
\caption{\label{fig:ang-fc4} Confidence level obtained from a likelihood scan (shaded blue) and from a Feldman-Cousins method (solid black).
The shaded blue and solid red vertical lines indicate the corresponding $68\%$ CL intervals obtained from the likelihood scan and the Feldman-Cousins method, respectively.}
\end{center}
\end{figure}

\begin{figure}[h]
\begin{center}
\includegraphics[width=0.37\textwidth]{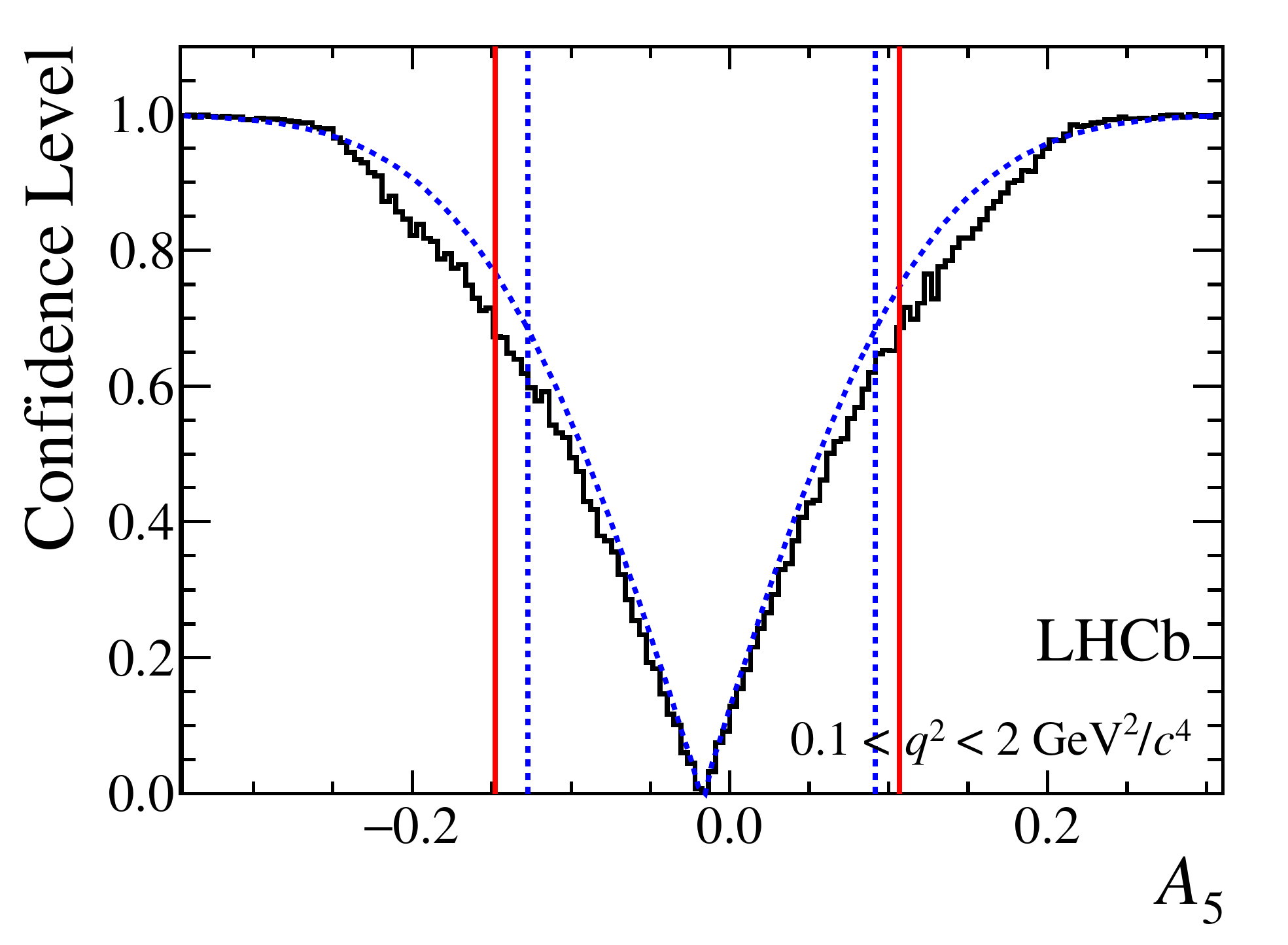}
\includegraphics[width=0.37\textwidth]{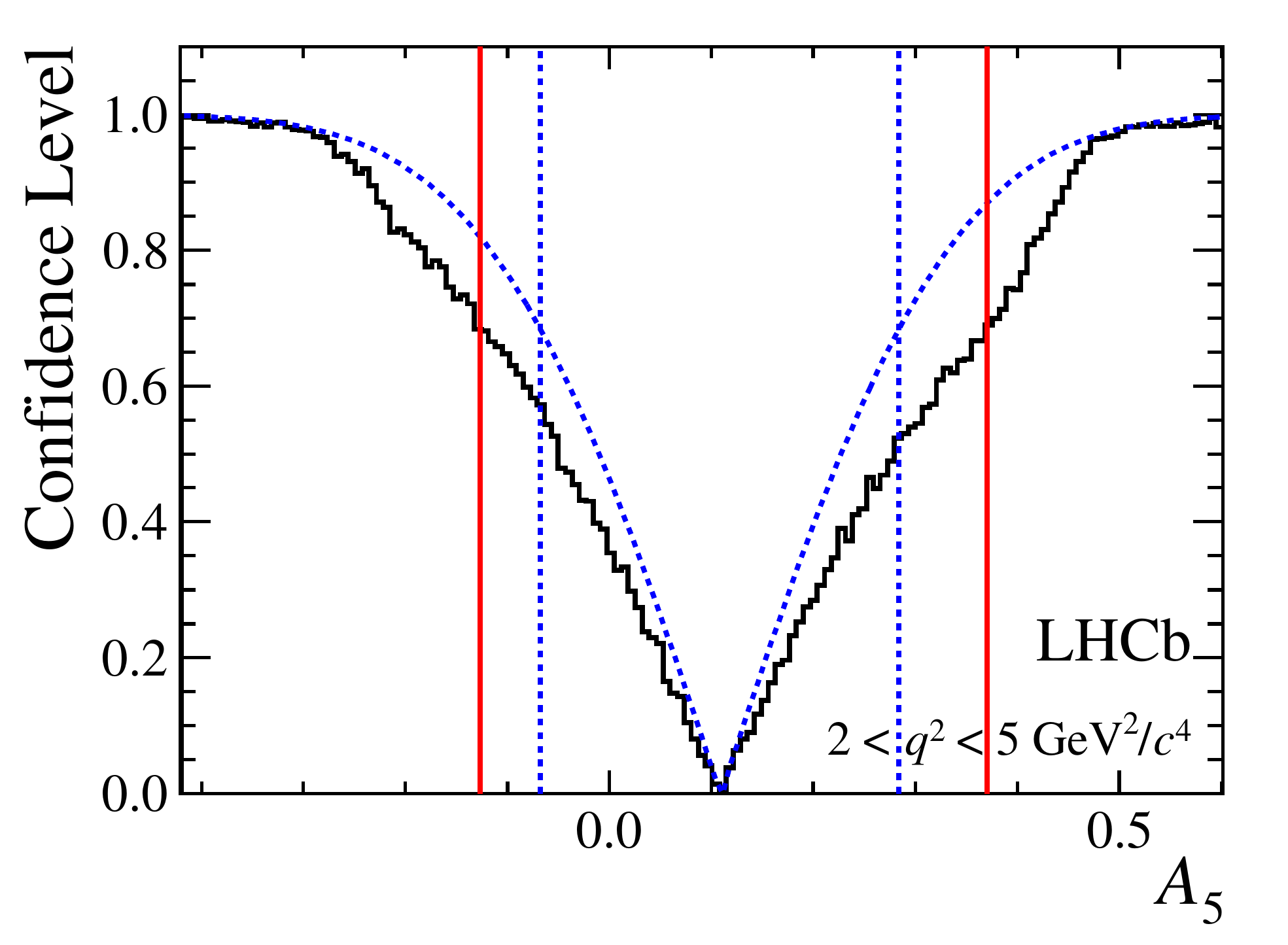}\\
\includegraphics[width=0.37\textwidth]{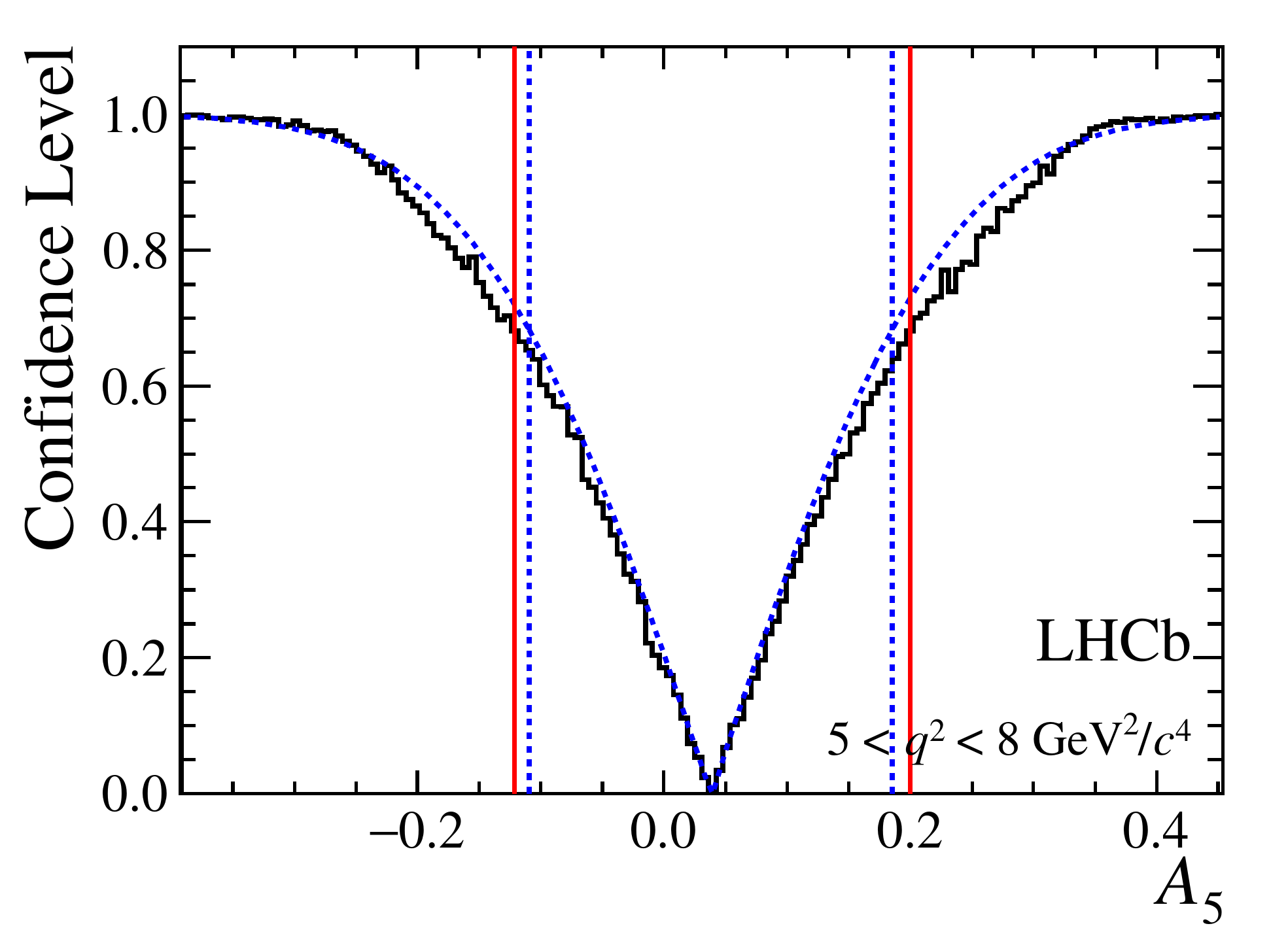}
\includegraphics[width=0.37\textwidth]{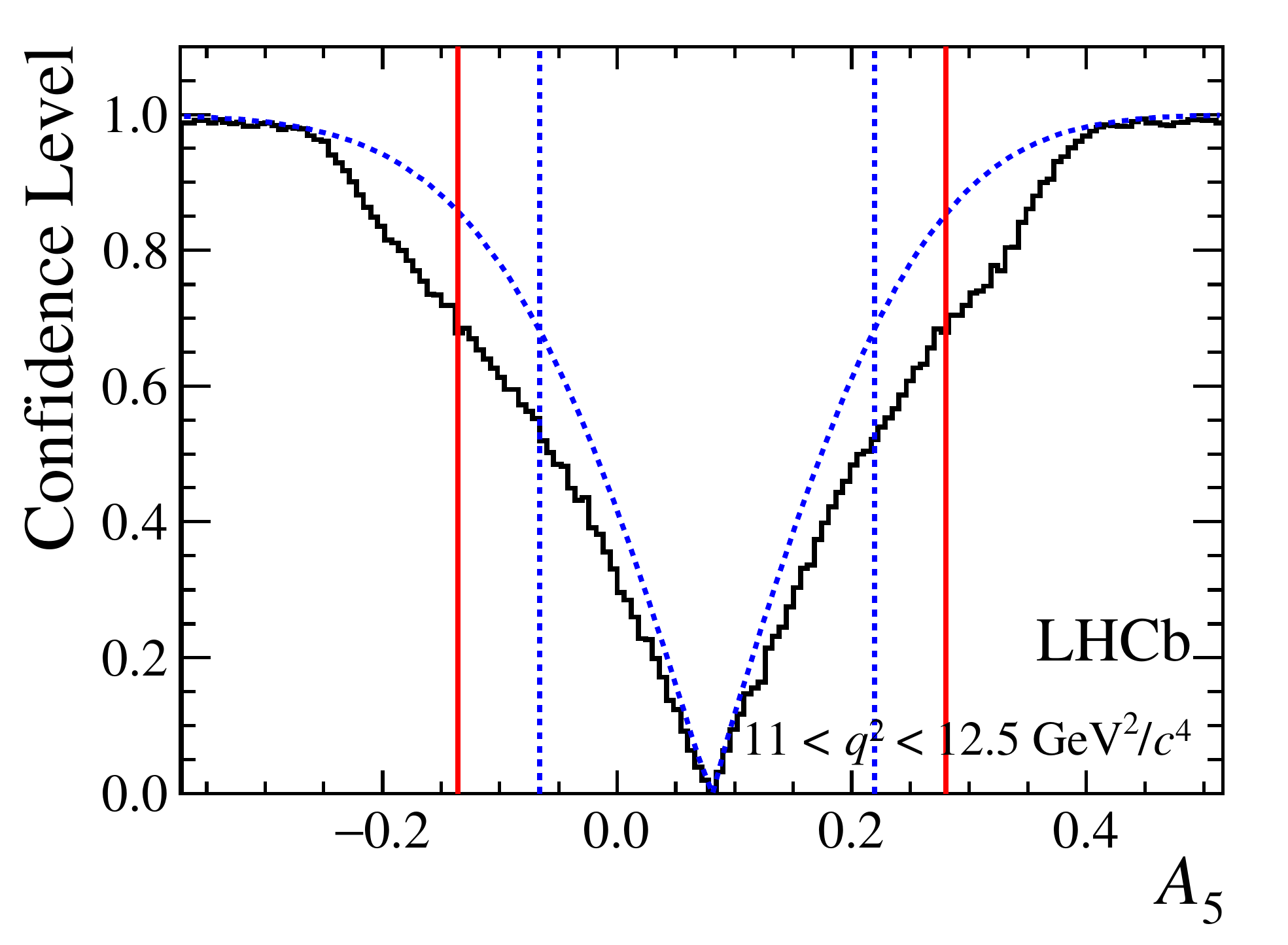}\\
\includegraphics[width=0.37\textwidth]{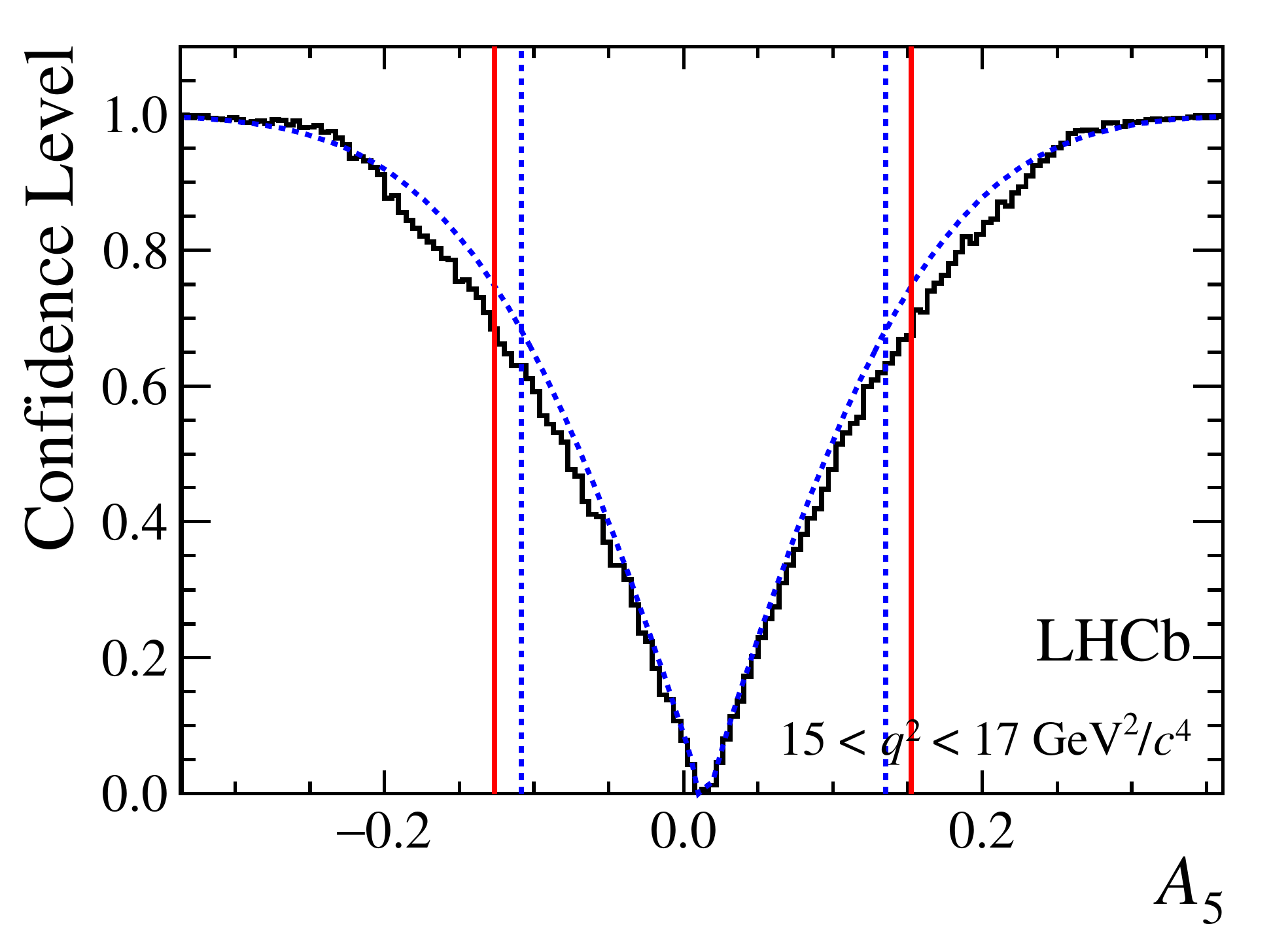}
\includegraphics[width=0.37\textwidth]{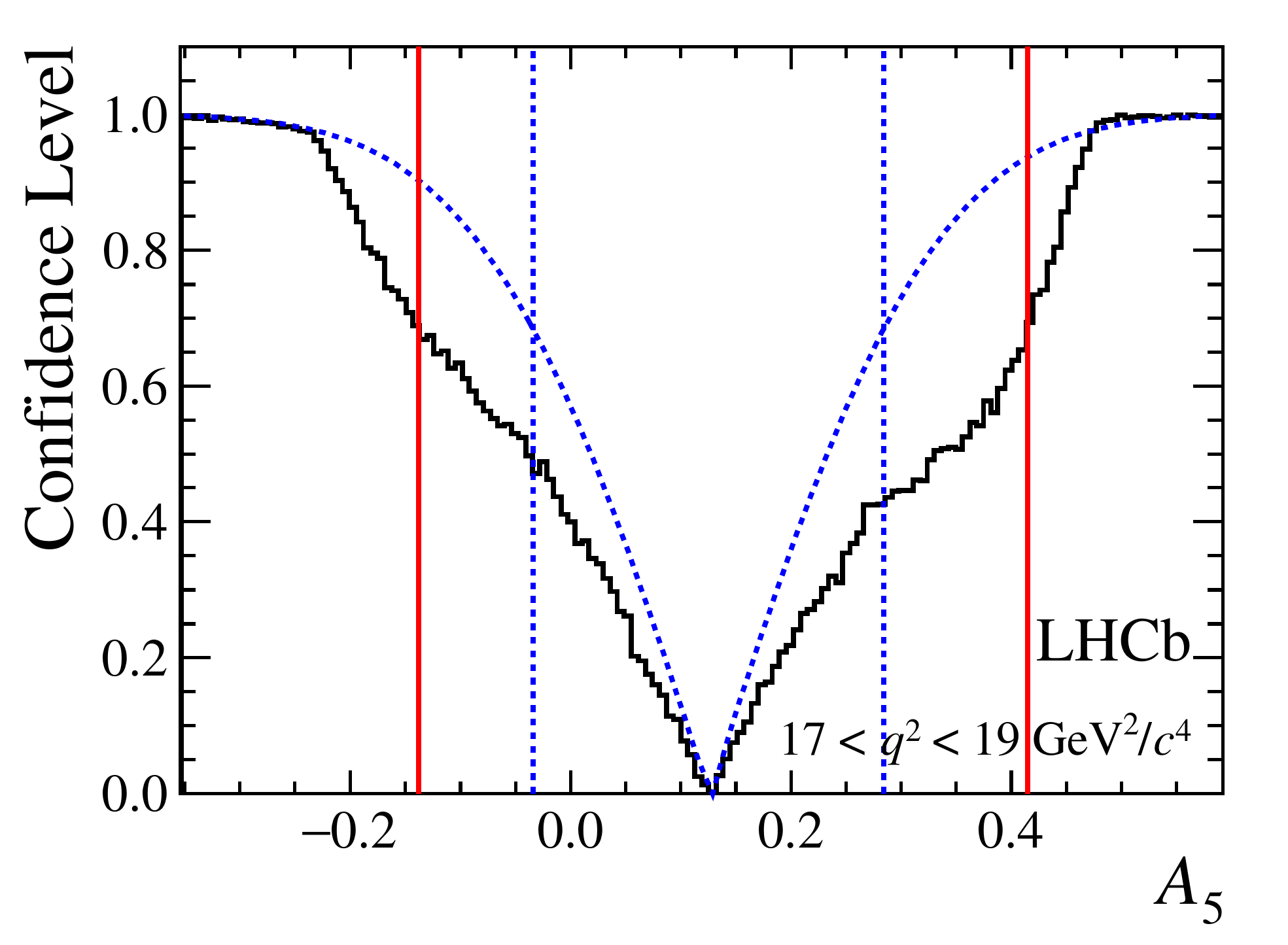}\\
\includegraphics[width=0.37\textwidth]{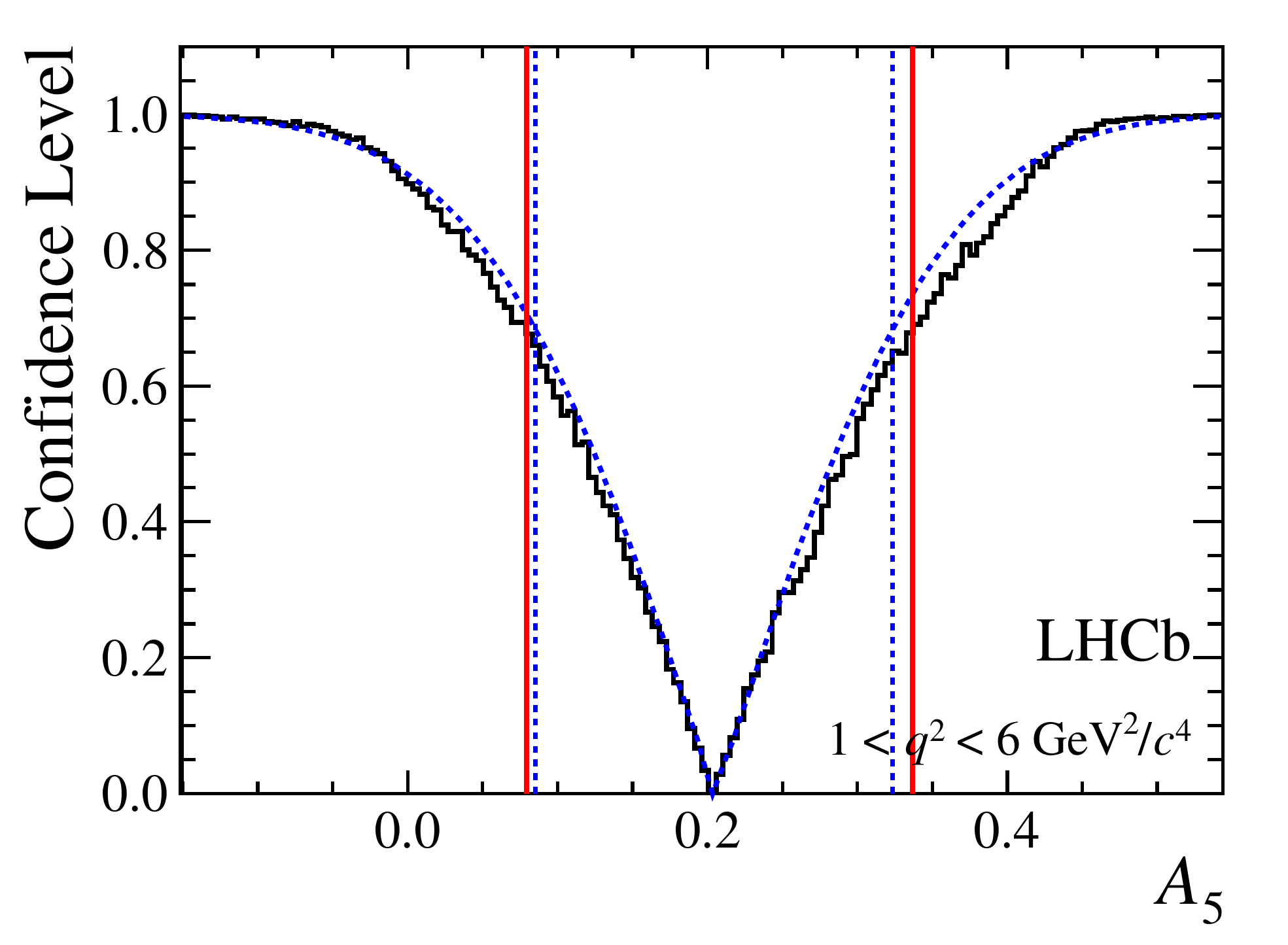}
\includegraphics[width=0.37\textwidth]{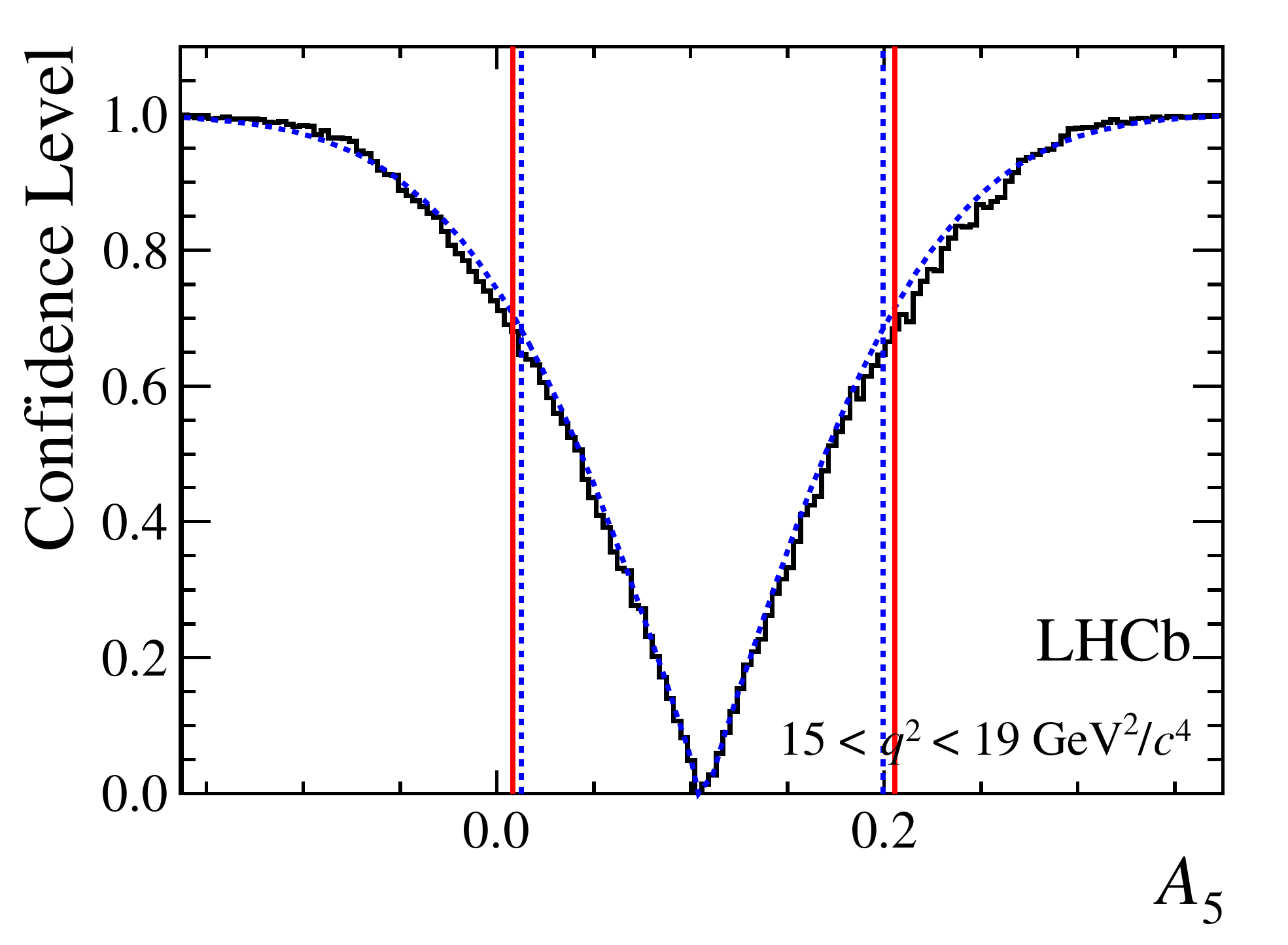}\\
\caption{\label{fig:ang-fc5} Confidence level obtained from a likelihood scan (shaded blue) and from a Feldman-Cousins method (solid black).
The shaded blue and solid red vertical lines indicate the corresponding $68\%$ CL intervals obtained from the likelihood scan and the Feldman-Cousins method, respectively.}
\end{center}
\end{figure}

\begin{figure}[h]
\begin{center}
\includegraphics[width=0.37\textwidth]{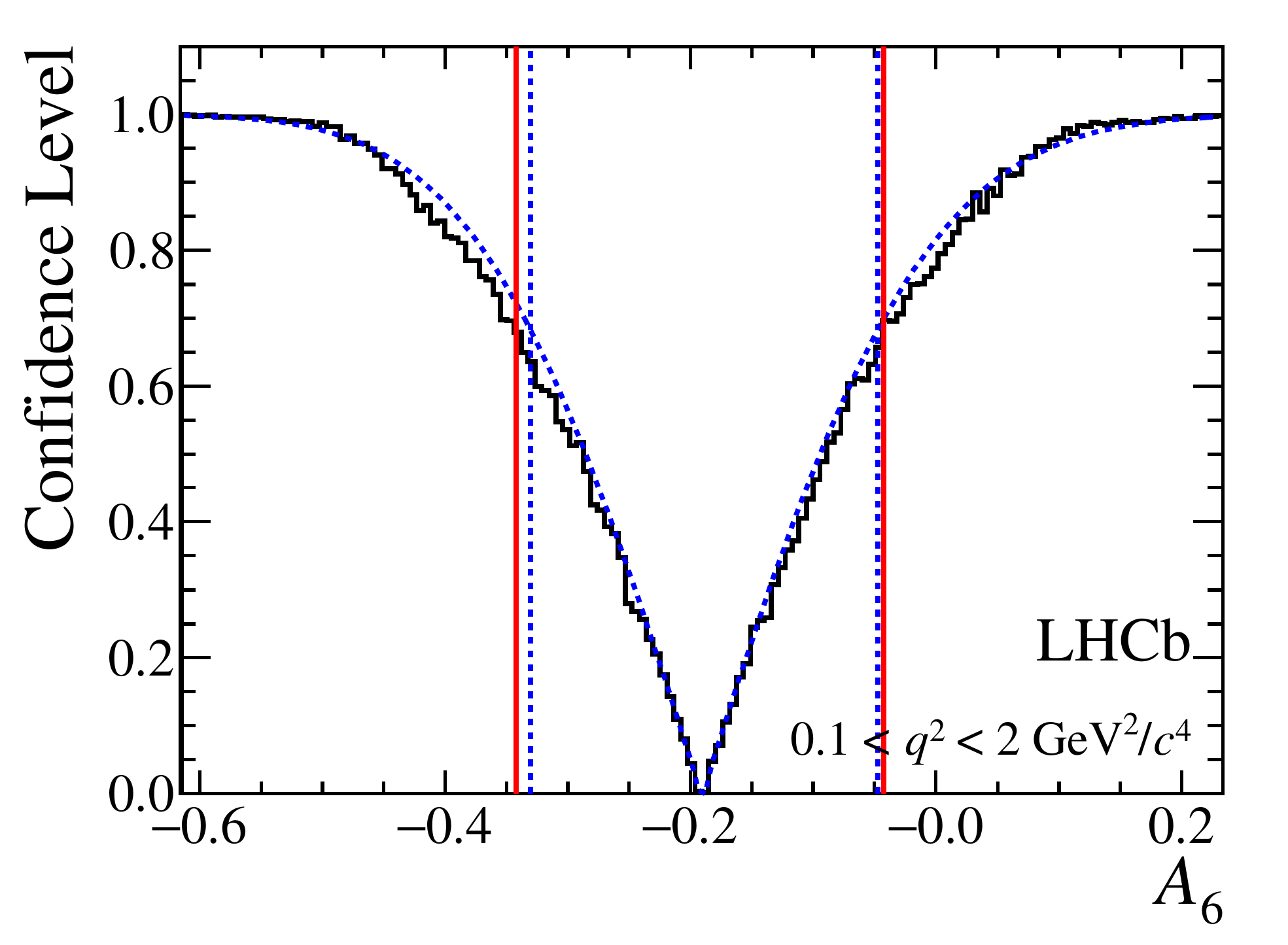}
\includegraphics[width=0.37\textwidth]{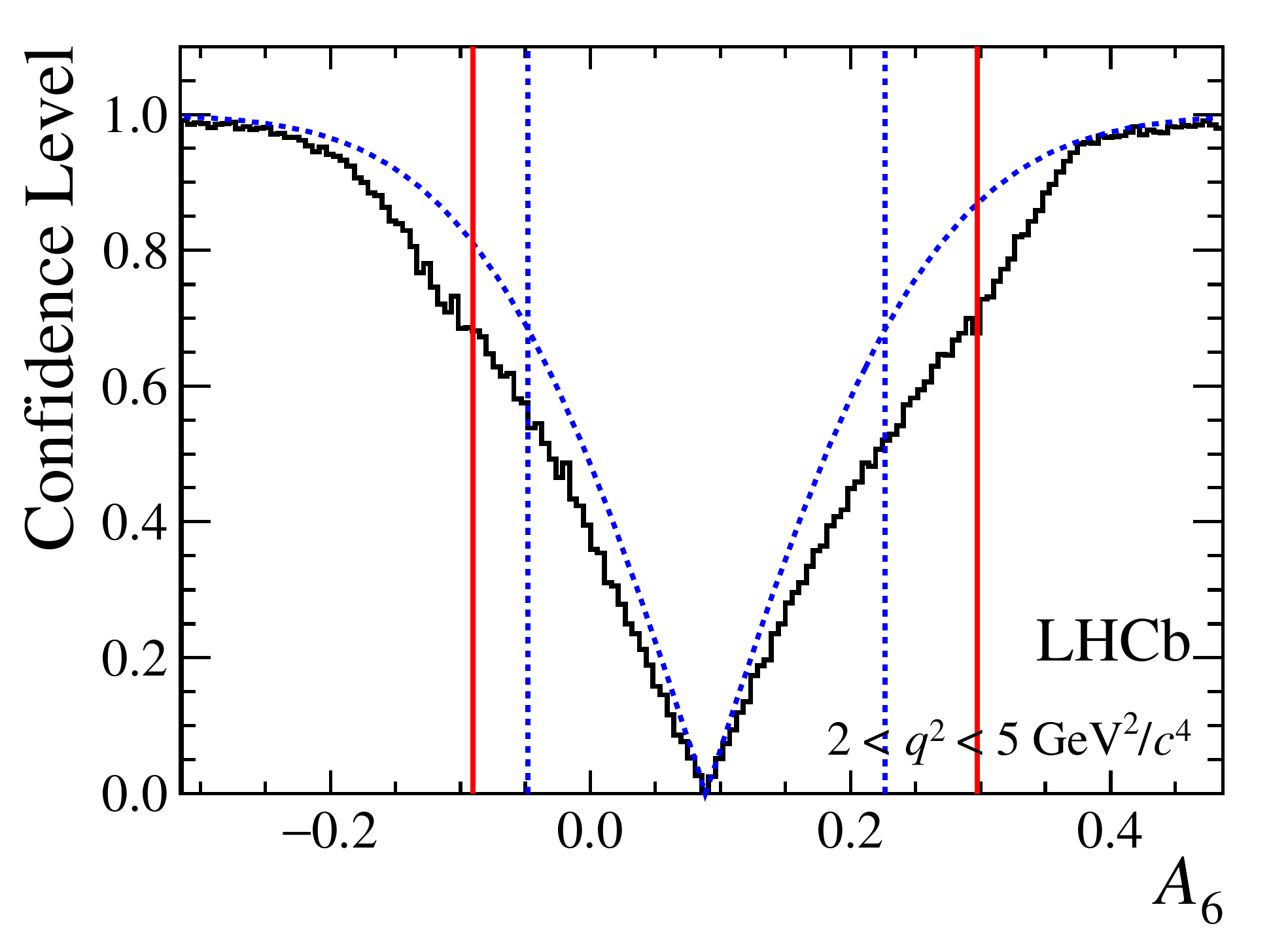}\\
\includegraphics[width=0.37\textwidth]{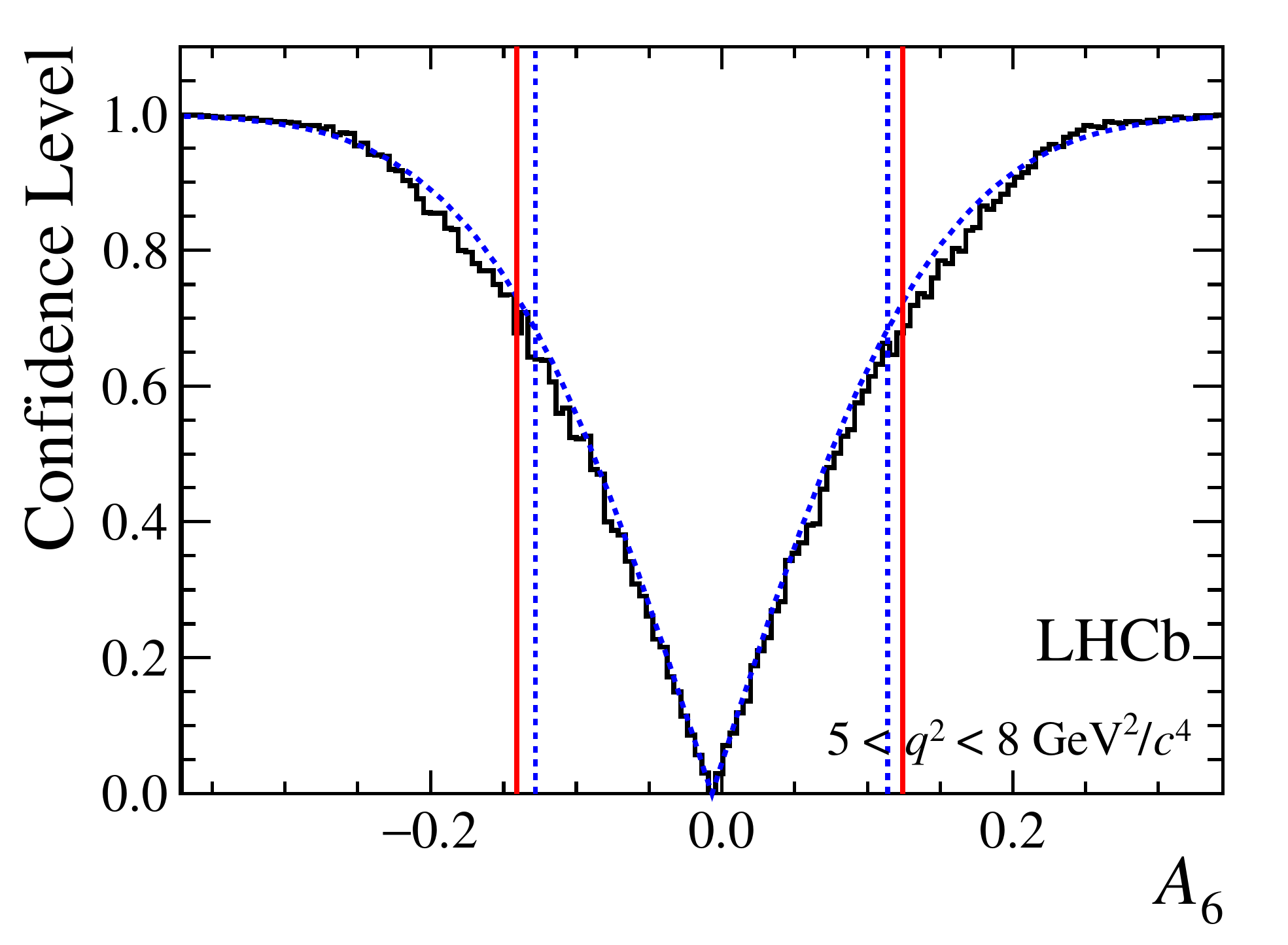}
\includegraphics[width=0.37\textwidth]{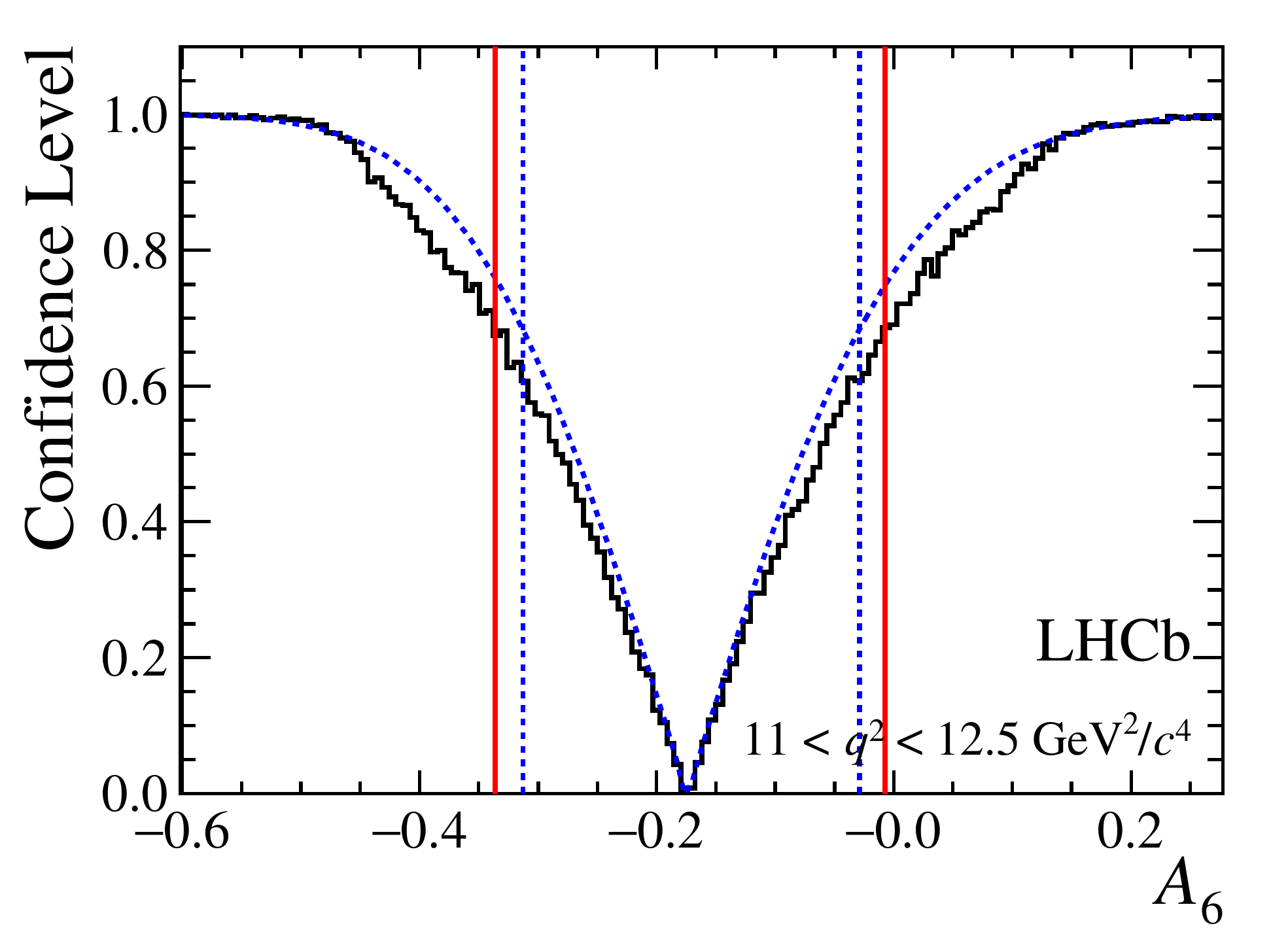}\\
\includegraphics[width=0.37\textwidth]{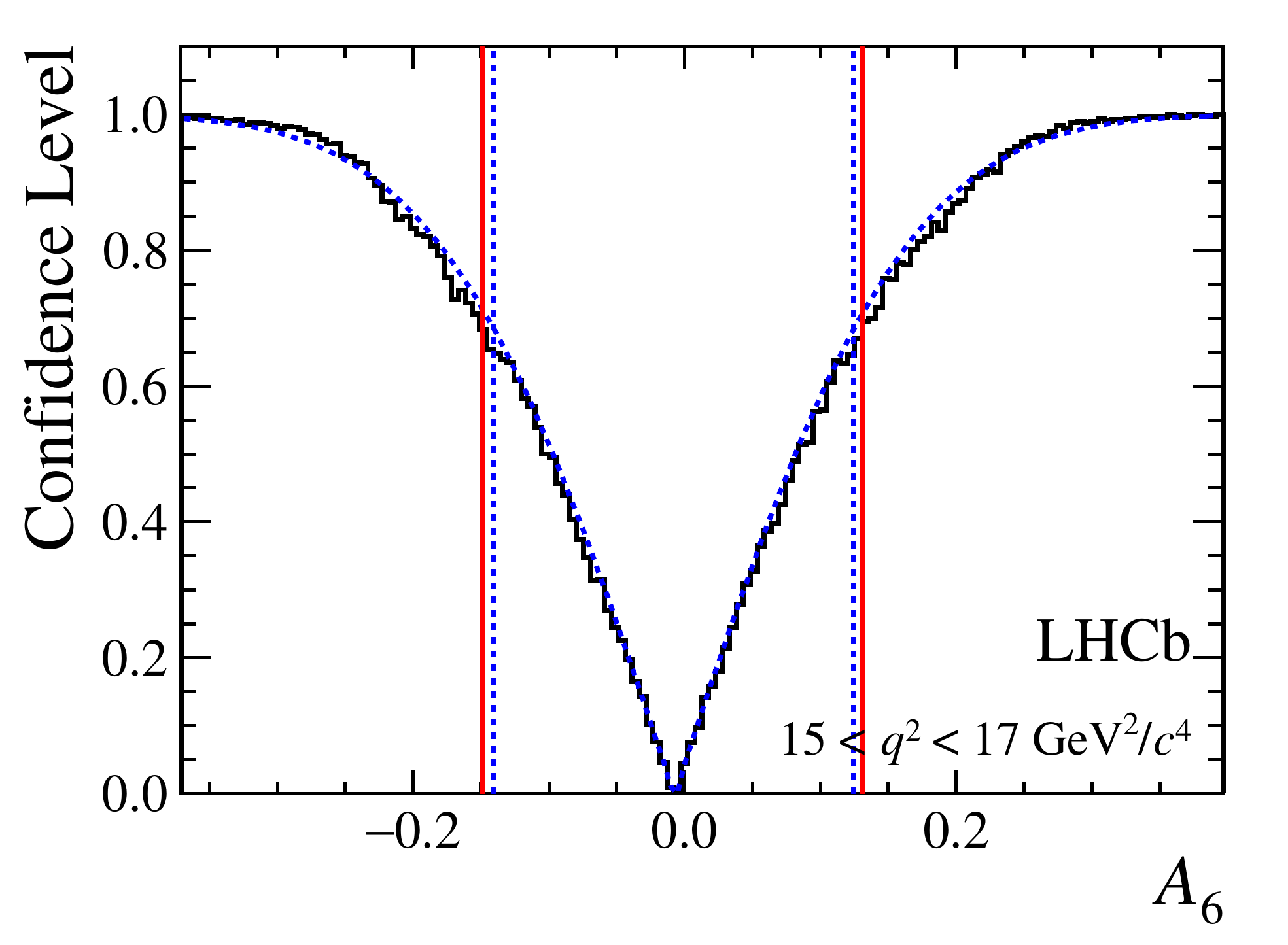}
\includegraphics[width=0.37\textwidth]{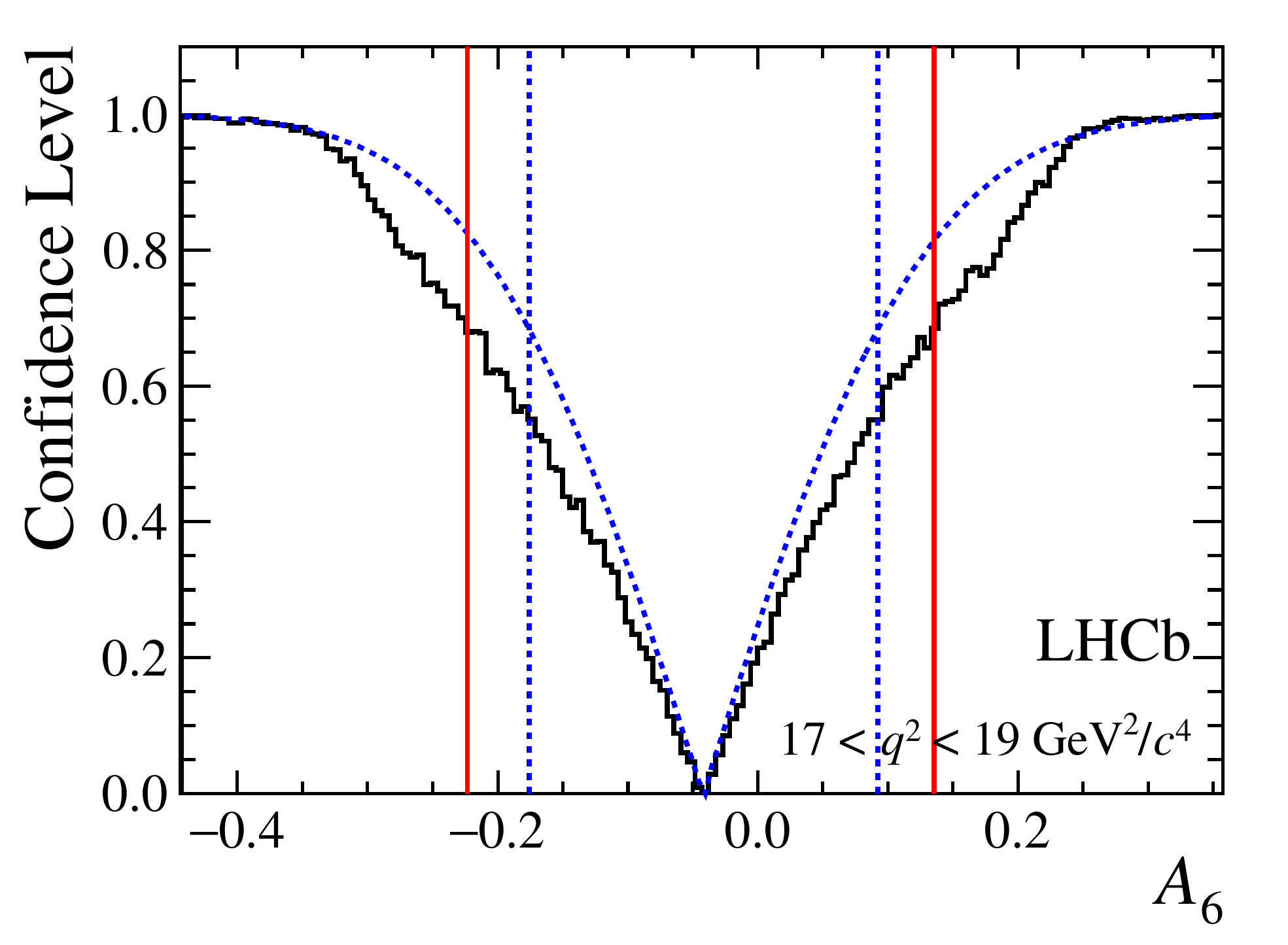}\\
\includegraphics[width=0.37\textwidth]{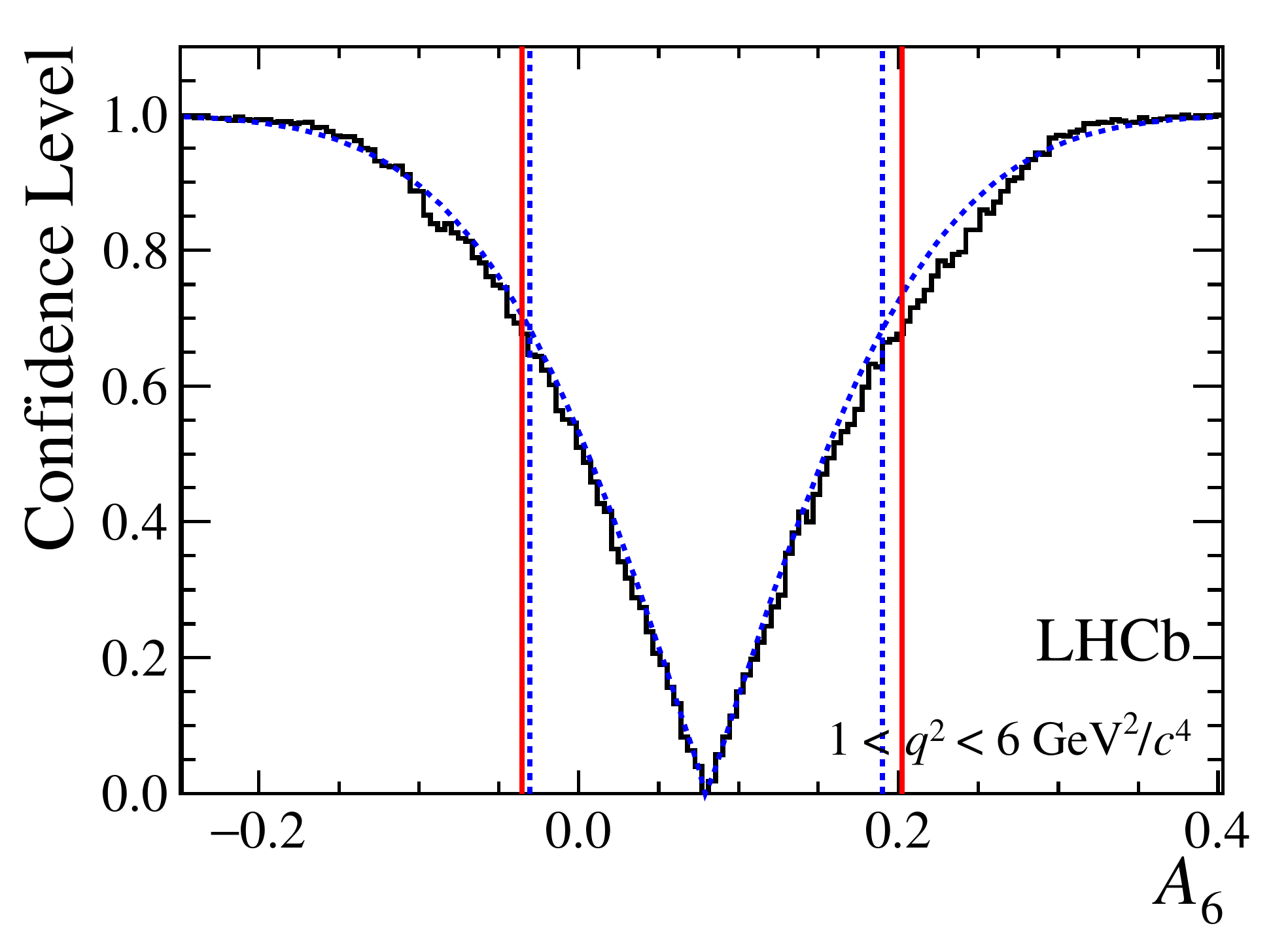}
\includegraphics[width=0.37\textwidth]{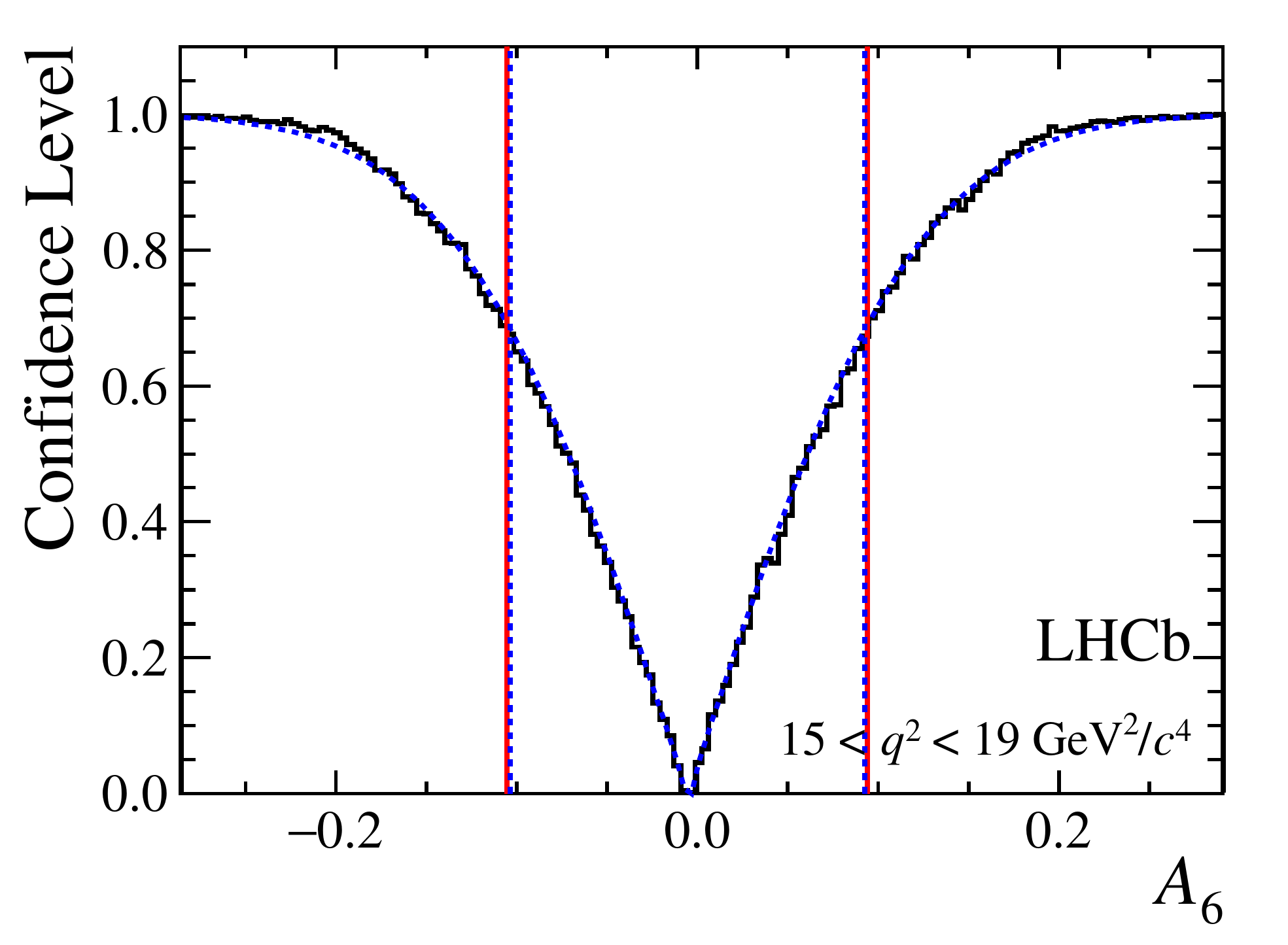}\\
\caption{\label{fig:ang-fc6} Confidence level obtained from a likelihood scan (shaded blue) and from a Feldman-Cousins method (solid black).
The shaded blue and solid red vertical lines indicate the corresponding $68\%$ CL intervals obtained from the likelihood scan and the Feldman-Cousins method, respectively.}
\end{center}
\end{figure}

\begin{figure}[h]
\begin{center}
\includegraphics[width=0.37\textwidth]{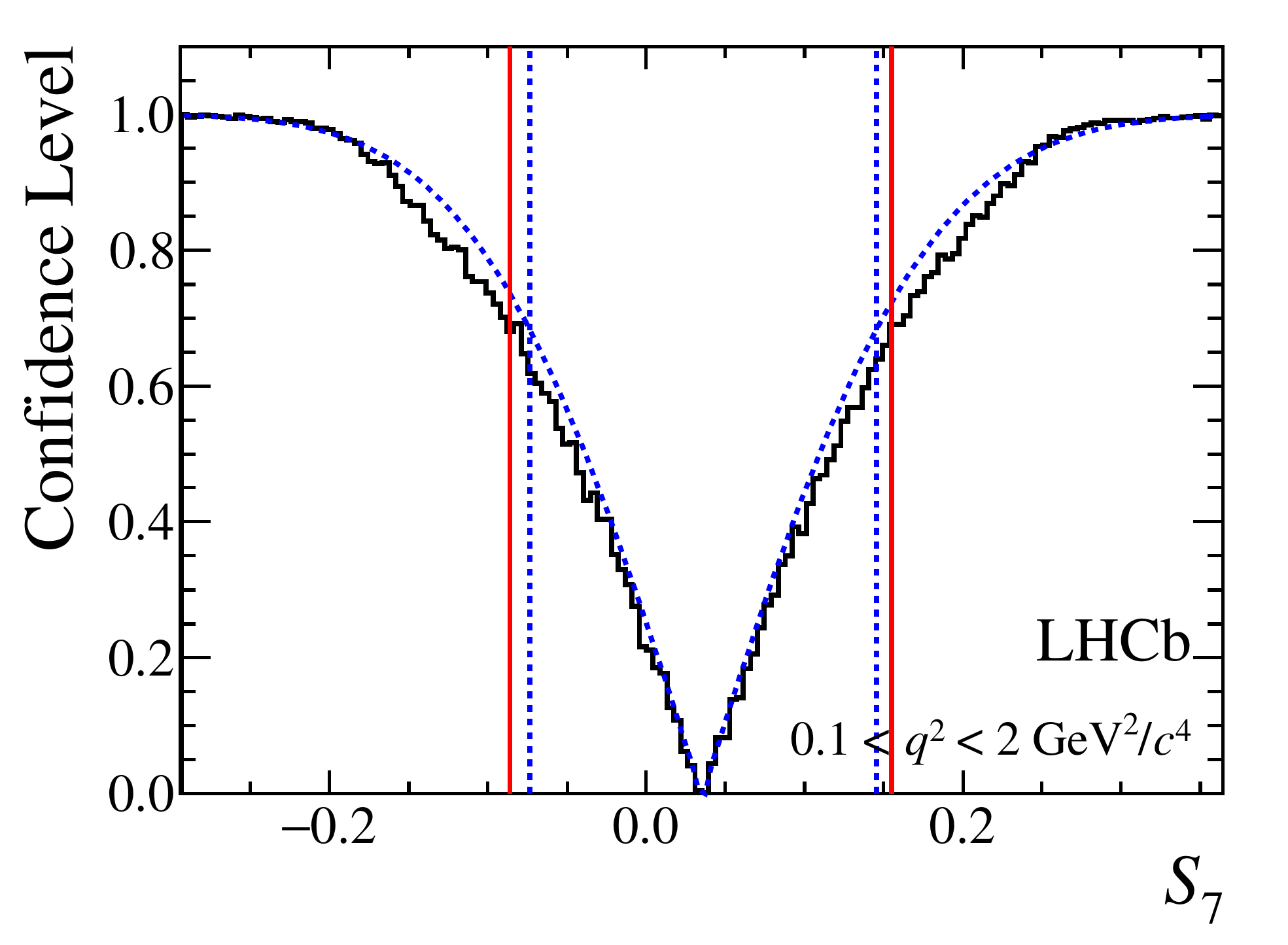}
\includegraphics[width=0.37\textwidth]{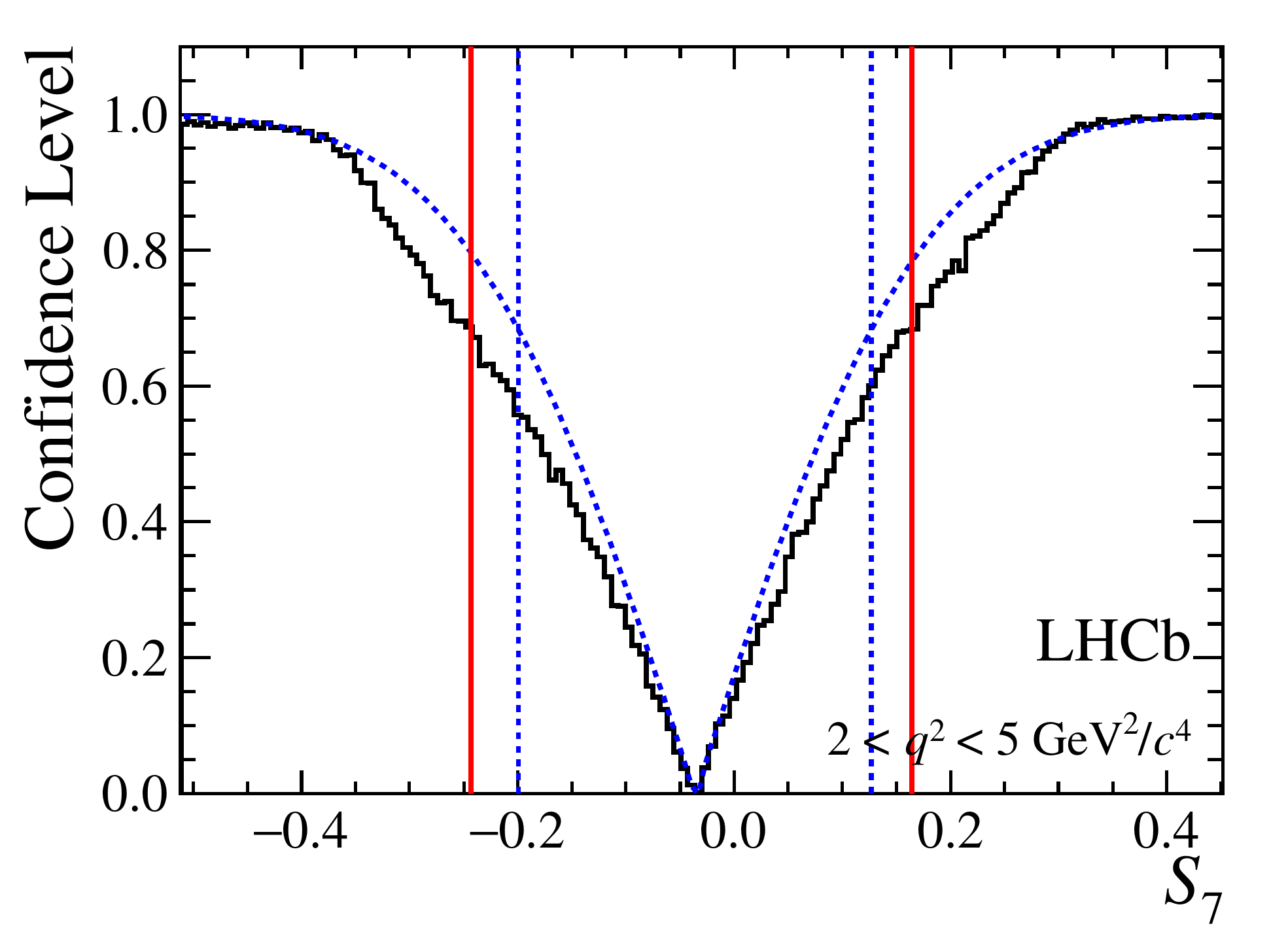}\\
\includegraphics[width=0.37\textwidth]{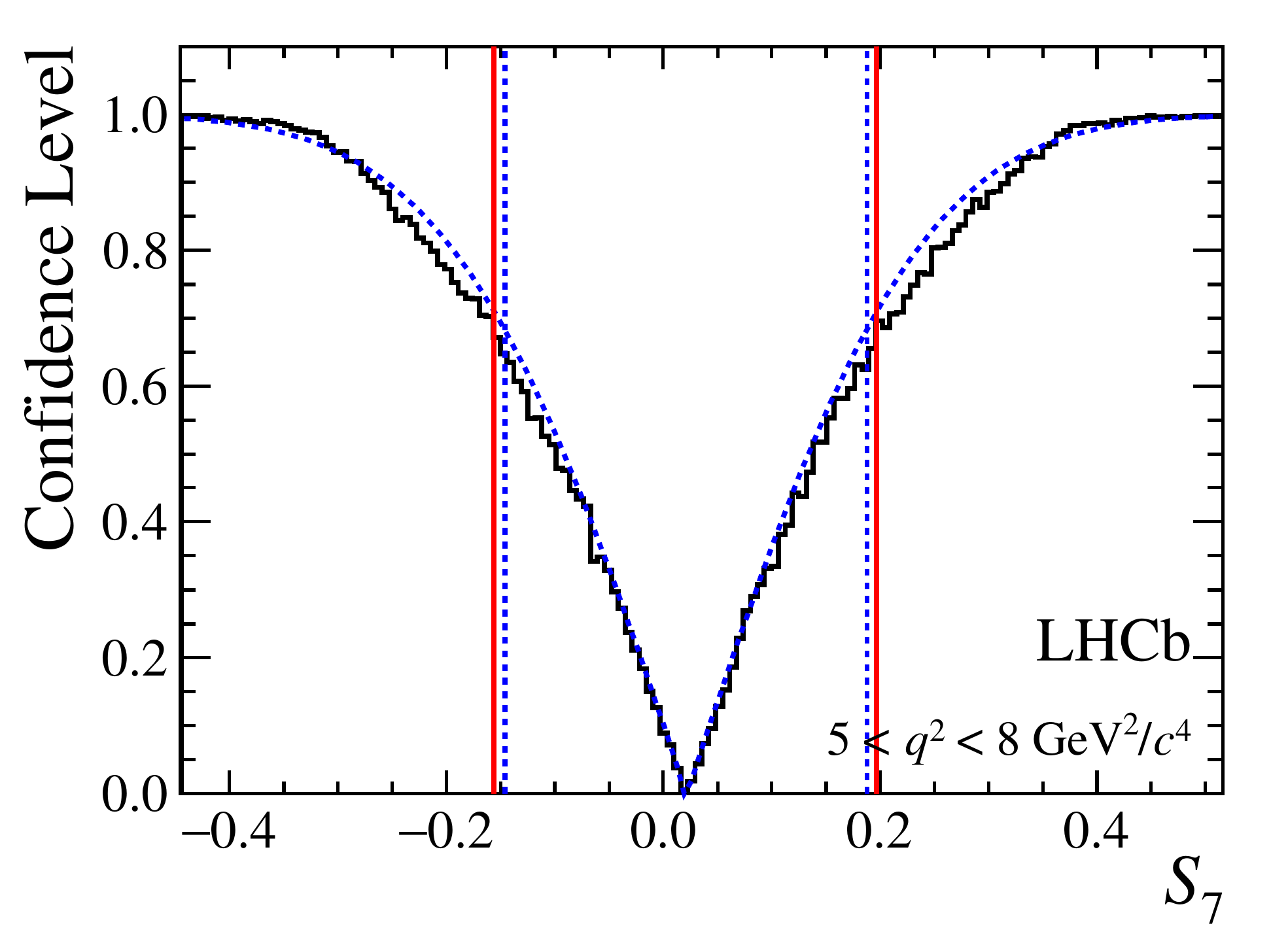}
\includegraphics[width=0.37\textwidth]{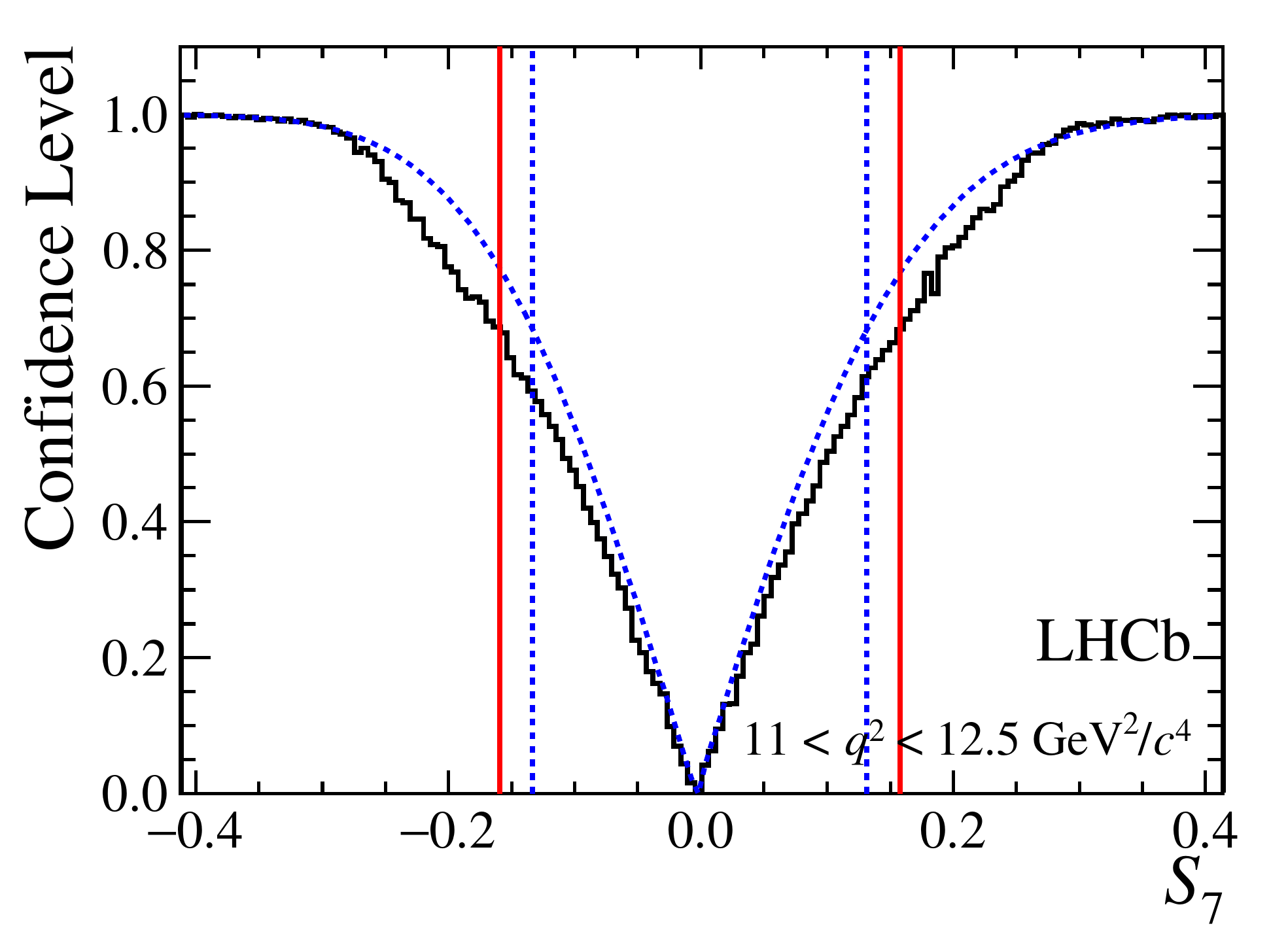}\\
\includegraphics[width=0.37\textwidth]{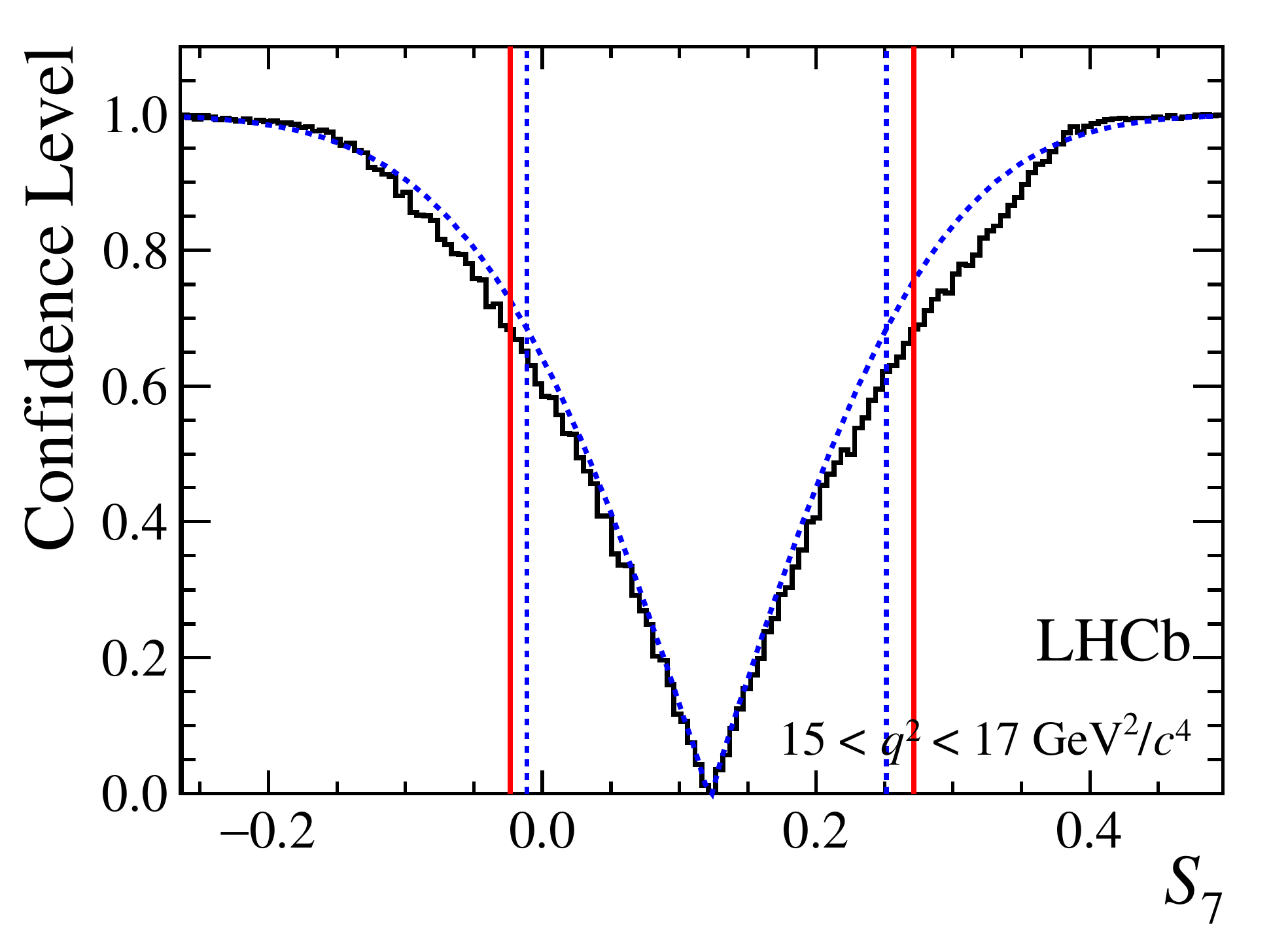}
\includegraphics[width=0.37\textwidth]{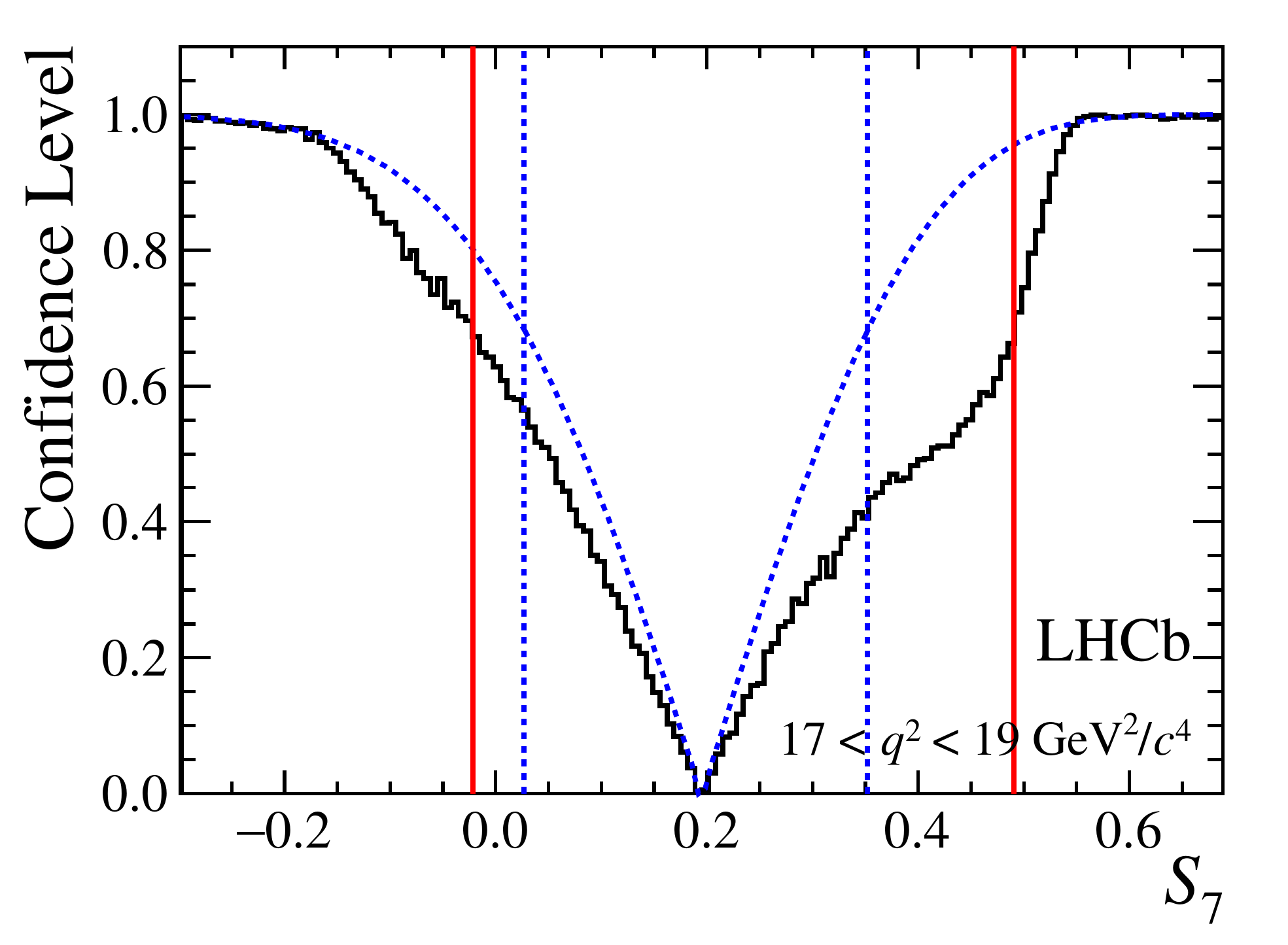}\\
\includegraphics[width=0.37\textwidth]{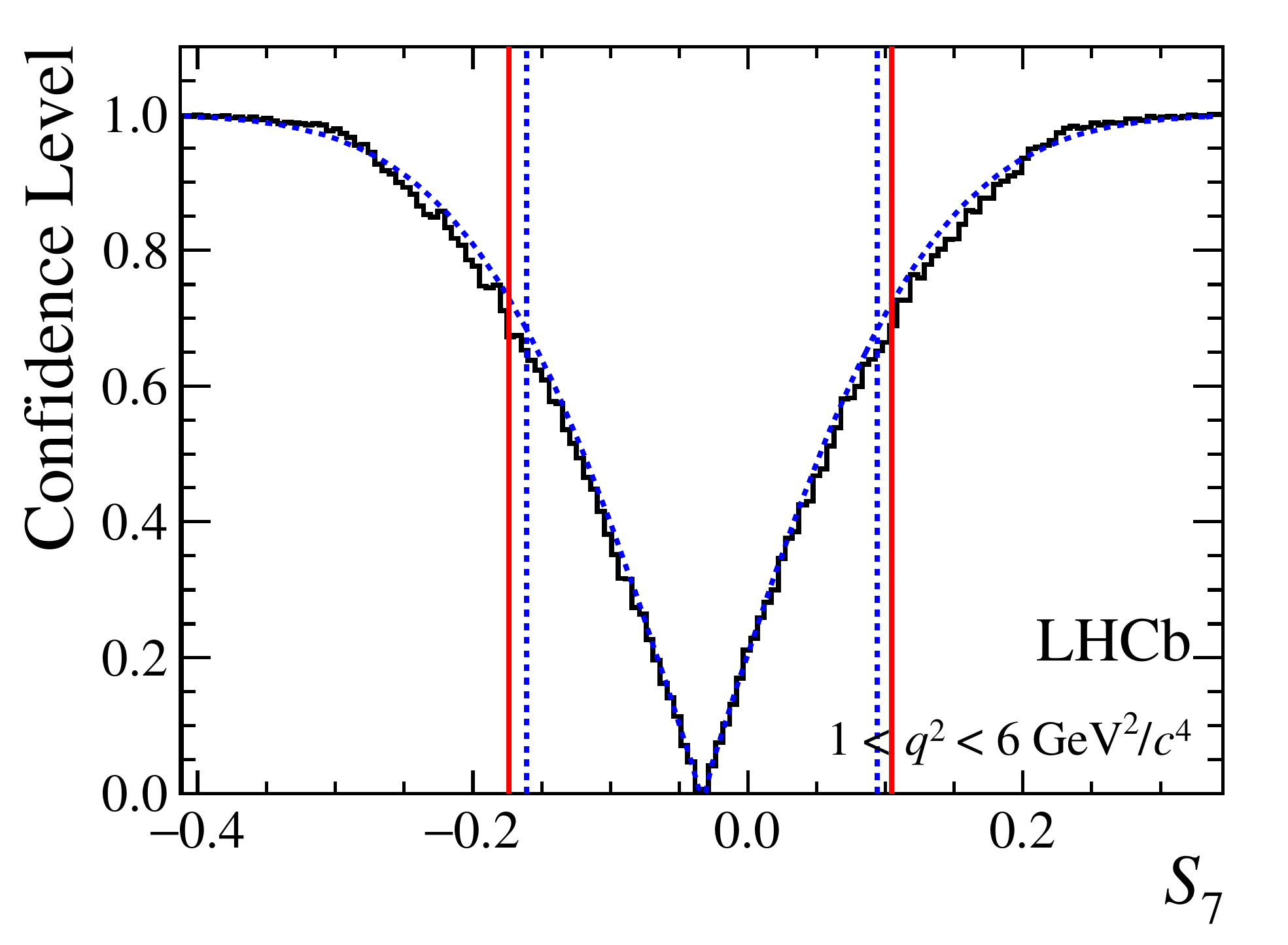}
\includegraphics[width=0.37\textwidth]{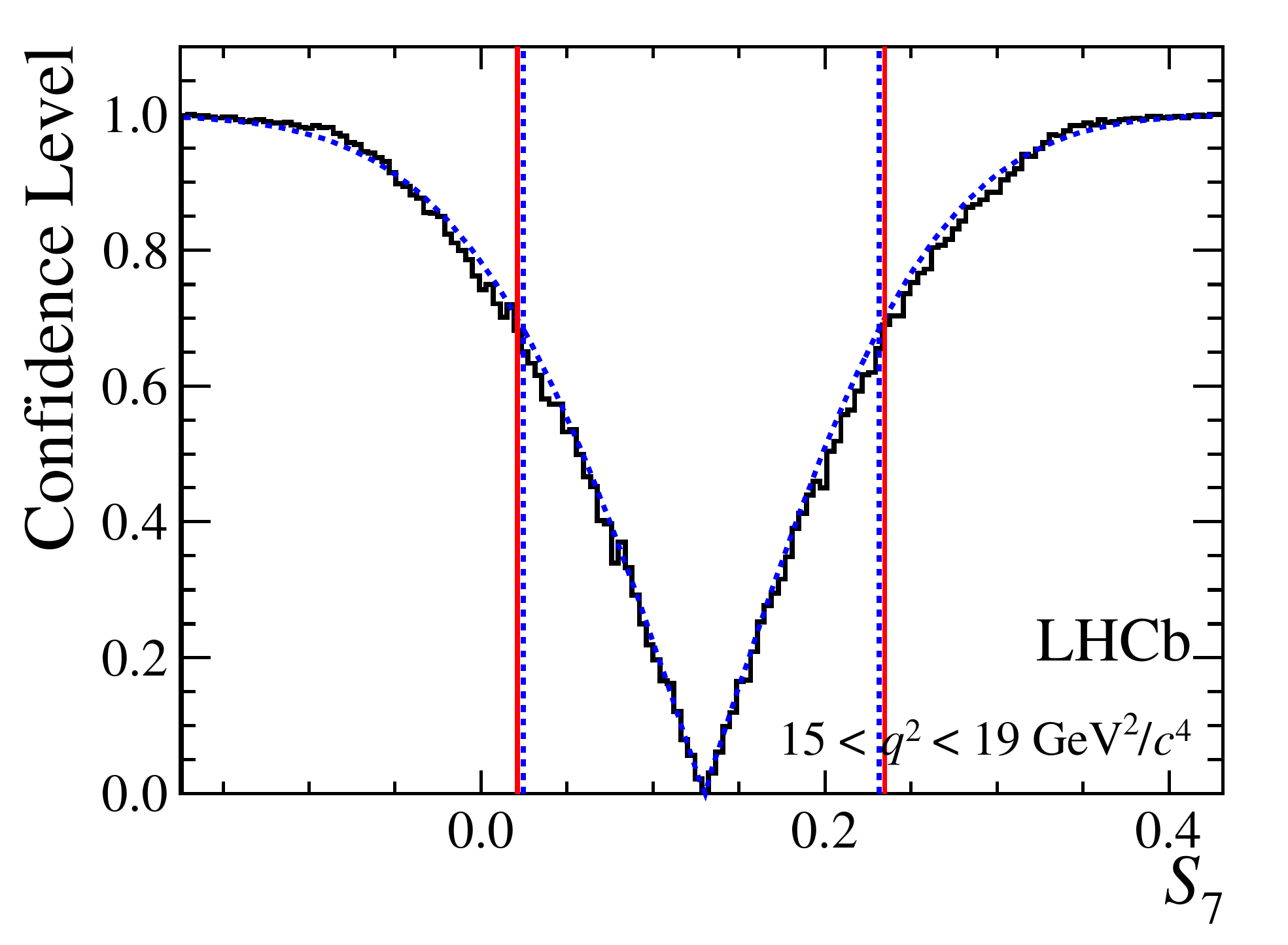}\\
\caption{\label{fig:ang-fc7} Confidence level obtained from a likelihood scan (shaded blue) and from a Feldman-Cousins method (solid black).
The shaded blue and solid red vertical lines indicate the corresponding $68\%$ CL intervals obtained from the likelihood scan and the Feldman-Cousins method, respectively.}
\end{center}
\end{figure}

\begin{figure}[h]
\begin{center}
\includegraphics[width=0.37\textwidth]{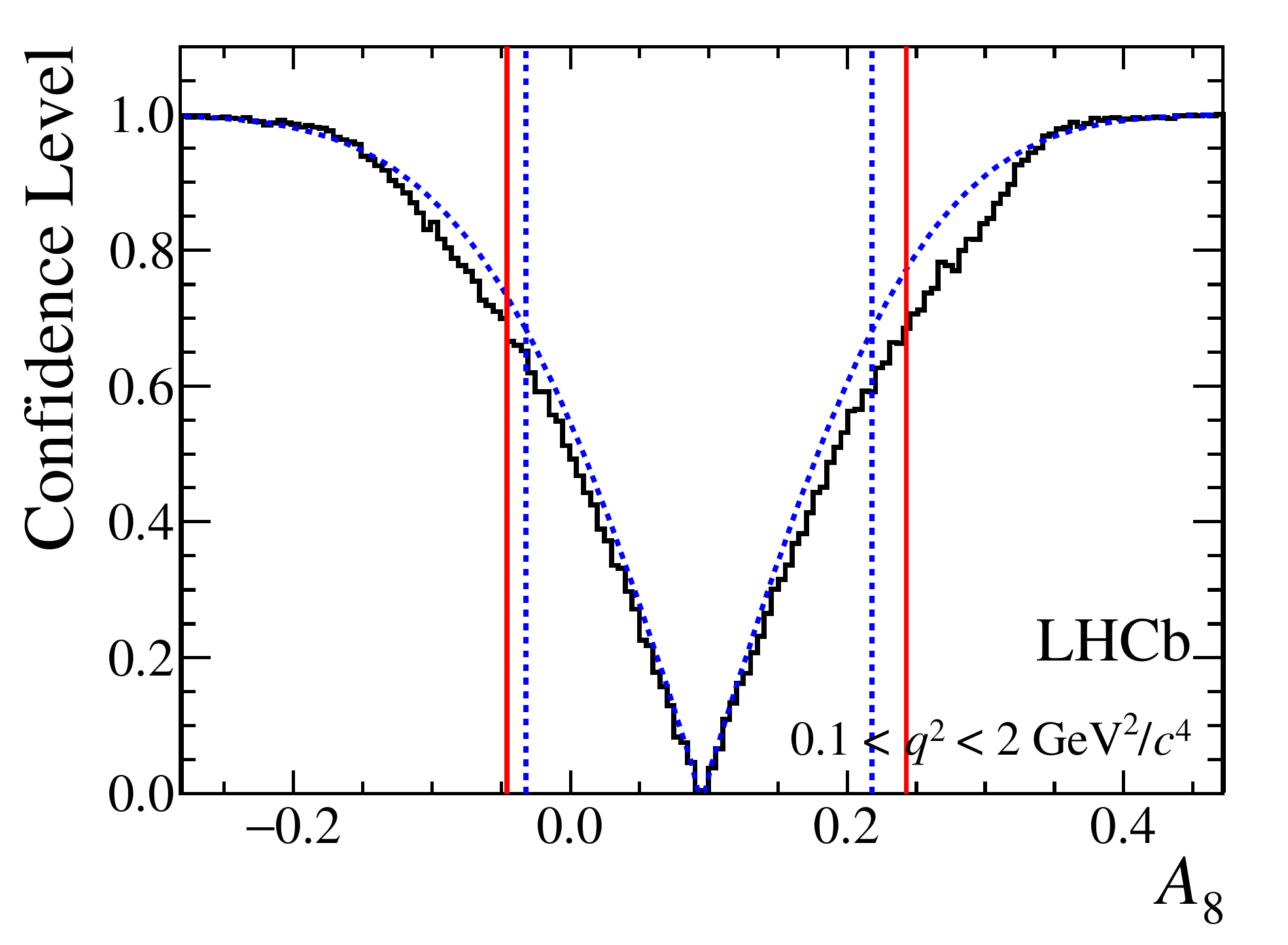}
\includegraphics[width=0.37\textwidth]{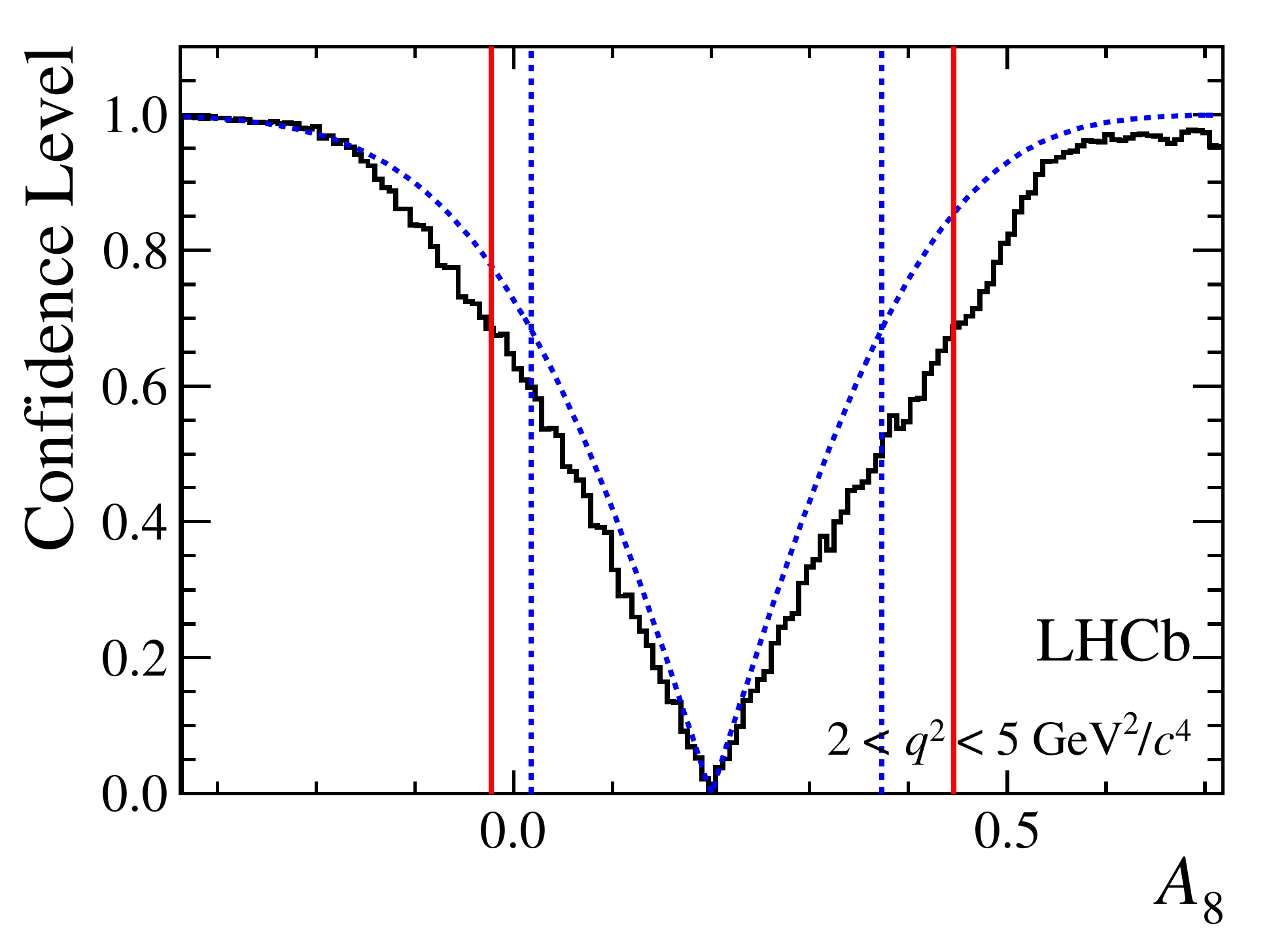}\\
\includegraphics[width=0.37\textwidth]{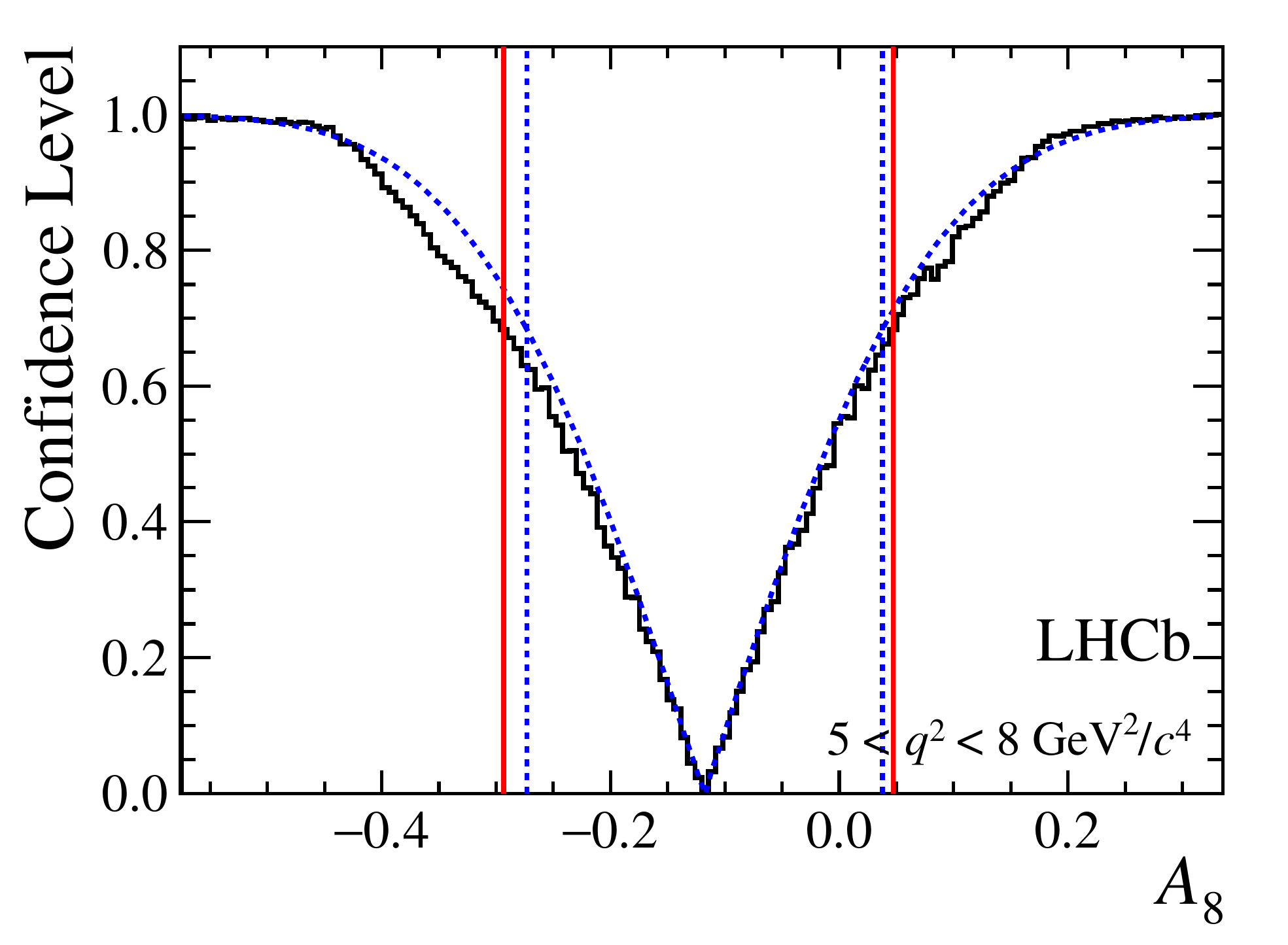}
\includegraphics[width=0.37\textwidth]{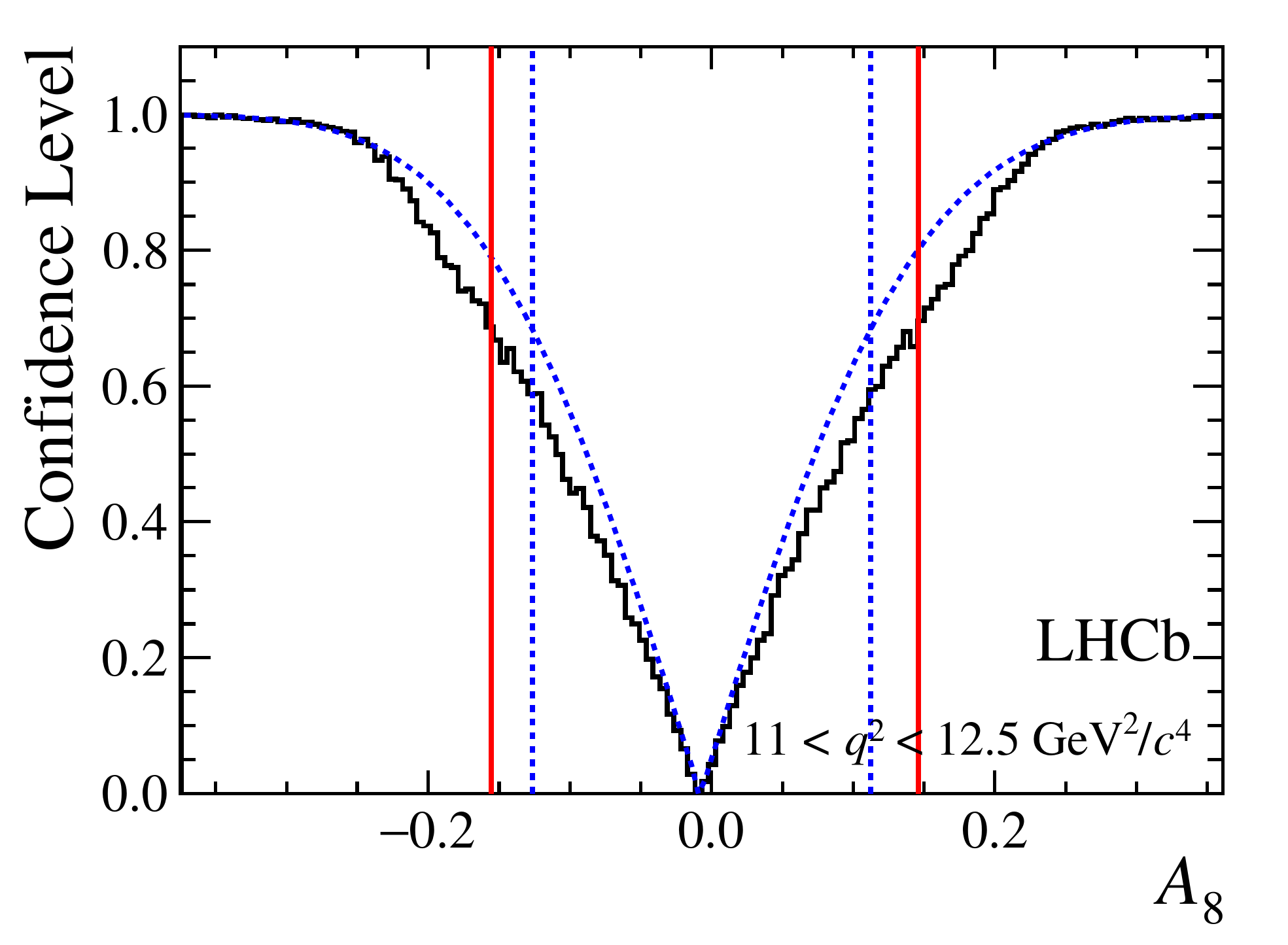}\\
\includegraphics[width=0.37\textwidth]{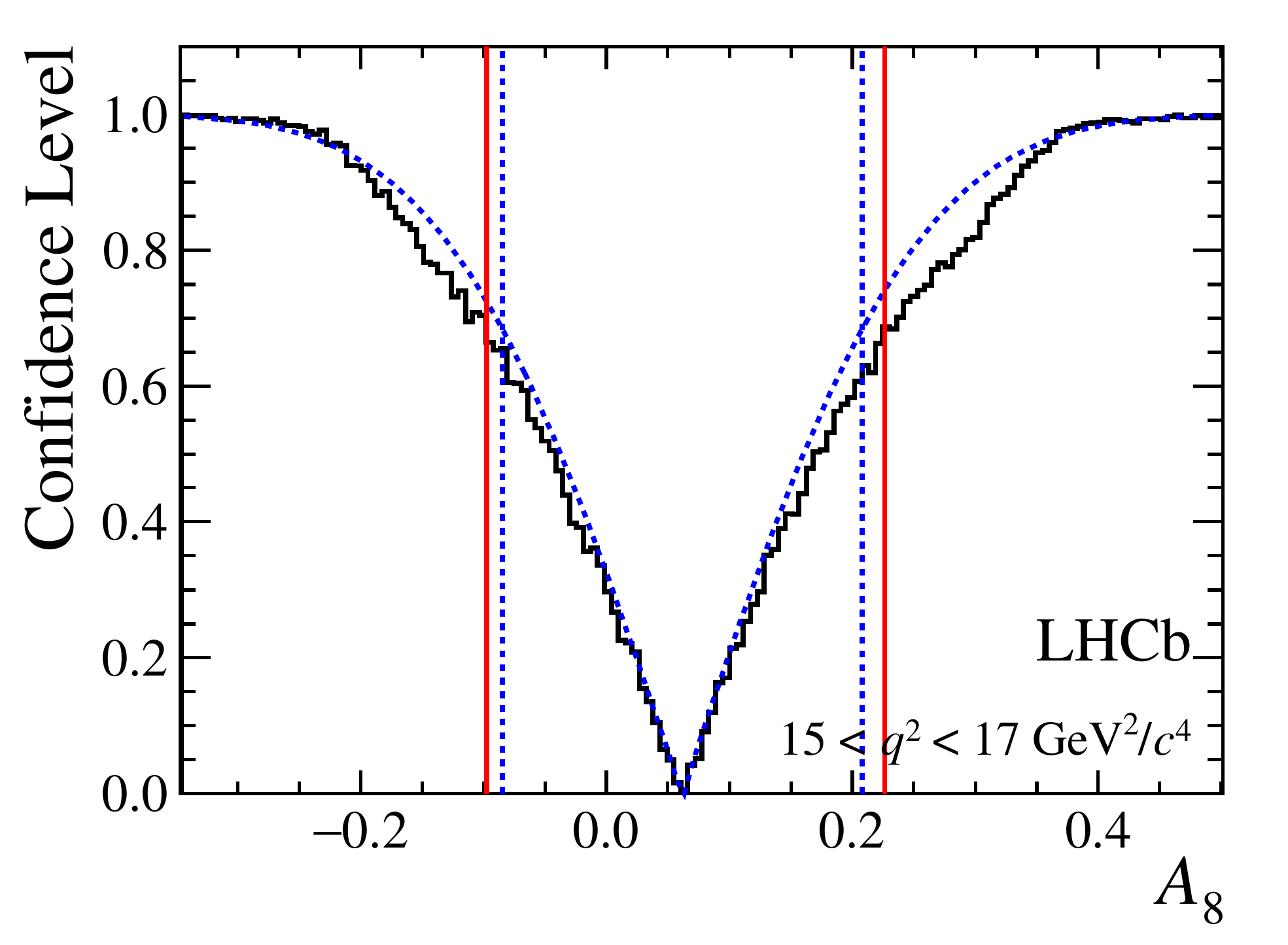}
\includegraphics[width=0.37\textwidth]{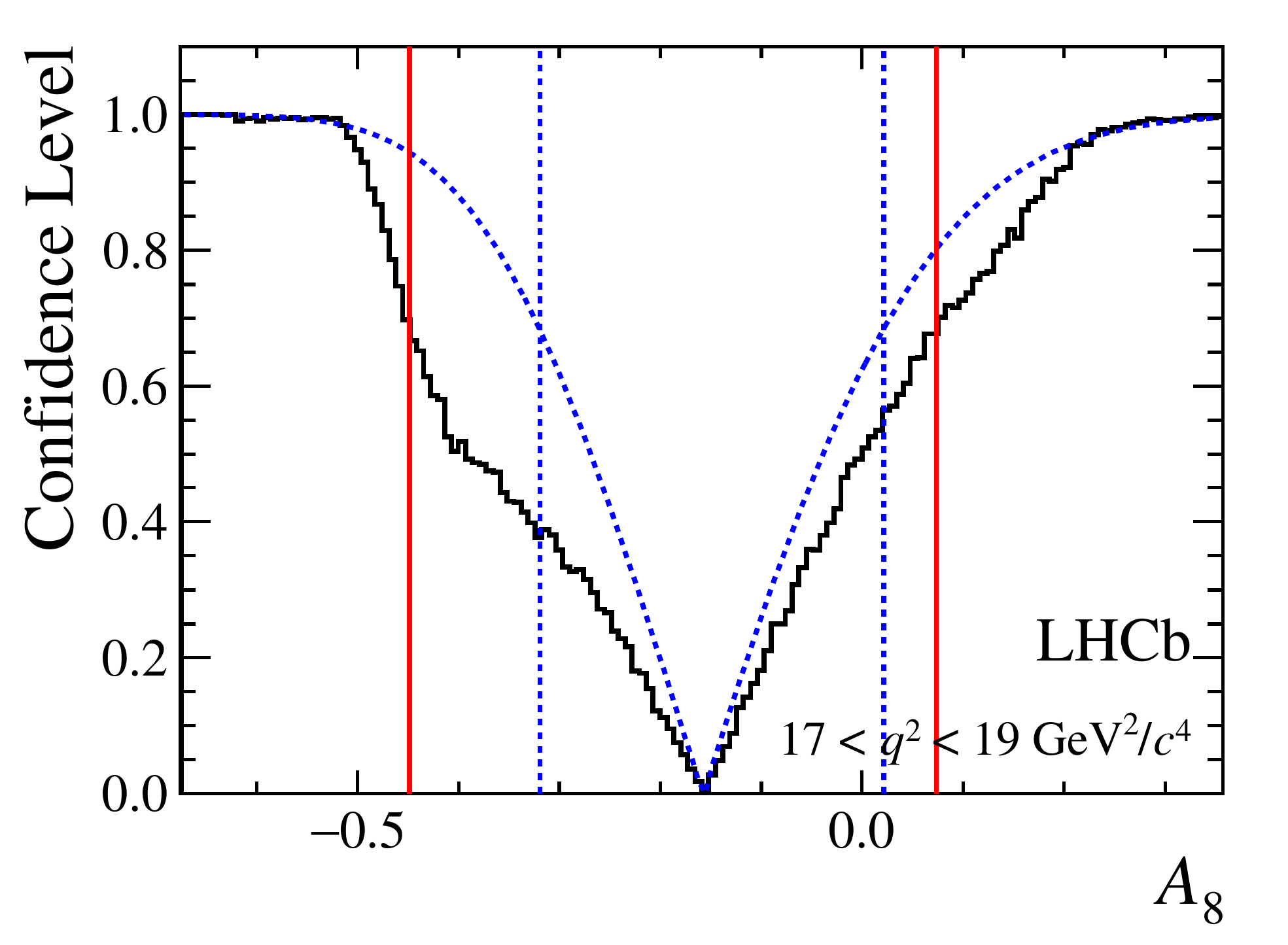}\\
\includegraphics[width=0.37\textwidth]{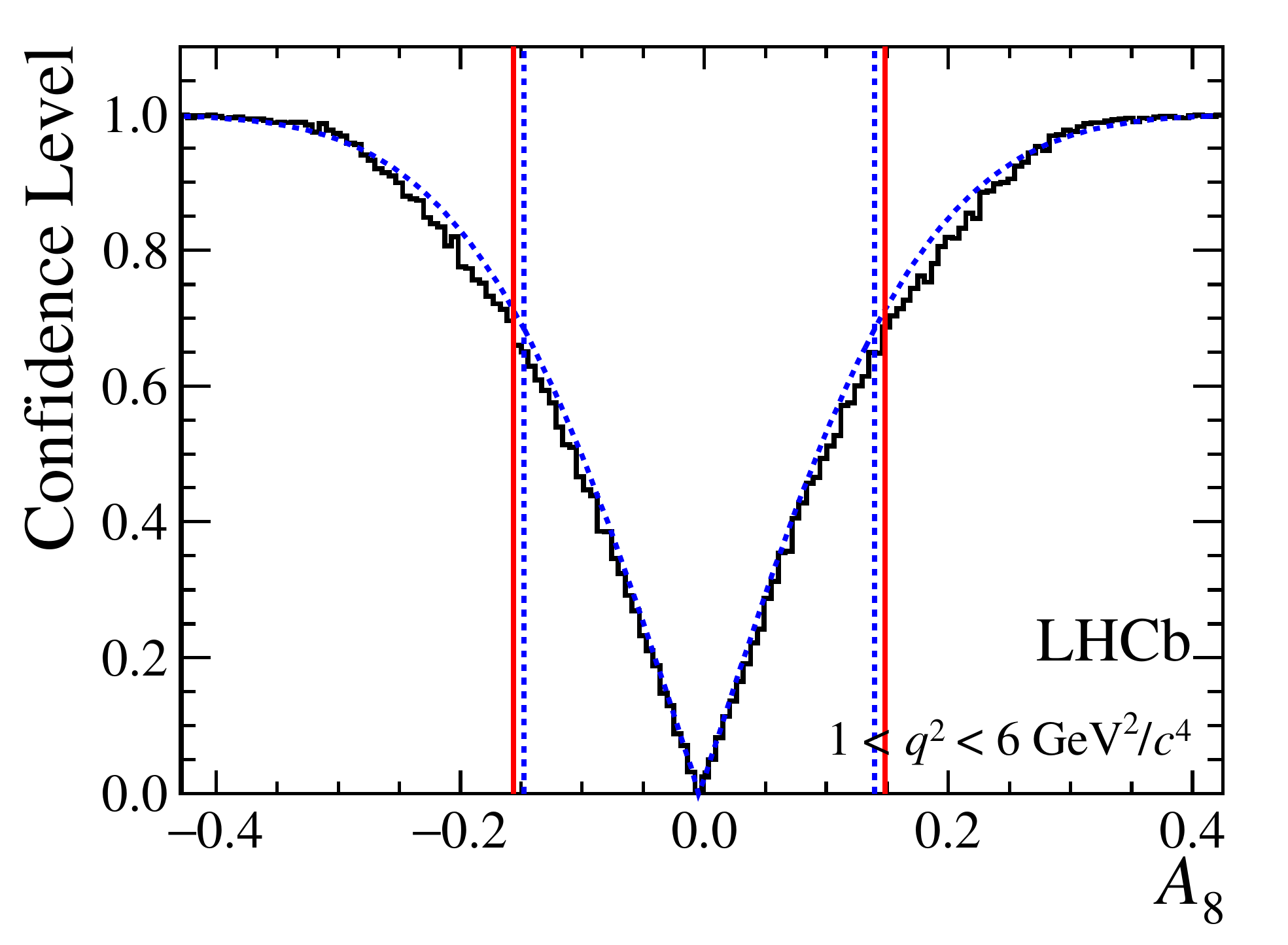}
\includegraphics[width=0.37\textwidth]{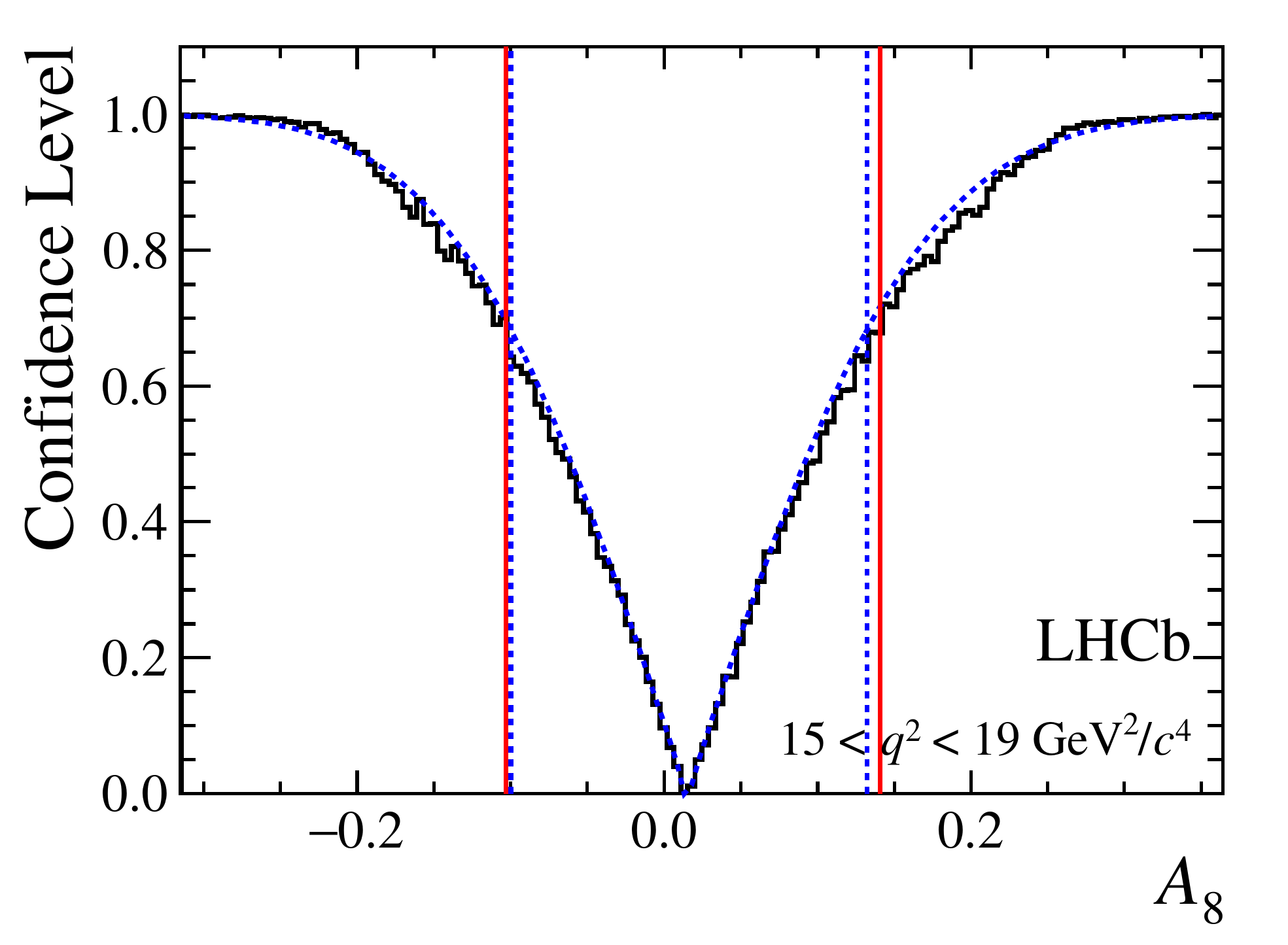}\\
\caption{\label{fig:ang-fc8} Confidence level obtained from a likelihood scan (shaded blue) and from a Feldman-Cousins method (solid black).
The shaded blue and solid red vertical lines indicate the corresponding $68\%$ CL intervals obtained from the likelihood scan and the Feldman-Cousins method, respectively.}
\end{center}
\end{figure}

\begin{figure}[h]
\begin{center}
\includegraphics[width=0.37\textwidth]{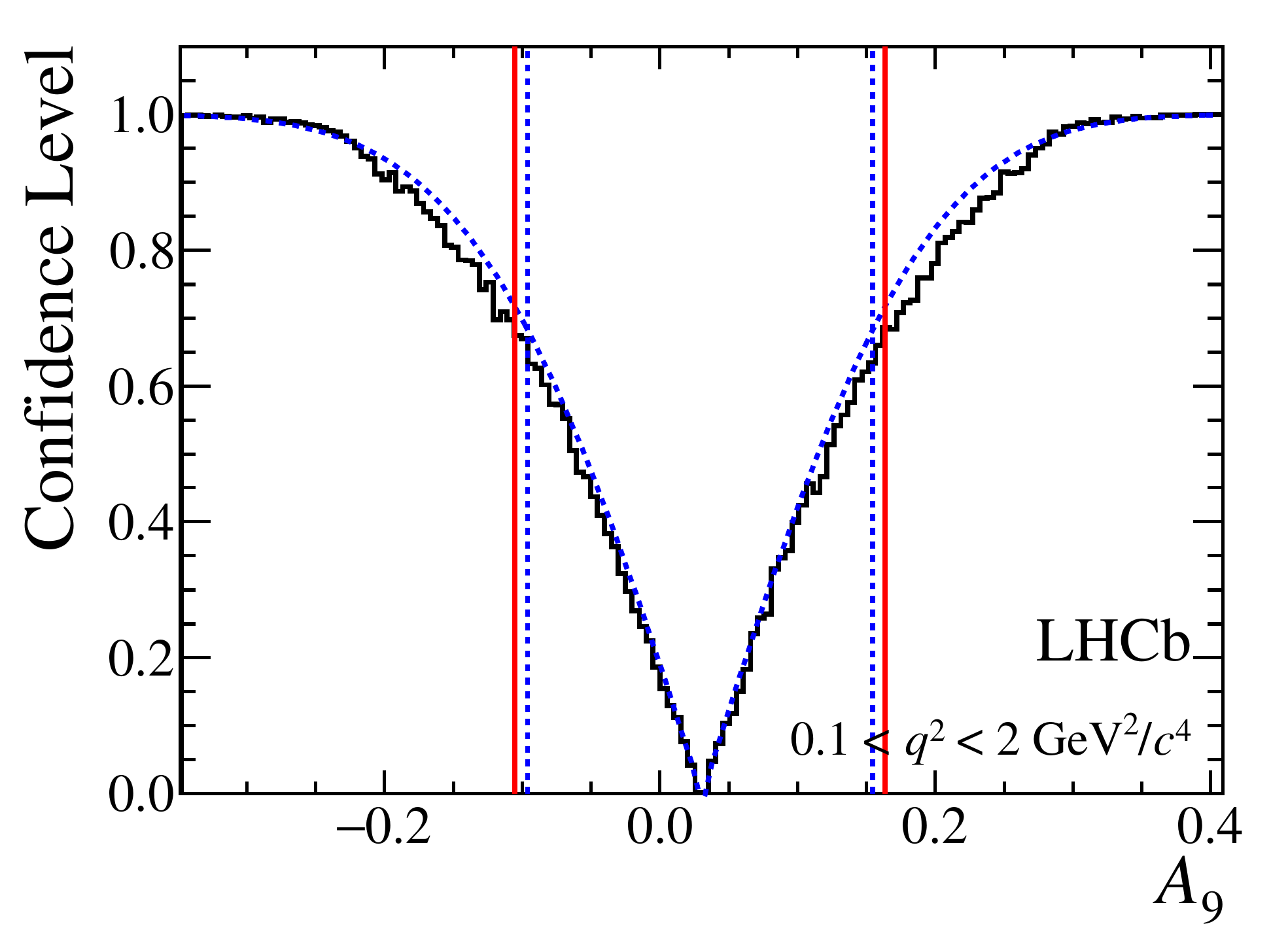}
\includegraphics[width=0.37\textwidth]{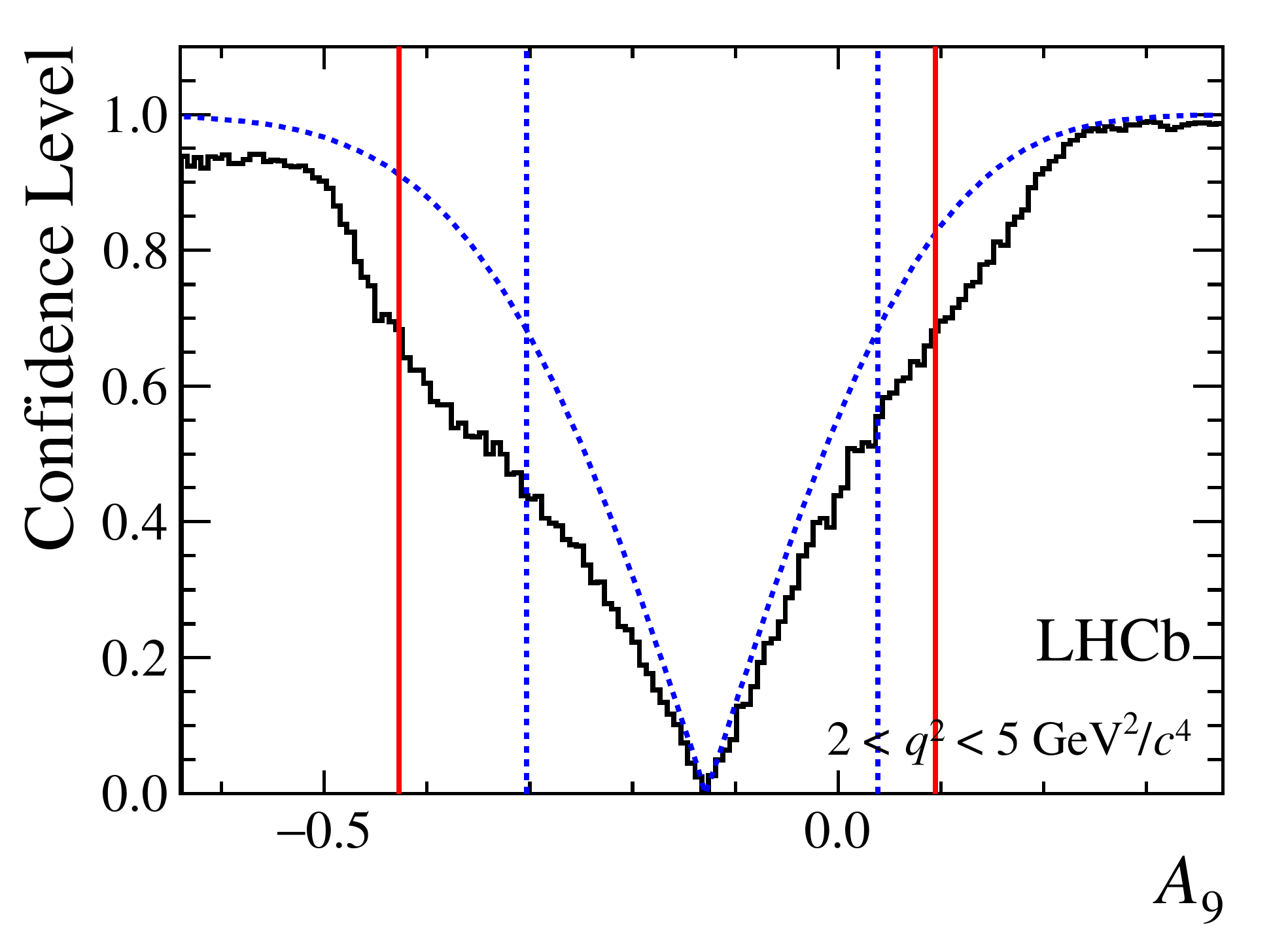}\\
\includegraphics[width=0.37\textwidth]{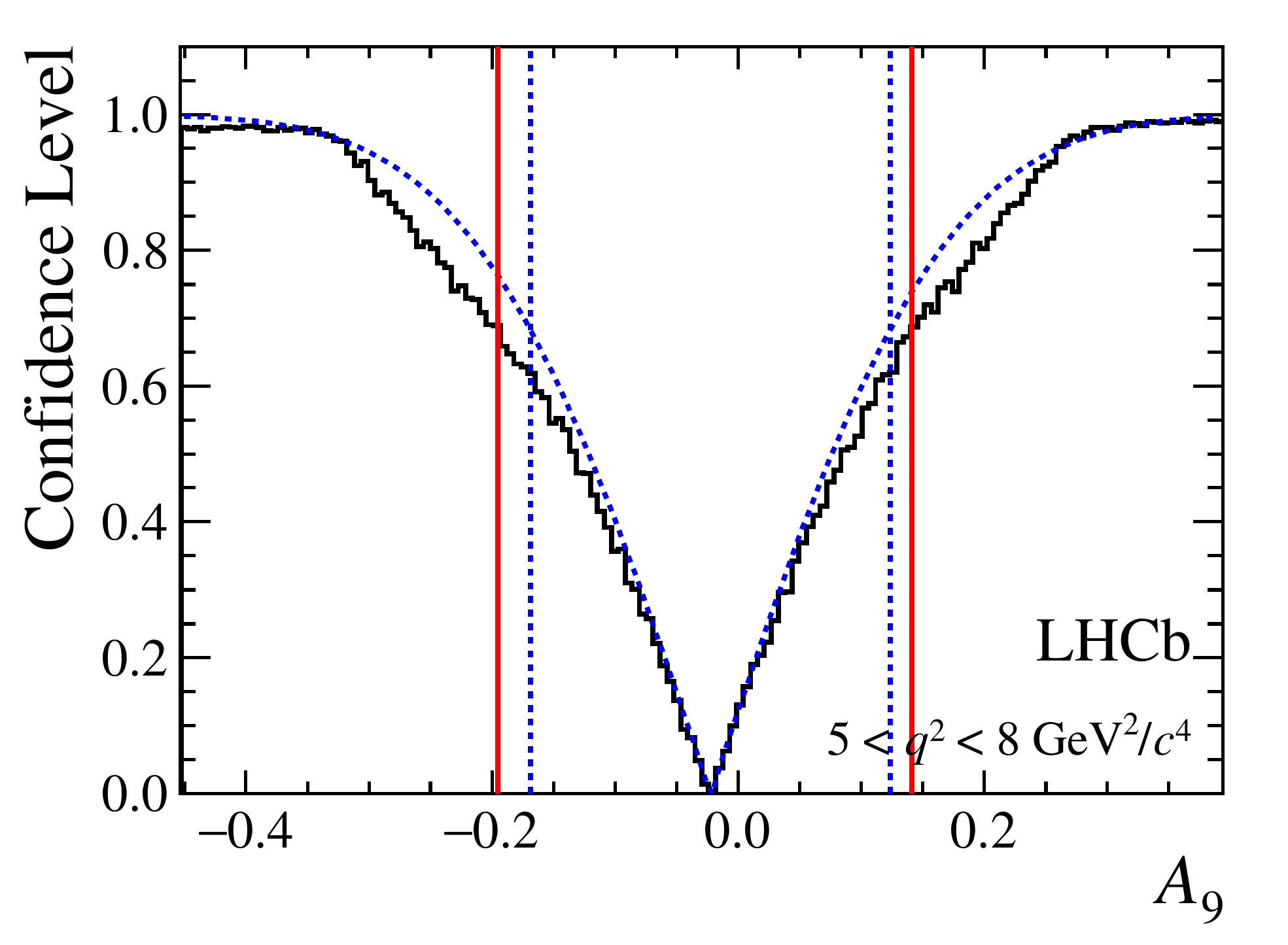}
\includegraphics[width=0.37\textwidth]{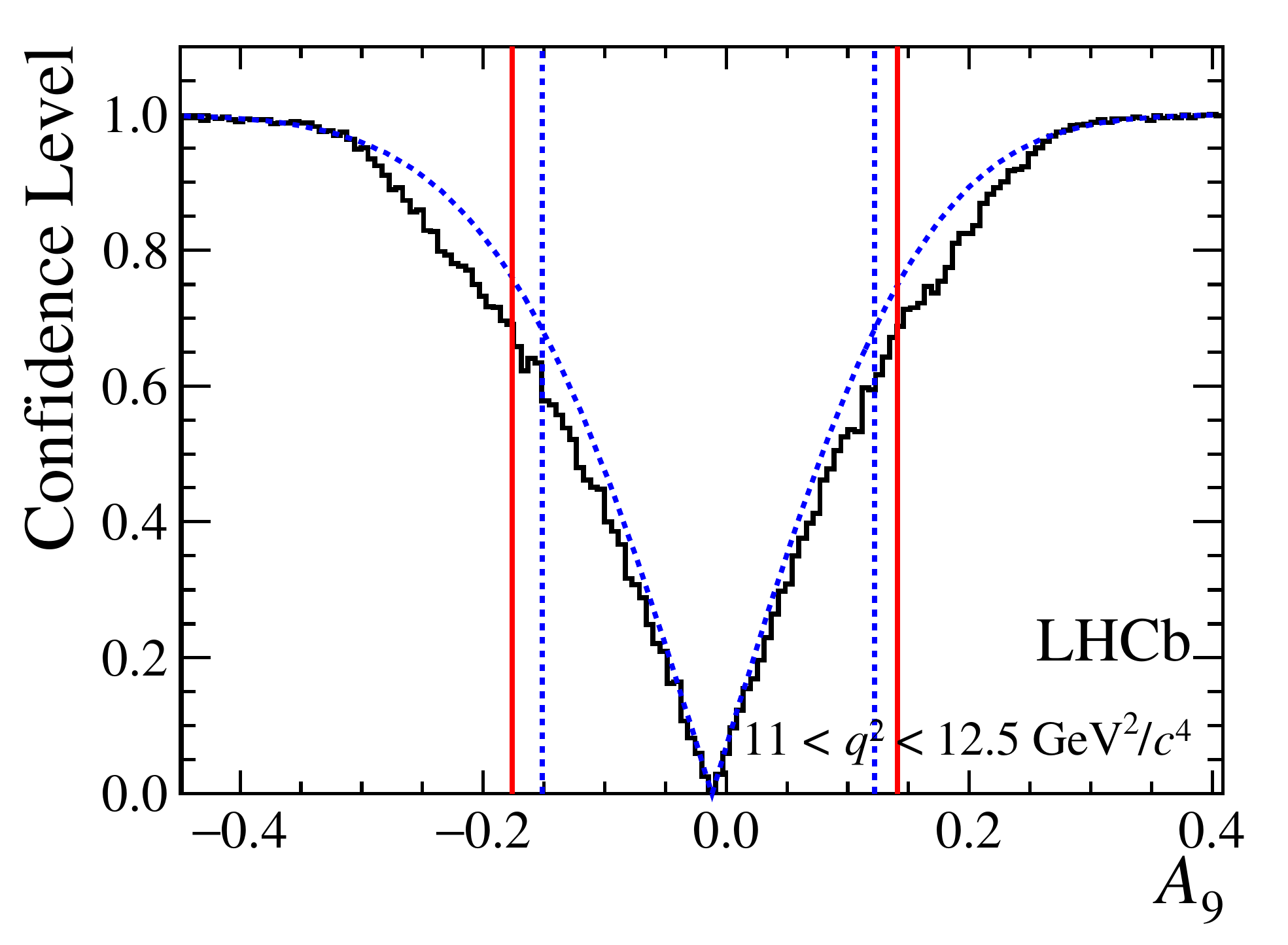}\\
\includegraphics[width=0.37\textwidth]{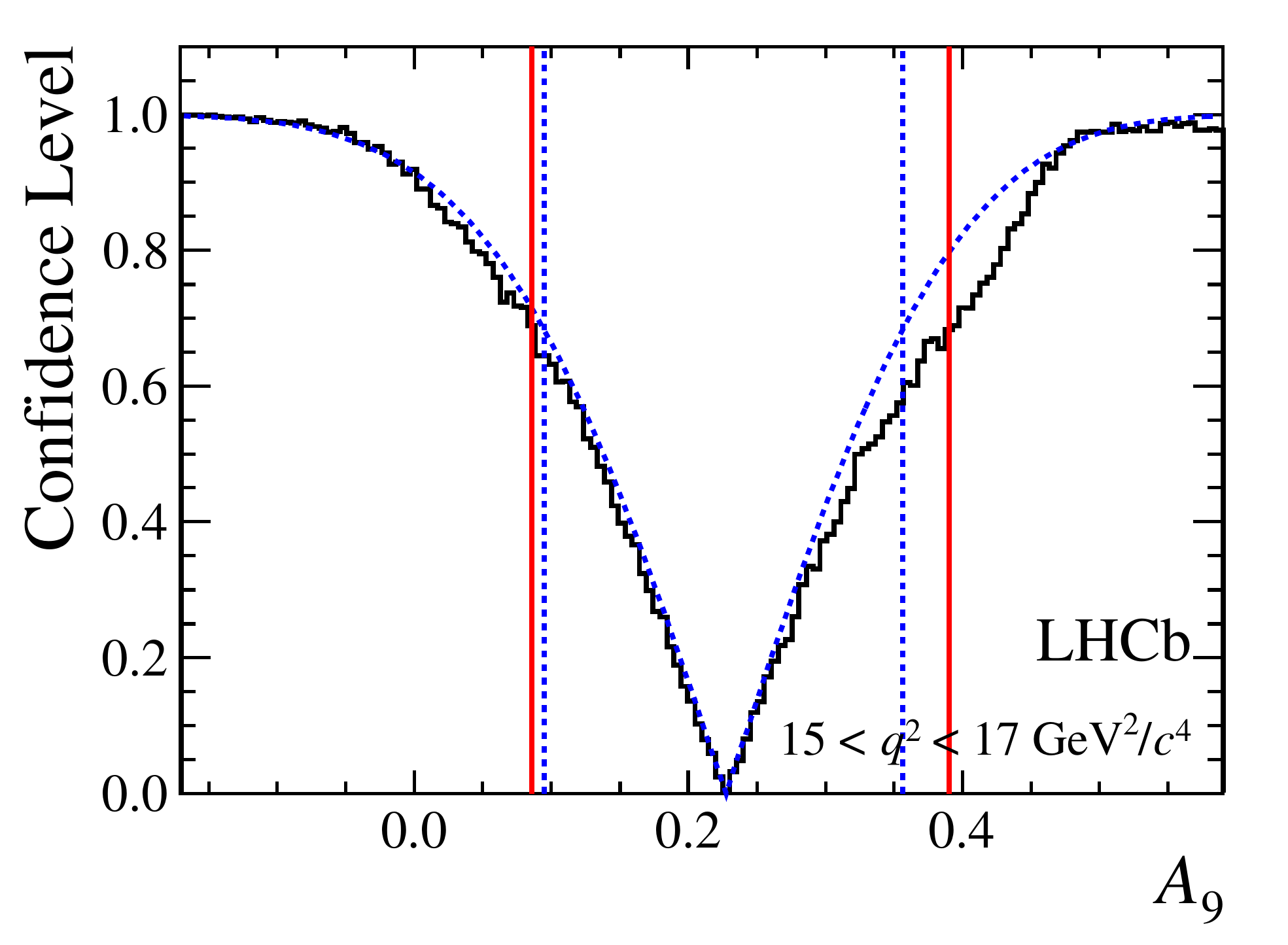}
\includegraphics[width=0.37\textwidth]{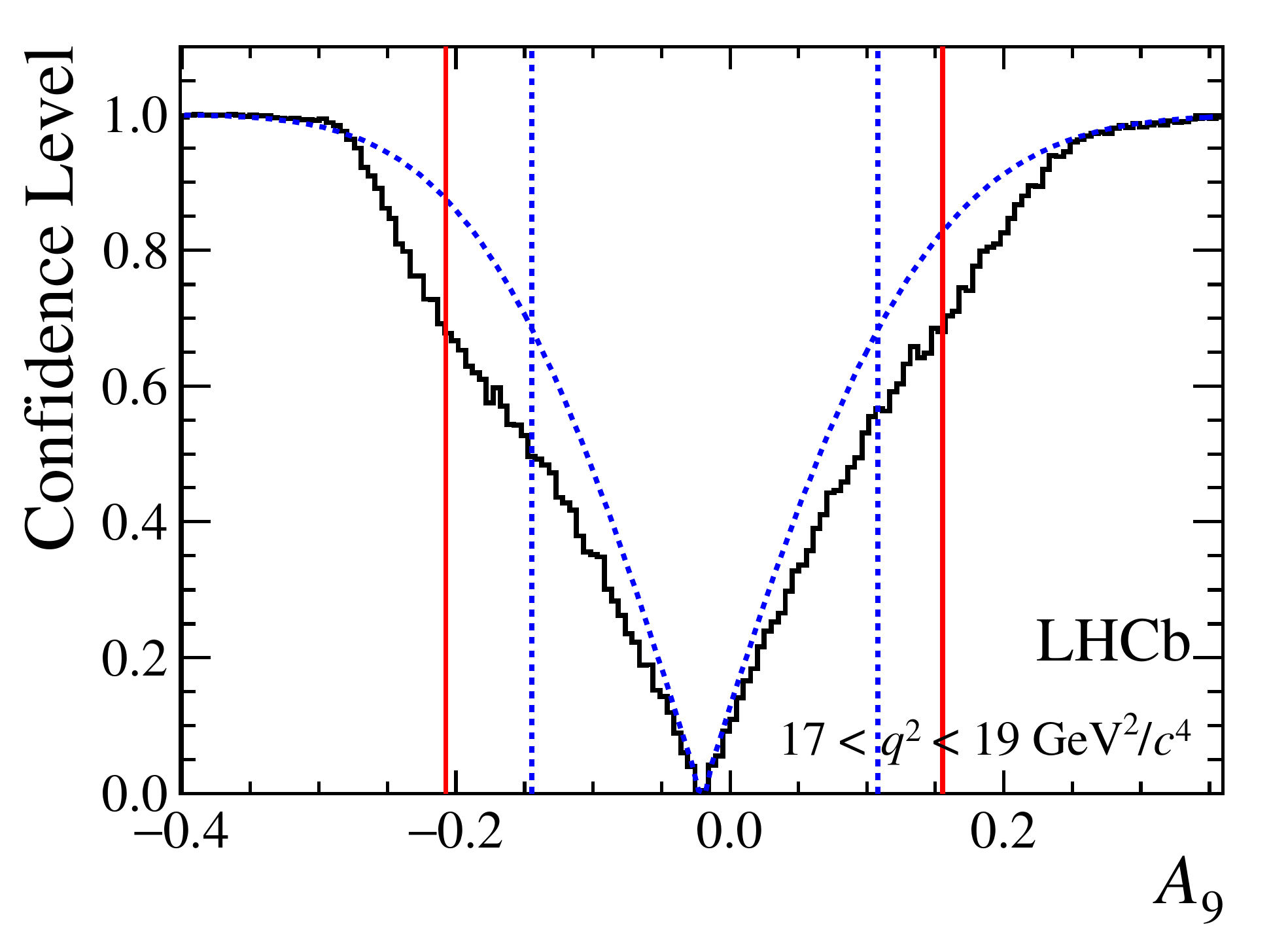}\\
\includegraphics[width=0.37\textwidth]{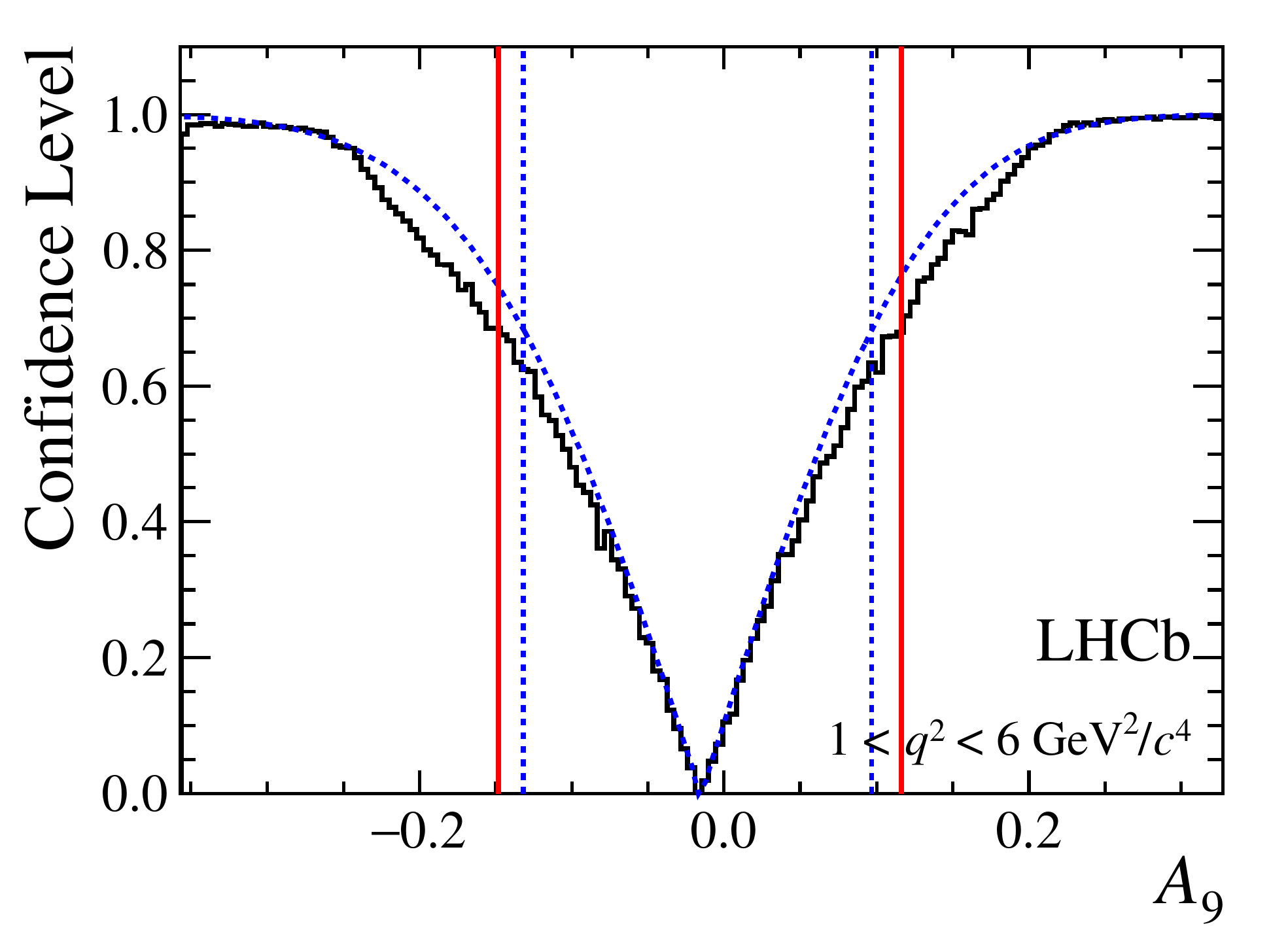}
\includegraphics[width=0.37\textwidth]{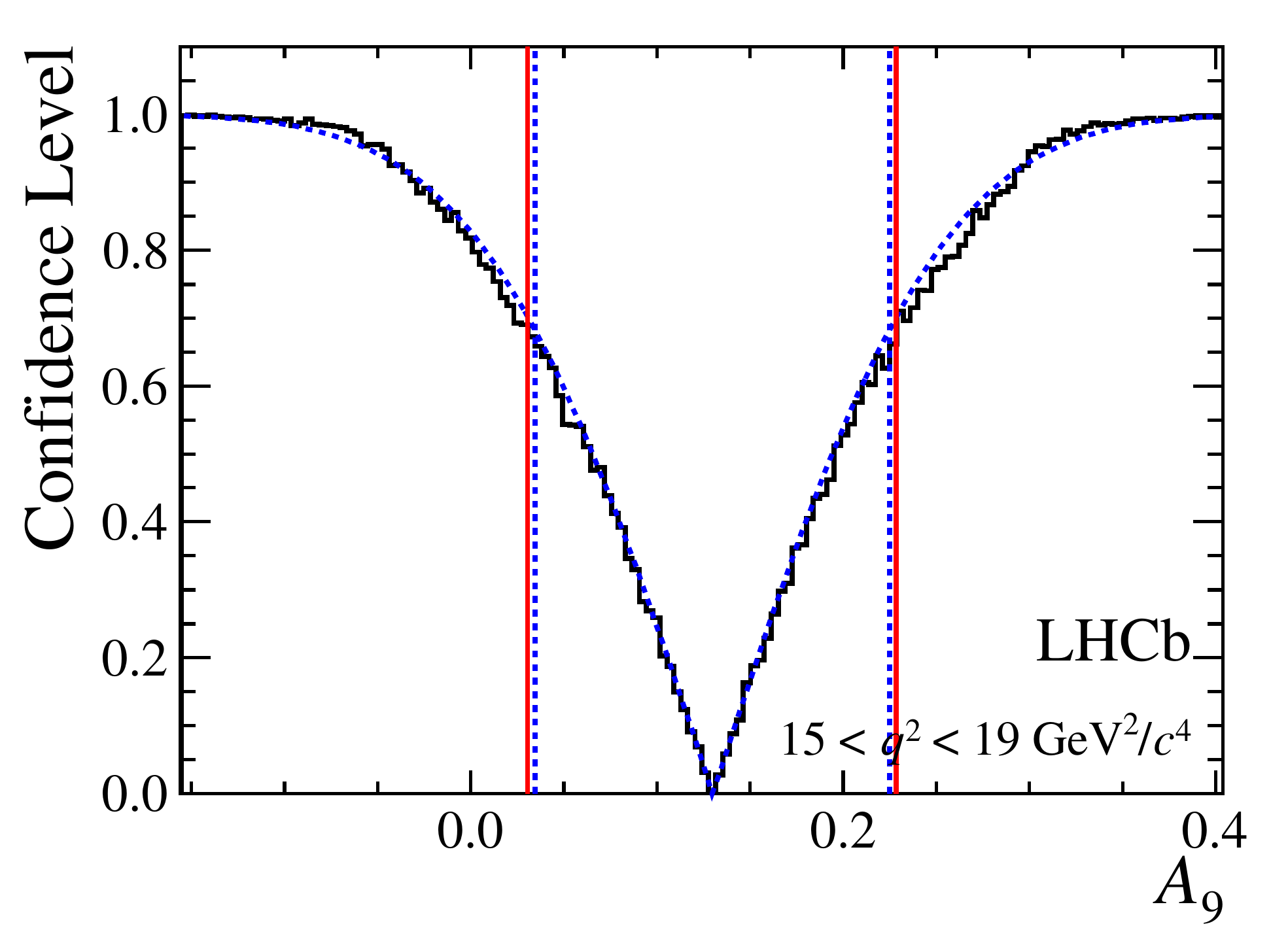}\\
\caption{\label{fig:ang-fc9} Confidence level obtained from a likelihood scan (shaded blue) and from a Feldman-Cousins method (solid black).
The shaded blue and solid red vertical lines indicate the corresponding $68\%$ CL intervals obtained from the likelihood scan and the Feldman-Cousins method, respectively.}
\end{center}
\end{figure}

\clearpage

\section{Correlation matrices}
\label{app:correlations}

\begin{table}[h]\small
\caption{Correlation matrices for the $q^2$ bins $0.1<q^2<2.0\gevgevcccc$, $2.0<q^2<5.0\gevgevcccc$ and $5.0<q^2<8.0\gevgevcccc$.}
    \begin{center}
  \begin{tabular}{lrrrrrrrr}\hline
 \multicolumn{9}{c}{Correlation matrix for $0.1<q^2<2.0\gevgevcccc$}\\
& $F_{\rm L}$ & $S_3$ & $S_4$ & $A_5$ & $A_6$ & $S_7$ & $A_8$ & $A_9$\\ \hline
        $F_{\rm L}$ & 1.00 & 0.03 & -0.15 & $-$ & 0.02 & 0.10 & 0.03 & $-$ \\ 
                   $S_{3}$ &   & 1.00 & 0.04 & 0.07 & $-$ & 0.05 & -0.18 & -0.05 \\ 
                   $S_{4}$ &   &   & 1.00 & -0.13 & -0.09 & -0.19 & 0.06 & -0.09 \\ 
                   $A_{5}$ &   &   &   & 1.00 & 0.11 & 0.06 & -0.14 & 0.10 \\ 
                   $A_{6}$ &   &   &   &   & 1.00 & 0.07 & -0.03 & -0.16 \\ 
                   $S_{7}$ &   &   &   &   &   & 1.00 & -0.30 & 0.03 \\ 
                   $A_{8}$ &   &   &   &   &   &   & 1.00 & 0.06 \\ 
                   $A_{9}$ &   &   &   &   &   &   &   & 1.00 \\ 
    \hline\end{tabular}\\[1cm]
     \begin{tabular}{lrrrrrrrr}\hline
 \multicolumn{9}{c}{Correlation matrix for $2.0<q^2<5.0\gevgevcccc$}\\
         & $F_{\rm L}$ & $S_{3}$ & $S_{4}$ & $A_{5}$ & $A_{6}$ & $S_{7}$ & $A_{8}$ & $A_{9}$\\ \hline
$                   F_{\rm L}$ & 1.00 & -0.05 & 0.27 & 0.04 & -0.09 & 0.02 & 0.02 & -0.16 \\ 
$                   S_{3}$ &   & 1.00 & -0.23 & -0.06 & -0.05 & 0.20 & -0.11 & 0.40 \\ 
$                   S_{4}$ &   &   & 1.00 & 0.11 & 0.16 & 0.14 & -0.41 & -0.33 \\ 
$                   A_{5}$ &   &   &   & 1.00 & -0.24 & -0.31 & 0.06 & 0.08 \\ 
$                   A_{6}$ &   &   &   &   & 1.00 & -0.03 & 0.05 & 0.11 \\ 
$                   S_{7}$ &   &   &   &   &   & 1.00 & -0.05 & -0.02 \\ 
$                   A_{8}$ &   &   &   &   &   &   & 1.00 & -0.16 \\ 
$                   A_{9}$ &   &   &   &   &   &   &   & 1.00 \\ 
    \hline\end{tabular}\\[1cm]
     \begin{tabular}{lrrrrrrrr}\hline
 \multicolumn{9}{c}{Correlation matrix for $5.0<q^2<8.0\gevgevcccc$}\\
& $F_{L}$ & $S_{3}$ & $S_{4}$ & $A_{5}$ & $A_{6}$ & $S_{7}$ & $A_{8}$ & $A_{9}$\\ \hline
$                   F_{L}$ & 1.00 & -0.03 & -0.01 & 0.11 & 0.01 & 0.03 & -0.07 & 0.09 \\ 
$                   S_{3}$ &   & 1.00 & -0.03 & -0.01 & -0.02 & -0.18 & $-$ & 0.05 \\ 
$                   S_{4}$ &   &   & 1.00 & -0.05 & -0.01 & -0.03 & 0.20 & -0.08 \\ 
$                   A_{5}$ &   &   &   & 1.00 & $-$ & 0.14 & -0.05 & -0.16 \\ 
$                   A_{6}$ &   &   &   &   & 1.00 & $-$ & 0.05 & -0.25 \\ 
$                   S_{7}$ &   &   &   &   &   & 1.00 & 0.04 & -0.14 \\ 
$                   A_{8}$ &   &   &   &   &   &   & 1.00 & -0.04 \\ 
$                   A_{9}$ &   &   &   &   &   &   &   & 1.00 \\ 
\hline \end{tabular}
\end{center}
\end{table}

\begin{table}[h]\small
\caption{Correlation matrices for the $q^2$ bins $11.0<q^2<12.5\gevgevcccc$, $15.0<q^2<17.0\gevgevcccc$ and $17.0<q^2<19.0\gevgevcccc$.}
    \begin{center}
     \begin{tabular}{lrrrrrrrr}\hline
 \multicolumn{9}{c}{Correlation matrix for $11.0<q^2<12.5\gevgevcccc$}\\
  & $F_{L}$ & $S_{3}$ & $S_{4}$ & $A_{5}$ & $A_{6}$ & $S_{7}$ & $A_{8}$ & $A_{9}$\\ \hline
$                   F_{L}$ & 1.00 & 0.21 & 0.11 & 0.20 & 0.06 & 0.21 & -0.05 & 0.03 \\ 
$                   S_{3}$ &   & 1.00 & 0.02 & 0.12 & 0.18 & 0.08 & -0.02 & 0.04 \\ 
$                   S_{4}$ &   &   & 1.00 & -0.11 & -0.37 & 0.26 & -0.01 & -0.09 \\ 
$                   A_{5}$ &   &   &   & 1.00 & -0.16 & -0.07 & 0.22 & 0.04 \\ 
$                   A_{6}$ &   &   &   &   & 1.00 & -0.04 & -0.11 & 0.14 \\ 
$                   S_{7}$ &   &   &   &   &   & 1.00 & -0.36 & 0.08 \\ 
$                   A_{8}$ &   &   &   &   &   &   & 1.00 & -0.23 \\ 
$                   A_{9}$ &   &   &   &   &   &   &   & 1.00 \\ 
\hline \end{tabular}\\[1cm]
     \begin{tabular}{lrrrrrrrr}\hline
 \multicolumn{9}{c}{Correlation matrix for $15.0<q^2<17.0\gevgevcccc$}\\
 & $F_{L}$ & $S_{3}$ & $S_{4}$ & $A_{5}$ & $A_{6}$ & $S_{7}$ & $A_{8}$ & $A_{9}$\\ \hline
$                   F_{L}$ & 1.00 & -0.01 & -0.01 & 0.07 & -0.05 & 0.06 & -0.06 & -0.06 \\ 
$                   S_{3}$ &   & 1.00 & 0.03 & -0.06 & 0.11 & -0.08 & -0.06 & 0.15 \\ 
$                   S_{4}$ &   &   & 1.00 & 0.01 & -0.07 & 0.04 & 0.22 & 0.04 \\ 
$                   A_{5}$ &   &   &   & 1.00 & -0.04 & 0.14 & 0.05 & 0.01 \\ 
$                   A_{6}$ &   &   &   &   & 1.00 & 0.05 & 0.01 & -0.09 \\ 
$                   S_{7}$ &   &   &   &   &   & 1.00 & -0.03 & 0.06 \\ 
$                   A_{8}$ &   &   &   &   &   &   & 1.00 & -0.11 \\ 
$                   A_{9}$ &   &   &   &   &   &   &   & 1.00 \\ 
\hline \end{tabular}\\[1cm]
       \begin{tabular}{lrrrrrrrr}\hline
 \multicolumn{9}{c}{Correlation matrix for $17.0<q^2<19.0\gevgevcccc$}\\
 & $F_{L}$ & $S_{3}$ & $S_{4}$ & $A_{5}$ & $A_{6}$ & $S_{7}$ & $A_{8}$ & $A_{9}$\\ \hline
$                   F_{L}$ & 1.00 & -0.12 & 0.08 & -0.32 & 0.06 & -0.04 & 0.15 & -0.01 \\ 
$                   S_{3}$ &   & 1.00 & -0.04 & 0.40 & 0.16 & 0.22 & $-$ & 0.02 \\ 
$                   S_{4}$ &   &   & 1.00 & 0.25 & 0.30 & -0.13 & 0.40 & 0.14 \\ 
$                   A_{5}$ &   &   &   & 1.00 & -0.05 & 0.16 & -0.06 & 0.28 \\ 
$                   A_{6}$ &   &   &   &   & 1.00 & -0.03 & 0.19 & -0.05 \\ 
$                   S_{7}$ &   &   &   &   &   & 1.00 & -0.02 & 0.18 \\ 
$                   A_{8}$ &   &   &   &   &   &   & 1.00 & -0.02 \\ 
$                   A_{9}$ &   &   &   &   &   &   &   & 1.00 \\ 
\hline \end{tabular}
\end{center}
\end{table}

\begin{table}[h]\small
\caption{Correlation matrices for the $q^2$ bins $1.0<q^2<6.0\gevgevcccc$ and $15.0<q^2<19.0\gevgevcccc$.}
    \begin{center}
     \begin{tabular}{lrrrrrrrr}\hline
     \multicolumn{9}{c}{Correlation matrix for $1.0<q^2<6.0\gevgevcccc$}\\
  & $F_{L}$ & $S_{3}$ & $S_{4}$ & $A_{5}$ & $A_{6}$ & $S_{7}$ & $A_{8}$ & $A_{9}$\\ \hline
$                   F_{L}$ & 1.00 & -0.02 & 0.08 & 0.08 & -0.04 & 0.07 & 0.03 & -0.05 \\ 
$                   S_{3}$ &   & 1.00 & -0.07 & 0.13 & -0.13 & 0.10 & -0.06 & 0.12 \\ 
$                   S_{4}$ &   &   & 1.00 & 0.17 & 0.10 & -0.06 & -0.01 & -0.13 \\ 
$                   A_{5}$ &   &   &   & 1.00 & -0.14 & -0.05 & -0.05 & 0.04 \\ 
$                   A_{6}$ &   &   &   &   & 1.00 & -0.04 & -0.09 & -0.06 \\ 
$                   S_{7}$ &   &   &   &   &   & 1.00 & 0.07 & 0.04 \\ 
$                   A_{8}$ &   &   &   &   &   &   & 1.00 & -0.10 \\ 
$                   A_{9}$ &   &   &   &   &   &   &   & 1.00 \\ 
\hline \end{tabular}\\[1cm]
     \begin{tabular}{lrrrrrrrr}\hline
     \multicolumn{9}{c}{Correlation matrix for $15.0<q^2<19.0\gevgevcccc$}\\
   & $F_{L}$ & $S_{3}$ & $S_{4}$ & $A_{5}$ & $A_{6}$ & $S_{7}$ & $A_{8}$ & $A_{9}$\\ \hline
$                   F_{L}$ & 1.00 & 0.01 & -0.05 & -0.03 & $-$ & 0.07 & -0.04 & -0.08 \\ 
$                   S_{3}$ &   & 1.00 & -0.03 & 0.16 & 0.09 & 0.05 & 0.02 & 0.07 \\ 
$                   S_{4}$ &   &   & 1.00 & 0.12 & 0.03 & 0.03 & 0.24 & 0.09 \\ 
$                   A_{5}$ &   &   &   & 1.00 & -0.09 & 0.08 & 0.04 & 0.14 \\ 
$                   A_{6}$ &   &   &   &   & 1.00 & 0.05 & 0.05 & -0.11 \\ 
$                   S_{7}$ &   &   &   &   &   & 1.00 & 0.01 & 0.10 \\ 
$                   A_{8}$ &   &   &   &   &   &   & 1.00 & -0.07 \\ 
$                   A_{9}$ &   &   &   &   &   &   &   & 1.00 \\ 
\hline \end{tabular}
\end{center}
\end{table}

\clearpage

\addcontentsline{toc}{section}{References}
\setboolean{inbibliography}{true}
\bibliographystyle{LHCb}
\bibliography{main,LHCb-PAPER,LHCb-CONF,LHCb-DP,LHCb-TDR}

\newpage

\newpage
\centerline{\large\bf LHCb collaboration}
\begin{flushleft}
\small
R.~Aaij$^{38}$, 
B.~Adeva$^{37}$, 
M.~Adinolfi$^{46}$, 
A.~Affolder$^{52}$, 
Z.~Ajaltouni$^{5}$, 
S.~Akar$^{6}$, 
J.~Albrecht$^{9}$, 
F.~Alessio$^{38}$, 
M.~Alexander$^{51}$, 
S.~Ali$^{41}$
% ...

{\footnotesize \it
% ...
$ ^{5}$Clermont Universit\'{e}, Universit\'{e} Blaise Pascal, CNRS/IN2P3, LPC, Clermont-Ferrand, France\\
$ ^{6}$CPPM, Aix-Marseille Universit\'{e}, CNRS/IN2P3, Marseille, France\\
% ...
$ ^{9}$Fakult\"{a}t Physik, Technische Universit\"{a}t Dortmund, Dortmund, Germany\\
% ...
$ ^{37}$Universidad de Santiago de Compostela, Santiago de Compostela, Spain\\
$ ^{38}$European Organization for Nuclear Research (CERN), Geneva, Switzerland\\
% ...
$ ^{41}$Nikhef National Institute for Subatomic Physics, Amsterdam, The Netherlands\\
% ...
$ ^{46}$H.H. Wills Physics Laboratory, University of Bristol, Bristol, United Kingdom\\
% ...
$ ^{51}$School of Physics and Astronomy, University of Glasgow, Glasgow, United Kingdom\\
$ ^{52}$Oliver Lodge Laboratory, University of Liverpool, Liverpool, United Kingdom\\
% ...
}
\end{flushleft}

\end{document}